\documentclass[11pt,letterpaper,onecolumn]{article}
\usepackage[T1]{fontenc}
\usepackage{mathpazo}
\usepackage[margin=1in]{geometry}
\usepackage{amsmath,amssymb,amsthm,mathtools,booktabs,array,microtype}
\usepackage{enumitem,etoolbox}
\usepackage{authblk}
\usepackage{algorithm,algpseudocode,caption}
\usepackage{tcolorbox}
\tcbuselibrary{breakable,skins}
\usepackage[colorlinks=true,linkcolor=blue,citecolor=blue,urlcolor=blue]{hyperref}
\patchcmd{\abstract}{\small}{\normalsize}{}{\PackageError{stoc-submission}{Could not set the abstract to the main text size}{Check the abstract environment before submitting.}}

\allowdisplaybreaks
\newtcolorbox{pseudocodebox}{
 enhanced,breakable,sharp corners,
 colback=white,colframe=black!65,boxrule=0.5pt,
 left=2mm,right=2mm,top=1mm,bottom=1mm,
 before skip=10pt,after skip=10pt,
 colbacktitle=black!5,coltitle=black,fonttitle=\normalsize\bfseries,
 title after break={Algorithm~\thealgorithm\ (continued)}
}
\numberwithin{equation}{section}
\newtheorem{theorem}{Theorem}[section]
\newtheorem{proposition}[theorem]{Proposition}
\newtheorem{lemma}[theorem]{Lemma}
\newtheorem{corollary}[theorem]{Corollary}
\theoremstyle{definition}
\newtheorem{definition}[theorem]{Definition}
\newtheorem{assumption}[theorem]{Assumption}
\theoremstyle{remark}

\newcounter{question}
\newcommand{\E}{\mathbb E}
\newcommand{\Prob}{\mathbb P}
\newcommand{\R}{\mathbb R}
\newcommand{\dd}{\,\mathrm d}
\newcommand{\norm}[1]{\left\lVert#1\right\rVert}

\newcommand{\KL}{D_{\mathrm{KL}}}
\newcommand{\dist}{\operatorname{dist}}
\newcommand{\clip}{\operatorname{clip}}
\newcommand{\OPT}{\operatorname{OPT}}
\newcommand{\Par}{\mathcal P_*}
\newcommand{\diag}{\operatorname{diag}}

\newcommand{\one}{\mathbf 1}

\newcommand{\egiamp}{EG-IAMP}
\newcommand{\gegiamp}{GEG-IAMP}
\newcommand{\etaacc}{\eta_{\mathrm{acc}}}
\title{A Fully Polynomial Approximation Algorithm for the Sherrington--Kirkpatrick Model}
\author[1]{Dylan Herman\thanks{Corresponding author: \href{mailto:dylan.a.herman@jpmchase.com}{dylan.a.herman@jpmchase.com}.}}
\author[1]{Shouvanik Chakrabarti}
\author[1]{Junhyung Lyle Kim}
\author[1]{Guneykan Ozgul}
\affil[1]{Global Technology Applied Research, JPMorganChase, New York, NY 10001, USA}
\date{}

\begin{document}
\maketitle
\begin{abstract}
Montanari's incremental approximate message passing (IAMP) algorithm [FOCS 2019; SIAM J. Comput., 2025] and the zero-temperature formulation of El Alaoui, Montanari and Sellke (AMS) [Ann. Probab., 2021] find a $(1-\varepsilon)$ approximation to the maximum Sherrington--Kirkpatrick (SK) energy in $C(\varepsilon)N^2$ operations for fixed accuracy, with probability tending to one as $N\to\infty$, under their respective full-support assumptions. However, it was unknown whether this approximation could be achieved with work polynomial in both $N$ and $1/\varepsilon$, together with quantitative guarantees at finite dimension. Under the full-support identity used by AMS and the unit-cost evaluation convention of Montanari and AMS, we introduce grouped IAMP with exact Gaussian query outputs (\gegiamp), which is the first SK approximation algorithm with total work polynomial in $N$ and $1/\varepsilon$ and a finite-size guarantee above a dimension threshold polynomial in $1/\varepsilon$ and $\log(1/\alpha)$. More precisely, for every $\alpha\in(0,1/2)$, it returns a $(1-\varepsilon)$ approximation with probability at least $1-\alpha$ whenever $N\ge \mathsf{K}\varepsilon^{-2}\log^2(e/\varepsilon)\log(\mathsf{K}/\alpha)$, using $O(N^2/\varepsilon)$ work and $O(N^2)$ storage. \gegiamp\ replaces the Onsager correction by Reeves's coupling construction [arXiv:2508.10782] and groups Euler--Maruyama steps into $O(1/\varepsilon)$ matrix--vector products. We also prove a relative guarantee for AMS IAMP with $O(N^2\varepsilon^{-3}\log^2(e/\varepsilon))$ work but a sufficient dimension exponential in inverse accuracy. By returning a fixed configuration below its dimension threshold, \gegiamp\ also provides an asymptotic fully polynomial randomized approximation scheme (AFPRAS), with an additive error independent of dimension.
\end{abstract}

\clearpage
\tableofcontents
\clearpage
\newpage
\section{Introduction}
\label{sec:introduction}

\subsection{Motivation}
\label{sec:motivation}

In the Sherrington--Kirkpatrick (SK) model~\cite{SherringtonKirkpatrick}, one seeks a spin configuration $\sigma\in\{-1,1\}^N$ maximizing the energy, or Hamiltonian, $H_N(\sigma)=\sigma^\top A\sigma/2$ for a symmetric Gaussian matrix $A\in\R^{N\times N}$. Let $\OPT_N$ denote this maximum. The Parisi formula expresses the limit $\Par=\lim_{N\to\infty}\E\OPT_N/N$ of the optimum per spin as the minimum of a deterministic variational problem~\cite{Talagrand,Panchenko,AuffingerChen}.

Several efficient algorithms for approximating the maximum energy build on approximate message passing (AMP), which was introduced for compressed sensing~\cite{DonohoMalekiMontanari}. AMP alternates multiplication by $A$ with nonlinear functions applied separately to each coordinate. Because $A$ is reused, each product is correlated with earlier iterates; the Onsager correction compensates for these correlations. Under regularity and initialization assumptions, the empirical distribution of the corrected products over the $N$ coordinates, taken jointly across iterations, converges to a Gaussian law as $N\to\infty$ with the number of iterations held fixed~\cite{FengAMP}. This description is called state evolution.

Montanari introduced incremental approximate message passing (IAMP) for SK using the finite-temperature Parisi solution~\cite{Montanari}. IAMP makes small updates to real-valued relaxations of the spins, called magnetizations, and converts them to spins at the end. El Alaoui, Montanari and Sellke (AMS) formulated it directly at zero temperature, where the Parisi formula describes the maximum itself~\cite{AMS}. To approach this maximum, they choose the nonlinear functions so that, by state evolution, each coordinate of an auxiliary state approximately follows a discretization of the diffusion $\dd X_t=\gamma(t)\partial_x\Phi(t,X_t)\dd t+\dd B_t$ for $0\le t<1$, with $X_0=0$. Here $X_t$ is the scalar state of one coordinate, and $B$ is a standard Brownian motion. The function $\gamma$, called the order parameter, is a minimizer of the zero-temperature Parisi variational problem. The function $\Phi(t,x)$ solves a partial differential equation (PDE) involving $\gamma$, with terminal value $\Phi(1,x)=|x|$.

AMS assume that $\gamma$ is strictly increasing on $[0,1)$, a property called full support. Montanari makes the analogous assumption at finite temperature. Under these assumptions, both prove a $(1-\varepsilon)$ approximation in $C(\varepsilon)N^2$ operations for each fixed accuracy $\varepsilon$, with probability tending to one as $N\to\infty$. These limit theorems neither bound $C(\varepsilon)$ nor say how large $N$ must be for a prescribed accuracy and failure probability, which leads to the following question.

\begin{center}
\refstepcounter{question}\label{q:finite-instance-work}
\textit{\textbf{Question~\thequestion.} How much work suffices to find a near-optimal spin configuration for a random SK instance, as a function of its dimension, the desired accuracy, and the allowed failure probability?}
\end{center}

In Question~\ref{q:finite-instance-work}, the \emph{work} of an algorithm is its number of operations, counted as in Definition~\ref{def:oracles}. Each exact real arithmetic operation, each standard Gaussian sample, and each evaluation of the supplied deterministic functions $\gamma$, $m=\partial_x\Phi$, and $a=\partial_{xx}\Phi$ at a point costs one unit. Constructing these functions is excluded from the operation count.

For IAMP, the answer to Question~\ref{q:finite-instance-work} depends on how finely the diffusion must be discretized. Each time step uses one product with $A$, which costs $O(N^2)$ operations. Since the terminal value $|x|$ is not differentiable at $x=0$, the derivatives of $\Phi$ grow as $t\to1$. Accurate discretization therefore requires many small steps and many products with $A$. A guarantee at finite dimension must also bound, over all these steps, the deviation of the empirical distribution of the corrected products from the Gaussian limit described by state evolution.

\subsection{Main results}
\label{sec:intro-results}

We answer Question~\ref{q:finite-instance-work} under the usual full-support hypothesis, in the form of the identity that AMS derive from it (Assumption~\ref{ass:frsb}). Our algorithm is based on a coupling construction that Reeves introduced to compare first-order methods with a Gaussian process~\cite{Reeves}. We use this construction as the matrix update of IAMP. To form the input to each matrix query, we subtract the magnetization's projection onto earlier query inputs and normalize the residual. Auxiliary Gaussian variables then correct the product within the span of the current and earlier inputs. With at most $N$ queries, the corrected outputs have independent standard Gaussian entries, although they still depend on $A$. With these operations in place of the Onsager term, we obtain IAMP with exact Gaussian query outputs (\egiamp).

We introduce grouped \egiamp\ (\gegiamp), which uses each corrected query output for several consecutive discretization steps.

\begin{theorem}[Relative approximation by GEG-IAMP; informal version of Theorem~\ref{thm:main}]
\label{thm:intro-main}
Under Assumption~\ref{ass:frsb} and Definition~\ref{def:oracles}, there is a universal constant $\mathsf{K}>0$ such that, for every $0<\varepsilon\le1$ and $0<\alpha<1/2$, \gegiamp\ returns a spin configuration $\sigma$ satisfying
\begin{equation*}
 \Prob\{H_N(\sigma)\ge(1-\varepsilon)\OPT_N\}\ge1-\alpha
 \quad\text{whenever}\quad
 N\ge \mathsf{K}\varepsilon^{-2}\log^2(e/\varepsilon)\log(\mathsf{K}/\alpha).
\end{equation*}
Total work is at most $\mathsf{K}N^2/\varepsilon$ and storage is $O(N^2)$.
\end{theorem}

\paragraph{Technical contributions.}
Our grouping technique reduces the number of matrix queries to $\Theta(\varepsilon^{-1})$, while retaining $\Theta(\varepsilon^{-2}\log^2(e/\varepsilon))$ Euler--Maruyama steps per coordinate. We reconstruct the increments for an entire group from one corrected query output and independent Gaussian variables, preserving their joint Gaussian law and their correlation with the magnetization updates (Section~\ref{sec:overview-grouping}).

Under the full-support identity, Proposition~\ref{prop:regularity} provides explicit bounds on the order parameter and the spatial derivatives of the Parisi solution near time one. These estimates quantify their growth toward the terminal time and sharpen the general inverse-linear bound on the order parameter.

We also sharpen the Euler--Maruyama discretization estimates used by Montanari and AMS~\cite{Montanari,AMS}. Weak estimates for the normalization bias and direct control of accumulated drift yield improved normalization and correlation bounds for our recursion, with explicit terminal-time dependence (Theorem~\ref{thm:scalar-error}). Including quadrature and setting the normalization factors to one, the final magnetization lies within $O(\varepsilon)$ of $[-1,1]$ in root mean square, and its correlation with the driving noise is at least $\Par-O(\varepsilon)$ (Lemma~\ref{lem:euler-quadrature} and Proposition~\ref{prop:unit-normalizers}).

At finite dimension, we adapt Reeves's energy identity to the grouped construction and bound the coordinate averages and projections that enter the output energy (Proposition~\ref{prop:energy-identity} and Lemma~\ref{lem:directional-qr}). Independence across coordinate paths and the conditional mean zero of later magnetization increments control these terms. Together with the clipping estimate of Lemma~\ref{lem:empirical-clipping}, these bounds establish the polynomial sufficient dimension in Theorem~\ref{thm:main}.

Our AMS IAMP analysis establishes a finite-dimensional relative guarantee with explicit bounds on sufficient dimension, failure probability, and complete work (Theorem~\ref{thm:onsager-main}).

Returning a fixed configuration below the threshold yields the first \emph{asymptotic fully polynomial randomized approximation scheme} (AFPRAS) for random SK instances under the stated assumptions (Corollary~\ref{thm:afptas}). Following the asymptotic schemes used in bin packing~\cite{FdlVL,KarmarkarKarp,Vazirani}, its guarantee permits an additive error $\mathfrak{E}_{\varepsilon,\alpha}$ independent of dimension:
\begin{equation}
 \Prob\{H_N(\sigma)\ge(1-\varepsilon)\OPT_N
							  -\mathfrak{E}_{\varepsilon,\alpha}\}\ge1-\alpha
 \qquad\text{for every }N,
 \label{eq:intro-afptas}
\end{equation}
where $\mathfrak{E}_{\varepsilon,\alpha}\le \mathsf{K}\varepsilon^{-2}\log^2(e/\varepsilon)\log(\mathsf{K}/\alpha)$ and work remains polynomial in $N,1/\varepsilon,\log(1/\alpha)$. Since $\E\OPT_N/N\to\Par>0$ and $\OPT_N/N$ concentrates around its expectation, the optimum grows linearly in $N$ with probability tending to one. For fixed $\varepsilon$ and $\alpha$, the additive error does not change with $N$, so it becomes negligible compared with the optimum.

\subsection{Related work}
\label{sec:related-work}

The accuracy dependence of IAMP has been analyzed for Montanari's construction. The authors of~\cite[Supplemental Material, Section~VIII, Theorem~4]{SpinBosonQAOA} state an asymptotic work bound $O(N^2\varepsilon^{-16}\exp(O(\varepsilon^{-4})))$ under his support assumption. Their bound relies on limiting state evolution and gives no dimension sufficient for a desired confidence, and they leave the accuracy dependence of AMS IAMP unresolved~\cite[Supplemental Material, Sections~VII--VIII]{SpinBosonQAOA}.

Our results, like those of AMS, use full support. Papers by H.-B. Chen~\cite[Theorem~1.3]{ChenFullSupport} and OpenAI~\cite[Theorem~1.1]{OpenAIFullSupport} claim that the pure, zero-field SK order parameter has full support. The OpenAI theorem also states \eqref{eq:frsb}, which we retain as a hypothesis without relying on those proofs.

Bayati and Montanari established state evolution for Gaussian sensing matrices~\cite{BayatiMontanari}. Finite-dimensional versions of state evolution have since been proved for Gaussian regression~\cite{RushVenkataramanan}, for iteration counts growing with $N$ in spiked matrix estimation~\cite{LiWeiSpiked}, and for entrywise dynamics and universality of first-order methods under regularity assumptions~\cite{HanGFOM}. For SK, the joint guarantee of Theorem~\ref{thm:main} on work and sufficient dimension does not follow directly from these results. Reeves proves the exact Gaussian law of his coupling construction and bounds the distance between the iterates of suitable first-order methods and those of the coupled Gaussian process~\cite[Definition~2 and Theorems~3--5]{Reeves}. El Alaoui and Montanari use a different projection in IAMP: they project increments of the corrected products off earlier iterates, with norm rescaling in their implementation~\cite[Eq.~(2.9) and Eqs.~(3.6)--(3.8)]{EAM}. \gegiamp\ instead orthogonalizes its queries, and its exact Gaussian law requires Reeves's corrections, as discussed in Appendix~\ref{sec:projected-iamp}.

Ivkov and Schramm simulate AMP by semidefinite programs, obtaining an SK approximation scheme that tolerates corruption of a small principal submatrix, with runtime $N^{O(d)}$ for a relaxation degree $d$ depending on accuracy~\cite[Corollary~1.7 and Remark~1.9]{IvkovSchrammSDP}. Their faster spectral construction simulates a fixed AMP iteration under such corruptions, with constants depending on the iteration count~\cite[Theorem~3.1 and Section~1.4]{IvkovSchrammFast}. These results concern robustness at fixed accuracy and iteration count.

Jekel, Sandhu and Shi~\cite{JSS} propose potential Hessian ascent, which chooses random increments through the Hessian, or matrix of second derivatives, of a regularized finite-temperature potential. Their oracle work bound is
\begin{equation*}
 \widetilde O\bigl(N^{2+4\zeta}
					\exp(O(\varepsilon^{-2}))\bigr),
 \qquad 0<\zeta\le1/19.
\end{equation*}
Their sufficient parameters include inverse temperature $\beta=10/\varepsilon$, step size $\eta=\exp(-\mathsf{K}\beta^2)$, and $N\ge\eta^{-90}$. Their guarantee is near-quadratic in $N$ at fixed accuracy but not jointly polynomial in $N$ and inverse accuracy. Their support hypothesis is at finite temperature. Both our work bounds and theirs exclude computing the minimizing order parameter and Parisi PDE solution.

\label{sec:quantum}
Among quantum algorithms, Farhi, Goldstone, Gutmann and Zhou~\cite{FarhiSK} derive the fixed-depth limiting energy of the quantum approximate optimization algorithm (QAOA) for SK and prove concentration. The authors of~\cite{BassoSK} relate it to high-degree MaxCut and conjecture that it converges to the Parisi value as the depth grows. A recent OpenAI paper claims convergence of the optimized, disorder-averaged expected energy per spin, with $N\to\infty$ before depth~\cite[Theorem~1.1]{OpenAIQAOASK}. Its deterministic angles are independent of dimension and disorder; it supplies no quantitative depth rate, efficient angle-selection procedure, or finite-size estimate.

The authors of~\cite{SpinBosonQAOA} prove a spin--boson representation of QAOA and report the numerical fit $p\approx\varepsilon^{-1.13}$ for the depth $p$ giving $O(\varepsilon)$ error per spin in the limiting expected energy. This numerical fit does not follow from the qualitative convergence claim. For a dense instance, one depth-$p$ circuit with fixed angles then uses $O(N^2/\varepsilon^{1.13})$ two-qubit gates and has physical depth $O(N/\varepsilon^{1.13})$ with all-to-all connectivity and parallel gates on disjoint pairs, excluding angle optimization and repeated runs. At the same error per spin, we prove total work $O(N^2/\varepsilon)$ for \gegiamp\ at confidence $1-\alpha$ above its dimension threshold.

\section*{AI methodology}
\addcontentsline{toc}{section}{AI methodology}
\label{sec:ai-methodology}
The authors initiated this research program to investigate whether sharper Euler--Maruyama estimates or higher-order discretizations of the Parisi stochastic differential equation could reduce the iteration complexity of SK IAMP. We supplied GPT-6.1 Sol with preliminary notes, attempted analyses, and specific research questions, and directed successive rounds of analysis and revision.

In the early scalar analyses, GPT-6.1 Sol developed an argument that replaced an exponential dependence on inverse accuracy in our earlier estimates by a sufficient $O(\varepsilon^{-6})$ step count. Following the authors' suggestion to investigate nonuniform time steps, the analysis improved to $O(\varepsilon^{-3})$ steps. These were intermediate scalar bounds, not complete finite-dimensional work guarantees.

For finite-dimensional bounds, GPT-6.1 Sol proposed running the Gaussian observation process in Reeves's coupling construction directly, rather than using it only to compare an Onsager trajectory with a Gaussian process. This led to \egiamp, whose corrected observations have an exact Gaussian law at finite dimension. Reeves's coupling construction, Gaussian observation law, and matrix identity are prior results. An early \egiamp\ analysis had a sufficient dimension of order $\varepsilon^{-6}$ up to logarithmic factors at fixed confidence.

Later work with GPT-6.1 Sol on nonuniform Euler--Maruyama steps and integrated drift coefficients reduced the step count to $\widetilde O(\varepsilon^{-2})$. While the authors were investigating It\^o--Taylor schemes of strong order $1.5$ (which was able to make work), GPT-6.1 Sol proposed grouping Euler--Maruyama steps between Gaussian observations. The resulting \gegiamp\ uses $K=\Theta(\varepsilon^{-2}\log^2(e/\varepsilon))$ time steps but only $\Theta(\varepsilon^{-1})$ products with $A$. This separation leads to $O(N^2/\varepsilon)$ total work for the AFPRAS in Corollary~\ref{thm:afptas}, including the $O(NK)$ coordinate updates, drift quadrature, Gaussian reconstruction, matrix operations, and rounding. As in that corollary, this work bound excludes computing the minimizing order parameter and Parisi PDE solution.

Grouping and subsequent GPT-6.1 Sol refinements of the matrix and clipping estimates led to the sufficient dimension $N\ge \mathsf{K}\varepsilon^{-2}\log^2(e/\varepsilon)\log(\mathsf{K}/\alpha)$ in Theorem~\ref{thm:main}. Additional GPT-6.1 Sol agents proposed revisions, and GPT-6 Sol and Claude Opus 5.5 assisted with independent checks of the grouped construction and later refinements. The AI-assisted work also included the AMS IAMP analysis in the appendix and the simplifications that set every normalization factor to one and add the final query only in the proof. Further attempts by the models did not improve the final exponents; this does not establish optimality.

Claude Opus 5.5 also assisted with readability and editing. We used these models to draft and revise proofs and to check calculations and references. The authors remain responsible for verifying the arguments and for the final mathematical claims.
\section{Model and assumptions}
\label{sec:setting}

\subsection{Gaussian input and the zero-temperature variational problem}

The input matrix $A\in\R^{N\times N}$ follows the Gaussian orthogonal ensemble (GOE): its entries on and above the diagonal are independent, with
\begin{equation}
 A_{ij}\sim\mathcal N(0,1/N)\quad(i<j),\qquad
 A_{ii}\sim\mathcal N(0,2/N).
 \label{eq:goe}
\end{equation}
We use the notation
\begin{equation}
 H_N(x)=\frac12x^\top Ax,\qquad
 \OPT_N=\max_{\sigma\in\{-1,1\}^N}H_N(\sigma).
 \label{eq:objective}
\end{equation}
Both $H_N$ and $\OPT_N$ are total energies. The accuracy $\varepsilon$ is relative to $\OPT_N$; dividing an energy difference by $N$ gives its error per spin.

The Parisi formula replaces the large random optimization problem by a deterministic variational problem over functions of one time variable. Its zero-temperature version describes the limiting maximum per spin~\cite{Talagrand,Panchenko,AuffingerChen}. Let $\mathcal U$ contain the nonnegative, nondecreasing, right-continuous functions on $[0,1)$ with finite integral. For $\gamma\in\mathcal U$, let $\Phi=\Phi_\gamma$ solve the partial differential equation
\begin{equation}
 \Phi_t+\frac12\Phi_{xx}+\frac12\gamma(t)\Phi_x^2=0,
                      \qquad \Phi(1,x)=|x|,
 \label{eq:pde}
\end{equation}
and define
\begin{equation}
 \mathcal P(\gamma)=\Phi_\gamma(0,0)
                  -\frac12\int_0^1t\gamma(t)\dd t,
 \qquad \Par=\min_{\gamma\in\mathcal U}\mathcal P(\gamma).
 \label{eq:parisi-functional}
\end{equation}
A minimizing function $\gamma$ exists~\cite[proof of Lemma~3]{AuffingerChen}, and $\E\OPT_N/N\to\Par$. We call $\gamma$ the order parameter. The algorithm uses it and its PDE solution as deterministic functions, which do not depend on the sampled matrix $A$.

Write $m=\Phi_x$ and $a=\Phi_{xx}$ for the first two spatial derivatives. Below time one they are continuous, with $|m|\le1$ and $a\ge0$~\cite[Section~6]{AMS}. For a standard real Brownian motion $B$, whose increment over time $h$ is centered Gaussian with variance $h$, the associated diffusion is
\begin{equation}
 \dd X_t=\gamma(t)m(t,X_t)\dd t+\dd B_t,\qquad X_0=0,
 \label{eq:true-diffusion}
\end{equation}
with a unique strong solution, continuous up to time one. The term $\gamma(t)m(t,X_t)\dd t$ is the drift, and $\dd B_t$ is the noise increment. The magnetization $M_t=m(t,X_t)\in[-1,1]$ is a real-valued relaxation of a spin and has continuous paths in time. It is a martingale: given its past, its expected future change is zero, and $\dd M_t=a(t,X_t)\dd B_t$. We call \eqref{eq:true-diffusion} the \emph{Parisi diffusion}; the algorithm approximates it at successive time nodes. Our analysis of this approximation applies It\^o's formula to the PDE solution and its spatial derivatives below time one, which the regularity results of Section~\ref{sec:prelim-parisi} justify.

\subsection{The full-support identity}

We fix a minimizing $\gamma$ and impose the following identity.

\begin{assumption}[Full-support identity]
\label{ass:frsb}
For every $0\le t<1$,
\begin{equation}
                         \E[m(t,X_t)^2]=t.
 \label{eq:frsb}
\end{equation}
\end{assumption}

Since $M_t=m(t,X_t)$ is a martingale, $\E[(M_t-M_s)^2]=\E M_t^2-\E M_s^2$ for $0\le s<t<1$, and \eqref{eq:frsb} makes this difference equal to $t-s$. The discretized magnetization must approximate this increment variance while approaching the spin values $\pm1$.

El Alaoui, Montanari and Sellke derive \eqref{eq:frsb} by variational stationarity when $\gamma$ is strictly increasing, with no constant interval~\cite[Lemma~6.15]{AMS}. Strict increase of the order parameter is the full-support condition, which Section~\ref{sec:prelim-support} expresses through its Stieltjes measure.

\subsection{Oracle arithmetic and the success criterion}

\begin{definition}[Parisi-data oracle model]
\label{def:oracles}
The algorithm receives the explicitly supplied matrix $A$, a target
accuracy, and a failure probability $\alpha\in(0,1/2)$. It can evaluate
\begin{equation}
 \gamma(t),\quad m(t,x),\quad a(t,x)
                         \qquad(0\le t<1,\ x\in\R).
 \label{eq:oracle-data}
\end{equation}
Each oracle evaluation, real arithmetic operation, comparison,
square root, logarithm of a positive real number, integer rounding,
and independent standard Gaussian draw costs one unit.
In particular, a dense matrix--vector product costs $O(N^2)$.
The output is an explicit spin vector in $\{-1,1\}^N$.
\end{definition}

Definition~\ref{def:oracles} formalizes the same unit-cost evaluation convention used by Montanari~\cite[Remark~1.2]{Montanari} and in the AMS work bounds~\cite[Remarks~2.1 and~2.4 in the journal version]{AMS}: point evaluations of the deterministic Parisi functions and real arithmetic have constant cost. Computing the minimizing order parameter and Parisi solution is excluded from our operation counts. Appendix~\ref{app:preprocessing} bounds the total work including this computation, under the hypothesis that exact evaluators can be constructed at specified costs. Numerical evaluation and a finite-bit implementation require additional error and stability estimates.

All success probabilities are joint over the GOE input and the algorithm's randomness. We use $\norm{\cdot}$ for the Euclidean norm of a vector and the induced operator norm of a matrix, $\one$ for the all-ones vector, and natural logarithms. Detailed notation is collected after the proof overview in Section~\ref{sec:notation}.
\section{The GEG-IAMP guarantee}
\label{sec:results}

\subsection{Relative approximation above a polynomial dimension}

Write
\begin{equation}
 L_\varepsilon=\log(e/\varepsilon),\qquad
 L_\alpha=\log(\mathsf{K}/\alpha).
 \label{eq:accuracy-logarithms}
\end{equation}
For other accuracy parameters $u$, write $L_u=\log(e/u)$. Constants in the following statements are universal after the algorithmic constants have been fixed.

In Theorem~\ref{thm:main}, each of the $B$ observation intervals groups consecutive discretization steps into one matrix query, and $\varepsilon$ denotes relative accuracy. In the detailed iteration and analysis, the same symbol denotes an internal additive error per spin, which the theorem sets to $\etaacc=c\varepsilon$.

\begin{theorem}[Relative approximation by GEG-IAMP]
\label{thm:main}
Under Assumption~\ref{ass:frsb} and Definition~\ref{def:oracles}, there are constants $c,\mathsf{K}>0$ with the following property. For $0<\varepsilon\le1$ and $0<\alpha<1/2$, set
\begin{equation}
 \etaacc=c\varepsilon,\qquad
 N_0=\left\lceil\mathsf{K}\etaacc^{-2}L_{\etaacc}^2L_\alpha\right\rceil.
 \label{eq:all-dimension-threshold}
\end{equation}
For every $N\ge N_0$, Algorithm~\ref{alg:egiamp} at internal accuracy $\etaacc$ returns $\sigma\in\{-1,1\}^N$ with
\begin{equation}
 \Prob\{H_N(\sigma)\ge(1-\varepsilon)\OPT_N\}\ge1-\alpha.
 \label{eq:main-relative}
\end{equation}
Total work is at most $\mathsf{K}N^2/\varepsilon$ and storage is $O(N^2)$. The algorithm uses $K=\Theta(\varepsilon^{-2}L_\varepsilon^2)$ discretization steps, grouped into $B=\Theta(\varepsilon^{-1})$ observation intervals. It executes $B$ matrix queries and one further dense product to initialize rounding, for $B+1$ products in total. Before simplification, total work is
\begin{equation}
 O\bigl(N^2B+NB^2+NK+K+\varepsilon^{-3}L_\varepsilon^2+N^2\bigr).
 \label{eq:afptas-work}
\end{equation}
\end{theorem}

In \eqref{eq:afptas-work}, $N^2B$ counts matrix products, $NB^2$ orthogonalization, and $NK$ coordinate updates and Gaussian reconstruction. The state update uses numerical quadrature of the integrated drift coefficient $\int_{t_k}^{t_{k+1}}\gamma(t)\dd t$. Computing these coefficients once for all coordinates costs $O(K+\varepsilon^{-3}L_\varepsilon^2)$; the final $N^2$ term includes rounding. For $N\ge N_0$, every term is at most $\mathsf{K}N^2B$.

\subsection{Comparison with AMS IAMP}
\label{sec:improvement-comparison}

\begin{theorem}[AMS IAMP; informal version of Theorem~\ref{thm:onsager-main}]
\label{thm:ams-informal}
Under Assumption~\ref{ass:frsb} and Definition~\ref{def:oracles}, there is a universal $\mathsf{K}>0$ such that, for every $0<\varepsilon\le1$ and $0<\alpha<1/2$, AMS IAMP with an internal increment cutoff and coefficients estimated from the iterates returns $\sigma\in\{-1,1\}^N$ satisfying $H_N(\sigma)\ge(1-\varepsilon)\OPT_N$ with probability at least $1-\alpha$ whenever
\begin{equation}
 N\ge\mathsf{K}\exp\!\left(\mathsf{K}\varepsilon^{-7/2}L_\varepsilon^{5/2}\right)L_\alpha^4.
 \label{eq:ams-dimension-comparison}
\end{equation}
Total work is $O(N^2\varepsilon^{-3}L_\varepsilon^2)$, including coefficient estimation, and storage is $O(N^2)$. There are $K=\Theta(\varepsilon^{-3}L_\varepsilon^2)$ steps and $K+2$ dense products. Probability is over the GOE input.
\end{theorem}

Our bounds on the deviation of the AMS IAMP iterates from the Gaussian process of state evolution and on the effect of empirical coefficient errors grow exponentially with the number of steps. As a result, the dimension bound \eqref{eq:ams-dimension-comparison} is exponential in a power of $1/\varepsilon$. Table~\ref{tab:results} compares both total work and the dimension sufficient for the relative guarantee. Both work bounds are quadratic in $N$ and polynomial in $1/\varepsilon$, but only the dimension bound for \gegiamp\ is polynomial. The dimension bounds are sufficient, not AMS lower bounds. Because the proved bounds for AMS IAMP are weaker, we defer its construction and analysis to Appendix~\ref{sec:iamp-comparison}.

\begin{table}[t]
\centering\normalsize
\begin{tabular}{@{}>{\raggedright\arraybackslash}p{0.22\textwidth}
>{\raggedright\arraybackslash}p{0.31\textwidth}
>{\raggedright\arraybackslash}p{0.37\textwidth}@{}}
\toprule
Method & Total work above threshold & Sufficient dimension for $(1-\varepsilon)$ approximation \\
\midrule
\gegiamp & $O(N^2\varepsilon^{-1})$ &
$\mathsf{K}\varepsilon^{-2}L_\varepsilon^2L_\alpha$ \\
AMS IAMP &
$O(N^2\varepsilon^{-3}L_\varepsilon^2)$ &
$\mathsf{K}\exp(\mathsf{K}\varepsilon^{-7/2}L_\varepsilon^{5/2})L_\alpha^4$ \\
\bottomrule
\end{tabular}
\caption{Proved work and sufficient dimensions at confidence $1-\alpha$, under Assumption~\ref{ass:frsb} and Definition~\ref{def:oracles}. Work includes quadrature and empirical coefficient estimation but excludes construction of the Parisi functions. Earlier stopping gives AMS work $O(N^2\varepsilon^{-8/3}L_\varepsilon^{7/3})$ and the still-exponential sufficient dimension $\mathsf{K}\exp(\mathsf{K}\varepsilon^{-3}L_\varepsilon^3)L_\alpha^4$ (Corollary~\ref{cor:onsager-earlier-stopping}).}
\label{tab:results}
\end{table}

\subsection{An AFPRAS as a corollary}

Extending the algorithm below its sufficient dimension permits an additive error independent of dimension.

\begin{definition}[AFPRAS for random SK instances]
\label{def:afptas}
A randomized family is an AFPRAS if, for every $N\ge1$, $0<\varepsilon\le1$, and $0<\alpha<1/2$, it returns $\sigma\in\{-1,1\}^N$ satisfying
\begin{equation}
 \Prob\{H_N(\sigma)\ge(1-\varepsilon)\OPT_N-\mathfrak{E}_{\varepsilon,\alpha}\}
                                                 \ge1-\alpha,
 \label{eq:afptas-definition}
\end{equation}
where $\mathfrak{E}_{\varepsilon,\alpha}$ is independent of $N$, and its work in Definition~\ref{def:oracles} is polynomial in $N,1/\varepsilon,\log(1/\alpha)$. Probability is joint over the GOE input and algorithmic randomness. The definition does not require $\mathfrak{E}_{\varepsilon,\alpha}$ to be polynomial in inverse accuracy.
\end{definition}

\begin{corollary}[An AFPRAS from GEG-IAMP]
\label{thm:afptas}
Under the assumptions of Theorem~\ref{thm:main}, Algorithm~\ref{alg:afptas} returns all-plus for $N<N_0$, before constructing the grid or quadrature, and otherwise runs Algorithm~\ref{alg:egiamp} at accuracy $\etaacc$. For every $N\ge1$,
\begin{equation}
 \Prob\{H_N(\sigma)\ge(1-\varepsilon)\OPT_N-4N_0\}\ge1-\alpha,
 \label{eq:afptas-result}
\end{equation}
with total work $O(N^2/\varepsilon)$, storage $O(N^2)$, and
\begin{equation}
 4N_0\le \mathsf{K}\varepsilon^{-2}L_\varepsilon^2L_\alpha.
 \label{eq:eg-afptas-allowance}
\end{equation}
For $N\ge N_0$, the additive term is omitted as in \eqref{eq:main-relative}.
\end{corollary}

The additive term is needed only for $N<N_0$, where the algorithm returns $\one$ and $(1-\varepsilon)\OPT_N-H_N(\one)\le N\norm A\le4N_0$ with probability at least $1-\alpha$. Corollary~\ref{thm:onsager-afptas} gives the corresponding AFPRAS for AMS IAMP, with an exponentially larger bound on the additive term.

\paragraph{Conversion to error per spin.}
On the event in \eqref{eq:afptas-definition},
\begin{equation}
 0\le\frac{\OPT_N-H_N(\sigma)}N
       \le\varepsilon\frac{\OPT_N}N
                            +\frac{\mathfrak{E}_{\varepsilon,\alpha}}N.
 \label{eq:per-spin-conversion}
\end{equation}
Since $\OPT_N/N\le\norm A/2$, Gaussian concentration bounds the relative contribution by $\mathsf{K}\varepsilon(1+\sqrt{L_\alpha/N})$ after dividing the failure probability. Above $N_0$, this is $O(\varepsilon)$ and there is no additive contribution.

\subsection{On exact optimization}
\label{sec:exact}

We compare the precision of Theorem~\ref{thm:main} with the precision that exact optimization would require. For fixed failure probability $\alpha$, the dimension threshold \eqref{eq:all-dimension-threshold} and Lemma~\ref{lem:eg-energy} permit a per-spin deficit $O(\log(eN)/\sqrt N)$.

The per-spin $N^{-1/2}$ scale comes from separately bounding empirical energy fluctuations, clipping and rounding losses, and the optimum's upper tail. These terms remain as the time step decreases; a direct comparison with the optimum might exploit cancellation between them.

For exact optimization, the relevant gap is the difference between the largest value of $H_N$ on spins and its largest strictly smaller value. An unproved extrapolation of the calculations of Franz and Parisi~\cite{FranzParisi} and Le Doussal, M\"uller and Wiese~\cite[Section~V]{LeDoussalMullerWiese} suggests that, for each fixed $\alpha$, this gap divided by $N$ is of order $N^{-3/2}$ with probability at least $1-\alpha$. Under that conjecture, an $o(N^{-3/2})$ per-spin deficit holding with probability tending to one would imply exact optimization with probability tending to one.

Without this conjecture, Lemma~\ref{lem:gap-lower-tail} bounds the probability of a gap at most $t$ by $Nt/2$. Consequently, by the reduction in Appendix~\ref{app:gap-lower-tail}, an algorithm guaranteeing $\OPT_N-H_N(\sigma)\le\varepsilon N$ for all dimensions and accuracies, with probability at least $1-\alpha$ and work polynomial in $N,1/\varepsilon,\log(1/\alpha)$, would yield an exact solver in the same oracle model, provided it has neither a dimension restriction nor an additional term $\mathfrak{E}_{\varepsilon,\alpha}$. Polynomial-time exact optimization is not expected; improving the approximation below $N^{-1/2}$ per spin remains a separate question.

Disorder chaos describes sensitivity to perturbing the couplings. Chatterjee proved it for SK at finite temperature~\cite{ChatterjeeChaos}; Chen, Handschy and Lerman proved the ground-state version~\cite[Theorem~2(i)]{ChenHandschyLerman}. For jointly Gaussian disorders with fixed correlation in $(0,1)$, their maximizers' normalized absolute inner product tends to zero in probability. An exact solver succeeding with probability tending to one must reproduce this sensitivity. Polynomial time does not imply stability under perturbations, so chaos does not prove hardness for arbitrary polynomial-time algorithms.
\section{AMP background and proof overview}
\label{sec:overview}

To prove Theorem~\ref{thm:main}, we compare both the output energy and the optimum with $N\Par$. Throughout this section, $\varepsilon$ denotes sufficiently small internal accuracy, a fixed multiple of relative accuracy in Algorithm~\ref{alg:afptas}.

\subsection{Why a matrix iteration can follow a one-dimensional diffusion}
\label{sec:overview-amp}

AMP alternates matrix multiplication and coordinatewise nonlinear updates. Reusing $A$ creates correlations with earlier iterates; the Onsager correction subtracts combinations of earlier inputs. When the nonlinear maps may depend on all earlier iterates, the update is
\begin{equation}
 z^{t+1}=A f_t(z^1,\ldots,z^t)
                      -\sum_{s\le t}b_{ts}f_{s-1}(z^1,\ldots,z^{s-1}).
 \label{eq:amp-background}
\end{equation}
Each iterate $z^t$ has $N$ components, $f_t$ acts on each coordinate's history, and $b_{ts}$ depends on derivatives or their limiting expectations. Under regularity and initialization assumptions, state evolution describes these histories at a fixed iteration count $r$: for suitable test functions $\psi$,
\begin{equation}
 \frac1N\sum_{i=1}^N\psi(z_i^1,\ldots,z_i^r)
   \xrightarrow[N\to\infty]{\Prob}
   \E\psi(\mathcal Z_1,\ldots,\mathcal Z_r),
 \label{eq:overview-state-evolution}
\end{equation}
where $z_i^t$ is coordinate $i$ of $z^t$ and $(\mathcal Z_1,\ldots,\mathcal Z_r)$ has the limiting Gaussian law~\cite{Montanari,AMS,FengAMP}.

AMS IAMP chooses the nonlinear maps so that, by state evolution, each coordinate of its auxiliary state approximately follows a discretization of the Parisi diffusion \eqref{eq:true-diffusion}. The magnetization $M_t=m(t,X_t)$ of this diffusion satisfies
\begin{equation}
 \dd M_t=a(t,X_t)\dd B_t,\qquad
 \E[(M_t-M_s)^2]=t-s\quad(0\le s<t<1).
 \label{eq:overview-martingale}
\end{equation}
The PDE identifies the correlation between the driving Brownian motion and terminal magnetization with the Parisi value, the limiting optimum per spin:
\begin{equation}
               \E[B_1M_1]=\int_0^1\E a(t,X_t)\dd t=\Par.
 \label{eq:overview-energy}
\end{equation}
These are the zero-temperature identities used by AMS~\cite{AMS}. Since $|M_1|\le1$ and $\E M_1^2=1$, $M_1\in\{-1,1\}$ almost surely. The correlation in \eqref{eq:overview-energy} becomes the leading energy term of the discretized algorithm. Its terminal magnetization should therefore approach $[-1,1]$ while retaining this correlation.

\subsection{Euler--Maruyama approximation of the Parisi diffusion}
\label{sec:overview-scalar}

The Euler--Maruyama scheme evaluates the spatial coefficients at the start of each time step and adds a Gaussian increment. For one coordinate, set $t_k=kh$, $T=Kh=1-\tau$, and take independent $G_k\sim\mathcal N(0,h)$. Numerical quadrature approximates $\int_{t_k}^{t_{k+1}}\gamma(t)\dd t$ by $\Gamma_k^{\rm q}$, using evaluations of $\gamma$. The state and magnetization, initialized at zero, satisfy
\begin{equation}
 \begin{aligned}
 Y_{k+1}&=Y_k+\Gamma_k^{\rm q}m(t_k,Y_k)+G_k,\\
 M_{k+1}&=M_k+a(t_k,Y_k)G_k,\qquad Y_0=M_0=0.
 \end{aligned}
 \label{eq:overview-scalar-update}
\end{equation}
Both updates use the same noise and previous state. Using the quadrature value in place of AMS's $h\gamma(t_k)$ controls the integration error even when $\gamma$ jumps. Including quadrature error, we choose
\begin{equation*}
 h\asymp\varepsilon^2/L_\varepsilon^2,\qquad
 \tau\asymp\varepsilon^{2/3}L_\varepsilon^{-1/3},\qquad
 K=\Theta(\varepsilon^{-2}L_\varepsilon^2).
\end{equation*}
For the one-coordinate recursion, we bound its terminal magnetization error, its correlation with the driving Gaussian increments, and the second moment defining $\rho_k$:
\begin{equation}
 \begin{aligned}
 \norm{M_K-m(T,Y_K)}_2&\le \mathsf{K}\varepsilon,&
 \E\left[\left(\sum_{k<K}G_k\right)M_K\right]&\ge\Par-\mathsf{K}\varepsilon,\\
 \max_{k<K}|\rho_k^2-1|&\le \mathsf{K}\varepsilon^{4/3}L_\varepsilon^{7/3}
                         =o(\varepsilon/L_\varepsilon),&
 \rho_k^2&=\E a(t_k,Y_k)^2.
 \end{aligned}
 \label{eq:overview-scalar-estimates}
\end{equation}
Dividing $a(t_k,Y_k)G_k$ by the normalization factor $\rho_k$ would give it variance $h$, matching \eqref{eq:overview-martingale}. By the last bound in \eqref{eq:overview-scalar-estimates}, \gegiamp\ can instead set all normalization factors to one. We control the resulting scalar error through the identity $\E[(\sum_k c_kG_k)^2]=h\sum_k\E[c_k^2]$, valid for square-integrable coefficients $c_k$ that are functions of $G_0,\ldots,G_{k-1}$. Relative-entropy and Hellinger comparisons of path laws transfer moments of $a$ from the Parisi diffusion to its Euler--Maruyama approximation and then bound the additional error that quadrature causes in the terminal magnetization.

\subsection{Reeves's coupling at finite dimension}
\label{sec:overview-gaussian}

Reeves's coupling construction compares a first-order method with a Gaussian process built from the same matrix~\cite{Reeves}.

\gegiamp\ instead uses Reeves's corrections as its matrix update. Each query input $q_s$, the vector multiplied by $A$, has length $\sqrt N$ and is orthogonal to earlier inputs. Conditional on earlier queries and products and on auxiliary variables used so far, the unobserved block on the orthogonal complement of earlier queries has independent centered upper-triangular Gaussian entries in an orthonormal basis, with variances $1/N$ off the diagonal and $2/N$ on it. Symmetry, however, determines the components of each new product along earlier queries through $q_r^\top Aq_s=(Aq_r)^\top q_s$, and auxiliary Gaussians correct the components within the query span. For $B\le N$, the corrected outputs obey the exact finite-dimensional law under the joint randomness of $A$ and the auxiliary variables:
\begin{equation}
             z_1,\ldots,z_B\ \text{are iid }\mathcal N(0,I_N).
 \label{eq:overview-exact-law}
\end{equation}
The outputs $z_s$ still depend on $A$ through the correction formula in the next subsection.

\subsection{The grouped algorithm}
\label{sec:overview-grouping}

For $N\ge N_0$ in Theorem~\ref{thm:main}, group the $K$ scalar steps into $B$ intervals of width $\Delta=m_{\rm f}h$; group $s$ contains $I_s=\{(s-1)m_{\rm f},\ldots,sm_{\rm f}-1\}$. Compute the quadrature coefficients and initialize $Y_0=0$, $\widehat M_0=\sqrt\Delta\one$. For $s=1,\ldots,B$, perform steps 1--4; step 5 follows the last group.
\begin{enumerate}[label=\arabic*.,leftmargin=2em,itemsep=3pt,parsep=0pt,topsep=4pt]
 \item \textbf{Form an orthogonal query.} Set $F_s=\widehat M_{(s-1)m_{\rm f}}$ and, for a nonzero residual, use
 \begin{equation}
  r_s=F_s-\sum_{r<s}\frac{q_r^\top F_s}{N}q_r,\qquad
  q_s=\sqrt N\,\frac{r_s}{\norm{r_s}}.
  \label{eq:overview-query-direction}
 \end{equation}
 If $r_s=0$, use instead the residual of the first coordinate vector outside the preceding query span, with the same normalization.
 \item \textbf{Multiply and correct.} Compute $v_s=Aq_s$ and
 \begin{equation}
  z_s=v_s+\frac12\left(J_{ss}-\frac{q_s^\top v_s}{N}\right)q_s
    +\sum_{r<s}\left(J_{rs}-\frac{q_r^\top v_s}{N}\right)q_r.
  \label{eq:overview-query-update}
 \end{equation}
 Here $J\in\R^{B\times B}$ is drawn once before the loop, independently of $A$ and the side variables. It is symmetric with independent centered upper-triangular Gaussian entries of variance $1/N$ off the diagonal and $2/N$ on it. Unlike Onsager coefficients in \eqref{eq:amp-background}, these corrections use projections, not derivatives.
 \item \textbf{Reconstruct the increments.} Fix an orthogonal $H_{\rm rec}\in\R^{m_{\rm f}\times m_{\rm f}}$ with first column $m_{\rm f}^{-1/2}\one$. In each coordinate $i$, draw side variables $\eta_{s,i}\sim\mathcal N(0,I_{m_{\rm f}-1})$ independently and set
 \begin{equation}
  (G_{j,i})_{j\in I_s}=\sqrt h\,H_{\rm rec}(z_{s,i},\eta_{s,i}^\top)^\top,
  \qquad \sum_{j\in I_s}G_{j,i}=\sqrt\Delta\,z_{s,i}.
  \label{eq:overview-reconstruction}
 \end{equation}
 After marginalizing over the side variables, all increments are independent $\mathcal N(0,h)$ variables. The centered coordinate paths $(Y_{j,i},\widehat M_{j,i}-\sqrt\Delta)$ are independent copies of \eqref{eq:overview-scalar-update}.
 \item \textbf{Advance the group.} For $j\in I_s$ in increasing order and every $i$, use the previous state in both updates:
 \begin{equation}
  \begin{aligned}
  Y_{j+1,i}&=Y_{j,i}+\Gamma_j^{\rm q}m(t_j,Y_{j,i})+G_{j,i},\\
  \widehat M_{j+1,i}&=\widehat M_{j,i}+a(t_j,Y_{j,i})G_{j,i}.
  \end{aligned}
  \label{eq:overview-vector-update}
 \end{equation}
 Each group uses one product with $A$ for all its time steps.
 \item \textbf{Return spins.} After all groups, set $f=\widehat M_K-\sqrt\Delta\one$ and $v=\clip(f,[-1,1]^N)$, moving each coordinate outside $[-1,1]$ to the nearest endpoint. Round $v$ sequentially; this requires one additional product $Av$, so the executed count is $B+1$.
\end{enumerate}
Section~\ref{sec:algorithm} specifies the grid and quadrature and shows how to reconstruct a group's increments in $O(m_{\rm f})$ operations per coordinate. Appendix~\ref{app:gegiamp-pseudocode} gives the pseudocode. To apply the energy identity, we need $f$ in the query span. The proof therefore appends a final query formed from $\widehat M_K$. This query drives no scalar update and is not executed.

Grouping preserves the energy correlation, not only the increment law. Writing $a_j=a(t_j,Y_j)$, \eqref{eq:overview-scalar-update} implies
\begin{equation}
 \E\left[\left(\sum_{j\text{ in group }s}G_j\right)
             \left(\sum_{k\text{ in group }s}a_kG_k\right)\right]
 =\sum_{j\text{ in group }s}h\,\E a_j.
 \label{eq:overview-group-correlation}
\end{equation}
After marginalizing over the side variables, for $j<k$ we condition before step $k$; for $j>k$, we use the independent centered $G_j$.

Every query must have length $\sqrt N$. A zero initial proposal could use the coordinate-vector rule of step~1, but would change the first query and the second-moment argument. We instead initialize at $\sqrt\Delta\one$, which gives $q_1=\one$, and subtract this vector at the end to match the scalar recursion started at zero. The first group is still executed and affects the later states. With normalization factors $\rho_j$, subtracting the initial vector omits the first group's correlation $c_1\le\Delta$ from the leading energy term, and the error from setting the factors to one is bounded separately. Thus, for this construction, $\Delta\asymp\varepsilon$ keeps this loss of the order of the accuracy and permits $B=\Theta(\varepsilon^{-1})$ queries.

\subsection{Bounding the output energy at finite dimension}
\label{sec:overview-finite}

From \eqref{eq:overview-query-update} and the symmetry of $A$, Reeves's energy identity expresses $H_N(f)/N$ exactly through coordinate averages of corrected outputs times magnetization increments, the coefficients expressing $F_s$ through $q_1,\ldots,q_s$, and $J$. We bound the averages using the independence of the coordinate paths, and we bound the projections of later increments onto earlier queries using their conditional mean zero. Only the combinations of query coefficients that enter the energy need to be controlled. For sufficiently small internal accuracy and $N\ge\mathsf{K}\varepsilon^{-2}L_\varepsilon^2L_\alpha$, these estimates hold on a common event of probability at least $1-\alpha$, on which
\begin{equation}
 \frac{H_N(f)}N\ge\Par-\mathsf{K}\varepsilon
   -\mathsf{K}\left\{L_\varepsilon\sqrt{\frac{L_\alpha}{N}}
             +\frac{L_\varepsilon^2(B+L_\alpha)}N\right\}.
 \label{eq:overview-finite-energy}
\end{equation}
Here $\mathsf{K}\varepsilon$ includes scalar error, normalization error, and the first-group loss.

To bound the clipping loss, we decompose $f_i-m(T,Y_{K,i})$ into a continuous martingale's final value and a uniformly $o(\varepsilon)$ remainder. The variance of its squared terminal value is bounded by its small second moment times a bound on quadratic variation, up to a universal constant. Since $|m|\le1$, clipping moves $f_i$ by at most this terminal error. With high probability, $v=\clip(f,[-1,1]^N)$ satisfies
\begin{equation}
 \frac{\norm{f-v}^2}{N}
 \le \mathsf{K}\varepsilon^2+\mathsf{K}\frac{L_\varepsilon^2L_\alpha}{N}.
 \label{eq:overview-clipping}
\end{equation}
The energy changes caused by clipping and sequential rounding satisfy
\begin{equation}
 \begin{aligned}
 \frac{|H_N(f)-H_N(v)|}{N}
 &\le\frac{\norm A}{2}\frac{\norm{f-v}}{\sqrt N}
                   \frac{\norm f+\norm v}{\sqrt N},\\
 H_N(\sigma)&\ge H_N(v)-\tfrac12\sum_i|A_{ii}|.
 \end{aligned}
 \label{eq:overview-rounding}
\end{equation}
The rounding step maximizes the zero-diagonal Hamiltonian one coordinate at a time. On the common success event, $\norm A\le4$, $\norm f/\sqrt N\le5/4$, and $N^{-1}\sum_i|A_{ii}|\le \mathsf{K}\sqrt{L_\alpha/N}$.

Finally, superadditivity and the Parisi limit imply the finite-$N$ expectation bound~\cite{GuerraToninelli,AuffingerChen}. Gaussian concentration bounds the upper tail of $\OPT_N/N$:
\begin{equation}
 \E\OPT_N/N\le\Par,\qquad
 \Prob\{\OPT_N/N>\Par+u\}\le e^{-Nu^2}\quad(u>0).
 \label{eq:overview-optimum}
\end{equation}
For $N\ge \mathsf{K}\varepsilon^{-2}L_\varepsilon^2L_\alpha$ and $B=\Theta(\varepsilon^{-1})$, \eqref{eq:overview-finite-energy}--\eqref{eq:overview-optimum} imply $\OPT_N-H_N(\sigma)\le \mathsf{K}\varepsilon N$ with probability at least $1-\alpha$. On the same event, $\OPT_N\ge H_N(\sigma)\ge N\Par/2>0$. We can therefore rescale the accuracy by a fixed factor, as in Theorem~\ref{thm:main}, to obtain a relative guarantee without using a rate of convergence in the thermodynamic limit.

\clearpage
\section{Preliminaries}
\label{sec:preliminaries}

We fix notation and recall the results needed to specify and analyze \gegiamp. The regularity of the Parisi solution allows us to apply It\^o's formula, and its stochastic-control representation helps us estimate the solution near time one. We also explain how full support implies \eqref{eq:frsb} and recall Reeves's construction of corrected query outputs with an exact Gaussian law. We use these results to specify the algorithm in Section~\ref{sec:algorithm}, before beginning the error analysis in Section~\ref{sec:scalar-analysis}.

\subsection{Notation}
\label{sec:notation}

For real vectors, $\langle x,y\rangle=x^\top y$ is the Euclidean inner product and $\norm{x}=(\sum_i x_i^2)^{1/2}$ is the Euclidean norm; $\top$ denotes transpose. For a matrix $U$, $\norm{U}=\sup_{\norm{x}=1}\norm{Ux}$ is the induced operator norm, and $\norm{U}_{\rm F}=(\sum_{i,j}U_{ij}^2)^{1/2}$ is the Frobenius norm. Norms are not normalized by dimension: factors such as $N^{-1/2}\norm{x}$ are written explicitly. We write $I_d$ for the $d\times d$ identity matrix, omitting the subscript when its size is clear, and $\one$ for the all-ones vector in the indicated dimension. The standard coordinate vector $\mathbf e_i$ has a one in coordinate $i$ and zeros elsewhere. The notation $U_{:,j}$ denotes column $j$ of $U$, and $\diag(U)$ is the diagonal matrix with the same diagonal as a square matrix $U$. For symmetric matrices, $U\preceq V$ means that $V-U$ is positive semidefinite.

We write $\Prob$ and $\E$ for probability and expectation under the law in question, and $\E[V\mid\mathcal F]$ for conditional expectation given a $\sigma$-field $\mathcal F$. The notation $V\sim\mathcal N(\mu,\sigma^2)$ specifies a real Gaussian variable with mean $\mu$ and variance $\sigma^2$; in several dimensions, $\mathcal N(\mu,\Sigma)$ specifies the mean vector and covariance matrix. The abbreviation iid means independent and identically distributed. For a scalar random variable $V$ and $1\le p<\infty$, $\norm{V}_p=(\E|V|^p)^{1/p}$, so $\norm{V}_2$ is its root mean square. We write $\norm{V}_{L^p(P)}$ when the law $P$ must be specified. For a bounded deterministic function, $\norm{f}_\infty$ is the supremum of $|f|$ on its stated domain; for a random variable it is the smallest almost sure bound on its absolute value under the specified law. An essential bound used to compare two laws must hold almost surely under both. Variance is centered: $\operatorname{Var}(V)=\E[(V-\E V)^2]$. Matrices of second moments use raw products without subtracting means; they need not be centered covariances. For a continuous martingale $Z$, $\langle Z\rangle_t$ denotes its quadratic variation through time $t$.

The notation $\one_E$ is the indicator of an event or set $E$, distinct from the all-ones vector. For $a\le b$, $\clip(x,[a,b])=\min\{b,\max\{a,x\}\}$ and $\dist(x,[a,b])=\inf_{y\in[a,b]}|x-y|$ for a real $x$; clipping to a cube is applied coordinatewise. We write $\lfloor x\rfloor$ and $\lceil x\rceil$ for the greatest integer at most $x$ and the least integer at least $x$. All logarithms are natural unless a base is displayed explicitly, and the unbolded symbol $e$ is their base. Universal constants $\mathsf{K},c>0$ can change between displays and are independent of $N$, accuracy, and confidence after the fixed algorithmic constants have been chosen. Any additional dependence is indicated by a subscript or stated explicitly.

On spins, the covariance of the SK Hamiltonian is $\E H_N(\sigma)H_N(\tau)=NR(\sigma,\tau)^2/2$, where $R(\sigma,\tau)=\sigma^\top\tau/N$ is the overlap.

\subsection{Regularity and representations of the Parisi solution}
\label{sec:prelim-parisi}

Below the terminal time, bounded spatial derivatives allow It\^o's formula to be applied to the Parisi solution and its diffusion. We use the regularity and representation results of El Alaoui, Montanari and Sellke \cite[Proposition~6.1, Lemmas~6.2, 6.4--6.5 and~6.7, and Corollary~6.6]{AMS}, with stochastic-control background in \cite{AuffingerChen,JT}.

\begin{proposition}[Parisi PDE and diffusion facts]
\label{prop:pde-facts}
For $\gamma\in\mathcal U$, the solution of \eqref{eq:pde} is continuous
and $\Phi(t,\cdot)$ is even, convex, and $1$-Lipschitz. It satisfies
$\Phi(t,x)\ge|x|$. For every $T_0<1$ and every fixed positive integer $j$, the spatial derivative $\partial_x^j\Phi$ has a jointly continuous representative on $[0,T_0]\times\R$ and is bounded there. Its weak time derivative is also bounded on this strip, so the representative is uniformly Lipschitz in time. The spatially differentiated PDEs hold
almost everywhere in time; no time differentiability of $\gamma$
is assumed. Write
\begin{equation}
 m=\Phi_x,\qquad a=\Phi_{xx},\qquad
 |m|\le1,\quad a\ge0,\quad m(t,0)=0.
 \label{eq:qualitative}
\end{equation}
The diffusion \eqref{eq:true-diffusion} has a unique strong solution, continuous up to time one. On compact
time intervals, It\^o's formula applies to $\Phi$ and its spatial derivatives with the
almost-everywhere time derivatives. In particular
$m(t,X_t)=\int_0^t a(s,X_s)\dd B_s$ is a bounded martingale.
\end{proposition}

The bound on the first spatial derivative is \cite[Lemma~6.2]{AMS}; the higher spatial and weak time derivative bounds are \cite[Lemma~6.4(a)--(b)]{AMS}. For each $j$, the bounded next spatial derivative makes $\partial_x^j\Phi$ Lipschitz in $x$, while its bounded weak time derivative gives a Lipschitz representative in time. Together these bounds justify the jointly continuous representatives in Proposition~\ref{prop:pde-facts}. In Appendix~\ref{app:scalar} we derive the additional quantitative bounds required as the terminal time is approached.

We use the zero-temperature stochastic-control formula to bound $\Phi(t,0)$ as $t\to1$ in the proof of Proposition~\ref{prop:identities}. This formula is equivalent to the construction that approximates $\gamma$ by step functions and applies Cole--Hopf transformations on their constant intervals, and it reads~\cite[Corollary~2 and the following paragraph]{AuffingerChen}
\begin{equation}
 \Phi(t,x)=\sup_{|u|\le1}\E\left[
 \left|x+B_1-B_t+\int_t^1\gamma(s)u_s\dd s\right|
                         -\frac12\int_t^1\gamma(s)u_s^2\dd s\right],
 \label{eq:control}
\end{equation}
where the controls are progressively measurable for the filtration generated by $B$ and bounded in absolute value by one. The cited result includes the extension from step functions to every $\gamma\in\mathcal U$, allowing an integrable divergence near time one. Jagannath and Tobasco prove the finite-temperature counterpart~\cite[Lemma~18]{JT}. The quadratic control cost produces the nonlinear term in \eqref{eq:pde}.

The limit $\E\OPT_N/N\to\Par$ stated in Section~\ref{sec:setting} follows from the almost-sure Parisi limit~\cite[Theorem~1]{AuffingerChen} and uniform integrability. Indeed, $Q_N=N^{-1}\max_\sigma|H_N(\sigma)|$ satisfies $\Prob\{Q_N>u\}\le\min\{1,2^{N+1}e^{-Nu^2}\}$ by a Gaussian union bound, since each $H_N(\sigma)/N$ has variance $1/(2N)$. For $u\ge\sqrt{2\log2}$, this is at most $2e^{-u^2/2}$, so $\sup_N\E Q_N^2<\infty$. The normalized maxima are therefore uniformly integrable, and their almost-sure limit also holds in expectation.

\subsection{Full support of the order parameter}
\label{sec:prelim-support}

Full support concerns where $\gamma$ increases, not merely where it is positive. Its locally finite Stieltjes measure $\mu_\gamma$ on $[0,1)$ is defined by
\begin{equation}
 \mu_\gamma([0,t])=\gamma(t),\qquad
 \mu_\gamma((s,t])=\gamma(t)-\gamma(s)
 \quad(0\le s<t<1).
 \label{eq:stieltjes-measure}
\end{equation}
Thus $\mu_\gamma(\{0\})=\gamma(0)$; its total mass need not be finite. Support is relative to $[0,1)$, and
\begin{equation}
 \operatorname{supp}\mu_\gamma=[0,1)
 \quad\Longleftrightarrow\quad
 \gamma(t)>\gamma(s)\quad(0\le s<t<1).
 \label{eq:full-stieltjes-support}
\end{equation}
Full support therefore excludes every constant interval. Infinitely many support points~\cite{ACZ}, or positivity of $\gamma$, do not suffice; full support alone also imposes no density assumption.

As recalled in Section~\ref{sec:setting}, full support implies \eqref{eq:frsb} by variational stationarity~\cite[Lemma~6.15]{AMS}. Montanari uses its finite-temperature counterpart~\cite[Lemma~3.3]{Montanari}. It\^o isometry then implies $\E[a(t,X_t)^2]=1$ and $\E[(M_t-M_s)^2]=t-s$ for $0\le s\le t<1$; Proposition~\ref{prop:identities} proves these diffusion identities. They are not exact identities for the Euler--Maruyama recursion.

\subsection{Reeves's coupling construction}
\label{sec:reeves-coupling}

Although \gegiamp\ uses only the zero-mean case of Reeves's coupling construction, we recall its general form because the AMS IAMP comparison in Appendix~\ref{sec:iamp-comparison} uses the same coupling on the same input matrix.

\begin{definition}[Coupling construction {\cite[Definition~2]{Reeves}}]
\label{def:reeves-coupling}
Let $1\le d\le N$, let $A$ have law \eqref{eq:goe}, and let $A'$ be an independent copy. Fix a deterministic upper triangular $\Omega\in\R^{d\times d}$ and measurable query and mean maps $f_t,\mu_t:(\R^N)^{t-1}\to\R^N$. At step $t$, project $F_t=f_t(y_{<t})$ off the preceding query span. Normalize a nonzero residual to length $\sqrt N$, obtaining $q_t$. If it is zero, project the first standard coordinate vector outside that span and normalize its projection instead. This choice changes only the query $q_t$, not the proposal $F_t$. Set
\begin{align}
 z_t&=Aq_t+\frac12\left(A'_{tt}-\frac{q_t^\top Aq_t}{N}\right)q_t
       +\sum_{s<t}\left(A'_{st}-\frac{q_s^\top Aq_t}{N}\right)q_s,
 \label{eq:reeves-coupling-z}\\
 w_t&=\sum_{s\le t}\Omega_{st}z_s,\qquad
 y_t=\mu_t(y_{<t})+w_t.
 \label{eq:reeves-coupling-fields}
\end{align}
\end{definition}

Reeves proves that the entries of $Z=[z_1|\cdots|z_d]$ are independent standard Gaussians for arbitrary measurable predictable query maps \cite[Theorem~3]{Reeves}. The centered fields $w_t$ in \eqref{eq:reeves-coupling-fields} therefore have covariance $(\Omega^\top\Omega)_{st}I_N$. This covariance statement concerns $w_t$, not the fields $y_t$, whose means can depend on earlier fields. The corrected query outputs $z_t$ still depend on $A$, and we use their exact relation to $A$ to express the energy in Proposition~\ref{prop:energy-identity}.

Only the leading $d\times d$ block of $A'$ occurs in \eqref{eq:reeves-coupling-z}; call it $J$:
\begin{equation}
 J_{st}\sim\mathcal N(0,1/N)\quad(s<t),\qquad
 J_{tt}\sim\mathcal N(0,2/N),\qquad J_{ts}=J_{st}.
 \label{eq:auxiliary-j}
\end{equation}
The matrix $J\in\R^{d\times d}$ is independent of $A$ and of the Gaussian side variables introduced below. It is the leading block of the ambient $N\times N$ matrix $A'$ in a fixed coordinate basis, not a projection onto the query directions. Its entry variances remain $1/N$ and $2/N$, not $1/d$ and $2/d$.
\section{Parameter choices and implementation of GEG-IAMP}
\label{sec:algorithm}

Section~\ref{sec:overview-grouping} explains how \gegiamp\ uses one matrix product for several scalar time steps. We now choose the parameters and describe the implementation needed for the accuracy and work bounds. We specify a uniform grid with an integer number of time steps in each group, compute the drift coefficients from point evaluations of $\gamma$, and reconstruct the Gaussian increments in each coordinate with work proportional to the number of steps in the group. The quadrature rule also applies when $\gamma$ has jumps.

We use the zero-mean case of Reeves's coupling from Section~\ref{sec:reeves-coupling} to define the matrix queries. We specify how to choose a query input when the magnetization lies in the span of earlier inputs, how to resolve ties when rounding to spins, and why the dimension must be checked before computing the grid or drift coefficients. Appendix~\ref{app:gegiamp-pseudocode} gives the pseudocode. The scalar expectations and additional query defined in Section~\ref{sec:complete-algorithms} are used only in the proofs and are not computed by the algorithm. Here $\varepsilon$ is the internal error tolerance per spin, which Theorem~\ref{thm:main} sets to a fixed multiple of the requested relative accuracy.

\subsection{Time grids and cutoff}
\label{sec:grids}

The recursion stops at $T=1-\tau$ to keep the bounds on $a$ and its derivatives finite. The fixed small constants $c_\tau,c_h,c_B>0$ control the terminal contribution, the Euler--Maruyama error, and the observation interval width, respectively; we choose them in Lemma~\ref{lem:partition} and the proof of Lemma~\ref{lem:eg-energy}. Starting with $\tau=1/2$, halve it until the first time
\begin{equation}
 \tau^3L_\varepsilon\le c_\tau^3\varepsilon^2,
 \qquad T=1-\tau,\qquad L_\tau=\log(e/\tau).
 \label{eq:grouped-cutoff}
\end{equation}
The logarithm $L_\varepsilon$ was defined in \eqref{eq:accuracy-logarithms}. Define
\begin{equation}
 \begin{aligned}
 B&=\lceil T/(c_B\varepsilon)\rceil,&\Delta&=T/B,
 &h_*&=c_h\varepsilon^2/L_\tau^2,\\
 m_{\rm f}&=\lceil\Delta/h_*\rceil,&h&=\Delta/m_{\rm f},
 &K&=Bm_{\rm f},\qquad d=B+1.
 \end{aligned}
 \label{eq:grid-parameters}
\end{equation}
Here $d=B+1$ counts the $B$ executed queries and the final query used only in the proof, described after \eqref{eq:recentered-output}. The exact state-evolution argument requires this number to satisfy $d\le N$.

Thus $t_k=kh$, $0\le k\le K$, are the time-grid nodes, and $s\Delta=t_{sm_{\rm f}}$, $0\le s\le B$, are the observation nodes. Group $s$ contains the indices
\begin{equation}
 I_s=\{(s-1)m_{\rm f},\ldots,sm_{\rm f}-1\}.
 \label{eq:grouped-indices}
\end{equation}
The choices in \eqref{eq:grouped-cutoff}--\eqref{eq:grid-parameters} use arithmetic, logarithms, comparisons, and integer rounding, without evaluating fractional powers of $\varepsilon$. For sufficiently small $\varepsilon$ they satisfy
\begin{equation}
 \begin{gathered}
 \tau\asymp\varepsilon^{2/3}L_\varepsilon^{-1/3},\quad
 L_\tau\asymp L_\varepsilon,\quad
 h\asymp\varepsilon^2/L_\varepsilon^2,\\
 K=\Theta(\varepsilon^{-2}L_\varepsilon^2),\qquad
 B=\Theta(\varepsilon^{-1}),\qquad \Delta\le c_B\varepsilon.
 \end{gathered}
 \label{eq:partition-size}
\end{equation}
In Lemma~\ref{lem:partition} we verify these parameter scales and the smallness assumptions used in Theorem~\ref{thm:scalar-error}, where the scalar error bounds are proved. The two widths control different errors: $h$ controls the diffusion approximation, while $\Delta$ bounds the first group's contribution to the deterministic leading energy term, omitted after the subtraction in \eqref{eq:recentered-output}. The distance $\tau$ from the terminal time in \eqref{eq:grouped-cutoff} also limits the growth of $a$, reducing the quadrature work, and makes the second-moment error small enough to set the normalization factors to one in Proposition~\ref{prop:unit-normalizers}.

\subsection{Point quadrature and the scalar recursion}

The algorithm uses $\Gamma_k^{\rm q}$ to approximate the integrated drift $\Gamma_k=\int_{t_k}^{t_{k+1}}\gamma(t)\dd t$ in each scalar update. To choose the quadrature tolerance, write $\Lambda(t)=\log(e/(1-t))/(1-t)$, the weight in \eqref{eq:derivative-weight} that measures the growth of the coefficient bounds near the terminal time, and fix
\begin{equation}
 R=\mathsf{K}_R\Lambda(T),\qquad \eta_{\rm q}=c_{\rm q}\varepsilon^2/R,
 \label{eq:quadrature-tolerance}
\end{equation}
where $\mathsf{K}_R$ is a fixed constant large enough that $a(t,x)^2\le \mathsf{K}_R\Lambda(t)$ by \eqref{eq:sharp-regularity}, and $c_{\rm q}>0$ is sufficiently small. Thus $R$ bounds $a(t,x)^2$ for $t\le T$. The choice of $\eta_{\rm q}$ makes $\sqrt{R\eta_{\rm q}}=\sqrt{c_{\rm q}}\varepsilon$ a small multiple of accuracy, as required for the terminal estimate in Lemma~\ref{lem:euler-quadrature}. For each time-grid cell set
\begin{align}
 r_k^{\rm q}&=\max\{1,\lceil[\gamma(t_{k+1})-\gamma(t_k)]/\eta_{\rm q}\rceil\},
 &s_{k,j}&=t_k+jh/r_k^{\rm q},\nonumber\\
 \Gamma_k^{\rm q}&=\frac{h}{2r_k^{\rm q}}
       \sum_{j=0}^{r_k^{\rm q}-1}
       \{\gamma(s_{k,j})+\gamma(s_{k,j+1})\}.
 \label{eq:drift-quadrature}
\end{align}
The algorithm stores the same $\Gamma_k^{\rm q}$ for every coordinate. The subdivision count uses the total increase $\gamma(t_{k+1})-\gamma(t_k)$: by monotonicity, the gap between the left- and right-endpoint sums is $h[\gamma(t_{k+1})-\gamma(t_k)]/r_k^{\rm q}\le h\eta_{\rm q}$. This bound on the gap controls the quadrature error even when $\gamma$ jumps, without requiring the increase on each quadrature subinterval to be at most $\eta_{\rm q}$. Lemma~\ref{lem:drift-quadrature} proves the error and work bounds. The exact $\Gamma_k$ is used only in the analysis.

To define the scalar law used in the analysis, let $G_k$ be independent $\mathcal N(0,h)$ variables. For deterministic normalization factors $\omega_k>0$, which divide the magnetization increments, set
\begin{align}
 Y_0&=0,&Y_{k+1}&=Y_k+\Gamma_k^{\rm q}m(t_k,Y_k)+G_k,
 \label{eq:scalar-y}\\
 M_0^\omega&=0,&M_{k+1}^\omega&=M_k^\omega+
                  a(t_k,Y_k)G_k/\omega_k.
 \label{eq:scalar-m}
\end{align}
Both updates use the same $Y_k$ and $G_k$, and only the magnetization depends on $\omega$. The implementation always uses $\omega_k=1$; other deterministic positive choices are introduced for the scalar comparisons and matrix bounds. The matrix algorithm obtains its increments by reconstruction from corrected query outputs, rather than drawing an additional independent scalar path for every coordinate.

\subsection{The Gaussian reconstruction map}

In each coordinate, the corrected query output determines the sum of the $m_{\rm f}$ increments in a group, while $m_{\rm f}-1$ independent standard Gaussian side variables determine their components orthogonal to $\one$. Let $v=m_{\rm f}^{-1/2}\one\in\R^{m_{\rm f}}$ be the normalized all-ones vector, used only to define the following reflection. When $m_{\rm f}>1$, put
\begin{equation}
 u=\mathbf e_1-v,\qquad H_{\rm rec}=I-2uu^\top/\norm u^2;
 \quad\text{when }m_{\rm f}=1,\quad H_{\rm rec}=I.
 \label{eq:householder}
\end{equation}
The Householder matrix $H_{\rm rec}$ is orthogonal and symmetric, and $H_{\rm rec}\mathbf e_1=v$. Writing $z_s$ for the entry of the corrected query output in one coordinate, and taking independent standard Gaussians $\eta_{s,1},\ldots,\eta_{s,m_{\rm f}-1}$, define
\begin{equation}
 (G_j)_{j\in I_s}=\sqrt h\,H_{\rm rec}
       (z_s,\eta_{s,1},\ldots,\eta_{s,m_{\rm f}-1})^\top.
 \label{eq:group-reconstruction}
\end{equation}
If $z_s$ and the $\eta_{s,j}$ are independent standard Gaussians, orthogonal invariance makes the $G_j$ independent $\mathcal N(0,h)$ variables. Furthermore $\one^\top H_{\rm rec}=\sqrt{m_{\rm f}}\mathbf e_1^\top$, so
\begin{equation}
                  \sum_{j\in I_s}G_j=\sqrt\Delta\,z_s.
 \label{eq:group-sum}
\end{equation}
We apply the reflection $H_{\rm rec}$ as one inner product and one rank-one update, in $O(m_{\rm f})$ operations, without storing or multiplying a dense $m_{\rm f}\times m_{\rm f}$ matrix. For fixed deterministic $\omega$ and $d\le N$, Proposition~\ref{prop:gaussian-law} establishes the stated joint law over all groups and coordinates after marginalizing over the side variables, although each vector $z_s$ is computed from $A$ and earlier corrected query outputs.

\subsection{Query maps and spin output}

Initialize vector states $Y_0=0$ and $\widehat M_0=\sqrt\Delta\one$, where the hat distinguishes the implemented vector from the centered scalar output in \eqref{eq:scalar-m}. This nonzero initial magnetization supplies the first query. At the start of each group $s=1,\ldots,B$, propose
\begin{equation}
 F_s=\widehat M_{(s-1)m_{\rm f}},\qquad
 P_{s-1}=\frac1N\sum_{r<s}q_rq_r^\top,\qquad
 r_s=(I-P_{s-1})F_s.
 \label{eq:query-proposals}
\end{equation}
If $r_s\ne0$, set $q_s=\sqrt N r_s/\norm{r_s}$. Otherwise take the smallest $i$ with $(P_{s-1})_{ii}<1$, subtract the orthogonal projection of $\mathbf e_i$ onto the earlier query span, and normalize the residual. Such an index exists for $s\le N$ because the diagonal entries of $P_{s-1}$ sum to $s-1<N$. The proposal $F_s$ is unchanged. With one product $v_s=Aq_s$, compute
\begin{equation}
 z_s=v_s+\frac12(J_{ss}-q_s^\top v_s/N)q_s
       +\sum_{r<s}(J_{rs}-q_r^\top v_s/N)q_r.
 \label{eq:gaussian-coupling}
\end{equation}
With all side variables fixed, the construction with the final query appended is the zero-mean case $\mu_t=0$ of Definition~\ref{def:reeves-coupling}, with $d=B+1$ and $A'_{rs}=J_{rs}$ in \eqref{eq:reeves-coupling-z}. Its first $B$ query maps return the magnetization at the start of each group. The algorithm draws only the leading $B\times B$ block of $J$; we add its last row and column independently in the proof. In the proof of Proposition~\ref{prop:gaussian-law}, we condition only on the entries used by earlier queries, so the unused entries remain independent with their original Gaussian laws.

Apply \eqref{eq:group-reconstruction} in each coordinate $i$, using the entry $z_{s,i}$ of the corrected query output and independent side variables, to obtain $(G_{j,i})_{j\in I_s}$. With every normalization factor set to one, perform successively for $j\in I_s$,
\begin{align}
 Y_{j+1,i}&=Y_{j,i}+\Gamma_j^{\rm q}m(t_j,Y_{j,i})+G_{j,i},\nonumber\\
 \widehat M_{j+1,i}&=\widehat M_{j,i}
                   +a(t_j,Y_{j,i})G_{j,i}.
 \label{eq:vector-scalar-update}
\end{align}
After the $B$th group, put $F_d=\widehat M_K$ for use in the proof and form the vector before clipping and rounding,
\begin{equation}
                   f=\widehat M_K-\sqrt\Delta\one=F_d-F_1.
 \label{eq:recentered-output}
\end{equation}
For the construction analyzed with fixed deterministic normalization factors $\omega_k>0$ and $d\le N$, Proposition~\ref{prop:gaussian-law} identifies the coordinates of $f$ with independent copies of $M_K^\omega$ in \eqref{eq:scalar-m}, which starts at zero. Thus, for the implemented choice $\omega_k=1$ and sufficiently small internal accuracy, \eqref{eq:scalar-final} controls their clipping error. In Lemma~\ref{lem:row-structure} we account for the subtraction in the deterministic leading energy term: it omits the first group's scalar correlation, which is at most $\Delta$. The finite-dimensional energy errors are bounded separately in Lemma~\ref{lem:finite-energy}.

Clip $f$ to $v\in[-1,1]^N$. Compute $\ell=Av-\diag(A)v$, with one further dense product. In order $i=1,\ldots,N$, replace $v_i$ by the sign of $\ell_i$, choosing $+1$ at a tie, and update $\ell$ by the coordinate change times column $i$ of $A-\diag(A)$. Lemma~\ref{lem:rounding} proves the energy bound. Since clipping can move $v$ outside the query span, the implementation computes $Av$ by a separate dense multiplication.

Algorithm~\ref{alg:afptas} first computes $N_0$ and $\etaacc$ from relative accuracy as in Theorem~\ref{thm:main}, then checks $N<N_0$ before constructing these grids or quadrature coefficients. It returns $\one$ below the threshold and otherwise runs Algorithm~\ref{alg:egiamp} at internal accuracy $\etaacc$. This order ensures that small dimensions incur no accuracy-dependent preprocessing and yields Corollary~\ref{thm:afptas}.

\subsection{Quantities used in the proofs}
\label{sec:complete-algorithms}

The scalar analysis compares the implemented choice $\omega_k=1$ with factors $\rho_k$ defined by the law of \eqref{eq:scalar-y}. With these factors, every magnetization increment has variance exactly $h$. We take the nonnegative square roots in the following definition; their positivity on the parameter range used here is proved in Lemma~\ref{lem:euler-quadrature}. The correlation $\mathcal E$ concerns $M_K^\rho$, rather than the implemented $M_K^\omega$:
\begin{align}
 \rho_k^2&=\E a(t_k,Y_k)^2,
 \label{eq:exact-normalizers}
 \\
 \mathcal E&=\E[W_TM_K^\rho]
       =\sum_{k<K}h\frac{\E a(t_k,Y_k)}{\rho_k},
 \qquad W_T=\sum_{k<K}G_k.
 \label{eq:scalar-energy}
\end{align}
The algorithm neither receives nor estimates these expectations. The tolerance $\nu$ bounds the relative squared mismatch $\max_{k<K}|\omega_k^2/\rho_k^2-1|$ allowed in \eqref{eq:fixed-normalizer-condition}. For a small universal $c_\nu>0$, set
\begin{equation}
                 \nu=\frac{c_\nu\varepsilon}{\log(4d)}.
 \label{eq:normalizer-tolerance}
\end{equation}
We retain the conservative logarithmic factor, so that $\nu\log(4d)=c_\nu\varepsilon$. The estimates of the errors of order $\nu$ in \eqref{eq:population-mismatch} and \eqref{eq:uncut-energy-bound} use only the weaker consequence $\nu\le c_\nu\varepsilon$, which keeps these errors within a small multiple of accuracy. Proposition~\ref{prop:unit-normalizers} proves that $\omega_k=1$ satisfies \eqref{eq:fixed-normalizer-condition} with this $\nu$ at sufficiently small internal accuracy. We obtain the terminal approximation and correlation bounds \eqref{eq:scalar-final} while retaining the deviation of $\rho_k$ from one under \eqref{eq:scalar-y}.

To apply Reeves's energy identity \eqref{eq:exact-energy}, we must express $f$ in the span of the queries. Every $F_s$ belongs to the span of $q_1,\ldots,q_s$, including when its residual is zero, but $f=F_d-F_1$ need not lie in the span of the $B$ executed queries. In the proof only, we therefore form a final query $q_d$ from $F_d$ by the rule after \eqref{eq:query-proposals} and compute its corrected query output using fresh entries $J_{rd}$ from \eqref{eq:auxiliary-j}. This output is not used in any scalar update, in $f$, or in rounding. Proposition~\ref{prop:virtual-final-query} shows that omitting the query from execution leaves the outputs and their probabilities unchanged. The proof still requires $d=B+1\le N$.

Algorithm~\ref{alg:egiamp} executes $B$ products with matrix queries and one product to initialize rounding, for $B+1$ dense products overall, and $K$ Euler--Maruyama steps per coordinate. In Section~\ref{sec:main-proofs} we also count reconstruction of the Gaussian increments, all $NK$ coordinate updates, and point quadrature; the final query added only in the proof contributes no executed work.

With the algorithm specified, we turn to its error analysis. In Section~\ref{sec:scalar-analysis}, we establish the terminal approximation and correlation bounds \eqref{eq:scalar-final}. In Section~\ref{sec:matrix-analysis}, we combine these estimates with Reeves's identity \eqref{eq:exact-energy}, the directional estimates of Lemma~\ref{lem:directional-qr}, and the clipping estimate \eqref{eq:empirical-clip} to bound the output energy at finite dimension. Finally, in Section~\ref{sec:main-proofs}, we verify the dimension threshold, count all operations, and convert the energy bound to a relative approximation guarantee. We handle dimensions below the threshold separately in Corollary~\ref{thm:afptas}.
\section{Euler--Maruyama estimates for the Parisi diffusion}
\label{sec:scalar-analysis}

We begin the proof of Theorem~\ref{thm:main} by analyzing the scalar recursion from Section~\ref{sec:algorithm}. Using the grid from Section~\ref{sec:grids}, we prove the two estimates in \eqref{eq:scalar-final} for a sufficiently small internal error tolerance. With $\omega_k=1$, the terminal magnetization is within $\varepsilon/1000$ of $m(T,Y_K)\in[-1,1]$ in root mean square, which controls the error caused by clipping. We also prove $\mathcal E\ge\Par-\varepsilon/16$ for the correlation in \eqref{eq:scalar-energy}. This correlation uses the factors $\rho_k$ and determines the leading energy term. We bound separately the effect of setting these factors to one in the algorithm.

We first derive the diffusion identities in Proposition~\ref{prop:identities} and state the coefficient bounds in Proposition~\ref{prop:regularity}, using the results recalled in Section~\ref{sec:prelim-parisi}. The coefficient bounds are proved in Appendix~\ref{app:scalar-endpoint}. To separate time discretization from quadrature error, we first use exact drift integrals. In Lemmas~\ref{lem:moments-euler} and~\ref{lem:euler-weak}, we establish the moment and weak-error estimates needed for Theorem~\ref{thm:scalar-error}. We then check the grid in Lemma~\ref{lem:partition}, bound the errors from quadrature and changes of normalization factors in Lemmas~\ref{lem:euler-quadrature} and~\ref{lem:fixed-normalizers}, and justify setting all normalization factors to one in Proposition~\ref{prop:unit-normalizers}.

We analyze independent $\mathcal N(0,h)$ increments. For fixed deterministic normalization factors and $d\le N$, the matrix construction produces increments with this same joint distribution (Proposition~\ref{prop:gaussian-law}). The scalar estimates therefore apply to each coordinate after subtracting the initial magnetization. In Section~\ref{sec:matrix-analysis}, we use them to bound the output energy at finite dimension.

\subsection{Diffusion identities and coefficient bounds}

\begin{proposition}[Zero-temperature Parisi diffusion identities, adapted from {\cite{AMS}}]
\label{prop:identities}
Under Assumption~\ref{ass:frsb}, write $\mathcal M_t=m(t,X_t)$ for the magnetization $M_t$ of Section~\ref{sec:setting}, distinguishing it from the discrete magnetizations, and let $g(t)=\E a(t,X_t)$. For $s\le t<1$,
\begin{gather}
 \dd\mathcal M_t=a(t,X_t)\dd B_t,\qquad
 \E a(t,X_t)^2=1,\qquad a(0,0)=1,
 \label{eq:martingale-identities}\\
 \E[(\mathcal M_t-\mathcal M_s)^2]=t-s,\qquad
 g(t)=1-\int_0^t\gamma(r)\dd r,\qquad 0\le g\le1.
 \label{eq:g-identities}
\end{gather}
The bounded martingale has an $L^2$ terminal limit $\mathcal M_1\in\{-1,1\}$. Moreover
\begin{gather}
 \Par=\int_0^1g(t)\dd t,
 \label{eq:parisi-integral}\\
 \E a_x(t,X_t)^2=2\gamma(t)\E a(t,X_t)^3
 \quad\text{for almost every }t<1.
 \label{eq:weighted-identity}
\end{gather}
\end{proposition}

\begin{proof}
The differentiated PDE and It\^o's formula imply
\begin{equation}
 \dd m(t,X_t)=a\dd B_t,\qquad
 \dd a(t,X_t)=a_x\dd B_t-\gamma a^2\dd t.
 \label{eq:true-curvature-sde}
\end{equation}
Lemma~\ref{lem:parisi-ito} justifies these formulas with measurable $\gamma$, and all coefficients are bounded on $[0,t]\times\R$ for every $t<1$. It\^o isometry and \eqref{eq:frsb} imply $\int_0^t\E a(s,X_s)^2\dd s=t$. The integrand is continuous in $s$, so it equals one at every $t<1$. At $t=0$ this reads $a(0,0)^2=1$, and $a\ge0$ selects $a(0,0)=1$. The same isometry proves the increment identity in \eqref{eq:g-identities}, and taking expectations in \eqref{eq:true-curvature-sde} proves the formula for $g$. Since $\E a(t,X_t)^2$ is constant, It\^o's formula for $a(t,X_t)^2$ shows that $\E a_x(t,X_t)^2-2\gamma(t)\E a(t,X_t)^3=0$ for almost every $t$, which is \eqref{eq:weighted-identity}. Finally, $\mathcal M_t$ is a martingale with values in $[-1,1]$, so it converges in $L^2$ to a limit $\mathcal M_1\in[-1,1]$ with $\E\mathcal M_1^2=\lim_{t\to1}\E\mathcal M_t^2=1$; hence $\mathcal M_1\in\{-1,1\}$ almost surely.

To prove \eqref{eq:parisi-integral}, we apply It\^o's formula to $\Phi(t,X_t)$, using \eqref{eq:pde}, and to the product $X_t\mathcal M_t$. Taking expectations and using $\E\mathcal M_s^2=s$, we obtain
\begin{equation}
 \E[\Phi(t,X_t)-X_t\mathcal M_t]
 =\Phi(0,0)-\frac12\int_0^ts\gamma(s)\dd s-\int_0^tg(s)\dd s.
 \label{eq:terminal-verification}
\end{equation}
Convexity and \eqref{eq:qualitative} imply $0\le\Phi(t,x)-xm(t,x)\le\Phi(t,0)$. From \eqref{eq:control} and $|u|-u^2/2\le1/2$,
\begin{equation}
 0\le\Phi(t,0)\le\sqrt{2(1-t)/\pi}+\frac12\int_t^1\gamma(s)\dd s\longrightarrow0.
\end{equation}
The left side of \eqref{eq:terminal-verification} therefore tends to zero as $t\to1$. Since $\gamma$ minimizes \eqref{eq:parisi-functional}, the right side tends to $\Par-\int_0^1g(s)\dd s$, which proves \eqref{eq:parisi-integral}.
\end{proof}

The coefficient bounds increase as $t\to1$. We measure their growth by $\Lambda(t)$ and use its integrals to bound errors accumulated over the time grid. Recalling $T$ and $L_\tau$ from \eqref{eq:grouped-cutoff}, write
\begin{equation}
 r_t=1-t,\qquad \ell_t=\log(e/r_t),\qquad
 \Lambda(t)=\ell_t/r_t,\qquad L_\tau=\log(e/\tau),\qquad T=1-\tau.
 \label{eq:derivative-weight}
\end{equation}

\begin{proposition}[Quantitative bounds near the terminal time]
\label{prop:regularity}
There is a universal $\mathsf{K}_\kappa\ge1$ such that
\begin{gather}
 \int_t^1g(s)\dd s\le \mathsf{K}_\kappa r_t^{3/2}\sqrt{\ell_t},\qquad
 g(t)\le \mathsf{K}_\kappa\sqrt{r_t\ell_t},
 \label{eq:terminal-g}\\
 \gamma(t),\ \norm{a(t,\cdot)}_\infty\le \mathsf{K}_\kappa\sqrt{\Lambda(t)},\qquad
 4\norm{a(t,\cdot)}_\infty^2\le \mathsf{K}_\kappa\Lambda(t),
 \label{eq:sharp-regularity}\\
 \norm{a_x(t,\cdot)}_\infty\le \mathsf{K}_\kappa\Lambda(t),\qquad
 \norm{a_{xx}(t,\cdot)}_\infty\le \mathsf{K}_\kappa\Lambda(t)^{3/2}.
 \label{eq:sharp-spatial-derivative}
\end{gather}
Also $g(1-)=0$, $\int_0^1\gamma=1$, and
\begin{gather}
 \int_0^T\Lambda(t)\dd t=(L_\tau^2-1)/2,\qquad
 \int_0^T\Lambda(t)^p\dd t\le \mathsf{K}_pL_\tau^p\tau^{1-p}\quad(p>1),
 \label{eq:singular-weights}\\
 \int_0^1r_t^{-p}\ell_t^j\dd t<\infty\quad (p<1,\ j\ge0).
 \label{eq:integrable-weights}
\end{gather}
\end{proposition}

Under the full-support identity, Proposition~\ref{prop:regularity} provides explicit bounds near the terminal time for the order parameter and the spatial derivatives of the Parisi solution. These bounds quantify the terminal dependence left unspecified by the regularity results used in AMS~\cite[Lemma~6.4]{AMS} and sharpen the general inverse-linear upper bound on the order parameter. For SK, the latter bound is $\gamma(t)=O((1-t)^{-1})$. It follows by taking the zero-temperature limit in~\cite[Eqs.~(15)--(16) and the proof of Lemma~3]{AuffingerChen} and using uniqueness of the minimizing order parameter~\cite[Theorem~4]{ChenHandschyLerman}. Equation~\eqref{eq:sharp-regularity} improves this bound to $\gamma(t)\le\mathsf{K}_\kappa\sqrt{\log(e/(1-t))/(1-t)}$.

We prove Proposition~\ref{prop:regularity} in Appendix~\ref{app:scalar-endpoint} in two steps. First, $\int_t^1g=\E[(B_1-B_t)\mathcal M_1]$ is the covariance of a Brownian increment with the terminal value $\mathcal M_1\in\{-1,1\}$, whose conditional mean given the path up to time $t$ is $\mathcal M_t$. By Lemma~\ref{lem:sign-covariance} and Jensen's inequality, we bound this covariance and obtain \eqref{eq:terminal-g} without additional regularity assumptions near $t=1$; since $\gamma$ is nondecreasing, the same bound controls $\gamma(t)$. Second, on a time interval after $t$ of length proportional to $1/\Lambda(t)$, the function $m$ solves a Burgers equation whose coefficient $\gamma$ is at most $\mathsf{K}\sqrt{\Lambda(t)}$, and we obtain the spatial bounds from the heat-semigroup estimate of Lemma~\ref{lem:heat-derivatives}. The derivative $a_{xx}$ is used only in Lemma~\ref{lem:euler-backward}, to justify and bound the spatial derivatives of $v$ in \eqref{eq:diffusion-curvature-expectation}, and in the energy part of the proof of Theorem~\ref{thm:scalar-error}. The algorithm evaluates neither $a_x$ nor $a_{xx}$.

\subsection{Moments of the Euler--Maruyama approximation}

To separate discretization from quadrature error, we first analyze the Euler--Maruyama approximation $Y^{\rm r}$ to the Parisi diffusion $X$ in \eqref{eq:true-diffusion}, using the exact integrals $\Gamma_k$ of $\gamma$ over the time-grid cells. The superscript $\mathrm r$ identifies the recursion using exact drift integrals, which occur only in the analysis; we bound the effect of the algorithm's quadrature values in Lemma~\ref{lem:euler-quadrature}. Let $W$ be a standard Brownian motion on $[0,T]$, and write its independent increments as $G_k=W_{t_{k+1}}-W_{t_k}\sim\mathcal N(0,h)$. Define
\begin{gather}
 Y_{k+1}^{\rm r}=Y_k^{\rm r}+\Gamma_km(t_k,Y_k^{\rm r})+G_k,\qquad
 Y_0^{\rm r}=0,\qquad \Gamma_k=\int_{t_k}^{t_{k+1}}\gamma(s)\dd s,
 \label{eq:reference-euler}\\
 (\rho_k^{\rm r})^2=\E a(t_k,Y_k^{\rm r})^2,\qquad
 F^{\rm r}=\sum_{k<K}\frac{a(t_k,Y_k^{\rm r})}{\rho_k^{\rm r}}G_k,
 \label{eq:reference-normalizers}\\
 \mathcal E^{\rm r}=\E[W_TF^{\rm r}]
       =\sum_{k<K}h\frac{\E a(t_k,Y_k^{\rm r})}{\rho_k^{\rm r}}.
 \label{eq:reference-correlation}
\end{gather}
The second expression in \eqref{eq:reference-correlation} follows by conditioning on the path before each increment: since $W_T=\sum_jG_j$, the products $G_ja(t_k,Y_k^{\rm r})G_k$ with $j\ne k$ have expectation zero, and $\E[a(t_k,Y_k^{\rm r})G_k^2]=h\E a(t_k,Y_k^{\rm r})$. Between the nodes we interpolate by
\begin{equation}
 \dd Y_t^{\rm r}=\gamma(t)m(t_k,Y_{t_k}^{\rm r})\dd t+\dd W_t,
 \qquad t_k\le t<t_{k+1}.
 \label{eq:interpolation}
\end{equation}
Integrating \eqref{eq:interpolation} over a cell recovers \eqref{eq:reference-euler}, so the interpolated process agrees with $Y^{\rm r}$ at the nodes. Within a cell, \eqref{eq:interpolation} evaluates $m$ at the left endpoint $(t_k,Y_{t_k}^{\rm r})$ but keeps the time dependence of $\gamma(t)$; the AMS IAMP scalar recursion \eqref{eq:uniform-iamp-scheme} evaluates $\gamma$ at the left endpoint as well. In the estimates for $Y^{\rm r}$ through Lemma~\ref{lem:partition}, write $a_t=a(t,Y_t^{\rm r})$, $b_t=a_x(t,Y_t^{\rm r})$, $c_t=a_{xx}(t,Y_t^{\rm r})$, and
\begin{equation}
 \delta_t=m(t,Y_t^{\rm r})-m(t_k,Y_{t_k}^{\rm r}),\qquad t_k\le t<t_{k+1}.
 \label{eq:within-cell-change}
\end{equation}
Here $b_t,c_t$ are spatial derivatives. In Lemma~\ref{lem:parisi-ito}, along a diffusion $\dd Y_t=\mu_t\dd t+\dd W_t$, we write the drift difference as $\mathfrak d_t=\gamma(t)m(t,Y_t)-\mu_t$. For the drift $\mu_t^{\rm r}=\gamma(t)m(t_k,Y_{t_k}^{\rm r})$ of \eqref{eq:interpolation}, this difference is $\mathfrak d_t=\gamma(t)\delta_t$, so
\begin{align}
 \dd\delta_t&=a_t\dd W_t-\gamma(t)a_t\delta_t\dd t,
 \label{eq:m-under-euler}\\
 \dd a_t&=b_t\dd W_t-\gamma(t)(a_t^2+b_t\delta_t)\dd t,
 \label{eq:a-under-euler}\\
 \dd b_t&=c_t\dd W_t-\gamma(t)(3a_tb_t+c_t\delta_t)\dd t.
 \label{eq:ax-under-euler}
\end{align}
The process $\delta$ restarts at zero at each cell's left endpoint, while $\delta_T$ at $T=t_K$ is its left limit from the last cell, using $k=K-1$ in \eqref{eq:within-cell-change}. Since $|m|\le1$, we have $|\delta_t|\le2$.

We compare moments along \eqref{eq:interpolation} with moments along the Parisi diffusion \eqref{eq:true-diffusion} through the relative entropy of their path laws. In the next lemma we convert a bound on relative entropy into bounds on differences of norms and means, which we use with \eqref{eq:entropy} and again in Lemma~\ref{lem:euler-quadrature}.

\begin{lemma}
\label{lem:entropy}
Let $P,Q$ be probability laws on a common measurable space, with densities $p,q$ relative to a common dominating measure. Let $V$ be a real-valued measurable function on this space, square-integrable under both laws. Write $\norm V_\infty$ for the smallest bound on $|V|$ that holds almost surely under both $P$ and $Q$. Define $d_{\rm Hel}^2(P,Q)=\int(\sqrt p-\sqrt q)^2$. Then
\begin{align}
 d_{\rm Hel}^2(P,Q)&\le\KL(P\Vert Q),
 \label{eq:hellinger-distance}\\
 |\norm V_{L^2(P)}-\norm V_{L^2(Q)}|&\le\norm V_\infty d_{\rm Hel}(P,Q),
 \label{eq:hellinger-norm}\\
 |\E_PV-\E_QV|&\le d_{\rm Hel}(P,Q)
       (\norm V_{L^2(P)}+\norm V_{L^2(Q)}).
 \label{eq:hellinger-mean}
\end{align}
The norm comparison is used for essentially bounded $V$; the mean comparison requires only finite second moments under both laws.
\end{lemma}

\begin{proof}
For $A_0=\int\sqrt{pq}$, Jensen implies $\KL(P\Vert Q)\ge-2\log A_0\ge2(1-A_0)=d_{\rm Hel}^2$. The reverse triangle inequality for $V\sqrt p,V\sqrt q$ proves \eqref{eq:hellinger-norm}. Factor $p-q=(\sqrt p-\sqrt q)(\sqrt p+\sqrt q)$ and apply Cauchy--Schwarz to prove \eqref{eq:hellinger-mean}.
\end{proof}

\begin{lemma}[Moment bounds for the Euler--Maruyama interpolation]
\label{lem:moments-euler}
Let $Y^{\rm r}$ be the continuous interpolation in \eqref{eq:interpolation}, and let $\delta_t$ be the difference in \eqref{eq:within-cell-change}. The moments below are evaluated along $Y^{\rm r}$, whereas $g(t)=\E[a(t,X_t)]$ is evaluated along the Parisi diffusion \eqref{eq:true-diffusion}. Suppose
\begin{equation}
             h\le\tau,\qquad hL_\tau^3/\tau\le c_0
 \label{eq:weighted-smallness}
\end{equation}
for a sufficiently small universal $c_0$. For every $t\le T$,
\begin{gather}
 3/4\le\norm{a(t,Y_t^{\rm r})}_2\le5/4,
 \label{eq:euler-norm-bounds}\\
 \E[a_x(t,Y_t^{\rm r})^2]\le \mathsf{K}\Lambda(t),\qquad
 \E[a(t,Y_t^{\rm r})^4]\le \mathsf{K}\Lambda(t).
 \label{eq:euler-derivative-moments}
\end{gather}
Moreover, the mean $\overline a(t)=\E[a(t,Y_t^{\rm r})]$ and the difference $\delta_t$ satisfy
\begin{equation}
 |\overline a(t)-g(t)|\le \mathsf{K}\sqrt h\,\ell_t,\qquad
 \E[\delta_t^2]\le \mathsf{K}(t-t_k).
 \label{eq:euler-local-moments}
\end{equation}
In particular the normalization factors $\rho_k^{\rm r}$ in \eqref{eq:reference-normalizers} are positive.
\end{lemma}

\begin{proof}
Fix $t\le T$ and let $P_t$ and $Q_t^{\rm r}$ be the laws on $C([0,t];\R)$ of the true path \eqref{eq:true-diffusion} and of the interpolated path \eqref{eq:interpolation}. Write $Z_s$ for the coordinate process and $\eta(s)=t_k$ for $t_k\le s<t_{k+1}$, with $\eta(T)=t_{K-1}$. Their drift functionals are
\begin{equation}
 \mu_s(Z)=\gamma(s)m(s,Z_s),\qquad
 \mu_s^{\rm r}(Z)=\gamma(s)m(\eta(s),Z_{\eta(s)}).
 \label{eq:reference-path-drifts}
\end{equation}
Both functionals are progressively measurable because $\eta(s)\le s$, and both are evaluated on the same coordinate path. Their difference is
\begin{equation}
 \Delta\mu_s=\mu_s(Z)-\mu_s^{\rm r}(Z)
 =\gamma(s)\{m(s,Z_s)-m(\eta(s),Z_{\eta(s)})\}.
 \label{eq:reference-drift-difference}
\end{equation}
Under $Q_t^{\rm r}$, the process $W_s^Q=Z_s-\int_0^s\mu_u^{\rm r}(Z)\dd u$ is Brownian. Since $|m|\le1$, we have $|\Delta\mu_s|\le2\gamma(s)$, and Novikov's condition follows from
\begin{equation}
 \E_{Q_t^{\rm r}}\exp\left(\frac12\int_0^t(\Delta\mu_s)^2\dd s\right)
 \le\exp\left(2\int_0^t\gamma(s)^2\dd s\right)<\infty.
 \label{eq:reference-novikov}
\end{equation}
Here $T<1$ ensures that $\gamma$ is bounded; no restriction on $h$ is needed. Girsanov's theorem changes the drift from $\mu^{\rm r}$ to $\mu$. The true drift is Lipschitz in space on the fixed strip, since $\partial_x\mu_s=\gamma(s)a(s,\cdot)$ is bounded by Proposition~\ref{prop:pde-facts}. Uniqueness in law therefore identifies the resulting measure with $P_t$, so
\begin{equation}
 \mathcal L_t=\frac{\dd P_t}{\dd Q_t^{\rm r}}
 =\exp\left(\int_0^t\Delta\mu_s\dd W_s^Q
                   -\frac12\int_0^t(\Delta\mu_s)^2\dd s\right).
 \label{eq:reference-girsanov-density}
\end{equation}
Under $P_t$, the process $B_s^P=W_s^Q-\int_0^s\Delta\mu_u\dd u$ is Brownian. Substituting $\dd W_s^Q=\dd B_s^P+\Delta\mu_s\dd s$ into \eqref{eq:reference-girsanov-density} yields
\begin{equation}
 \log\mathcal L_t=\int_0^t\Delta\mu_s\dd B_s^P
                         +\frac12\int_0^t(\Delta\mu_s)^2\dd s.
 \label{eq:reference-log-density}
\end{equation}
The stochastic integral in \eqref{eq:reference-log-density} is square-integrable and has mean zero under $P_t$, so
\begin{equation}
 \KL(P_t\Vert Q_t^{\rm r})=\frac12\E_{P_t}\int_0^t(\Delta\mu_s)^2\dd s.
 \label{eq:reference-entropy-identity}
\end{equation}
The expectation in \eqref{eq:reference-entropy-identity} is under the true law, where $\mathcal M_s=m(s,Z_s)$ satisfies $\dd\mathcal M_s=a(s,Z_s)\dd B_s^P$. For $t_k\le s<t_{k+1}$, It\^o isometry and \eqref{eq:martingale-identities} show explicitly the increment identity in \eqref{eq:g-identities}:
\begin{equation}
 \E_{P_t}[(\mathcal M_s-\mathcal M_{t_k})^2]
 =\int_{t_k}^s\E_{P_t}[a(u,Z_u)^2]\dd u=s-t_k.
 \label{eq:reference-true-increment}
\end{equation}
Substituting \eqref{eq:reference-drift-difference} and \eqref{eq:reference-true-increment} into \eqref{eq:reference-entropy-identity}, we obtain
\begin{align}
 \KL(P_t\Vert Q_t^{\rm r})
 &=\frac12\sum_{k:t_k<t}\int_{t_k}^{\min(t_{k+1},t)}
         \gamma(s)^2(s-t_k)\dd s\nonumber\\
 &\le\frac h2\int_0^t\gamma(s)^2\dd s
 \le \mathsf{K}h\int_0^t\Lambda(s)\dd s\le \mathsf{K}h\ell_t^2.
 \label{eq:entropy}
\end{align}
The last line uses \eqref{eq:sharp-regularity} and $\int_0^t\Lambda(s)\dd s=(\ell_t^2-1)/2$, obtained by the substitution $\ell_s=\log(e/(1-s))$. There is no variation term for $\gamma$: both drifts in \eqref{eq:reference-path-drifts} use its value at $s$.

Equations~\eqref{eq:hellinger-distance} and~\eqref{eq:entropy} imply $d_{\rm Hel}(P_t,Q_t^{\rm r})\le \mathsf{K}\sqrt h\,\ell_t$. Apply \eqref{eq:hellinger-norm} to the path observable $V=a(t,Z_t)$. Its $L^2(P_t)$ norm is one by \eqref{eq:martingale-identities}, so
\begin{align}
 |\norm{a(t,Y_t^{\rm r})}_2-1|
 &\le\norm{a(t,\cdot)}_\infty d_{\rm Hel}(P_t,Q_t^{\rm r})\nonumber\\
 &\le \mathsf{K}\sqrt{h\ell_t^3/r_t}\le \mathsf{K}\sqrt{hL_\tau^3/\tau}\le1/4,
 \label{eq:reference-curvature-norm-transfer}
\end{align}
where \eqref{eq:sharp-regularity} bounds the supremum norm, and \eqref{eq:weighted-smallness} with sufficiently small $c_0$ proves the last inequality. This establishes \eqref{eq:euler-norm-bounds}.

For the derivative moment under the true law, \eqref{eq:weighted-identity}, $a\ge0$, and \eqref{eq:martingale-identities} imply, for almost every $t$,
\begin{align}
 \E[a_x(t,X_t)^2]
 &=2\gamma(t)\E[a(t,X_t)^3]\nonumber\\
 &\le2\gamma(t)\norm{a(t,\cdot)}_\infty
                \E[a(t,X_t)^2]\le \mathsf{K}\Lambda(t).
 \label{eq:true-curvature-derivative-moment}
\end{align}
Joint continuity and boundedness of $a_x$ on the fixed strip allow dominated convergence, so the expectation is continuous in $t$ and the upper bound extends to every time. Apply \eqref{eq:hellinger-norm} to $V=a_x(t,Z_t)$ and use \eqref{eq:sharp-spatial-derivative}:
\begin{align}
 \norm{a_x(t,Y_t^{\rm r})}_2
 &\le\norm{a_x(t,X_t)}_2
       +\norm{a_x(t,\cdot)}_\infty d_{\rm Hel}(P_t,Q_t^{\rm r})\nonumber\\
 &\le \mathsf{K}\sqrt{\Lambda(t)}+\mathsf{K}\sqrt h\,\ell_t\Lambda(t).
 \label{eq:reference-derivative-norm-transfer}
\end{align}
Squaring with $(u+v)^2\le2u^2+2v^2$ yields
\begin{equation}
 \E[a_x(t,Y_t^{\rm r})^2]\le \mathsf{K}\Lambda(t)+\mathsf{K}h\ell_t^2\Lambda(t)^2
       =\mathsf{K}\Lambda(t)\{1+h\ell_t^3/r_t\}\le \mathsf{K}\Lambda(t),
 \label{eq:reference-derivative-second-moment}
\end{equation}
where the final inequality uses \eqref{eq:weighted-smallness}. The fourth-moment bound in \eqref{eq:euler-derivative-moments} follows from the second-moment bound \eqref{eq:euler-norm-bounds} and the supremum bound in \eqref{eq:sharp-regularity}:
\begin{equation}
 \E[a(t,Y_t^{\rm r})^4]\le\norm{a(t,\cdot)}_\infty^2\E[a(t,Y_t^{\rm r})^2]
 \le\frac{25}{16}\norm{a(t,\cdot)}_\infty^2\le \mathsf{K}\Lambda(t).
 \label{eq:reference-curvature-fourth-moment}
\end{equation}
For the mean, \eqref{eq:hellinger-mean} uses the second moments under both laws rather than the supremum norm:
\begin{align}
 |\overline a(t)-g(t)|
 &\le d_{\rm Hel}(P_t,Q_t^{\rm r})\{1+\norm{a(t,Y_t^{\rm r})}_2\}\nonumber\\
 &\le\frac94d_{\rm Hel}(P_t,Q_t^{\rm r})\le \mathsf{K}\sqrt h\,\ell_t.
 \label{eq:reference-curvature-mean-transfer}
\end{align}
This proves the first bound in \eqref{eq:euler-local-moments}.

Finally, on a cell $[t_k,t_{k+1})$, recall that $a_s=a(s,Y_s^{\rm r})$. For $f(x)=x^2$, we have $f'(x)=2x$ and $f''(x)=2$, so It\^o's formula applied to \eqref{eq:m-under-euler} reads
\begin{align}
 \dd(\delta_s^2)
 &=2\delta_s\dd\delta_s+a_s^2\dd s\nonumber\\
 &=2a_s\delta_s\dd W_s
              +\{a_s^2-2\gamma(s)a_s\delta_s^2\}\dd s.
 \label{eq:reference-within-cell-square}
\end{align}
The correction $a_s^2\dd s$ comes from the square of the Brownian coefficient $a_s$, while $-2\gamma(s)a_s\delta_s^2\dd s$ comes from the drift in \eqref{eq:m-under-euler}. Integrating \eqref{eq:reference-within-cell-square} from $t_k$ to $t$, where $\delta_{t_k}=0$, produces the Brownian integral $2\int_{t_k}^t a_s\delta_s\dd W_s$. In \eqref{eq:interpolation}, $W$ is Brownian and the integrand $2a_s\delta_s$ is adapted and bounded, since $|\delta_s|\le2$ and $a$ is bounded on $[0,T]\times\R$. This integral is therefore a square-integrable martingale increment with expectation zero. Taking expectations in the integral form of \eqref{eq:reference-within-cell-square} yields
\begin{equation}
 \E\delta_t^2+2\int_{t_k}^t\gamma(s)\E[a_s\delta_s^2]\dd s
       =\int_{t_k}^t\E a_s^2\dd s.
 \label{eq:reference-within-cell-second-moment}
\end{equation}
Since $\gamma,a\ge0$, the second term on the left is nonnegative. By \eqref{eq:euler-norm-bounds}, $\E\delta_t^2\le(25/16)(t-t_k)$, proving the last bound in \eqref{eq:euler-local-moments}. At $T$, the same inequality holds by the last-cell limit specified after \eqref{eq:ax-under-euler}.
\end{proof}

\subsection{Weak error of the normalization factors}

We estimate $\E[a(t,Y_t^{\rm r})^2]-1$ directly to verify the normalization tolerance \eqref{eq:normalizer-tolerance}. By Lemma~\ref{lem:moments-euler}, the factors $\rho_k^{\rm r}$ in \eqref{eq:reference-normalizers} lie between $3/4$ and $5/4$, but its bound $|\norm{a(t,Y_t^{\rm r})}_2-1|\le \mathsf{K}\sqrt{h\ell_t^3/r_t}$ has scale $(\varepsilon L_\varepsilon)^{2/3}$ at $t=T$ on the grid \eqref{eq:partition-size}. This upper bound is too large to establish an error of order $\nu$. Instead, for fixed $t$, we use the expected value of $a(t,X_t)^2$ for the Parisi diffusion started at $(s,x)$, and express the weak error through its spatial derivative in \eqref{eq:euler-weak-identity}.

For $0\le s\le t<1$, let $X^{s,x}$ solve the same SDE as in \eqref{eq:true-diffusion}, but with starting condition $X_s^{s,x}=x$ at time $s$. Write $\E_{s,x}$ for expectation under this law, omitting the superscripts on $X$ inside this expectation. For any bounded measurable function $f:\R\to\R$, define the transition operator by
\begin{equation}
 P_{s,t}f(x):=\E[f(X_t^{s,x})]=\E_{s,x}[f(X_t)].
 \label{eq:diffusion-transition-operator}
\end{equation}
Thus $P_{s,t}$ maps a function of the state at time $t$ to its expected value when the diffusion starts from $x$ at time $s$. Taking $f(y)=a(t,y)^2$, define
\begin{equation}
 v(s,x)=P_{s,t}[a(t,\cdot)^2](x)=\E[a(t,X_t^{s,x})^2],\qquad p=v_x.
 \label{eq:diffusion-curvature-expectation}
\end{equation}
The terminal time $t$ is held fixed when differentiating $v$ with respect to $s$ or $x$. To bound $p$ and $p_x$, we use in Appendix~\ref{app:euler-backward} the identity \eqref{eq:euler-weighted-flow} for $\partial_xX_u^{s,x}$ weighted by $a$. This avoids a quantitative bound on the unweighted flow derivative.

\begin{lemma}[Spatial derivatives of the expected value of $a^2$]
\label{lem:euler-backward}
For $v$ and $p$ in \eqref{eq:diffusion-curvature-expectation} and $0\le s\le t<1$,
\begin{equation}
 |p(s,x)|\le \mathsf{K}\Lambda(t)a(s,x),\qquad
 |p_x(s,x)|\le \mathsf{K}\Lambda(t)\Lambda(s).
 \label{eq:euler-backward-bounds}
\end{equation}
The expectation $v$ satisfies the Kolmogorov backward equation with the prescribed terminal value:
\begin{equation}
 v_s+\tfrac12v_{xx}+\gamma(s)m(s,x)v_x=0,\qquad v(t,x)=a(t,x)^2.
 \label{eq:diffusion-kolmogorov-equation}
\end{equation}
The derivatives in the first equation are evaluated at $(s,x)$. Equation~\eqref{eq:diffusion-kolmogorov-equation} and its spatially differentiated equations hold almost everywhere in $s$, and It\^o's formula for products is valid with these time derivatives.
\end{lemma}

The proof of Lemma~\ref{lem:euler-backward} is in Appendix~\ref{app:euler-backward}.

\begin{lemma}[Weak error of $\E a^2$ and $\rho_k^{\rm r}$]
\label{lem:euler-weak}
Let $\mathfrak r(t)=\E a(t,Y_t^{\rm r})^2-1$ be the signed difference from the value $\E a(t,X_t)^2=1$ along the Parisi diffusion in \eqref{eq:martingale-identities}. Under \eqref{eq:weighted-smallness},
\begin{equation}
 |\mathfrak r(t)|\le \mathsf{K}h\ell_t^4/r_t,\qquad
 \left(\sum_{k<K}h(\rho_k^{\rm r}-1)^2\right)^{1/2}\le \mathsf{K}hL_\tau^4/\sqrt\tau.
 \label{eq:euler-weak-error}
\end{equation}
\end{lemma}

\begin{proof}
Fix $0\le t\le T$ and hold this terminal time fixed in $v$ from \eqref{eq:diffusion-curvature-expectation}. On a cell $t_k\le s<t_{k+1}$, the identity $m(t_k,Y_{t_k}^{\rm r})=m(s,Y_s^{\rm r})-\delta_s$ follows from \eqref{eq:within-cell-change}. The interpolation \eqref{eq:interpolation} has Brownian coefficient one, so the second-order term in It\^o's formula is $v_{xx}\dd s/2$. With $v_s$, $v_{xx}$, and $p=v_x$ evaluated at $(s,Y_s^{\rm r})$, the formula reads
\begin{align}
 \dd v(s,Y_s^{\rm r})
 &=v_s\dd s+p\dd Y_s^{\rm r}+\tfrac12v_{xx}\dd s\nonumber\\
 &=p\dd W_s+\{v_s+\tfrac12v_{xx}+\gamma(s)m(s,Y_s^{\rm r})p\}\dd s
       -\gamma(s)\delta_sp\dd s\nonumber\\
 &=p\dd W_s-\gamma(s)\delta_sp\dd s.
 \label{eq:euler-weak-ito}
\end{align}
The last equality uses \eqref{eq:diffusion-kolmogorov-equation}, which holds almost everywhere in $s$; Lemma~\ref{lem:euler-backward} justifies It\^o's formula with these time derivatives. Integrating over the cells up to $t$, the endpoint values telescope because $v(s,Y_s^{\rm r})$ is continuous. Since $v(t,x)=a(t,x)^2$ and $v(0,0)=\E a(t,X_t)^2=1$, we obtain
\begin{equation}
 a(t,Y_t^{\rm r})^2-1
 =\int_0^t p(s,Y_s^{\rm r})\dd W_s
       -\int_0^t\gamma(s)\delta_sp(s,Y_s^{\rm r})\dd s.
 \label{eq:euler-weak-integrated}
\end{equation}
On this fixed time interval, $p$ is bounded by \eqref{eq:euler-backward-bounds}, so the stochastic integral is a square-integrable martingale with expectation zero. The drift is integrable, so taking expectations in \eqref{eq:euler-weak-integrated} and applying Fubini's theorem yields
\begin{equation}
 \mathfrak r(t)=-\int_0^t\gamma(s)\E[\delta_sp(s,Y_s^{\rm r})]\dd s.
 \label{eq:euler-weak-identity}
\end{equation}
On a time-grid cell $t_k\le s<t_{k+1}$, up to the fixed terminal time $t$, apply It\^o's formula to $p(s,Y_s^{\rm r})$. With $p,p_s,p_x,p_{xx}$ evaluated at $(s,Y_s^{\rm r})$, the backward equation \eqref{eq:euler-backward-p} implies
\begin{align}
 \dd p(s,Y_s^{\rm r})
 &=p_s\dd s+p_x\dd Y_s^{\rm r}+\tfrac12p_{xx}\dd s\nonumber\\
 &=p_x\dd W_s+\{p_s+\tfrac12p_{xx}+\gamma(s)m(s,Y_s^{\rm r})p_x\}\dd s
       -\gamma(s)\delta_sp_x\dd s\nonumber\\
 &=p_x\dd W_s-\gamma(s)\{a_sp+\delta_sp_x\}\dd s.
 \label{eq:euler-weak-p-ito}
\end{align}
Together with $\dd\delta_s=a_s\dd W_s-\gamma(s)a_s\delta_s\dd s$ from \eqref{eq:m-under-euler}, the It\^o product rule yields
\begin{align}
 \dd\{\delta_sp(s,Y_s^{\rm r})\}
 &=\delta_s\dd p(s,Y_s^{\rm r})+p\dd\delta_s+a_sp_x\dd s\nonumber\\
 &=(\delta_sp_x+a_sp)\dd W_s\nonumber\\
 &\quad+\{a_sp_x-2\gamma(s)a_s\delta_sp-\gamma(s)\delta_s^2p_x\}\dd s.
 \label{eq:euler-weak-product-ito}
\end{align}
The correction $a_sp_x\dd s$ is the product of the two Brownian coefficients. The drift terms from $\delta_s\dd p$ and $p\dd\delta_s$ each contribute $-\gamma(s)a_s\delta_sp\dd s$, which explains the factor two. On the fixed interval, $a,p,p_x,\delta$ are bounded, so the stochastic integral in \eqref{eq:euler-weak-product-ito} is square-integrable and has expectation zero. Integrating from $t_k$ to $s$, taking expectations, and applying Fubini expresses $\E[\delta_sp(s,Y_s^{\rm r})]$ as the integral of the expected drift, since $\delta_{t_k}=0$. This expectation is absolutely continuous on the cell, so the fundamental theorem of calculus yields, for almost every $s$,
\begin{align}
 \frac\dd{\dd s}\E[\delta_sp(s,Y_s^{\rm r})]
 =\E[a_sp_x]-2\gamma(s)\E[a_s\delta_sp]
                  -\gamma(s)\E[\delta_s^2p_x].
 \label{eq:euler-weak-product}
\end{align}
The three expectations in \eqref{eq:euler-weak-product} have the separate bounds
\begin{align}
 |\E[a_sp_x]|&\le \mathsf{K}\Lambda(t)\Lambda(s)\overline a(s),\nonumber\\
 2\gamma(s)|\E[a_s\delta_sp]|&\le \mathsf{K}\gamma(s)\Lambda(t)\E a_s^2
       \le \mathsf{K}\gamma(s)\Lambda(t),\nonumber\\
 \gamma(s)|\E[\delta_s^2p_x]|&\le \mathsf{K}h\gamma(s)\Lambda(t)\Lambda(s).
 \label{eq:euler-weak-three-terms}
\end{align}
Here the first bound uses \eqref{eq:euler-backward-bounds} and $a\ge0$; the second uses $|p|\le \mathsf{K}\Lambda(t)a$, $|\delta|\le2$, and the bounded second moment of $a$; and the last uses $\E\delta_s^2\le \mathsf{K}h$. Thus the absolute value of the derivative in \eqref{eq:euler-weak-product} is at most
\begin{equation}
 \mathsf{K}\Lambda(t)\{\Lambda(s)\overline a(s)+\gamma(s)+h\gamma(s)\Lambda(s)\}.
\end{equation}
The expectation $\E[\delta_sp(s,Y_s^{\rm r})]$ vanishes at each left endpoint, because $\delta_{t_k}=0$. Write $q(s)$ for its almost-everywhere derivative in \eqref{eq:euler-weak-product}. Substituting $\E[\delta_sp(s,Y_s^{\rm r})]=\int_{t_k}^sq(u)\dd u$ into \eqref{eq:euler-weak-identity} and exchanging the order of integration, we obtain the exact expression
\begin{equation}
 \mathfrak r(t)=-\sum_{k:t_k<t}\int_{t_k}^{\min(t_{k+1},t)}
       \left(\int_s^{\min(t_{k+1},t)}\gamma(u)\dd u\right)q(s)\dd s.
 \label{eq:euler-weak-fubini}
\end{equation}
The deterministic weight in \eqref{eq:euler-weak-fubini} is
\begin{equation}
 R_{k,t}(s)=\int_s^{\min(t_{k+1},t)}\gamma(u)\dd u
       \le \mathsf{K}h\sqrt{\Lambda(s)}.
 \label{eq:euler-cell-weight}
\end{equation}
Indeed, since $h\le\tau$, the values of $\Lambda$ on one time-grid cell are comparable, and \eqref{eq:sharp-regularity} bounds $\gamma$. Thus
\begin{equation}
 |\mathfrak r(t)|\le \mathsf{K}h\Lambda(t)\left\{
 \int_0^t\Lambda(s)^{3/2}\overline a(s)\dd s
 +\int_0^t\Lambda(s)\dd s+h\int_0^t\Lambda(s)^2\dd s\right\}.
 \label{eq:euler-weak-integrals}
\end{equation}
By \eqref{eq:terminal-g} and \eqref{eq:euler-local-moments}, the mean $\overline a(s)$ in the first integral is at most $\mathsf{K}\sqrt{r_s\ell_s}+\mathsf{K}\sqrt h\ell_s$. Consequently
\begin{align}
 \int_0^t\Lambda(s)^{3/2}\overline a(s)\dd s
 &\le \mathsf{K}\int_0^t\ell_s^2/r_s\dd s
       +\mathsf{K}\sqrt h\int_0^t\ell_s^{5/2}/r_s^{3/2}\dd s\nonumber\\
 &\le \mathsf{K}\ell_t^3+\mathsf{K}\sqrt{h/r_t}\ell_t^{5/2}\le \mathsf{K}\ell_t^3.
\end{align}
The other integrals in \eqref{eq:euler-weak-integrals} are at most $\mathsf{K}\ell_t^2$, since $h\le r_t$. This proves the first bound in \eqref{eq:euler-weak-error}. Since $(\rho_k^{\rm r})^2=1+\mathfrak r(t_k)$ and $\rho_k^{\rm r}>0$ by Lemma~\ref{lem:moments-euler}, we have $|\rho_k^{\rm r}-1|\le|\mathfrak r(t_k)|\le \mathsf{K}h\ell_{t_k}^4/r_{t_k}$. Because $\ell_t^8/r_t^2$ increases with $t$, the left sum is at most the corresponding integral, and
\begin{equation}
 \sum_kh(\rho_k^{\rm r}-1)^2
       \le \mathsf{K}h^2\int_0^T\ell_t^8/r_t^2\dd t
       \le \mathsf{K}h^2L_\tau^8/\tau.
\end{equation}
Taking square roots proves the second bound. This argument does not require the pointwise bound $\mathsf{K}h\ell_t^4/r_t$ to be small.
\end{proof}

\subsection{Terminal magnetization and energy correlation}

We now bound the terminal error $F^{\rm r}-m(T,Y_K^{\rm r})$ and the correlation $\mathcal E^{\rm r}$. The terminal error splits into the three terms of \eqref{eq:terminal-coupling} below: the error from the normalization factors $\rho_k^{\rm r}$ in \eqref{eq:reference-normalizers}, which Lemma~\ref{lem:euler-weak} controls; the change of $a$ within each time-grid cell; and the drift error from evaluating $m$ at the left endpoint of each cell. The next lemma bounds the last two terms.

\begin{lemma}
\label{lem:local-errors}
Let $a_t=a(t,Y_t^{\rm r})$ for the interpolation \eqref{eq:interpolation}, and let $\delta_t$ be as in \eqref{eq:within-cell-change}. Under \eqref{eq:weighted-smallness},
\begin{align}
 \sum_k\int_{t_k}^{t_{k+1}}\E|a_t-a_{t_k}|^2\dd t
       &\le \mathsf{K}\{hL_\tau^2+h^2L_\tau^2/\tau\}\le \mathsf{K}hL_\tau^2,
 \label{eq:euler-curvature-variation}\\
 \left\|\int_0^T\gamma(t)a_t\delta_t\dd t\right\|_2
       &\le \mathsf{K}hL_\tau^{3/2}/\sqrt\tau.
 \label{eq:cumulative-drift}
\end{align}
\end{lemma}

\begin{proof}
For $t_k\le t\le t_{k+1}$, integrate \eqref{eq:a-under-euler} to write
\begin{equation}
 a_t-a_{t_k}=\int_{t_k}^t b_s\dd W_s
       -\int_{t_k}^t\gamma(s)(a_s^2+b_s\delta_s)\dd s.
 \label{eq:euler-curvature-cell-decomposition}
\end{equation}
The process $b$ is adapted and bounded on the fixed interval, so It\^o isometry applies to the stochastic integral. Applying the isometry and then Tonelli's theorem to the nonnegative time integrals, we obtain
\begin{align}
 \sum_k\int_{t_k}^{t_{k+1}}\E\left|\int_{t_k}^t b_s\dd W_s\right|^2\dd t
 &=\sum_k\int_{t_k}^{t_{k+1}}\int_{t_k}^t\E b_s^2\dd s\dd t\nonumber\\
 &=\sum_k\int_{t_k}^{t_{k+1}}(t_{k+1}-s)\E b_s^2\dd s\nonumber\\
 &\le \mathsf{K}h\int_0^T\Lambda(s)\dd s.
 \label{eq:euler-curvature-stochastic}
\end{align}
The last inequality uses $t_{k+1}-s\le h$ and the bound on $\E b_s^2$ in \eqref{eq:euler-derivative-moments}.

For the drift, $|\delta_s|\le2$ and \eqref{eq:euler-derivative-moments} imply $\E[(a_s^2+b_s\delta_s)^2]\le2\E a_s^4+8\E b_s^2\le \mathsf{K}\Lambda(s)$. Cauchy--Schwarz on $[t_k,t]$ yields the pathwise inequality
\begin{equation}
 \left|\int_{t_k}^t\gamma(s)(a_s^2+b_s\delta_s)\dd s\right|^2
 \le(t-t_k)\int_{t_k}^t\gamma(s)^2(a_s^2+b_s\delta_s)^2\dd s.
 \label{eq:euler-curvature-drift-cell}
\end{equation}
Taking expectations, using $t-t_k\le h$, and changing the order of integration over each cell, we obtain
\begin{align}
 \sum_k\int_{t_k}^{t_{k+1}}\E\left|\int_{t_k}^t\gamma(s)(a_s^2+b_s\delta_s)\dd s\right|^2\dd t
 &\le \mathsf{K}h\sum_k\int_{t_k}^{t_{k+1}}\int_{t_k}^t\gamma(s)^2\Lambda(s)\dd s\dd t\nonumber\\
 &=\mathsf{K}h\sum_k\int_{t_k}^{t_{k+1}}(t_{k+1}-s)\gamma(s)^2\Lambda(s)\dd s\nonumber\\
 &\le \mathsf{K}h^2\int_0^T\gamma(s)^2\Lambda(s)\dd s
 \le \mathsf{K}h^2\int_0^T\Lambda(s)^2\dd s.
 \label{eq:euler-curvature-drift}
\end{align}
One factor $h$ comes from Cauchy--Schwarz and $t-t_k\le h$; the second comes from $t_{k+1}-s\le h$ after changing the order of integration. The last inequality uses $\gamma(s)^2\le \mathsf{K}\Lambda(s)$ from \eqref{eq:sharp-regularity}. Finally, apply $|x-y|^2\le2|x|^2+2|y|^2$ to \eqref{eq:euler-curvature-cell-decomposition} and use \eqref{eq:singular-weights} to obtain the first bound in \eqref{eq:euler-curvature-variation}. The second follows from $h\le\tau$.

To prove the accumulated-drift bound \eqref{eq:cumulative-drift}, apply the It\^o product rule to \eqref{eq:m-under-euler} and \eqref{eq:a-under-euler} on each cell. The Brownian coefficients of $\delta_t$ and $a_t$ are $a_t$ and $b_t$, so their product contributes $a_tb_t\dd t$:
\begin{align}
 \dd(a_t\delta_t)
 &=a_t\dd\delta_t+\delta_t\dd a_t+a_tb_t\dd t\nonumber\\
 &=a_t^2\dd W_t-\gamma(t)a_t^2\delta_t\dd t
       +\delta_tb_t\dd W_t\nonumber\\
 &\quad-\gamma(t)\delta_t(a_t^2+b_t\delta_t)\dd t+a_tb_t\dd t.
 \label{eq:euler-drift-product-rule}
\end{align}
The two terms $-\gamma(t)a_t^2\delta_t\dd t$ combine, so
\begin{equation}
 \dd(a_t\delta_t)=(a_t^2+b_t\delta_t)\dd W_t
       +(a_tb_t-2\gamma a_t^2\delta_t-\gamma b_t\delta_t^2)\dd t.
 \label{eq:euler-drift-product}
\end{equation}
Since $\delta_{t_k}=0$, the product starts at zero on each cell. Integrating \eqref{eq:euler-drift-product} from $t_k$ to $t<t_{k+1}$ yields
\begin{align}
 a_t\delta_t
 ={}&\int_{t_k}^t(a_s^2+b_s\delta_s)\dd W_s\nonumber\\
 &+\int_{t_k}^t
       (a_sb_s-2\gamma(s)a_s^2\delta_s-\gamma(s)b_s\delta_s^2)\dd s.
 \label{eq:euler-drift-cell-integral}
\end{align}
Multiply this identity by $\gamma(t)$ and integrate over $t\in[t_k,t_{k+1}]$. The double integrals range over $t_k\le s\le t\le t_{k+1}$. After reversing the order, $t$ runs from $s$ to $t_{k+1}$, so define the deterministic weight $R_k(s)=\int_s^{t_{k+1}}\gamma(u)\dd u$. All coefficients are bounded on the fixed interval, which justifies both stochastic and ordinary Fubini. For the stochastic term,
\begin{align}
 &\int_{t_k}^{t_{k+1}}\gamma(t)
       \left\{\int_{t_k}^t(a_s^2+b_s\delta_s)\dd W_s\right\}\dd t\nonumber\\
 &\quad=\int_{t_k}^{t_{k+1}}
       \left(\int_s^{t_{k+1}}\gamma(t)\dd t\right)
       (a_s^2+b_s\delta_s)\dd W_s\nonumber\\
 &\quad=\int_{t_k}^{t_{k+1}}R_k(s)(a_s^2+b_s\delta_s)\dd W_s.
 \label{eq:euler-drift-stochastic-fubini}
\end{align}
Ordinary Fubini applied to the time integral similarly yields
\begin{align}
 &\int_{t_k}^{t_{k+1}}\gamma(t)\int_{t_k}^t
       (a_sb_s-2\gamma(s)a_s^2\delta_s-\gamma(s)b_s\delta_s^2)\dd s\dd t\nonumber\\
 &\quad=\int_{t_k}^{t_{k+1}}R_k(s)
       (a_sb_s-2\gamma(s)a_s^2\delta_s-\gamma(s)b_s\delta_s^2)\dd s.
 \label{eq:euler-drift-deterministic-fubini}
\end{align}
Summing over the cells, with $k(s)$ the cell containing $s$, we obtain the exact identity
\begin{align}
 \int_0^T\gamma a\delta\dd t
 ={}&\int_0^TR_{k(s)}(s)(a_s^2+b_s\delta_s)\dd W_s\nonumber\\
 &+\int_0^TR_{k(s)}(s)
       (a_sb_s-2\gamma a_s^2\delta_s-\gamma b_s\delta_s^2)\dd s.
 \label{eq:euler-drift-fubini}
\end{align}
The weight $R_k(s)$ integrates $\gamma$ over only the remainder of one cell. Its bound $R_k(s)\le \mathsf{K}h\sqrt{\Lambda(s)}$ in \eqref{eq:euler-cell-weight} supplies the factor $h$. Keeping the stochastic terms in one integral over $[0,T]$ allows It\^o isometry to combine their second moments rather than summing their norms. As in \eqref{eq:euler-curvature-stochastic}--\eqref{eq:euler-curvature-drift}, the moment bounds in \eqref{eq:euler-derivative-moments} and $|\delta_s|\le2$ imply $\E[(a_s^2+b_s\delta_s)^2]\le \mathsf{K}\Lambda(s)$. Combining this with the bound on $R_{k(s)}(s)$, It\^o isometry yields
\begin{align}
 \left\|\int_0^TR_{k(s)}(s)(a_s^2+b_s\delta_s)\dd W_s\right\|_2^2
 &=\int_0^TR_{k(s)}(s)^2\E[(a_s^2+b_s\delta_s)^2]\dd s\nonumber\\
 &\le \mathsf{K}h^2\int_0^T\Lambda(s)^2\dd s.
 \label{eq:euler-drift-isometry}
\end{align}
For the time integral, \eqref{eq:sharp-regularity}, \eqref{eq:euler-derivative-moments}, and $|\delta_s|\le2$ bound its three drift terms separately:
\begin{align}
 \|a_sb_s\|_2
 &\le\|a(s,\cdot)\|_\infty\|b_s\|_2\le \mathsf{K}\Lambda(s),\nonumber\\
 \|2\gamma(s)a_s^2\delta_s\|_2
 &\le4\gamma(s)(\E a_s^4)^{1/2}\le \mathsf{K}\Lambda(s),\nonumber\\
 \|\gamma(s)b_s\delta_s^2\|_2
 &\le4\gamma(s)\|b_s\|_2\le \mathsf{K}\Lambda(s).
 \label{eq:euler-drift-integrand-norms}
\end{align}
Minkowski's integral inequality, followed by the triangle inequality for these three terms, therefore yields
\begin{align}
 &\left\|\int_0^TR_{k(s)}(s)
       (a_sb_s-2\gamma(s)a_s^2\delta_s-\gamma(s)b_s\delta_s^2)\dd s\right\|_2\nonumber\\
 &\quad\le\int_0^TR_{k(s)}(s)
       \bigl(\|a_sb_s\|_2+\|2\gamma(s)a_s^2\delta_s\|_2
                           +\|\gamma(s)b_s\delta_s^2\|_2\bigr)\dd s\nonumber\\
 &\quad\le \mathsf{K}h\int_0^T\Lambda(s)^{3/2}\dd s.
 \label{eq:euler-drift-minkowski}
\end{align}
Taking the square root in \eqref{eq:euler-drift-isometry} and combining it with \eqref{eq:euler-drift-minkowski} by the triangle inequality, the $L^2$ norm of \eqref{eq:euler-drift-fubini} is at most
\begin{equation}
 \mathsf{K}h\left\{\left(\int_0^T\Lambda(s)^2\dd s\right)^{1/2}
                  +\int_0^T\Lambda(s)^{3/2}\dd s\right\}
       \le \mathsf{K}hL_\tau^{3/2}/\sqrt\tau.
\end{equation}
Here \eqref{eq:singular-weights} with $p=2$ and $p=3/2$ bounds the two integrals, and $L_\tau\ge1$ allows the first contribution to be absorbed into the second. This proves \eqref{eq:cumulative-drift}.
\end{proof}

The preceding lemmas sharpen the discretization estimates used by Montanari~\cite[Lemma~2.4]{Montanari} and AMS~\cite[Propositions~5.3--5.4]{AMS}. Those analyses use strong couplings with the continuous diffusion. At a fixed terminal time, AMS bounds the normalization bias in its Eq.~(5.7) and the energy approximation in its Proposition~5.4 by $O(\sqrt h)$. For our integrated-drift recursion, weak normalization estimates and the signed mean-error calculation below give $O(h)$ normalization bias and correlation discretization error at fixed $T<1$. We bound the accumulated drift directly and track the dependence on $\tau=1-T$ in the following theorem.

\begin{theorem}
\label{thm:scalar-error}
Under \eqref{eq:weighted-smallness}, the quantities $Y^{\rm r}$, $F^{\rm r}$, and $\mathcal E^{\rm r}$ in \eqref{eq:reference-euler}--\eqref{eq:reference-correlation} satisfy
\begin{align}
 \norm{F^{\rm r}-m(T,Y_K^{\rm r})}_2
       &\le \mathsf{K}(\sqrt h\,L_\tau+hL_\tau^4/\sqrt\tau),
 \label{eq:scalar-clipping}\\
 \Par-\mathcal E^{\rm r}
       &\le \mathsf{K}(\tau^{3/2}\sqrt{L_\tau}+hL_\tau^3+hL_\tau^4/\sqrt\tau).
 \label{eq:scalar-energy-bound}
\end{align}
The first bound also controls $\norm{\dist(F^{\rm r},[-1,1])}_2$.
\end{theorem}

\begin{proof}
Integrating \eqref{eq:m-under-euler} over all cells and subtracting the result from $F^{\rm r}$ in \eqref{eq:reference-normalizers}, we obtain
\begin{align}
 F^{\rm r}-m(T,Y_K^{\rm r})
 ={}&\sum_k(1/\rho_k^{\rm r}-1)a_{t_k}G_k
       +\int_0^T(a_{t_{k(t)}}-a_t)\dd W_t
       +\int_0^T\gamma a\delta\dd t.
 \label{eq:terminal-coupling}
\end{align}
The three terms on the right come from normalization by $\rho_k^{\rm r}$ in \eqref{eq:reference-normalizers}, from the change of $a$ within each time-grid cell, and from evaluating $m$ in the drift at the left endpoint of each cell. The summands of the first term are orthogonal and $\E a_{t_k}^2=(\rho_k^{\rm r})^2$, so its variance is exactly $\sum_kh(\rho_k^{\rm r}-1)^2$. Equations~\eqref{eq:euler-weak-error}, \eqref{eq:euler-curvature-variation}, and \eqref{eq:cumulative-drift} therefore prove \eqref{eq:scalar-clipping}. We used $L_\tau^{3/2}\le L_\tau^4$ to absorb the drift term.

For the correlation bound \eqref{eq:scalar-energy-bound}, let $\chi(t)=\E[b_t\delta_t]$ and, as in Lemma~\ref{lem:euler-weak}, $\mathfrak r(t)=\E a_t^2-1$. Equations~\eqref{eq:m-under-euler} and \eqref{eq:ax-under-euler} imply
\begin{equation}
 \dd(b_t\delta_t)=(a_tb_t+c_t\delta_t)\dd W_t
       +(a_tc_t-4\gamma a_tb_t\delta_t-\gamma c_t\delta_t^2)\dd t.
 \label{eq:euler-energy-product}
\end{equation}
The three drift expectations have absolute bounds $\mathsf{K}\Lambda^{3/2}$, $\mathsf{K}\Lambda$, and $\mathsf{K}h\Lambda^2$, respectively. For the first, use \eqref{eq:sharp-spatial-derivative} and $\E|a_t|\le \mathsf{K}$. For the middle bound, use $\gamma\norm b_\infty\E|a\delta|\le \mathsf{K}\Lambda^{3/2}\sqrt h\le \mathsf{K}\Lambda$, since $h\Lambda(T)\le c_0$. For the last, use $\E\delta^2\le \mathsf{K}h$. Also $h\sqrt{\Lambda(T)}$ is bounded, so the last term is at most $\mathsf{K}\Lambda^{3/2}$. Since $\delta_{t_k}=0$, integrating the expected drift from the left endpoint of the cell proves
\begin{equation}
                         |\chi(t)|\le \mathsf{K}h\Lambda(t)^{3/2}.
 \label{eq:euler-energy-local}
\end{equation}
Integrate \eqref{eq:a-under-euler} from $0$ to $t$. The stochastic integral $\int_0^t b_s\dd W_s$ is a square-integrable martingale, since $b$ is bounded on $[0,T]$, and therefore has expectation zero. The drift is integrable, so Fubini's theorem allows expectation to pass through its time integral. Using $a(0,0)=1$, we obtain
\begin{align}
 \overline a(t)
 &=1-\int_0^t\gamma(s)\{\E a_s^2+\E[b_s\delta_s]\}\dd s\nonumber\\
 &=1-\int_0^t\gamma(s)\{1+\mathfrak r(s)+\chi(s)\}\dd s.
 \label{eq:euler-mean-integral}
\end{align}
Thus $\overline a$ is absolutely continuous, and the fundamental theorem of calculus implies $\overline a'(t)=-\gamma(t)\{1+\mathfrak r(t)+\chi(t)\}$ for almost every $t\in[0,T]$. Subtracting $g(t)=1-\int_0^t\gamma(s)\dd s$ in \eqref{eq:g-identities} yields $\overline a(t)-g(t)=-\int_0^t\gamma(s)\{\mathfrak r(s)+\chi(s)\}\dd s$ for every $t\le T$. Fubini and $T-s\le r_s$ then yield
\begin{align}
 \left|\int_0^T(\overline a(t)-g(t))\dd t\right|
 &\le\int_0^Tr_s\gamma(s)(|\mathfrak r(s)|+|\chi(s)|)\dd s\nonumber\\
 &\le \mathsf{K}h\int_0^1r_s^{-1/2}\ell_s^{9/2}\dd s
          +\mathsf{K}h\int_0^T\ell_s^2/r_s\dd s
 \le \mathsf{K}hL_\tau^3.
 \label{eq:euler-integrated-mean}
\end{align}
The first integral is finite by \eqref{eq:integrable-weights}. Furthermore $|\chi|\le(\E b^2\E\delta^2)^{1/2}\le \mathsf{K}\sqrt{h\Lambda}\le \mathsf{K}$ and $\E a^2\le \mathsf{K}$, so $|\overline a'|\le \mathsf{K}\gamma$ almost everywhere. By absolute continuity, the left Riemann sum for $\overline a$ differs from its integral by at most $\mathsf{K}h\int_0^T\gamma\le \mathsf{K}h$. Finally $0\le\E a_{t_k}\le\rho_k^{\rm r}$ implies
\begin{equation}
 \left|\sum_kh\frac{\E a_{t_k}}{\rho_k^{\rm r}}-\sum_kh\E a_{t_k}\right|
       \le\sum_kh|\rho_k^{\rm r}-1|\le \mathsf{K}hL_\tau^4/\sqrt\tau.
\end{equation}
Since $\Par=\int_0^Tg+\int_T^1g$ by \eqref{eq:parisi-integral}, this inequality, \eqref{eq:euler-integrated-mean}, the Riemann-sum bound above, and the tail bound \eqref{eq:terminal-g} prove \eqref{eq:scalar-energy-bound}.
\end{proof}

We next check that the grid \eqref{eq:grouped-cutoff}--\eqref{eq:grid-parameters} satisfies \eqref{eq:weighted-smallness} and makes the right sides of \eqref{eq:scalar-clipping} and \eqref{eq:scalar-energy-bound} small multiples of $\varepsilon$.

\begin{lemma}[Choice of the grid constants]
\label{lem:partition}
For every fixed $c_B>0$ and every prescribed pair of positive clipping and correlation tolerances, $c_\tau,c_h>0$ and $\varepsilon_0>0$ can be chosen so that \eqref{eq:partition-size} and \eqref{eq:weighted-smallness} hold and the right sides of \eqref{eq:scalar-clipping} and \eqref{eq:scalar-energy-bound} are at most the corresponding tolerances times $\varepsilon$ for every $0<\varepsilon\le\varepsilon_0$. These choices and the implicit constants in \eqref{eq:partition-size} may depend on the fixed $c_B$ and the prescribed tolerances.
\end{lemma}

\begin{proof}
The accepted value $\tau$ passes the test in \eqref{eq:grouped-cutoff}, while the preceding dyadic value $2\tau$ fails it, so
\begin{equation}
 (c_\tau/2)\varepsilon^{2/3}L_\varepsilon^{-1/3}
       <\tau\le c_\tau\varepsilon^{2/3}L_\varepsilon^{-1/3}.
\end{equation}
After fixing $c_\tau$ and restricting $\varepsilon_0\le c_\tau$, the cutoff bounds imply
\begin{equation}
 \frac23L_\varepsilon\le L_\tau
 <1+\log(2/c_\tau)+\frac23\log(1/\varepsilon)+\frac13\log L_\varepsilon
 \le2L_\varepsilon.
\end{equation}
For the last inequality use $\log(1/c_\tau)\le\log(1/\varepsilon)$, $\log L_\varepsilon\le\log(1/\varepsilon)$, and $1+\log2<2$. In particular the tail in \eqref{eq:scalar-energy-bound} is at most $\mathsf{K}c_\tau^{3/2}\varepsilon$. Choose $c_\tau$ first. Since $\Delta/h_*\to\infty$, the ceiling in $m_{\rm f}$ changes $h$ by at most a factor two for sufficiently small accuracy. Then
\begin{equation}
 \begin{aligned}
 \sqrt hL_\tau&\le\sqrt{c_h}\varepsilon,&
 hL_\tau^4/\sqrt\tau&\le \mathsf{K}\varepsilon^{5/3}L_\varepsilon^{13/6},\\
 hL_\tau^3&\le \mathsf{K}\varepsilon^2L_\varepsilon,&
 hL_\tau^3/\tau&\le \mathsf{K}\varepsilon^{4/3}L_\varepsilon^{4/3}.
 \end{aligned}
\end{equation}
Choose $c_h$ small, then $\varepsilon_0$ small to meet the remaining error tolerances and \eqref{eq:weighted-smallness}. The ceilings defining $B$ and $m_{\rm f}$ now prove \eqref{eq:partition-size}. The constant $c_B$ is chosen separately for the first-block energy loss.
\end{proof}

\subsection{Drift quadrature and normalization factors}

Theorem~\ref{thm:scalar-error} bounds $F^{\rm r}$ and $\mathcal E^{\rm r}$ in \eqref{eq:reference-normalizers}--\eqref{eq:reference-correlation}, which use the state recursion \eqref{eq:reference-euler} with exact drift integrals. The algorithm instead computes $\Gamma_k^{\rm q}$ by \eqref{eq:drift-quadrature} and sets every normalization factor $\omega_k$ in \eqref{eq:scalar-m} to one. Lemma~\ref{lem:euler-quadrature} first compares \eqref{eq:reference-euler} and \eqref{eq:scalar-y}, using the factors $\rho_k^{\rm r}$ and $\rho_k$ defined by their respective state laws. Lemma~\ref{lem:fixed-normalizers} then bounds the effect of changing these factors. Proposition~\ref{prop:unit-normalizers} shows that setting every $\omega_k=1$ changes the estimates by less than the permitted error.

\begin{lemma}[Quadrature for $\Gamma_k$ under Definition~\ref{def:oracles}]
\label{lem:drift-quadrature}
For the quadrature rule \eqref{eq:drift-quadrature} and every time-grid cell $[t_k,t_{k+1}]$, $0\le k<K$,
\begin{equation}
 |\Gamma_k^{\rm q}-\Gamma_k|\le\eta_{\rm q}h/2,\qquad
 0\le\Gamma_k^{\rm q}\le h\gamma(t_{k+1}).
 \label{eq:quadrature-error}
\end{equation}
Work is $O(K+\gamma(T)/\eta_{\rm q})=O(K+\varepsilon^{-3}L_\varepsilon^2)$ and storage is $O(K)$ at the chosen parameters.
\end{lemma}

\begin{proof}
On each quadrature subinterval $[s_{k,j},s_{k,j+1}]$ from \eqref{eq:drift-quadrature}, monotonicity implies $\gamma(s_{k,j})\le\gamma(u)\le\gamma(s_{k,j+1})$ for every $u$ in that subinterval. Integrating these inequalities and summing over $j$ shows that the left-endpoint sum is a lower bound for $\Gamma_k$, while the right-endpoint sum is an upper bound:
\begin{align}
 \Gamma_k^-:=\frac{h}{r_k^{\rm q}}\sum_{j=0}^{r_k^{\rm q}-1}\gamma(s_{k,j})
 &\le\Gamma_k,\nonumber\\
 \Gamma_k^+:=\frac{h}{r_k^{\rm q}}\sum_{j=0}^{r_k^{\rm q}-1}\gamma(s_{k,j+1})
 &\ge\Gamma_k.
 \label{eq:quadrature-endpoint-sums}
\end{align}
These bounds do not require continuity of $\gamma$ and remain valid when it jumps. The quadrature value in \eqref{eq:drift-quadrature} is the average $\Gamma_k^{\rm q}=(\Gamma_k^-+\Gamma_k^+)/2$. The difference of the two sums telescopes, so
\begin{equation}
 \Gamma_k^+-\Gamma_k^-
 =\frac{h}{r_k^{\rm q}}[\gamma(t_{k+1})-\gamma(t_k)]
 \le h\eta_{\rm q},
 \label{eq:quadrature-endpoint-gap}
\end{equation}
where the last inequality follows from the choice of $r_k^{\rm q}$ in \eqref{eq:drift-quadrature}. Since $\Gamma_k^-\le\Gamma_k\le\Gamma_k^+$ and $\Gamma_k^{\rm q}$ is the midpoint of these bounds, $|\Gamma_k^{\rm q}-\Gamma_k|\le(\Gamma_k^+-\Gamma_k^-)/2\le h\eta_{\rm q}/2$. Also $0\le\Gamma_k^{\rm q}\le\Gamma_k^+\le h\gamma(t_{k+1})$, because $\gamma$ is nonnegative and nondecreasing. This proves \eqref{eq:quadrature-error}. The number of subintervals satisfies
\begin{equation}
 \sum_kr_k^{\rm q}\le K+[\gamma(T)-\gamma(0)]/\eta_{\rm q}.
\end{equation}
By \eqref{eq:partition-size} and \eqref{eq:sharp-regularity}, $R=O(\varepsilon^{-2/3}L_\varepsilon^{4/3})$ and $\gamma(T)\le \mathsf{K}\sqrt R$. Substitution of \eqref{eq:quadrature-tolerance} proves the work assertion.
\end{proof}

\begin{lemma}[Effect of drift quadrature]
\label{lem:euler-quadrature}
Use the normalization factors $\rho_k^{\rm r}$ from \eqref{eq:reference-normalizers} and $\rho_k$ from \eqref{eq:exact-normalizers}, defined using \eqref{eq:reference-euler} and \eqref{eq:scalar-y}, respectively. Under \eqref{eq:weighted-smallness}, if $\sqrt R\eta_{\rm q}$ is at most a sufficiently small universal constant, the factors $\rho_k$ satisfy $1/2\le\rho_k\le3/2$, and
\begin{align}
 |\rho_k-\rho_k^{\rm r}|&\le \mathsf{K}\sqrt R\eta_{\rm q},
 \label{eq:quadrature-normalizers}\\
 \norm{M_K^\rho-m(T,Y_K)}_2
 &\le\left(\norm{F^{\rm r}-m(T,Y_K^{\rm r})}_2^2+\mathsf{K}R\eta_{\rm q}\right)^{1/2}
        +\mathsf{K}\sqrt R\eta_{\rm q},
 \label{eq:quadrature-strong}\\
 |\mathcal E-\mathcal E^{\rm r}|&\le \mathsf{K}\sqrt R\eta_{\rm q}.
 \label{eq:quadrature-energy}
\end{align}
\end{lemma}

\begin{proof}
Let $Q^{\rm r}$ and $Q^{\rm q}$ be the laws of the node sequence $(Y_0,\ldots,Y_K)$ under \eqref{eq:reference-euler} and \eqref{eq:scalar-y}, respectively. Conditional on the same node history ending at $Y_k=y$, the next node is Gaussian with variance $h$ under both laws, with respective means
\begin{equation}
 \mu_k^{\rm r}(y)=y+\Gamma_km(t_k,y),\qquad
 \mu_k^{\rm q}(y)=y+\Gamma_k^{\rm q}m(t_k,y).
 \label{eq:quadrature-conditional-means}
\end{equation}
For two Gaussian laws with the same variance, the normalizing factors in their densities cancel. Thus, if $Z\sim\mathcal N(\mu,h)$, the expected logarithm of the density ratio is
\begin{align}
 \KL\bigl(\mathcal N(\mu,h)\Vert\mathcal N(\widetilde\mu,h)\bigr)
 &=\frac{\E[(Z-\widetilde\mu)^2-(Z-\mu)^2]}{2h}\nonumber\\
 &=\frac{(\mu-\widetilde\mu)^2}{2h},
 \label{eq:quadrature-gaussian-entropy}
\end{align}
where the last equality uses $\E(Z-\mu)=0$. Applying this formula to the conditional transitions, \eqref{eq:quadrature-error} and $|m|\le1$ imply, for every $y$,
\begin{align}
 \KL\bigl(\mathcal N(\mu_k^{\rm r}(y),h)\Vert\mathcal N(\mu_k^{\rm q}(y),h)\bigr)
 &=\frac{(\Gamma_k-\Gamma_k^{\rm q})^2m(t_k,y)^2}{2h}\nonumber\\
 &\le\frac{\eta_{\rm q}^2h}{8}.
 \label{eq:quadrature-transition-entropy}
\end{align}
For joint laws $P_{U,V}$ and $Q_{U,V}$ of real vectors $U,V$, with the first law absolutely continuous with respect to the second, the chain rule for relative entropy states
\begin{align}
 \KL(P_{U,V}\Vert Q_{U,V})
 &=\KL(P_U\Vert Q_U)\nonumber\\
 &\quad+\E_{u\sim P_U}\left[
       \KL(P_{V\mid U=u}\Vert Q_{V\mid U=u})\right].
 \label{eq:relative-entropy-chain-rule}
\end{align}
Here $P_U,Q_U$ are the marginal laws of $U$. Both node laws start at the same deterministic $Y_0=0$, so the initial marginal term is zero. Applying \eqref{eq:relative-entropy-chain-rule} successively with $U=(Y_0,\ldots,Y_k)$ and $V=Y_{k+1}$, and using \eqref{eq:quadrature-gaussian-entropy} and \eqref{eq:quadrature-transition-entropy}, yields
\begin{align}
 \KL(Q^{\rm r}\Vert Q^{\rm q})
 &=\sum_{k<K}\E_{Q^{\rm r}}\left[
       \frac{\{\mu_k^{\rm r}(Y_k)-\mu_k^{\rm q}(Y_k)\}^2}{2h}\right]\nonumber\\
 &\le\frac{\eta_{\rm q}^2}{8}\sum_{k<K}h
       =\frac{\eta_{\rm q}^2T}{8}\le\frac{\eta_{\rm q}^2}{8}.
 \label{eq:quadrature-path-entropy}
\end{align}
Here both conditional means are evaluated at the same node $Y_k$, and $Kh=T\le1$. Since $a^2\le R$, \eqref{eq:hellinger-norm} implies $|\rho_k-\rho_k^{\rm r}|\le \mathsf{K}\sqrt R\eta_{\rm q}$. Together with $3/4\le\rho_k^{\rm r}\le5/4$ from Lemma~\ref{lem:moments-euler}, this proves the bounds on $\rho_k$.

We compare $F^{\rm r}$ in \eqref{eq:reference-normalizers} with $M_K^\rho$ in \eqref{eq:scalar-m} by recovering the increments from \eqref{eq:reference-euler} and \eqref{eq:scalar-y}, respectively. On any deterministic node sequence $y=(y_0,\ldots,y_K)$ with $y_0=0$, this defines
\begin{align}
 F^{\rm r}(y)
 &=\sum_{k<K}\frac{a(t_k,y_k)}{\rho_k^{\rm r}}
       \{y_{k+1}-y_k-\Gamma_km(t_k,y_k)\},\nonumber\\
 M_K^\rho(y)
 &=\sum_{k<K}\frac{a(t_k,y_k)}{\rho_k}
       \{y_{k+1}-y_k-\Gamma_k^{\rm q}m(t_k,y_k)\}.
 \label{eq:quadrature-output-functions}
\end{align}
The factors $\rho_k^{\rm r}$ and $\rho_k$ remain the fixed constants from \eqref{eq:reference-normalizers} and \eqref{eq:exact-normalizers}; they are not recomputed from $y$. Under $Q^{\rm r}$, $F^{\rm r}(Y_0,\ldots,Y_K)$ is the random variable defined in \eqref{eq:reference-normalizers}. Under $Q^{\rm q}$, $M_K^\rho(Y_0,\ldots,Y_K)$ is the terminal value of \eqref{eq:scalar-m} with $\omega=\rho$. We evaluate $F^{\rm r}(y)$ and $M_K^\rho(y)$ at the same $y$, without assuming that paths sampled under the two laws coincide.

To estimate $F^{\rm r}-M_K^\rho$, let the node sequence have law $Q^{\rm q}$, and write $a_k=a(t_k,Y_k)$ and $m_k=m(t_k,Y_k)$. By \eqref{eq:scalar-y}, its centered Gaussian increment is $G_k=Y_{k+1}-Y_k-\Gamma_k^{\rm q}m_k$. The residual used in $F^{\rm r}$ in \eqref{eq:quadrature-output-functions} is $G_k^{\rm r}=Y_{k+1}-Y_k-\Gamma_km_k$. Hence
\begin{equation}
 G_k^{\rm r}=G_k+(\Gamma_k^{\rm q}-\Gamma_k)m(t_k,Y_k).
\end{equation}
Under $Q^{\rm q}$, the conditional mean of $G_k$ is zero, while that of $G_k^{\rm r}$ is $(\Gamma_k^{\rm q}-\Gamma_k)m_k$. Subtracting $M_K^\rho$ from $F^{\rm r}$ in \eqref{eq:quadrature-output-functions} on this sequence therefore yields
\begin{equation}
 F^{\rm r}-M_K^\rho
 =\sum_ka_k\left(\frac1{\rho_k^{\rm r}}-\frac1{\rho_k}\right)G_k
       +\sum_k\frac{a_km_k}{\rho_k^{\rm r}}(\Gamma_k^{\rm q}-\Gamma_k).
 \label{eq:quadrature-output-difference}
\end{equation}
Under $Q^{\rm q}$, each coefficient of $G_k$ in \eqref{eq:quadrature-output-difference} is determined by $Y_0,\ldots,Y_k$, while $G_k$ is independent of these nodes and has variance $h$. The summands are therefore martingale differences and are orthogonal in $L^2(Q^{\rm q})$. Since $\E_{Q^{\rm q}}a_k^2=\rho_k^2$, their sum satisfies
\begin{align}
 \left\|\sum_{k<K}a_k\left(\frac1{\rho_k^{\rm r}}-\frac1{\rho_k}\right)G_k\right\|_{L^2(Q^{\rm q})}^2
 &=\sum_{k<K}h\rho_k^2
       \left(\frac1{\rho_k^{\rm r}}-\frac1{\rho_k}\right)^2\nonumber\\
 &=\sum_{k<K}h\left(\frac{\rho_k}{\rho_k^{\rm r}}-1\right)^2
 \le \mathsf{K}R\eta_{\rm q}^2.
 \label{eq:quadrature-difference-martingale}
\end{align}
The last bound uses $\rho_k^{\rm r}\ge3/4$, \eqref{eq:quadrature-normalizers}, and $\sum_{k<K}h=T\le1$. For the drift sum, $|a_k|\le\sqrt R$, $|m_k|\le1$, and \eqref{eq:quadrature-error} imply the deterministic bound
\begin{equation}
 \left|\sum_{k<K}\frac{a_km_k}{\rho_k^{\rm r}}(\Gamma_k^{\rm q}-\Gamma_k)\right|
 \le \mathsf{K}\sqrt R\sum_{k<K}|\Gamma_k^{\rm q}-\Gamma_k|
 \le \mathsf{K}\sqrt R\eta_{\rm q}.
 \label{eq:quadrature-difference-drift}
\end{equation}
Taking square roots in \eqref{eq:quadrature-difference-martingale} and using \eqref{eq:quadrature-difference-drift} in \eqref{eq:quadrature-output-difference}, the triangle inequality yields $\|F^{\rm r}-M_K^\rho\|_{L^2(Q^{\rm q})}\le \mathsf{K}\sqrt R\eta_{\rm q}$.

We next bound the fourth moment of $F^{\rm r}-m(T,Y_K)$ under both node laws. Define two distinct partial sums for $0\le\ell\le K$:
\begin{equation}
 S_\ell^{\rm r}=\sum_{k<\ell}\frac{a_k}{\rho_k^{\rm r}}G_k^{\rm r},\qquad
 S_\ell^{\rm q}=\sum_{k<\ell}\frac{a_k}{\rho_k^{\rm r}}G_k.
 \label{eq:quadrature-martingale-sums}
\end{equation}
Both sums retain the factors $\rho_k^{\rm r}$; in particular, $S_K^{\rm q}$ is not $M_K^\rho$, which uses $\rho_k$ in \eqref{eq:scalar-m}. With $\mathcal F_k=\sigma(Y_0,\ldots,Y_k)$, the process $(S_\ell^{\rm r})$ is a martingale under $Q^{\rm r}$, because $G_k^{\rm r}$ is centered Gaussian conditional on $\mathcal F_k$ under that law. The process $(S_\ell^{\rm q})$ is a martingale under $Q^{\rm q}$ for the same reason with $G_k$. By \eqref{eq:quadrature-output-functions} and the relation between the two residuals, the following identities hold on every node sequence:
\begin{align}
 F^{\rm r}&=S_K^{\rm r},\nonumber\\
 F^{\rm r}&=S_K^{\rm q}
       +\sum_{k<K}\frac{a_km(t_k,Y_k)}{\rho_k^{\rm r}}(\Gamma_k^{\rm q}-\Gamma_k).
 \label{eq:quadrature-martingale-decomposition}
\end{align}
We use the first identity to bound moments under $Q^{\rm r}$ and the second under $Q^{\rm q}$, where its drift sum is bounded by \eqref{eq:quadrature-difference-drift}.

For $j\in\{{\rm r},{\rm q}\}$, the increment $S_{k+1}^j-S_k^j$ is centered Gaussian conditional on $\mathcal F_k$ under $Q^j$, with variance $ha_k^2/(\rho_k^{\rm r})^2$. Thus, for every real $\lambda$,
\begin{align}
 \E_{Q^j}\left[e^{\lambda(S_{k+1}^j-S_k^j)}\mid\mathcal F_k\right]
 &=\exp\left\{\frac{\lambda^2ha_k^2}{2(\rho_k^{\rm r})^2}\right\}
 \le\exp\{\mathsf{K}\lambda^2hR\},\nonumber\\
 \E_{Q^j} e^{\lambda S_K^j}&\le\exp\{\mathsf{K}\lambda^2R\}.
 \label{eq:quadrature-martingale-mgf}
\end{align}
The second bound follows by successive conditioning and $Kh\le1$. Chernoff's inequality, optimized separately for $S_K^j$ and $-S_K^j$, and integration of the resulting tail yield
\begin{align}
 Q^j\{|S_K^j|>u\}&\le2e^{-u^2/(\mathsf{K}R)},\qquad u>0,\nonumber\\
 \E_{Q^j}|S_K^j|^4
 &=\int_0^\infty4u^3Q^j\{|S_K^j|>u\}\dd u\le \mathsf{K}R^2.
 \label{eq:quadrature-martingale-fourth}
\end{align}
Using the respective representations in \eqref{eq:quadrature-martingale-decomposition}, $|m(T,Y_K)|\le1$, $R\ge1$, and the bound on $\eta_{\rm q}$ from the smallness of $\sqrt R\eta_{\rm q}$, we obtain for each $j\in\{{\rm r},{\rm q}\}$
\begin{equation}
 \E_{Q^j}|F^{\rm r}-m(T,Y_K)|^4
 \le \mathsf{K}\{\E_{Q^j}|S_K^j|^4+R^2\eta_{\rm q}^4+1\}
 \le \mathsf{K}R^2.
 \label{eq:quadrature-terminal-fourth}
\end{equation}

Apply the Hellinger comparison to the single function $V(y)=\{F^{\rm r}(y)-m(T,y_K)\}^2$. Equation~\eqref{eq:quadrature-terminal-fourth} implies $\|V\|_{L^2(Q^{\rm r})}+\|V\|_{L^2(Q^{\rm q})}\le \mathsf{K}R$, and \eqref{eq:hellinger-distance} and \eqref{eq:quadrature-path-entropy} imply $d_{\rm Hel}(Q^{\rm r},Q^{\rm q})\le \mathsf{K}\eta_{\rm q}$. Therefore \eqref{eq:hellinger-mean} yields
\begin{align}
 |\E_{Q^{\rm q}}V-\E_{Q^{\rm r}}V|
 &\le d_{\rm Hel}(Q^{\rm r},Q^{\rm q})
       \{\|V\|_{L^2(Q^{\rm r})}+\|V\|_{L^2(Q^{\rm q})}\}\nonumber\\
 &\le \mathsf{K}R\eta_{\rm q}.
 \label{eq:quadrature-terminal-hellinger}
\end{align}
Under $Q^{\rm r}$, $\E_{Q^{\rm r}}V=\|F^{\rm r}-m(T,Y_K)\|_{L^2(Q^{\rm r})}^2$. Combining \eqref{eq:quadrature-terminal-hellinger} with the bound on $F^{\rm r}-M_K^\rho$ above, the triangle inequality under $Q^{\rm q}$ gives
\begin{align}
 \|M_K^\rho-m(T,Y_K)\|_{L^2(Q^{\rm q})}
 &\le\|F^{\rm r}-m(T,Y_K)\|_{L^2(Q^{\rm q})}
       +\|M_K^\rho-F^{\rm r}\|_{L^2(Q^{\rm q})}\nonumber\\
 &\le\left(\|F^{\rm r}-m(T,Y_K)\|_{L^2(Q^{\rm r})}^2
                      +\mathsf{K}R\eta_{\rm q}\right)^{1/2}
       +\mathsf{K}\sqrt R\eta_{\rm q}.
 \label{eq:quadrature-terminal-transfer}
\end{align}
The $L^2(Q^{\rm r})$ norm in the last line equals $\|F^{\rm r}-m(T,Y_K^{\rm r})\|_2$ in \eqref{eq:quadrature-strong}, so this proves that bound.

For the correlation estimate, apply \eqref{eq:hellinger-mean} to the function $y\mapsto a(t_k,y_k)$. Its $L^2$ norms under $Q^{\rm q}$ and $Q^{\rm r}$ are $\rho_k$ and $\rho_k^{\rm r}$, respectively, and both are bounded. Hence
\begin{equation}
 |\E_{Q^{\rm q}}a_k-\E_{Q^{\rm r}}a_k|
 \le d_{\rm Hel}(Q^{\rm r},Q^{\rm q})(\rho_k+\rho_k^{\rm r})
 \le \mathsf{K}\eta_{\rm q}.
 \label{eq:quadrature-mean-transfer}
\end{equation}
Use the sum representations \eqref{eq:scalar-energy} and \eqref{eq:reference-correlation}. Subtracting their $k$th summands, adding and subtracting $\E_{Q^{\rm r}}a_k/\rho_k$, and using $\rho_k\ge1/2$ and $\rho_k^{\rm r}\ge3/4$, we obtain
\begin{align}
 |\mathcal E-\mathcal E^{\rm r}|
 &\le\sum_{k<K}h\left\{
       \frac{|\E_{Q^{\rm q}}a_k-\E_{Q^{\rm r}}a_k|}{\rho_k}
       +|\E_{Q^{\rm r}}a_k|\left|\frac1{\rho_k}-\frac1{\rho_k^{\rm r}}\right|
                          \right\}\nonumber\\
 &\le \mathsf{K}\sum_{k<K}h\{\eta_{\rm q}+|\rho_k-\rho_k^{\rm r}|\}
 \le \mathsf{K}\sqrt R\eta_{\rm q}.
 \label{eq:quadrature-correlation-transfer}
\end{align}
For the second inequality, Cauchy--Schwarz and the definition \eqref{eq:reference-normalizers} imply $|\E_{Q^{\rm r}}a_k|\le(\E_{Q^{\rm r}}a_k^2)^{1/2}=\rho_k^{\rm r}\le5/4$, where the last bound is \eqref{eq:euler-norm-bounds} at $t=t_k$. The last step uses \eqref{eq:quadrature-normalizers}, $\sum_{k<K}h\le1$, and $R\ge1$. This proves \eqref{eq:quadrature-energy}.
\end{proof}

\begin{lemma}[Effect of fixed normalization factors]
\label{lem:fixed-normalizers}
Let $0<\nu\le1/2$ and let $\omega$ be a deterministic vector of normalization factors satisfying
\begin{equation}
 \frac12\le\omega_k\le2,\qquad
 \max_{k<K}|\omega_k^2/\rho_k^2-1|\le\nu.
 \label{eq:fixed-normalizer-condition}
\end{equation}
Define $M^\omega$ and $M^\rho$ by \eqref{eq:scalar-y}--\eqref{eq:scalar-m} using the same Gaussian increments $G_k$. Then
\begin{equation}
 \norm{M_K^\omega-M_K^\rho}_2^2
       =\sum_{k<K}h(\rho_k/\omega_k-1)^2\le \mathsf{K}\nu^2.
 \label{eq:normalizer-error-variance}
\end{equation}
Consequently, replacing $\rho$ by $\omega$ changes the terminal norm $\norm{M_K^\rho-m(T,Y_K)}_2$ in \eqref{eq:quadrature-strong} and the clipping norm $\norm{\dist(M_K^\rho,[-1,1])}_2$ by at most $\mathsf{K}\nu$, using \eqref{eq:normalizer-error-variance}. It also changes the correlation $\mathcal E=\E[W_TM_K^\rho]$ in \eqref{eq:scalar-energy} by at most $\mathsf{K}\nu$.
\end{lemma}

\begin{proof}
Starting from $Y_0=0$, the recursion \eqref{eq:scalar-y} computes $Y_{k+1}=Y_k+\Gamma_k^{\rm q}m(t_k,Y_k)+G_k$ without using either normalization vector. Thus the same Gaussian increments $G_k$ produce the same state sequence $(Y_k)$ for $M^\omega$ and $M^\rho$. Only the denominator in \eqref{eq:scalar-m} changes, from $\rho_k$ to $\omega_k$. Subtracting these updates and using $M_0^\omega=M_0^\rho=0$, we obtain
\begin{equation}
 M_K^\omega-M_K^\rho
 =\sum_{k<K}a(t_k,Y_k)
       \left(\frac1{\omega_k}-\frac1{\rho_k}\right)G_k.
 \label{eq:normalizer-output-difference}
\end{equation}
Let $\mathcal F_k=\sigma(G_0,\ldots,G_{k-1})$. Since $Y_k$ is $\mathcal F_k$-measurable and both normalization factors are deterministic, the coefficient of $G_k$ in \eqref{eq:normalizer-output-difference} is $\mathcal F_k$-measurable. The increment $G_k$ is independent of $\mathcal F_k$, with mean zero and variance $h$. Hence the summands have conditional mean zero and are pairwise orthogonal in $L^2$. By \eqref{eq:exact-normalizers}, the second moment of the $k$th summand is
\begin{equation}
 h\E a(t_k,Y_k)^2
       \left(\frac1{\omega_k}-\frac1{\rho_k}\right)^2
       =h(\rho_k/\omega_k-1)^2.
\end{equation}
Summing these second moments proves the equality in \eqref{eq:normalizer-error-variance}. Condition~\eqref{eq:fixed-normalizer-condition} implies $|\rho_k/\omega_k-1|\le \mathsf{K}\nu$, and $Kh=T\le1$ proves its upper bound. The triangle inequality controls the terminal approximation. Distance to $[-1,1]$ is $1$-Lipschitz, and Cauchy--Schwarz with $\norm{W_T}_2\le1$ controls the correlation change.
\end{proof}

In this scalar comparison, the Gaussian increments $G_k$ are held fixed when $\rho$ is replaced by $\omega$. In the matrix algorithm, the current magnetization vector is used in \eqref{eq:query-proposals} to choose the next vector $q_s$ to be multiplied by $A$. Changing the normalization factors can change the magnetization, and hence $q_s$, the corrected query output $z_s$ in \eqref{eq:gaussian-coupling}, and the increments reconstructed by \eqref{eq:group-reconstruction}. In Section~\ref{sec:matrix-analysis} we therefore prove the energy and clipping bounds directly for each fixed $\omega$ satisfying \eqref{eq:fixed-normalizer-condition}. The implemented algorithm uses only $\omega_k=1$, and Proposition~\ref{prop:unit-normalizers} verifies this choice for sufficiently small internal accuracy.

\begin{proposition}
\label{prop:unit-normalizers}
After fixing the grid, quadrature, and tolerance constants, there is a universal $\varepsilon_0>0$ such that $\omega_k=1$ satisfies \eqref{eq:fixed-normalizer-condition}, with $\nu$ from \eqref{eq:normalizer-tolerance}, for every $0<\varepsilon\le\varepsilon_0$. This accuracy threshold is independent of $N$ and $\alpha$.
\end{proposition}

\begin{proof}
For $\omega_k=1$, the relative-error requirement in \eqref{eq:fixed-normalizer-condition} is $\max_{k<K}|1/\rho_k^2-1|\le\nu$, where $\nu=c_\nu\varepsilon/\log(4d)$ by \eqref{eq:normalizer-tolerance}. At the grid nodes, \eqref{eq:euler-weak-error} and \eqref{eq:quadrature-normalizers}, together with the bounds on the positive normalization factors in Lemma~\ref{lem:euler-quadrature}, imply
\begin{align}
 \delta_\varepsilon:=\max_{k<K}|\rho_k^2-1|
 &\le \mathsf{K}\{hL_\tau^4/\tau+\sqrt R\eta_{\rm q}\}\nonumber\\
 &\le \mathsf{K}\{\varepsilon^{4/3}L_\varepsilon^{7/3}
                 +\varepsilon^{7/3}L_\varepsilon^{-2/3}\}.
 \label{eq:unit-normalizer-deviation}
\end{align}
Indeed \eqref{eq:partition-size} gives $h=O(\varepsilon^2L_\varepsilon^{-2})$, $\tau\asymp\varepsilon^{2/3}L_\varepsilon^{-1/3}$, and $L_\tau\asymp L_\varepsilon$. Also $R\asymp\varepsilon^{-2/3}L_\varepsilon^{4/3}$ and \eqref{eq:quadrature-tolerance} gives $\sqrt R\eta_{\rm q}=c_{\rm q}\varepsilon^2/\sqrt R$.

The lower bound $\rho_k^2\ge1/4$ from Lemma~\ref{lem:euler-quadrature} implies $\max_{k<K}|1/\rho_k^2-1|\le4\delta_\varepsilon$. It therefore suffices to prove $4\delta_\varepsilon\le\nu$, or equivalently $\delta_\varepsilon\log(4d)/\varepsilon\le c_\nu/4$ by the definition of $\nu$ in \eqref{eq:normalizer-tolerance}. The definition of $d$ in \eqref{eq:grid-parameters} and the bound \eqref{eq:partition-size} imply $\log(4d)\le \mathsf{K} L_\varepsilon$. Multiplying \eqref{eq:unit-normalizer-deviation} by $\log(4d)/\varepsilon$, we obtain
\begin{equation}
 \frac{\delta_\varepsilon\log(4d)}{\varepsilon}
 \le \mathsf{K}\{\varepsilon^{1/3}L_\varepsilon^{10/3}
                 +\varepsilon^{4/3}L_\varepsilon^{1/3}\}
 \longrightarrow0.
 \label{eq:unit-normalizer-tolerance}
\end{equation}
For all sufficiently small $\varepsilon$, the left-hand side of \eqref{eq:unit-normalizer-tolerance} is at most $c_\nu/4$, so $4\delta_\varepsilon\le\nu$ as required. The other bounds in \eqref{eq:fixed-normalizer-condition} are immediate for $\omega_k=1$.
\end{proof}

After fixing $c_B>0$ and choosing the remaining constants as described above, there is an accuracy threshold $\varepsilon_0>0$ such that, for every $0<\varepsilon\le\varepsilon_0$, the correlation $\mathcal E$ of \eqref{eq:scalar-energy} and every deterministic normalization vector $\omega$ satisfying \eqref{eq:fixed-normalizer-condition}, with $\nu$ from \eqref{eq:normalizer-tolerance}, obey
\begin{equation}
 \norm{M_K^\omega-m(T,Y_K)}_2
       \le\varepsilon/1000,\qquad
                         \mathcal E\ge\Par-\varepsilon/16.
 \label{eq:scalar-final}
\end{equation}
By Lemma~\ref{lem:partition}, we first choose the constants so that the errors in \eqref{eq:scalar-clipping} and \eqref{eq:scalar-energy-bound} are small fixed multiples of $\varepsilon$. By \eqref{eq:quadrature-tolerance}, $\mathsf{K}R\eta_{\rm q}=\mathsf{K}c_{\rm q}\varepsilon^2$, so a small $c_{\rm q}$ keeps the additional error in Lemma~\ref{lem:euler-quadrature} within the required bounds. By Lemma~\ref{lem:fixed-normalizers}, replacing $\rho$ by $\omega$ changes the terminal norm by at most $\mathsf{K}\nu$, and Proposition~\ref{prop:unit-normalizers} verifies its hypotheses for $\omega_k=1$ on this accuracy range. Since $m(T,Y_K)\in[-1,1]$, the first bound in \eqref{eq:scalar-final} also bounds the root-mean-square clipping loss. The distance to $m(T,Y_K)$ is the stronger estimate used in Lemma~\ref{lem:empirical-clipping} to control the second moment of a martingale. The correlation $\mathcal E$ is defined with $\rho_k$ from \eqref{eq:exact-normalizers} and is independent of $\omega$; in the matrix proof we retain the $O(\nu)$ second-moment mismatch caused by using $\omega_k=1$. We then estimate the output energy directly, rather than compare its coordinate trajectories with the Parisi diffusion.
\section{Finite-dimensional analysis of GEG-IAMP}
\label{sec:matrix-analysis}
\label{app:matrix}

The main energy estimate, Lemma~\ref{lem:finite-energy}, bounds the deviation of $H_N(f)/N$ from $\mathcal E-c_1$ by \eqref{eq:uncut-energy-bound}, under the normalization and dimension hypotheses stated there. Here $f$ is the magnetization vector before clipping, $\mathcal E$ is the scalar correlation \eqref{eq:scalar-energy}, and $c_1\le\Delta$ is the first group's correlation in \eqref{eq:block-increments}. We first establish the exact coordinate-path law in Section~\ref{sec:matrix-state-evolution}. In Section~\ref{sec:matrix-energy}, we derive the energy identity of Proposition~\ref{prop:energy-identity}, compute its leading term in Lemma~\ref{lem:row-structure}, and bound its errors in Lemmas~\ref{lem:directional-qr}--\ref{lem:auxiliary-events}. We then bound the losses from clipping in Section~\ref{sec:matrix-clipping} and sequential rounding in Section~\ref{sec:matrix-rounding}, before comparing the optimum with $N\Par$ in Section~\ref{sec:matrix-optimum}. In the energy expansion, the triangular factor of query orthogonalization occurs only through its action on the deterministic output weights. We bound that action using the conditional mean zero of projections onto earlier queries, rather than an operator-norm estimate for the whole factor.

The implemented algorithm uses $\omega_j=1$ for every $j$. We prove the bounds below for any fixed deterministic vector $\omega$ satisfying \eqref{eq:fixed-normalizer-condition}, with magnetization increment $a(t_j,Y_j)G_j/\omega_j$ at time step $j$. By Proposition~\ref{prop:unit-normalizers}, this includes $\omega_j=1$ for $0<\varepsilon\le\varepsilon_0$, after the algorithmic constants are fixed. All matrices below include the final query $q_d$, $d=B+1$, and its corrected query output $z_d$, appended after the last group as described after \eqref{eq:recentered-output}. This construction requires $d\le N$, so that the coordinate-vector rule after \eqref{eq:query-proposals} remains available for a zero residual. By Proposition~\ref{prop:virtual-final-query}, its output agrees with that of the executed algorithm, which performs only $B$ products with matrix queries and the product $Av$ that initializes rounding.

\subsection{Exact state evolution at finite dimension}
\label{sec:matrix-state-evolution}

The vector magnetization $\widehat M_j\in\R^N$ is initialized at $\widehat M_0=\sqrt\Delta\one$ before \eqref{eq:query-proposals}; this nonzero vector supplies the first query. With the fixed normalization factors considered in this section, its increment in \eqref{eq:vector-scalar-update} is $a(t_j,Y_{j,i})G_{j,i}/\omega_j$, with $\omega_j=1$ in the implemented algorithm. Consequently, for $0\le j\le K$,
\begin{equation}
 \widehat M_{j,i}-\sqrt\Delta
 =\sum_{\ell<j}\frac{a(t_\ell,Y_{\ell,i})}{\omega_\ell}G_{\ell,i},
 \qquad 1\le i\le N.
 \label{eq:centered-vector-magnetization}
\end{equation}
Subtracting $\sqrt\Delta$ removes the deterministic initial value, matching $M_0^\omega=0$ in \eqref{eq:scalar-m}. We call the joint coordinate-path law stated in Proposition~\ref{prop:gaussian-law} \emph{exact state evolution at finite dimension}.

The nonzero initialization aligns the first proposal with the first normalized query: $F_1=\sqrt\Delta\one$ and $q_1=\one$. A zero vector cannot itself be a query because queries have norm $\sqrt N$. A zero proposal $F_1=0$ would invoke the coordinate-vector rule after \eqref{eq:query-proposals}, producing $q_1=\sqrt N\mathbf e_1$ while the first normalized proposal column would be zero. The second-moment calculation and the identity $U_{:,1}=q_1=\one$ used below would then change. With our initialization, subtracting $F_1$ sets the first output weight to zero and omits the first group's correlation $c_1\le\Delta$ from the leading energy term in Lemma~\ref{lem:row-structure}. The group is still executed and affects all later states; the omitted correlation is a cost of this construction, not a query lower bound.

We directly prove the zero-mean special case of \cite[Theorem~3]{Reeves} for \eqref{eq:gaussian-coupling}. Recall the matrix of corrected query outputs $Z=[z_1|\cdots|z_d]\in\R^{N\times d}$, introduced after Definition~\ref{def:reeves-coupling}. We first condition on the side variables in \eqref{eq:group-reconstruction} for all groups and coordinates and show that the conditional law of $Z$ is the same product standard Gaussian law for every fixed choice. We then marginalize over those variables to obtain the coordinate-path law. In the later scalar martingale calculations, we instead condition only on earlier increments of one coordinate, as specified in \eqref{eq:fine-and-block-filtrations}.

The conditional law of $Z$ follows from a basic property of GOE matrices: in an orthonormal basis fixed by the conditioning, the block on the remaining basis vectors keeps its original entry variances and is independent of the first columns.

\begin{lemma}[GOE blocks in a conditionally fixed orthonormal basis]
\label{lem:conditional-goe-blocks}
Conditional on a sigma-algebra $\mathcal F$, suppose a symmetric $m\times m$ matrix $G$ has independent centered Gaussian entries on and above its diagonal, with variance $2/N$ on the diagonal and $1/N$ above it. For any $\mathcal F$-measurable orthogonal matrix $\mathsf U=[\mathsf U_1|\mathsf U_2]$, the first columns $\mathsf U^\top G\mathsf U_1$ are independent of $\mathsf U_2^\top G\mathsf U_2$ under this conditioning. The remaining block has the same entry variances, not variances rescaled by its smaller dimension.
\end{lemma}

\begin{proof}
Conditional on $\mathcal F$, the matrix $\mathsf U$ is fixed. Orthogonal invariance implies that $\mathsf U^\top G\mathsf U$ again has independent centered Gaussian entries on and above its diagonal, with variance $2/N$ on the diagonal and $1/N$ above it. The first columns $\mathsf U^\top G\mathsf U_1$ are determined by the entries on and above the diagonal in the first rows of $\mathsf U^\top G\mathsf U$, while $\mathsf U_2^\top G\mathsf U_2$ is determined by those in the remaining rows. These sets of independent entries are disjoint, proving the conditional independence, and the remaining block keeps the same entry variances.
\end{proof}

We apply this GOE block property to the successive corrected query outputs in the next proposition.

\begin{proposition}[Exact state evolution]
\label{prop:gaussian-law}
For fixed deterministic normalization factors $\omega_k>0$ and $d\le N$, conditional on the independent side variables of all groups and coordinates, the entries of $Z$ generated by \eqref{eq:gaussian-coupling}, including the final corrected query output appended in the proof, are independent standard Gaussians. This conditional law is the same for every value of the side variables. Consequently, $Z$ is independent of these variables, and the pairs formed by row $i$ of $Z$ and the side variables of coordinate $i$ are independent copies over $i$. After marginalizing over the side variables, the reconstructed $G_{j,i}$ are independent $\mathcal N(0,h)$ variables, and the full paths $(Y_{j,i},\widehat M_{j,i}-\sqrt\Delta)_{j=0}^K$, $1\le i\le N$, have the joint law of independent copies of $(Y_j,M_j^\omega)_{j=0}^K$ in \eqref{eq:scalar-y}--\eqref{eq:scalar-m}, under the joint randomness of $A$ and the algorithm. In particular, the coordinates of $f$ in \eqref{eq:recentered-output} are independent copies of $M_K^\omega$ in \eqref{eq:scalar-m}. No independence of $Z$ from $A$ or the full $J$ is asserted.
\end{proposition}

\begin{proof}
We first fix the deterministic coefficients and the side variables of all groups and coordinates. The query at the start of group $s$ is then a measurable function of $z_1,\ldots,z_{s-1}$, and the final query $q_d$ is a measurable function of $z_1,\ldots,z_B$. Thus each query is determined before its corresponding observation is generated. We prove the Gaussian law directly by tracking the remaining GOE block and the entries of $J$ not yet used.

Let $\mathcal F_s$ be generated by the fixed side variables, the products $Aq_1,\ldots,Aq_s$, and the entries $J_{rt}$ with $r\le t\le s$. Write $u_s=q_s/\sqrt N$ and $P_s=\sum_{r\le s}u_ru_r^\top$. Conditional on $\mathcal F_{s-1}$, the query $q_s$ is fixed, so $u_s$ is a fixed unit vector orthogonal to $u_1,\ldots,u_{s-1}$.

We complete $u_s$ to an orthonormal basis of $\operatorname{span}\{u_1,\ldots,u_{s-1}\}^\perp$, which has dimension $N-s+1$. The additional basis vectors must lie in $\operatorname{span}\{u_1,\ldots,u_s\}^\perp$, a space of dimension $N-s$. We process the standard coordinate vectors $\mathbf e_1,\ldots,\mathbf e_N$ in increasing order. For each $\mathbf e_i$, we subtract its components along $u_1,\ldots,u_s$ and along the additional basis vectors already selected. The residual is its orthogonal projection onto the subspace orthogonal to all these vectors. We skip a zero residual and normalize each nonzero residual. To see that this supplies enough vectors, every $x\in\operatorname{span}\{u_1,\ldots,u_s\}^\perp$ satisfies
\begin{equation}
 x=(I-P_s)x=\sum_{i=1}^N x_i(I-P_s)\mathbf e_i.
 \label{eq:coordinate-projection-span}
\end{equation}
Thus the projected coordinate vectors $(I-P_s)\mathbf e_i$ span this $(N-s)$-dimensional space, so at least $N-s$ of them are nonzero. Gram--Schmidt retains exactly $N-s$ linearly independent nonzero residuals. Together with $u_s$, their normalized versions form the required $N-s+1$ basis vectors. Since $u_1,\ldots,u_s$ are $\mathcal F_{s-1}$-measurable and the coordinate order is fixed, the resulting basis is also $\mathcal F_{s-1}$-measurable.

Let $\mathsf V_s\in\R^{N\times(N-s)}$ collect the additional orthonormal basis vectors, and set $\mathsf U_s=[u_s|\mathsf V_s]\in\R^{N\times(N-s+1)}$. Then $\mathsf U_s\mathsf U_s^\top=I-P_{s-1}$ and $\mathsf V_s\mathsf V_s^\top=I-P_s$. The matrices used below are
\begin{equation}
 \mathsf U_s^\top A\mathsf U_s
 =\begin{pmatrix}
            u_s^\top Au_s & u_s^\top A\mathsf V_s\\
            \mathsf V_s^\top Au_s & \mathsf V_s^\top A\mathsf V_s
       \end{pmatrix},
 \label{eq:conditional-goe-basis-matrix}
\end{equation}
where the lower-right block $\mathsf V_s^\top A\mathsf V_s$ has size $(N-s)\times(N-s)$.

We argue by induction on $s$ that, conditional on $\mathcal F_{s-1}$, the matrix $\mathsf U_s^\top A\mathsf U_s$ has independent centered Gaussian entries on and above its diagonal, with variances $1/N$ off the diagonal and $2/N$ on it. For $s=1$, this follows from the GOE law of $A$ and orthogonal invariance, since $\mathsf U_1$ is fixed under the conditioning. By Lemma~\ref{lem:conditional-goe-blocks}, the first column of \eqref{eq:conditional-goe-basis-matrix}, consisting of $u_s^\top Au_s$ and $\mathsf V_s^\top Au_s$, is conditionally independent of $\mathsf V_s^\top A\mathsf V_s$. This lower-right block retains the same entry variances.

The components of $Aq_s$ along earlier query directions are already determined by the past, because $q_r^\top Aq_s=(Aq_r)^\top q_s$ by symmetry. Thus, given $\mathcal F_{s-1}$, observing $Aq_s$ reveals only the first column of \eqref{eq:conditional-goe-basis-matrix}. The unused entries $J_{rs}$, $r\le s$, are conditionally independent of this column and the lower-right block. Consequently, after adding $Aq_s$ and these entries to obtain $\mathcal F_s$, the conditional law of $\mathsf V_s^\top A\mathsf V_s$ is unchanged. For $s<N$, the columns of $\mathsf V_s$ and $\mathsf U_{s+1}$ are two $\mathcal F_s$-measurable orthonormal bases of the same space. Orthogonal invariance therefore proves the induction claim for $\mathsf U_{s+1}^\top A\mathsf U_{s+1}$. For $s=N$, only the scalar $u_s^\top Au_s$ remains and the lower-right block is empty.

Adding and subtracting $P_sAq_s=\sum_{r\le s}(q_r^\top Aq_s/N)q_r$ in \eqref{eq:gaussian-coupling}, we can rewrite the observation as
\begin{equation}
 z_s=(I-P_s)Aq_s+\frac12(q_s^\top Aq_s/N+J_{ss})q_s
                         +\sum_{r<s}J_{rs}q_r.
 \label{eq:coupling-decomposition}
\end{equation}
We now read off the coefficients of $z_s$ in the orthonormal basis $[u_1|\cdots|u_s|\mathsf V_s]$. For a column $v$ of $\mathsf V_s$, the coefficient is $\sqrt N\,v^\top Au_s$, with variance $N(1/N)=1$ by \eqref{eq:conditional-goe-basis-matrix}. Along $u_s$, it is $\frac{\sqrt N}2(u_s^\top Au_s+J_{ss})$, with variance $N(2/N+2/N)/4=1$. Along $u_r$, $r<s$, it is $\sqrt NJ_{rs}$, with variance $N(1/N)=1$. These coefficients are mutually independent conditional on $\mathcal F_{s-1}$, and the three terms of \eqref{eq:coupling-decomposition} have conditional covariance matrices $I-P_s$, $u_su_s^\top$, and $P_{s-1}$, whose sum is $I$. Thus $\mathcal L(z_s\mid\mathcal F_{s-1})=\mathcal N(0,I_N)$, independent of the realized past. Since $z_1,\ldots,z_{s-1}$ are $\mathcal F_{s-1}$-measurable, the entries of $Z$ are independent standard Gaussians given the side variables. The conditional law of $z_s$ does not depend on the rule that produced $q_s$, so it also holds for the query selected when $r_s=0$ by the rule after \eqref{eq:query-proposals}.

The conditional law of $Z$ just obtained is the same for every value of the side variables, so $Z$ is independent of the side variables. Their joint law is therefore the product standard Gaussian law. For each group and coordinate, \eqref{eq:group-reconstruction} applies $\sqrt h$ times the fixed Householder reflection $H_{\rm rec}$ to one entry of $Z$ and $m_{\rm f}-1$ side variables, and different groups and coordinates use disjoint sets of these variables. Hence, after marginalizing over the side variables, the per-step Gaussian increments $G_{j,i}$ are independent $\mathcal N(0,h)$ variables. In coordinate $i$, the output $f_i=\sum_{j<K}a(t_j,Y_{j,i})G_{j,i}/\omega_j$ follows \eqref{eq:scalar-y}--\eqref{eq:scalar-m} with these increments, so it has the law of $M_K^\omega$. The path of coordinate $i$ is a function of row $i$ of $Z$, the side variables of coordinate $i$, and the deterministic coefficients. These inputs are independent and identically distributed over $i$, which proves the remaining assertions.
\end{proof}

For the scalar martingale calculations below, we consider one coordinate after marginalizing over the side variables, and define
\begin{equation}
 \mathcal A_j=\sigma(G_0,\ldots,G_{j-1}),\qquad
 \mathcal B_s=\mathcal A_{sm_{\rm f}},\qquad 0\le s\le B,
 \label{eq:fine-and-block-filtrations}
\end{equation}
so $\mathcal A_j$ is generated by the per-step Gaussian increments before step $j$, and $\mathcal B_s$ is generated by the increments of groups $1,\ldots,s$. The algorithm computes $z_s$, and hence the sum $\sum_{j\in I_s}G_j=\sqrt\Delta z_s$ in \eqref{eq:group-sum}, before it runs the time steps of group $s$. For $j$ inside group $s$, however, $\mathcal A_j$ contains only $G_0,\ldots,G_{j-1}$, including the earlier increments of the same group, but not $\sum_{k\in I_s}G_k$. Conditioning on $\mathcal A_j$ is nevertheless justified. After marginalizing over the side variables, the increments $G_j$ are independent $\mathcal N(0,h)$ variables by Proposition~\ref{prop:gaussian-law}, and the recursion \eqref{eq:scalar-y}--\eqref{eq:scalar-m} computes $Y_j$ and $M_j^\omega$ from $G_0,\ldots,G_{j-1}$. Thus, conditional on $\mathcal A_j$, the increment $G_j$ has law $\mathcal N(0,h)$. With the side variables held fixed, the increments of group $s$ would instead be affine functions of the single corrected query output $z_s$. We therefore use conditional laws of per-step Gaussian increments only after marginalizing over the side variables.

\subsection{The exact energy identity and its error terms}
\label{sec:matrix-energy}

We first express the energy of $f$, the magnetization vector before clipping in \eqref{eq:recentered-output}, in terms of the magnetization increments and their empirical second moments. For one coordinate, with the state $Y_j$ of the implemented recursion \eqref{eq:scalar-y}, define
\begin{equation}
 \begin{gathered}
 \Delta M_j^\rho=\frac{a(t_j,Y_j)}{\rho_j}G_j,\qquad
 \delta_s M^\rho=\sum_{j\in I_s}\Delta M_j^\rho,\\
 \delta_s M^\omega=\sum_{j\in I_s}\frac{a(t_j,Y_j)}{\omega_j}G_j,
 \qquad c_s=\E\left[\left(\sum_{j\in I_s}G_j\right)\delta_s M^\rho\right].
 \end{gathered}
 \label{eq:block-increments}
\end{equation}
Here $\Delta M_j^\rho=M_{j+1}^\rho-M_j^\rho$ is the random increment of \eqref{eq:scalar-m} with the factors $\rho_j$ from \eqref{eq:exact-normalizers}. The magnetization increments $\delta_s M^\rho$ and $\delta_s M^\omega$ over group $s$ equal $M_{sm_{\rm f}}^\rho-M_{(s-1)m_{\rm f}}^\rho$ and $M_{sm_{\rm f}}^\omega-M_{(s-1)m_{\rm f}}^\omega$, respectively. For the algorithm, $\omega_j=1$ and $\delta_s M^\omega=\sum_{j\in I_s}a(t_j,Y_j)G_j$ is the change of the magnetization over group $s$. The quantity $c_s$ is the correlation between the sum $\sum_{j\in I_s}G_j$ of the Gaussian increments of group $s$ and the magnetization change $\delta_s M^\rho$ normalized by $\rho_j$; by \eqref{eq:block-correlation}, these correlations add up to $\mathcal E$.

To apply the energy identity of Proposition~\ref{prop:energy-identity}, we write $f=Uw$ as a fixed combination of the magnetization changes over the groups. The columns of $U$ after the first will be these changes divided by $\sqrt\Delta$, and $w$ will contain their deterministic output weights. Let $F=[F_1|\cdots|F_d]$ collect the proposals $F_s$ of \eqref{eq:query-proposals} and $F_d=\widehat M_K$, and define
\begin{equation}
 \Omega_{rs}=\sqrt\Delta\one_{\{r\le s\}},\qquad
 U=F\Omega^{-1},\qquad
 w=\sqrt\Delta(0,1,\ldots,1)^\top.
 \label{eq:whitening-matrix}
\end{equation}
Since $F=U\Omega$, each column of $F$ is $F_s=\sqrt\Delta\sum_{r\le s}U_{:,r}$. Hence the columns of $U$ are successive differences of the columns of $F$, divided by $\sqrt\Delta$: $U_{:,1}=F_1/\sqrt\Delta=\one$, and $U_{:,s+1}=(\widehat M_{sm_{\rm f}}-\widehat M_{(s-1)m_{\rm f}})/\sqrt\Delta$ is the change of the magnetization over group $s$, divided by $\sqrt\Delta$. Therefore $Uw=\sqrt\Delta\sum_{s=2}^dU_{:,s}=\widehat M_K-\sqrt\Delta\one=f$, and $\norm w^2=B\Delta=T$. In the expansion $f=Uw$, the entry $w_1=0$ is the coefficient of the first column $U_{:,1}=\one$; it removes the initial magnetization $\widehat M_0=\sqrt\Delta\one$ from the output. For one coordinate, the corresponding row of $U$ is $u=(1,\delta_1 M^\omega/\sqrt\Delta,\ldots,\delta_B M^\omega/\sqrt\Delta)$, and $u^0=(1,\delta_1 M^\rho/\sqrt\Delta,\ldots,\delta_B M^\rho/\sqrt\Delta)$ is the same row normalized by $\rho_j$. Let $\zeta$ denote the corresponding row of $Z$, and set
\begin{equation}
 S=\E[u^\top u],\quad D=\E[\zeta^\top u],\quad
 D^0=\E[\zeta^\top u^0],\qquad
 \widehat S=U^\top U/N,\quad \widehat D=Z^\top U/N.
 \label{eq:empirical-matrices}
\end{equation}
The deterministic matrices $S$, $D$, and $D^0$ are second moments, not centered covariances, because the first coordinate of $u$ is one. The random empirical matrices $\widehat S$ and $\widehat D$ are the averages of $u^\top u$ and $\zeta^\top u$ over the $N$ rows.

The queries are obtained by orthogonalizing the columns of $F$ in order, so $F=QR_0$, where $Q=[q_1|\cdots|q_d]$, $Q^\top Q=NI$, and $R_0$ is upper triangular with nonnegative diagonal. Define $L_0=R_0\Omega^{-1}$, which is also upper triangular. Then $U=F\Omega^{-1}=QL_0$ and $L_0^\top L_0=U^\top U/N=\widehat S$.

For $1\le s\le d$, recall the residual $r_s=(I-P_{s-1})F_s$ from \eqref{eq:query-proposals}, with the same rule used for the final query. The diagonal entry $(R_0)_{ss}=q_s^\top F_s/N$ equals $\norm{r_s}/\sqrt N$ when $r_s\ne0$, because $q_s=\sqrt N r_s/\norm{r_s}$. If $r_s=0$, the rule after \eqref{eq:query-proposals} forms $q_s$ by projecting and normalizing the first coordinate vector outside the preceding query span. Since $F_s$ lies in that span and $q_s$ is orthogonal to it, $(R_0)_{ss}=0$. With $\Omega_{ss}=\sqrt\Delta$ from \eqref{eq:whitening-matrix}, in both cases
\begin{equation}
 (R_0)_{ss}=\frac{\norm{r_s}}{\sqrt N},\qquad
 (L_0)_{ss}=\frac{(R_0)_{ss}}{\sqrt\Delta}
           =\frac{\norm{r_s}}{\sqrt{N\Delta}}.
 \label{eq:query-residual-diagonal}
\end{equation}
If $\widehat S$ is positive definite, the identity $L_0^\top L_0=\widehat S$ implies that $L_0$ is invertible. Every diagonal entry of this triangular matrix is therefore nonzero. By \eqref{eq:query-residual-diagonal}, all residuals $r_s$ are nonzero. The diagonal entries of $L_0$ are then positive. The directional estimates below, and their cutoff counterpart for the AMS IAMP comparator in Lemma~\ref{lem:onsager-comparator-moments}, do not require positive definiteness.

\begin{proposition}[Energy identities, adapted from {\cite[Eqs.~(10) and (21)]{Reeves}}]
\label{prop:energy-identity}
For any predictable queries from Definition~\ref{def:reeves-coupling} and any invertible upper triangular $\Omega$, with $F=QR_0$, $U=F\Omega^{-1}$, and $L_0=R_0\Omega^{-1}$, the following identities hold on every outcome:
\begin{equation}
 AQ=Z+QZ^\top Q/N-QJ,\qquad
 U=QL_0,\qquad L_0^\top L_0=\widehat S.
 \label{eq:AQ}
\end{equation}
For any vector $w$ with $f=Uw$,
\begin{equation}
 \frac{H_N(f)}N=w^\top L_0^\top\widehat Dw
                     -\frac12w^\top L_0^\top JL_0w.
 \label{eq:exact-energy}
\end{equation}
\end{proposition}

\begin{proof}
We take the inner product of \eqref{eq:gaussian-coupling} with $q_r$. Since the queries are orthogonal and $q_r^\top q_r=N$, we obtain
\begin{equation}
 \frac{q_r^\top z_s}{N}=
 \begin{cases}
 J_{rs},&r<s,\\
 (q_s^\top Aq_s/N+J_{ss})/2,&r=s,\\
 q_r^\top Aq_s/N,&r>s.
 \end{cases}
 \label{eq:query-inner-products}
\end{equation}
By \eqref{eq:gaussian-coupling}, $Aq_s-z_s$ lies in the span of $q_1,\ldots,q_s$, so the difference between the two sides of the first identity in \eqref{eq:AQ} lies in the span of $Q$. In column $s$, its coefficient along $q_r$ is $q_r^\top Aq_s/N-q_r^\top z_s/N-q_s^\top z_r/N+J_{rs}$. This coefficient vanishes in each of the three cases of \eqref{eq:query-inner-products}, because $A$ and $J$ are symmetric. The other two identities follow from $F=QR_0$ and $U=F\Omega^{-1}$. Multiplying the first identity on the right by $L_0$, and using $QL_0=U$ and $Z^\top U/N=\widehat D$, we obtain
\begin{equation}
 AU=ZL_0+Q\widehat D-QJL_0.
\end{equation}
Since $f=Uw=QL_0w$, we have $H_N(f)/N=f^\top AUw/(2N)$. We multiply the last identity on the right by $w$ and take the inner product with $f/(2N)=QL_0w/(2N)$. The first two terms contribute the same scalar $w^\top L_0^\top\widehat Dw/2$, because $(QL_0w)^\top ZL_0w=(ZL_0w)^\top Uw$ and $Q^\top Q=NI$. The last term contributes $-\tfrac12w^\top L_0^\top JL_0w$. This proves \eqref{eq:exact-energy}. These algebraic identities hold on every outcome; they require neither positive residuals nor any independence between the matrices.
\end{proof}

The energy identity \eqref{eq:exact-energy} uses $f=QL_0w$, which in general requires all $d$ query directions, including the final query $q_d$. The next proposition shows that this query need not be executed.

\begin{proposition}[A final query used only in the proof]
\label{prop:virtual-final-query}
Suppose $d=B+1\le N$ and fix deterministic normalization factors $\omega_k>0$. Run the executed construction and the construction with the final query $q_d$ appended using the same $A$, the same leading $B\times B$ block of $J$, and the same side variables, adding independent entries $J_{rd}$ only for the final query. They produce the same vector $f$ before clipping and the same rounded spin configuration. All energy estimates for the construction with $q_d$ appended therefore apply to the executed algorithm, which computes $B+1$ dense products, including rounding initialization.
\end{proposition}

\begin{proof}
We construct both algorithms on one probability space, using the same input $A$, the same leading $B\times B$ block of $J$, and the same side variables for the $B$ groups. Their last executed query $q_B$ is formed from $F_B=\widehat M_{(B-1)m_{\rm f}}$. Its observation $z_B$ supplies the Gaussian increments of group $B$ through \eqref{eq:group-reconstruction}. After the updates for this group, both constructions have the same final magnetization $F_d=\widehat M_{Bm_{\rm f}}=\widehat M_K$ and the same output $f=F_d-F_1$ in \eqref{eq:recentered-output}. Thus $F_d$ and $f$ are known before $q_d$ or $z_d$ is formed.

We extend the probability space by independent Gaussian entries $J_{rd}$, $r\le d$, with the law \eqref{eq:auxiliary-j}. We form $q_d$ from the final magnetization $F_d$, projecting off all $B$ preceding queries: its residual is $r_d=(I-P_B)F_d$. If $r_d\ne0$, set $q_d=\sqrt N r_d/\norm{r_d}$; otherwise use the coordinate-vector rule after \eqref{eq:query-proposals}, which is available because $d\le N$. In the proof only, we use the product $Aq_d$ and the new entries $J_{rd}$ in \eqref{eq:gaussian-coupling} to form $z_d$. No scalar update follows $z_d$.

The query $q_d$ is determined before its observation, so Proposition~\ref{prop:gaussian-law} applies to the extended construction. Adding $q_d$ ensures that $F_d$, and hence $f=F_d-F_1$, lies in the span of $q_1,\ldots,q_d$. We can therefore write $f=QL_0w$ and apply the energy identity \eqref{eq:exact-energy}. The additional observation $z_d$ enters $Z$ and hence $\widehat D=Z^\top U/N$ in that identity; it is not used to construct the output.

Rounding is the same deterministic function of $A$ and $\clip(f,[-1,1]^N)$ in both constructions. The added entries $J_{rd}$ change neither $A$ nor $f$ nor the rounded spins, so marginalizing over them leaves the probability of every event determined by these outputs unchanged. Execution requires only the first $B$ observation products and the product $Av$ used to initialize rounding, for a total of $B+1$ dense products. This rounding product is not the additional product $Aq_d$ used only in the proof.
\end{proof}

To isolate the leading term $w^\top D^0w$ in the energy identity \eqref{eq:exact-energy}, define
\begin{equation}
 E_D=\widehat D-D^0,\qquad P=L_0-I,\qquad
 x=D^0w.
 \label{eq:energy-perturbations}
\end{equation}
Substituting $L_0=I+P$ and $\widehat D=D^0+E_D$ in \eqref{eq:exact-energy}, and using the symmetry of $J$, we obtain the exact expansion
\begin{equation}
 \frac{H_N(f)}N=w^\top D^0w+w^\top E_Dw
       +x^\top Pw-\tfrac12w^\top Jw+\mathcal R_N,
 \label{eq:energy-expansion}
\end{equation}
where
\begin{equation}
 \mathcal R_N=(Pw)^\top E_Dw-(Pw)^\top Jw
                                             -\tfrac12(Pw)^\top J(Pw).
 \label{eq:energy-remainder}
\end{equation}
The leading term $w^\top D^0w$ depends only on the scalar law. The remaining terms in \eqref{eq:energy-expansion}--\eqref{eq:energy-remainder} require bounds on $Pw$, $x^\top Pw$, $E_Dw$, $w^\top E_Dw$, and $J$. We first compute the leading term and bound the deterministic differences in second moments caused by replacing $\rho_j$ with $\omega_j$. We then bound the empirical fluctuations and the terms involving $J$.

\paragraph{The leading term and normalization errors.}

The next lemma identifies $w^\top D^0w$ in \eqref{eq:energy-expansion} with the scalar correlation $\mathcal E$ except for the first-group contribution $c_1$. Its bound on $D-D^0$ controls the deterministic part of $E_Dw=(\widehat D-D)w+(D-D^0)w$: since $\norm w\le1$, $\norm{(D-D^0)w}\le \mathsf{K}\nu$. The resulting contribution to $w^\top E_Dw$ in \eqref{eq:energy-expansion} is at most $\mathsf{K}\nu$, and its contribution to $(Pw)^\top E_Dw$ in \eqref{eq:energy-remainder} is at most $\mathsf{K}\nu\norm{Pw}$.

The bound on $S-I$ controls the deterministic variance error in the later estimates of $Pw$ and $x^\top Pw$. Indeed, $L_0^\top L_0-I=(\widehat S-S)+(S-I)$ separates empirical fluctuations of the second moments from their deterministic difference from the identity. Lemma~\ref{lem:directional-qr} combines this variance bound with bounds on the projections onto earlier query directions, without estimating $\norm P$.

\begin{lemma}[Leading term and normalization errors]
\label{lem:row-structure}
With the factors $\rho_j$ of \eqref{eq:exact-normalizers}, $\E[(u^0)^\top u^0]=I_d$. The entries of $D^0$ vanish unless their column index is one larger than their row index, and
\begin{equation}
 D^0_{s,s+1}=c_s/\Delta,\qquad 0\le c_s\le\Delta,\qquad
 \norm{D^0}\le1,\qquad w^\top D^0w=\mathcal E-c_1.
 \label{eq:population-cross}
\end{equation}
For every deterministic vector satisfying \eqref{eq:fixed-normalizer-condition},
\begin{equation}
 \norm{S-I}\le \mathsf{K}\nu,\qquad \norm{D-D^0}\le \mathsf{K}\nu,\qquad
 \left\|\E[(u,\zeta)^\top(u,\zeta)]\right\|\le3.
 \label{eq:population-mismatch}
\end{equation}
\end{lemma}

\begin{proof}
Each increment $\Delta M_j^\rho$ is a martingale difference for the filtration $(\mathcal A_j)$, with variance $h\E a(t_j,Y_j)^2/\rho_j^2=h$ by \eqref{eq:exact-normalizers}. Hence $\delta_s M^\rho$ has mean zero and variance $m_{\rm f}h=\Delta$ for every $s$, and increments over distinct groups are orthogonal. Since the first coordinate of $u^0$ is one and each $\delta_s M^\rho$ has mean zero, $\E[(u^0)^\top u^0]=I$. For the cross moments, $\E[G_j\Delta M_\ell^\rho]=0$ whenever $j\ne\ell$. If $j<\ell$, we condition on $\mathcal A_\ell$, which determines $G_j$ and $Y_\ell$, and use $\E[G_\ell\mid\mathcal A_\ell]=0$. If $j>\ell$, then $\Delta M_\ell^\rho$ is $\mathcal A_j$-measurable, while $G_j$ is centered and independent of $\mathcal A_j$. For $j=\ell$, conditioning on $\mathcal A_j$ shows that $\E[G_j\Delta M_j^\rho]=\E[a(t_j,Y_j)G_j^2]/\rho_j=h\E a(t_j,Y_j)/\rho_j$. Therefore
\begin{equation}
 c_s=\sum_{j\in I_s}h\frac{\E a(t_j,Y_j)}{\rho_j},\qquad
                         \sum_sc_s=\mathcal E.
 \label{eq:block-correlation}
\end{equation}
By \eqref{eq:group-sum}, $z_s=\sum_{j\in I_s}G_j/\sqrt\Delta$ for $s\le B$, so $D^0_{s,s+1}=\E[z_s\delta_s M^\rho]/\sqrt\Delta=c_s/\Delta$. The cross moments of $z_s$ with $\delta_r M^\rho$, $r\ne s$, vanish by the same argument, and $\E z_s=0$, so its cross moment with the constant first coordinate also vanishes. The last entry of $\zeta$ is the observation of the final query $q_d$. It drives no per-step Gaussian increment, and it is centered and independent of $u^0$, so its cross moments vanish. This proves the structure of $D^0$. Since $a\ge0$ by \eqref{eq:qualitative}, $c_s\ge0$. By Cauchy--Schwarz, $c_s\le(\E[(\sum_{j\in I_s}G_j)^2]\,\E[(\delta_s M^\rho)^2])^{1/2}=\Delta$. For any vector $x$, $\norm{D^0x}^2=\sum_{s<d}(c_s/\Delta)^2x_{s+1}^2\le\norm x^2$, so $\norm{D^0}\le1$. Finally, $w^\top D^0w=\sum_{s=1}^Bw_sw_{s+1}c_s/\Delta$. Since $w_1=0$ and $w_j=\sqrt\Delta$ for $j\ge2$, this sum omits exactly the term $s=1$ and equals $\sum_{s=2}^Bc_s=\mathcal E-c_1$, which is the last identity in \eqref{eq:population-cross}.

We now compare these population moments with the moments for the fixed coefficients $\omega$. Put $\chi_j=\rho_j/\omega_j$. Condition~\eqref{eq:fixed-normalizer-condition} implies $|\chi_j-1|\le \mathsf{K}\nu$, and the same martingale calculation shows that
\begin{equation}
 S_{s+1,s+1}=\frac h\Delta\sum_{j\in I_s}\chi_j^2,\qquad
 \E|\delta_s M^\omega-\delta_s M^\rho|^2=h\sum_{j\in I_s}(\chi_j-1)^2\le \mathsf{K}\Delta\nu^2.
\end{equation}
Since $|\chi_j^2-1|\le \mathsf{K}\nu$ and $I_s$ has $m_{\rm f}=\Delta/h$ elements, the first identity implies $|S_{s+1,s+1}-1|\le \mathsf{K}\nu$. The entry $S_{11}$ equals one, and the off-diagonal entries of $S$ vanish because $\delta_s M^\omega$ has mean zero and increments over distinct groups are orthogonal. Hence $S-I$ is diagonal with entries at most $\mathsf{K}\nu$ in absolute value. The entries of $D$ still vanish except at positions $(s,s+1)$, and there $|D_{s,s+1}-D^0_{s,s+1}|=|\E[z_s(\delta_s M^\omega-\delta_s M^\rho)]|/\sqrt\Delta\le \mathsf{K}\nu$ by Cauchy--Schwarz and the second identity. A matrix whose nonzero entries lie only at positions $(s,s+1)$ has norm equal to its largest absolute entry, so the first two assertions in \eqref{eq:population-mismatch} follow.

For the joint second-moment bound, condition~\eqref{eq:fixed-normalizer-condition} implies $\chi_j^2\le(1-\nu)^{-1}\le2$ for every $0<\nu\le1/2$. Since $S$ is diagonal, $S_{11}=1$, and each $S_{s+1,s+1}$ is the average of the $\chi_j^2$ over $j\in I_s$, we have $\norm S\le2$. For deterministic $x,y\in\R^d$, the triangle inequality in $L^2$, the identity $\E[(\zeta y)^2]=\norm y^2$, and Cauchy--Schwarz imply
\begin{align*}
 \E[(ux+\zeta y)^2]
 &\le\bigl(\sqrt{\norm S}\norm x+\norm y\bigr)^2\\
 &\le(\norm S+1)(\norm x^2+\norm y^2)
 \le3(\norm x^2+\norm y^2).
\end{align*}
Taking the supremum over $\norm x^2+\norm y^2=1$ proves the last assertion of \eqref{eq:population-mismatch}.

The ratios $\chi_j$ can vary within a group, so $\delta_s M^\omega$ need not be a multiple of $\delta_s M^\rho$. In particular, setting $\omega_j=1$ makes $S$ equal to $I$ only up to an error of norm at most $\mathsf{K}\nu$, not exactly.
\end{proof}

The bound $c_1\le\Delta$ in \eqref{eq:population-cross} explains the observation width chosen in \eqref{eq:grid-parameters}. The deterministic leading term $w^\top D^0w$ equals $\mathcal E-c_1$, so its omitted first-group contribution is at most $\Delta$. We may therefore take $\Delta$ proportional to accuracy, independently of the much smaller time-step width $h$. This calculation concerns the leading term; the actual output energy also includes the errors in \eqref{eq:energy-expansion}--\eqref{eq:energy-remainder}.

To bound $x^\top Pw$ in \eqref{eq:energy-expansion}, we will use the following bounds on the deterministic vectors $w$ in \eqref{eq:whitening-matrix} and $x=D^0w$. By \eqref{eq:population-cross}, $x_s=c_s/\sqrt\Delta$ for $s\le B$ and $x_d=0$. Since also $\norm w^2=T\le1$ and $\norm{D^0}\le1$,
\begin{equation}
 w_1=0,\quad w_j=\sqrt\Delta\ (j\ge2),\quad
 0\le x_i\le\sqrt\Delta,\quad \norm x,\norm w\le1.
 \label{eq:energy-flat-weights}
\end{equation}

\paragraph{Concentration estimates.}

We use the moment generating function form of Bernstein's inequality for the unbounded scalar variables in this section, Gaussian concentration for Lipschitz functions, and an operator-norm estimate for GOE matrices. We also state the bounded scalar Bernstein inequality used for the AMS IAMP clipping estimate in Appendix~\ref{sec:iamp-comparison}.

\begin{lemma}[Bernstein and Gaussian bounds {\cite{BLM,Tropp}}]
\label{lem:probability-tools}
For independent centered scalars bounded by $R$ with sum variance at most $v$,
\begin{equation}
 \Prob\{|\sum_iX_i|\ge t\}\le2\exp\{-t^2/[2(v+Rt/3)]\}.
 \label{eq:scalar-bernstein}
\end{equation}
More generally, independent centered scalars $X_i$ need not be bounded. Suppose $b>0$, $v_i\ge0$, and $v=\sum_i v_i$, with
\begin{equation}
 \log\E e^{\lambda X_i}
 \le\frac{v_i\lambda^2}{2(1-b|\lambda|)},
 \qquad |\lambda|<b^{-1}.
 \label{eq:moment-bernstein-mgf}
\end{equation}
Then, for $t>0$,
\begin{equation}
 \Prob\left\{\sum_iX_i\ge\sqrt{2vt}+bt\right\}\le e^{-t},\qquad
 \Prob\left\{\left|\sum_iX_i\right|\ge\sqrt{2vt}+bt\right\}\le2e^{-t}.
 \label{eq:moment-bernstein-tail}
\end{equation}
A sufficient condition for \eqref{eq:moment-bernstein-mgf} is
\begin{equation}
 \E|X_i|^p\le\frac{p!}{2}v_i b^{p-2},
 \qquad p\ge2\text{ an integer}.
 \label{eq:moment-bernstein-condition}
\end{equation}
A function of independent standard Gaussians with Euclidean Lipschitz constant $L$ has upper tail $\exp[-t^2/(2L^2)]$ above its mean. If $G$ is symmetric Gaussian with independent upper entries of variance $s^2$ off the diagonal and $2s^2$ on it, then
\begin{equation}
                  \Prob\{\norm G>2s\sqrt m+2s\sqrt u\}\le2e^{-u}.
 \label{eq:gaussian-matrix-norms}
\end{equation}
\end{lemma}

The bounded scalar Bernstein inequality is \cite[Theorem~1.6.1]{Tropp}. For \eqref{eq:moment-bernstein-mgf}, centering removes the linear term in the exponential series, and \eqref{eq:moment-bernstein-condition} bounds the remaining series by $v_i\lambda^2/[2(1-b|\lambda|)]$; then $\log(1+z)\le z$ proves the moment generating function bound. Independence permits these bounds to be added for $\sum_iX_i$. For $v>0$, Chernoff's inequality at $\lambda=\sqrt{2t/v}/(1+b\sqrt{2t/v})$ proves the one-sided bound in \eqref{eq:moment-bernstein-tail}. Applying the same argument to $-X_i$ and taking a union bound proves its two-sided form. If $v=0$, Jensen's inequality and \eqref{eq:moment-bernstein-mgf} force every $X_i=0$ almost surely, so the tail bounds are immediate.

For completeness, we prove the GOE norm estimate. For unit vectors $x,y\in\R^m$, the increment $x^\top Gx-y^\top Gy$ has variance $4s^2[1-(x^\top y)^2]$, which is at most the variance $8s^2(1-x^\top y)$ of the increment $2sg^\top(x-y)$, where $g$ is a standard Gaussian vector in $\R^m$. The Sudakov--Fernique inequality therefore implies $\E\lambda_{\max}(G)\le2s\E\norm g\le2s\sqrt m$. As a function of the independent standard Gaussian variables that generate the entries on and above the diagonal, $\lambda_{\max}(G)$ is $\sqrt2s$-Lipschitz. Gaussian concentration for $G$ and $-G$ proves \eqref{eq:gaussian-matrix-norms}.

We next obtain concentration estimates for the magnetization increments $\delta_s M^\omega$ entering the empirical matrices \eqref{eq:empirical-matrices}. The normalization vector $\omega$ remains deterministic, and every conditional statement uses only the filtrations \eqref{eq:fine-and-block-filtrations}; we do not condition on $A$ or on events defined by later increments. Conditional on $\mathcal A_j$, the increment $a(t_j,Y_j)G_j/\omega_j$ is centered Gaussian with variance $ha(t_j,Y_j)^2/\omega_j^2$. Choose a sufficiently large universal $\mathsf{K}_x$ so that $\omega_j\ge1/2$ and \eqref{eq:sharp-regularity} bound this variance by $\mathsf{K}_xh\Lambda(t_j)$ for every state. Define
\begin{equation}
 x_j^{\mathrm{var}}=\mathsf{K}_xh\Lambda(t_j),\quad V_s=\sum_{j\in I_s}x_j^{\mathrm{var}},\quad
 V=\sum_{j<K}x_j^{\mathrm{var}}\le \mathsf{K}L_\tau^2,\quad v_s=V_s/\Delta.
 \label{eq:clipping-coefficient-sum}
\end{equation}
Thus $x_j^{\mathrm{var}}$ bounds the conditional variance of one time-step increment; it is distinct from the entry $x_j$ of $x=D^0w$. The sums $V_s$ and $V$ bound the sums of conditional variances over group $s$ and over all steps. They determine the exponential-moment bounds in Lemma~\ref{lem:gaussian-sum}, while $v_s=V_s/\Delta$ is the variance parameter for the group increment divided by $\sqrt\Delta$. The bound on $V$ holds because $\Lambda$ is increasing, so the left sum $\sum_{j<K}h\Lambda(t_j)$ is at most $\int_0^T\Lambda(t)\dd t=(L_\tau^2-1)/2$ by \eqref{eq:singular-weights}. Write $\overline V=1+V$ to include a constant together with the total variance bound.

The next lemma bounds the conditional exponential moments of the per-step magnetization increments. Iterating these bounds over one group or over the whole path, we obtain tail bounds for $\delta_s M^\omega$ and for $M_K^\omega$. Lemma~\ref{lem:directional-cross} uses the tail of $M_K^\omega$, and the proof of Lemma~\ref{lem:directional-qr} iterates \eqref{eq:predictable-mgf} over single groups.

\begin{lemma}[Exponential moments of magnetization increments]
\label{lem:gaussian-sum}
For every real $\lambda$,
\begin{equation}
 \E[\exp\{\lambda a(t_j,Y_j)G_j/\omega_j\}\mid\mathcal A_j]
        \le e^{\lambda^2x_j^{\mathrm{var}}/2}.
 \label{eq:predictable-mgf}
\end{equation}
Hence, for $a_0>0$,
\begin{equation}
 \Prob\{|\delta_s M^\omega|>\sqrt{2V_sa_0}\mid\mathcal B_{s-1}\}\le2e^{-a_0},\qquad
 \Prob\{|M_K^\omega|>\sqrt{2Va_0}\}\le2e^{-a_0},
 \label{eq:predictable-tail}
\end{equation}
and $\E[(\delta_s M^\omega/\sqrt\Delta)^4]\le16v_s^2$.
\end{lemma}

\begin{proof}
Conditional on $\mathcal A_j$, the coefficient $a(t_j,Y_j)/\omega_j$ is fixed and $G_j\sim\mathcal N(0,h)$ is independent of it; here $\mathcal A_j$ contains all earlier per-step Gaussian increments, including those of the current group. The conditional exponential moment in \eqref{eq:predictable-mgf} is therefore exactly $\exp\{\lambda^2ha(t_j,Y_j)^2/(2\omega_j^2)\}$, which is at most $e^{\lambda^2x_j^{\mathrm{var}}/2}$ by the choice of $\mathsf{K}_x$. We condition successively on $\mathcal A_j$ for the time steps of one group, starting from the last step, and multiply these bounds to obtain $\E[e^{\lambda\delta_s M^\omega}\mid\mathcal B_{s-1}]\le e^{\lambda^2V_s/2}$. Repeating this over all time steps, we obtain $\E e^{\lambda M_K^\omega}\le e^{\lambda^2V/2}$. Chernoff's inequality in both signs then proves the two tail bounds. With $a_0=t^2/(2v_s)$, the first tail bound reads $\Prob\{|\delta_s M^\omega/\sqrt\Delta|>t\mid\mathcal B_{s-1}\}\le2e^{-t^2/(2v_s)}$. Integrating this tail against $4t^3\dd t$ proves $\E[(\delta_s M^\omega/\sqrt\Delta)^4]\le16v_s^2$. The increments $\delta_s M^\omega$ need not be Gaussian.
\end{proof}

For $j\ge2$, the entries of column $j$ of $U$ are copies of $\delta_{j-1}M^\omega/\sqrt\Delta$ from \eqref{eq:block-increments}, and $\widehat S_{jj}$ averages their squares. To control its fluctuations, we use a fourth-moment bound for $U_{\ell j}$ proportional to its second moment $S_{jj}=\E[U_{\ell j}^2]$. The bound $16v_{j-1}^2$ in Lemma~\ref{lem:gaussian-sum} does not retain this second-moment factor. We first prove a general moment bound of the required form for continuous martingales and then represent $\delta_s M^\omega$ as a terminal value of such a martingale.

\begin{lemma}[Squares of continuous martingales with bounded quadratic variation]
\label{lem:bracket-moments}
\label{app:bracket-concentration}
Let $Z_0=0$ be a continuous square-integrable martingale with $\langle Z\rangle_T\le v$ almost surely, for deterministic $v>0$, and put $\mu_Z=\E Z_T^2$. For every integer $p\ge2$,
\begin{equation}
 \E|Z_T|^{2p}\le p!\mathsf{K}^p\mu_Zv^{p-1}.
 \label{eq:bracket-moments}
\end{equation}
For $N$ independent copies and $t_{\rm dev}\ge1$, with failure at most $e^{-t_{\rm dev}}$,
\begin{equation}
 \frac1N\sum_{i=1}^NZ_{T,i}^2
 \le\mu_Z+\mathsf{K}\sqrt{\mu_Zvt_{\rm dev}/N}+\mathsf{K}vt_{\rm dev}/N
 \le2\mu_Z+\mathsf{K}vt_{\rm dev}/N.
 \label{eq:bracket-bernstein}
\end{equation}
\end{lemma}

\begin{proof}
Let $\sigma=\inf\{t\le T:|Z_t|=\sqrt v\}$, with $\sigma=\infty$ if the level is not reached. Optional stopping and the martingale isometry imply
\begin{equation}
 v\Prob\{\sigma\le T\}\le\E Z_{\sigma\wedge T}^2
       =\E\langle Z\rangle_{\sigma\wedge T}\le\mu_Z.
 \label{eq:bracket-hitting-probability}
\end{equation}
After $\sigma$, the remaining quadratic variation is at most $v$. The exponential supermartingale inequality, conditional on the stopped sigma-algebra, gives
\begin{equation}
 \Prob\{|Z_T-Z_\sigma|>u\mid\mathcal F_\sigma\}
       \le2\exp\{-u^2/(2v)\}\quad\text{on }\{\sigma\le T\}.
\end{equation}
By continuity, $|Z_T|>u\ge2\sqrt v$ requires $\sigma\le T$ and $|Z_T-Z_\sigma|>u/2$. Hence
\begin{equation}
 \Prob\{|Z_T|>u\}\le\frac{2\mu_Z}{v}e^{-u^2/(8v)},
                           \qquad u\ge2\sqrt v.
 \label{eq:bracket-variance-tail}
\end{equation}
On $|Z_T|\le2\sqrt v$, the $2p$-th power is at most $(4v)^{p-1}Z_T^2$. On its complement, integrate \eqref{eq:bracket-variance-tail}; the boundary term is at most $\mathsf{K}^p\mu_Zv^{p-1}$ and the remaining integral is at most $2p!(8v)^p\mu_Z/v$. This proves \eqref{eq:bracket-moments}.

For $X=Z_T^2$, centering changes only the universal constants, since $\mu_Z\le v$. Expanding the exponential series gives, for $|\lambda|<(\mathsf{K}v)^{-1}$,
\begin{equation}
 \log\E e^{\lambda(X-\E X)}
       \le\frac{\mathsf{K}\mu_Zv\lambda^2}{1-\mathsf{K}v|\lambda|}.
 \label{eq:bracket-centered-mgf}
\end{equation}
Applying \eqref{eq:moment-bernstein-tail} to the independent copies proves the first inequality in \eqref{eq:bracket-bernstein}; Young's inequality proves the second. The same argument in both signs gives a two-sided deviation with failure $2e^{-t_{\rm dev}}$. If $v=0$, the martingale is identically zero and the assertions are immediate.
\end{proof}

Lemma~\ref{lem:bracket-moments} applies to martingales with continuous paths, whereas the algorithm specifies values only on a discrete time grid. To use it for $\delta_s M^\omega$ in \eqref{eq:block-increments}, in the next lemma we embed the Gaussian increments in Brownian motion and the magnetization increments in a continuous-time martingale. Independent Brownian bridges supply the path between consecutive grid points without changing any $G_k$ or any grid value of the magnetization. This extension is used only in the proof; the algorithm remains discrete and generates no bridges.

\begin{lemma}[A continuous martingale representation of magnetization increments]
\label{lem:martingale-representations}
Fix deterministic $\omega$ satisfying \eqref{eq:fixed-normalizer-condition}, and let $(Y_k,M_k^\omega)$ follow \eqref{eq:scalar-y}--\eqref{eq:scalar-m} with the quadrature coefficients \eqref{eq:drift-quadrature}. On a probability space enlarged by independent Brownian bridges inside the time-grid cells, each magnetization increment $\delta_s M^\omega$, $1\le s\le B$, is the terminal value of a continuous square-integrable martingale starting at zero, with quadratic variation at most $V_s$ from \eqref{eq:clipping-coefficient-sum}. For independent coordinate paths, the bridges and resulting martingales can be constructed independently across coordinates.
\end{lemma}

\begin{proof}
On an enlarged probability space, let $\beta_k$, $0\le k<K$, be independent Brownian bridges on $[0,h]$, independent of all the Gaussian increments $G_j$ in \eqref{eq:scalar-y}. Each bridge has endpoints $\beta_k(0)=\beta_k(h)=0$, mean zero, and covariance $\E[\beta_k(u)\beta_k(v)]=\min(u,v)-uv/h$. Define
\begin{equation}
 W_{t_k+u}=\sum_{j<k}G_j+\frac{u}{h}G_k+\beta_k(u),
 \qquad 0\le u\le h,\quad 0\le k<K.
 \label{eq:brownian-bridge-interpolation}
\end{equation}
The bridge vanishes at both endpoints, so $W_{t_{k+1}}-W_{t_k}=G_k$. Linear interpolation alone would produce a continuous path, but not a Brownian path. The term $\beta_k(u)$ supplies the Brownian fluctuations between the endpoints. Since $\E G_k^2=h$ and $G_k$ is independent of $\beta_k$,
\begin{equation}
 \E\big[(W_{t_k+u}-W_{t_k})(W_{t_k+v}-W_{t_k})\big]
 =\frac{uv}{h}+\min(u,v)-\frac{uv}{h}=\min(u,v).
 \label{eq:bridge-cell-covariance}
\end{equation}
The processes $u\mapsto W_{t_k+u}-W_{t_k}$ for distinct cells are independent because they depend on independent pairs $(G_k,\beta_k)$. Consequently, $W$ is a continuous centered Gaussian process with covariance $\E[W_tW_r]=\min(t,r)$, hence a Brownian motion in its natural filtration. We use this filtration, which reveals the path only up to the current time, rather than a filtration containing all $G_k$ at time zero. Choosing the bridges independently across coordinates preserves independence of the coordinate paths.

Let $k(r)$ be the time-step index with $t_{k(r)}\le r<t_{k(r)+1}$. For group $s$, define the continuous martingale
\begin{equation}
 \begin{gathered}
 \widetilde M_{s,t}^\omega
 =\int_{(s-1)\Delta}^{(s-1)\Delta+t}
       \frac{a(t_{k(r)},Y_{k(r)})}{\omega_{k(r)}}\dd W_r,
 \qquad 0\le t\le\Delta,\\
 \widetilde M_{s,\Delta}^\omega=\delta_s M^\omega,\qquad
 \langle\widetilde M_s^\omega\rangle_\Delta
 =\sum_{k\in I_s}\frac{ha(t_k,Y_k)^2}{\omega_k^2}\le V_s.
 \end{gathered}
 \label{eq:group-martingale-representation}
\end{equation}
On a whole cell, the stochastic integral of the constant coefficient $a(t_k,Y_k)/\omega_k$ is $a(t_k,Y_k)(W_{t_{k+1}}-W_{t_k})/\omega_k=a(t_k,Y_k)G_k/\omega_k$. Thus the bridges supply the path between grid points without changing the terminal value $\delta_s M^\omega$ in \eqref{eq:group-martingale-representation}. The coefficient on cell $k$ is determined by the increments before that cell. The process is therefore adapted to the chronological Brownian filtration, starts at zero, and is square-integrable by the bound in \eqref{eq:group-martingale-representation}. The quadratic-variation bound follows from the definition of $x_k^{\mathrm{var}}$ in \eqref{eq:clipping-coefficient-sum}.
\end{proof}

\paragraph{Bounds on the energy error terms.}

We now apply the two preceding lemmas to the empirical second moments $\widehat S_{jj}=N^{-1}\sum_{\ell=1}^N U_{\ell j}^2$. For $j\ge2$, the entries $U_{\ell j}$ are independent copies of $\delta_{j-1}M^\omega/\sqrt\Delta$, with second moment $S_{jj}$. Lemma~\ref{lem:martingale-representations} represents each entry as the terminal value of a continuous martingale with quadratic variation at most $v_{j-1}=V_{j-1}/\Delta$. Applying Lemma~\ref{lem:bracket-moments} with $p=2$ therefore implies $\operatorname{Var}(U_{\ell j}^2)\le \mathsf{K} S_{jj}v_{j-1}$, and independence across coordinates implies $\operatorname{Var}(\widehat S_{jj})\le \mathsf{K} S_{jj}v_{j-1}/N$. The higher-moment bounds in the same lemma yield concentration of $\widehat S_{jj}$ about $S_{jj}$.

These empirical second moments control the diagonal entries of $L_0$ through $(L_0)_{jj}^2=\widehat S_{jj}-\sum_{i<j}(L_0)_{ij}^2$, which follows from the factorization $L_0^\top L_0=\widehat S$ in \eqref{eq:AQ}. Above the diagonal, $(L_0)_{ij}=q_i^\top U_{:,j}/N$ is a projection onto a query fixed before the group producing $U_{:,j}$ is generated, so it has conditional mean zero. The query $q_j$ itself depends on $U_{:,j}$. In the next lemma we therefore combine concentration of $\widehat S_{jj}$ about $S_{jj}$ with estimates for the projections $q_i^\top U_{:,j}/N$, $i<j$, and their squared sum to bound $Pw$ and $x^\top Pw$ in \eqref{eq:energy-expansion}--\eqref{eq:energy-remainder}. These bounds require neither an operator-norm bound on $P$ nor positive definiteness of $\widehat S$, and include outcomes on which a query residual vanishes.

\begin{lemma}[Bounds on $Pw$ and $x^\top Pw$]
\label{lem:directional-qr}
\label{app:directional-qr}
Fix $\omega$ satisfying \eqref{eq:fixed-normalizer-condition} and $t_{\rm dev}\ge1$. Let $w=\sqrt\Delta(0,1,\ldots,1)^\top\in\R^d$ and $x=D^0w$, where $D^0$ is defined in \eqref{eq:empirical-matrices}. Suppose $d\le N$ and
\begin{equation}
 N\ge \mathsf{K}R\{t_{\rm dev}+\log(\mathsf{K}d)\},
 \label{eq:directional-dimension}
\end{equation}
where $R=\mathsf{K}_R\Lambda(T)$ is the bound on $a^2$ from \eqref{eq:quadrature-tolerance}, and define
\begin{equation}
 r_N^2:=\nu^2+\overline V(d+t_{\rm dev})/N.
 \label{eq:directional-radius}
\end{equation}
With failure at most $\mathsf{K}e^{-t_{\rm dev}}$,
\begin{align}
 \norm{Pw}^2&\le \mathsf{K}r_N^2,
 \label{eq:directional-qr-norm}\\
 |x^\top Pw|&\le \mathsf{K}\{\nu+\sqrt{\overline Vt_{\rm dev}/N}
                         +\overline V(d+t_{\rm dev})/N\}.
 \label{eq:directional-qr-scalar}
\end{align}
\end{lemma}

\begin{proof}
The vector $\omega$ is deterministic, and we marginalize over the side variables. The matrices $F,Q,L_0$ include the final query appended in the proof. Let
\begin{equation}
 \mathcal H_s=\sigma(G_{k,i}:k<sm_{\rm f},\ 1\le i\le N),
 \qquad
 \upsilon_j=V_{j-1}/\Delta\quad(2\le j\le d).
\end{equation}
Then $\Delta\sum_{j=2}^d\upsilon_j=V$ and $\max_j\upsilon_j\le \mathsf{K}R$. Write $U_j=U_{:,j}$. Since
\begin{equation}
 F_j=\sqrt\Delta\sum_{r\le j}U_r,
 \qquad
 (I-P_{j-1})F_j=\sqrt\Delta(I-P_{j-1})U_j,
 \label{eq:query-qr-predictability}
\end{equation}
orthogonalizing the columns of $F$ and of $U$ produces the same query directions. When a residual is zero, both constructions use the rule after \eqref{eq:query-proposals}, projecting and normalizing the first coordinate vector outside the preceding query span. After marginalizing over the side variables, Proposition~\ref{prop:gaussian-law} shows that the per-step Gaussian increments $G_{k,i}$ of all steps and coordinates are independent $\mathcal N(0,h)$ variables. The query input $F_j=\widehat M_{(j-1)m_{\rm f}}$ is a function of the Gaussian increments in groups $1,\ldots,j-1$, and the deterministic rule after \eqref{eq:query-proposals} makes $q_j$ a function of $F_1,\ldots,F_j$, hence of those increments alone. The matrices $A$ and $J$ that generate the observations need not be independent of $q_j$ for this conclusion. In particular, $q_i$ is $\mathcal H_{j-2}$-measurable for every $i<j$. This measurability holds on every outcome because of the form of the query rule; it does not follow from the independence of rows alone.

Conditional on $\mathcal H_{j-2}$, the $N$ entries of $U_j$ are independent and have mean zero. Iterating \eqref{eq:predictable-mgf} over the time steps of group $j-1$ shows that each entry has conditional exponential moments bounded with variance parameter $\upsilon_j$. For deterministic $a\in\R^{j-1}$, the vector $\sum_{i<j}a_iq_i$ is $\mathcal H_{j-2}$-measurable and has squared norm $N\norm a^2$, because the earlier queries are orthogonal. Since $(L_0)_{ij}=q_i^\top U_j/N$, for every real $\vartheta$ we obtain
\begin{equation}
 \E\left[\left.\exp\left\{\vartheta\sum_{i<j}a_i(L_0)_{ij}\right\}
                      \right|\mathcal H_{j-2}\right]
 \le\exp\{\vartheta^2\upsilon_j\norm a^2/(2N)\}.
 \label{eq:predictable-qr-mgf}
\end{equation}
Let $P_{\rm str}$ retain only the strictly upper entries of $P=L_0-I$. Iterating \eqref{eq:predictable-qr-mgf} with weights $w_j=\sqrt\Delta$ yields, for every deterministic $a\in\R^d$,
\begin{equation}
 \E e^{\vartheta a^\top P_{\rm str}w}
               \le e^{\vartheta^2V\norm a^2/(2N)}.
\end{equation}
We use a deterministic $1/2$-net to pass from fixed directions to the Euclidean norm. For each $m\ge1$, choose a finite set $\mathcal N_m$ on the unit sphere in $\R^m$ such that every unit vector is within distance $1/2$ of some point of $\mathcal N_m$. Such a set can be chosen with at most $5^m$ points: take a maximal set whose distinct points have distance greater than $1/2$. Maximality ensures the covering property, and the disjoint balls of radius $1/4$ centered at these points lie inside the ball of radius $5/4$. Comparing their volumes proves the cardinality bound. For $y\ne0$, choose $\theta\in\mathcal N_m$ within distance $1/2$ of $y/\norm y$. Cauchy--Schwarz implies $\norm y\le|\theta^\top y|+\norm y/2$, so
\begin{equation}
 |\mathcal N_m|\le5^m,\qquad
 \norm y\le2\max_{\theta\in\mathcal N_m}|\theta^\top y|
 \quad(y\in\R^m).
 \label{eq:euclidean-net-bound}
\end{equation}
The case $y=0$ is immediate. The nets are fixed independently of all random variables.

For every deterministic unit vector $a$, Chernoff's inequality applied to the preceding exponential-moment bound implies $\Prob\{|a^\top P_{\rm str}w|>\sqrt{2Vt_{\rm tail}/N}\}\le2e^{-t_{\rm tail}}$ for $t_{\rm tail}>0$. Put $r=\sqrt{2Vt_{\rm tail}/N}$. Applying \eqref{eq:euclidean-net-bound} with $y=P_{\rm str}w$, we see that $\norm{P_{\rm str}w}>2r$ requires $|\theta^\top P_{\rm str}w|>r$ for at least one $\theta\in\mathcal N_d$. Since $(2r)^2=8Vt_{\rm tail}/N$,
\[
 \begin{aligned}
 \Prob\!\left\{\norm{P_{\rm str}w}^2>\frac{8Vt_{\rm tail}}N\right\}
 &\le\Prob\!\left\{\max_{\theta\in\mathcal N_d}
                          |\theta^\top P_{\rm str}w|>r\right\}\\
 &\le\sum_{\theta\in\mathcal N_d}
                  \Prob\{|\theta^\top P_{\rm str}w|>r\}
 \le2|\mathcal N_d|e^{-t_{\rm tail}}.
 \end{aligned}
\]
The first inequality uses the norm estimate, the second is the union bound, and the last applies the fixed-direction Chernoff bound. Taking $t_{\rm tail}=t_{\rm dev}+d\log5$ and using $|\mathcal N_d|\le5^d$ bounds the right-hand side by $2e^{-t_{\rm dev}}$.
For the deterministic direction $x$, apply the exponential-moment bound directly with $a=x$ and use $\norm x\le1$; no net is needed. Its two-sided tail at $\sqrt{2Vt_{\rm dev}/N}$ is at most $2e^{-t_{\rm dev}}$. Combining the two events, we obtain
\begin{equation}
 \norm{P_{\rm str}w}^2\le \mathsf{K}V(d+t_{\rm dev})/N,
 \qquad |x^\top P_{\rm str}w|\le \mathsf{K}\sqrt{Vt_{\rm dev}/N}
 \label{eq:strict-qr-directions}
\end{equation}
with total failure at most $\mathsf{K}e^{-t_{\rm dev}}$.

For the diagonal, put $s_j=\widehat S_{jj}$, $b_j=\sum_{i<j}(L_0)_{ij}^2$, $\lambda_j=(L_0)_{jj}$, and $t_{{\rm dev},d}=t_{\rm dev}+\log(\mathsf{K}d)$. The number $b_j$ is the squared norm of the vector $((L_0)_{ij})_{i<j}\in\R^{j-1}$. The projection onto a deterministic unit vector $a\in\R^{j-1}$ is the scalar $\sum_{i<j}a_i(L_0)_{ij}$. Equation~\eqref{eq:predictable-qr-mgf} states that $\E[\exp\{\vartheta\sum_{i<j}a_i(L_0)_{ij}\}\mid\mathcal H_{j-2}]\le\exp\{\vartheta^2\upsilon_j/(2N)\}$ for every real $\vartheta$. Conditional Chernoff bounds in both signs therefore imply, for $t_{\rm tail}>0$,
\[
 \Prob\!\left\{\left|\sum_{i<j}a_i(L_0)_{ij}\right|
             >\sqrt{\frac{2\upsilon_j t_{\rm tail}}N}
             \,\middle|\,\mathcal H_{j-2}\right\}\le2e^{-t_{\rm tail}}.
\]

Apply \eqref{eq:euclidean-net-bound} with $m=j-1$ and $y=((L_0)_{ij})_{i<j}$. Since $\norm y^2=b_j$, the norm estimate reads
\[
 \sqrt{b_j}\le2\max_{a\in\mathcal N_{j-1}}
                         \left|\sum_{i<j}a_i(L_0)_{ij}\right|.
\]
Thus $b_j>8\upsilon_j t_{\rm tail}/N$ requires a projection larger than $\sqrt{2\upsilon_j t_{\rm tail}/N}$ at some point of $\mathcal N_{j-1}$. This implication holds on every outcome, so it remains valid after conditioning on $\mathcal H_{j-2}$. Consequently,
\[
 \begin{aligned}
 \Prob\!\left\{b_j>\frac{8\upsilon_j t_{\rm tail}}N
                       \,\middle|\,\mathcal H_{j-2}\right\}
 &\le\Prob\!\left\{\max_{a\in\mathcal N_{j-1}}
             \left|\sum_{i<j}a_i(L_0)_{ij}\right|
             >\sqrt{\frac{2\upsilon_j t_{\rm tail}}N}
             \,\middle|\,\mathcal H_{j-2}\right\}\\
 &\le\sum_{a\in\mathcal N_{j-1}}
       \Prob\!\left\{\left|\sum_{i<j}a_i(L_0)_{ij}\right|
             >\sqrt{\frac{2\upsilon_j t_{\rm tail}}N}
             \,\middle|\,\mathcal H_{j-2}\right\}\\
 &\le2|\mathcal N_{j-1}|e^{-t_{\rm tail}}\le2\cdot5^{j-1}e^{-t_{\rm tail}}.
 \end{aligned}
\]
Here the first inequality uses the norm estimate, the second is the conditional union bound, and the last line uses the conditional Chernoff bound at each deterministic net point and $|\mathcal N_{j-1}|\le5^{j-1}$.
Take $t_{\rm tail}=(j-1)\log5+t_{{\rm dev},d}$ and let $E_j=\{b_j>8\upsilon_j t_{\rm tail}/N\}$. The preceding conditional bound and the tower property imply
\[
 \Prob(E_j)=\E\!\left[\Prob(E_j\mid\mathcal H_{j-2})\right]
          \le2e^{-t_{{\rm dev},d}}.
\]
Summing these unconditional probabilities over $2\le j\le d$ bounds the total failure by $2(d-1)e^{-t_{{\rm dev},d}}\le \mathsf{K}e^{-t_{\rm dev}}$. Thus
\begin{equation}
 b_j\le \mathsf{K}\upsilon_j(j+t_{{\rm dev},d})/N,
 \qquad \Delta\sum_{j=2}^db_j\le \mathsf{K}V(d+t_{\rm dev})/N
 \label{eq:qr-projection-squares}
\end{equation}
simultaneously. To prove the second inequality, multiply each bound on $b_j$ by $\Delta$ and sum over $2\le j\le d$:
\[
 \begin{aligned}
 \Delta\sum_{j=2}^d b_j
 &\le\frac{\mathsf{K}\Delta}{N}\sum_{j=2}^d\upsilon_j(j+t_{{\rm dev},d})\\
 &\le\frac{\mathsf{K}(d+t_{\rm dev})}{N}\Delta\sum_{j=2}^d\upsilon_j
       =\frac{\mathsf{K}V(d+t_{\rm dev})}{N}.
 \end{aligned}
\]
The second inequality uses $j+t_{{\rm dev},d}\le \mathsf{K}(d+t_{\rm dev})$ for $j\le d$, and the equality uses $\Delta\sum_{j=2}^d\upsilon_j=V$.

For one coordinate, $u_j=\delta_{j-1}M^\omega/\sqrt\Delta$. By Lemma~\ref{lem:martingale-representations}, it is the terminal value of a continuous martingale starting at zero with quadratic variation at most $V_{j-1}/\Delta=\upsilon_j$; its second moment is $S_{jj}\le2$. By Lemma~\ref{lem:bracket-moments},
\begin{equation}
 \E|u_j|^{2p}\le p!\mathsf{K}^p\upsilon_j^{p-1},\qquad p\ge2.
 \label{eq:group-variance-moments}
\end{equation}
The exponential-series argument in \eqref{eq:bracket-centered-mgf} and independence across rows, followed by \eqref{eq:moment-bernstein-tail} and a union bound over columns, therefore give simultaneously
\begin{equation}
 |s_j-S_{jj}|\le \mathsf{K}\{\sqrt{\upsilon_jt_{{\rm dev},d}/N}
                                             +\upsilon_jt_{{\rm dev},d}/N\}.
\end{equation}
Under $N\ge \mathsf{K}Rt_{{\rm dev},d}$, we square the preceding bound and sum with weight $\Delta$. Using $|S_{jj}-1|\le \mathsf{K}\nu$ and $\Delta(d-1)=T\le1$ from \eqref{eq:grid-parameters} for the $\mathsf{K}\nu^2$ term, we obtain
\begin{equation}
 \Delta\sum_{j=2}^d(s_j-1)^2\le \mathsf{K}\nu^2+\mathsf{K}Vt_{{\rm dev},d}/N.
 \label{eq:qr-diagonal-squares}
\end{equation}
The diagonal entry $\lambda_j$ is nonnegative because the query points along the residual when that residual is nonzero. Recall $r_j=(I-P_{j-1})F_j$ from \eqref{eq:query-proposals}. If $r_j\ne0$, then $q_j=\sqrt N r_j/\norm{r_j}$ and $q_j$ is orthogonal to $P_{j-1}F_j$. As in \eqref{eq:query-residual-diagonal},
\[
 \lambda_j=(L_0)_{jj}
 =\frac{q_j^\top F_j}{N\sqrt\Delta}
 =\frac{q_j^\top r_j}{N\sqrt\Delta}
 =\frac{\norm{r_j}}{\sqrt{N\Delta}}\ge0.
\]
If $r_j=0$, then $F_j$ lies in the preceding query span and the chosen $q_j$ is orthogonal to that span, so $q_j^\top F_j=0$ and $\lambda_j=0$. The factorization $L_0^\top L_0=\widehat S$ in \eqref{eq:AQ} implies
\begin{equation}
 \lambda_j^2=s_j-b_j,\qquad
 \lambda_j-1=\tfrac12(s_j-1)-\tfrac12b_j-\tfrac12(\lambda_j-1)^2.
 \label{eq:qr-diagonal-identity}
\end{equation}
Indeed, $s_j=\sum_{i\le j}(L_0)_{ij}^2$ proves the first identity, and substituting it into $\lambda_j-1=\tfrac12(\lambda_j^2-1)-\tfrac12(\lambda_j-1)^2$ proves the second. Since $\lambda_j\ge0$, the first identity also implies
\begin{equation}
 |\lambda_j-1|\le|\sqrt{s_j}-1|+\sqrt{b_j}
                     \le|s_j-1|+\sqrt{b_j},
\end{equation}
even if $\lambda_j=0$. Equations~\eqref{eq:qr-projection-squares}--\eqref{eq:qr-diagonal-squares} imply
\begin{equation}
 \Delta\sum_{j=2}^d(\lambda_j-1)^2
                      \le \mathsf{K}\nu^2+\mathsf{K}\overline V(d+t_{\rm dev})/N.
 \label{eq:qr-factor-diagonal-squares}
\end{equation}
Since $U_1=q_1=\one$, the first column of $P$ is zero. Combining \eqref{eq:strict-qr-directions} and \eqref{eq:qr-factor-diagonal-squares} proves \eqref{eq:directional-qr-norm}.

To bound $x^\top Pw$, put $\pi_j=x_jw_j$, where $x=D^0w$. The correlations $c_s\ge0$ in \eqref{eq:population-cross} make these weights nonnegative. Equations~\eqref{eq:energy-flat-weights} and \eqref{eq:population-cross} therefore imply
\begin{equation}
 0\le\pi_j\le\Delta,\qquad \pi_1=0,\qquad
 \sum_j\pi_j\le1,\qquad
 \sum_{j=2}^d\pi_j\upsilon_j\le\Delta\sum_{j=2}^d\upsilon_j=V.
\end{equation}
For one row, put $X=\sum_{j=2}^d\pi_ju_j^2$. By Minkowski's inequality, \eqref{eq:group-variance-moments}, and the weighted H\"older inequality,
\begin{align}
 \norm X_p
 &\le \mathsf{K}(p!)^{1/p}\sum_{j=2}^d\pi_j\upsilon_j^{1-1/p}\nonumber\\
 &\le \mathsf{K}(p!)^{1/p}\left(\sum_j\pi_j\right)^{1/p}
                      \left(\sum_j\pi_j\upsilon_j\right)^{1-1/p}
 \le \mathsf{K}(p!)^{1/p}\overline V^{1-1/p}.
\end{align}
Thus its centered moments satisfy \eqref{eq:moment-bernstein-condition} with variance parameter $\mathsf{K}\overline V$ and scale $\mathsf{K}\overline V$. Its mean differs from $\sum_j\pi_j$ by at most $\mathsf{K}\nu$. Applying \eqref{eq:moment-bernstein-tail} to the independent rows proves
\begin{equation}
 \left|\sum_j\pi_j(s_j-1)\right|
       \le \mathsf{K}\{\nu+\sqrt{\overline Vt_{\rm dev}/N}+\overline Vt_{\rm dev}/N\}.
 \label{eq:qr-weighted-diagonal}
\end{equation}
Finally, \eqref{eq:qr-diagonal-identity} gives the exact scalar expansion
\begin{equation}
 x^\top Pw=x^\top P_{\rm str}w+\frac12\sum_j\pi_j(s_j-1)
        -\frac12\sum_j\pi_jb_j-\frac12\sum_j\pi_j(\lambda_j-1)^2.
\end{equation}
Use $\pi_j\le\Delta$, \eqref{eq:strict-qr-directions}, \eqref{eq:qr-projection-squares}, \eqref{eq:qr-factor-diagonal-squares}, and \eqref{eq:qr-weighted-diagonal} to prove \eqref{eq:directional-qr-scalar}. All conditional calculations condition only on completed groups of Gaussian increments, never on the side variables of all groups and coordinates or on events defined by later groups.
\end{proof}

The empirical cross-moment matrix $\widehat D$ enters the expansion through $E_Dw$ and $w^\top E_Dw$. The next lemma bounds these quantities using the independence of the rows of $(U,Z)$ from Proposition~\ref{prop:gaussian-law}.

\begin{lemma}[Bounds on $E_Dw$ and $w^\top E_Dw$]
\label{lem:directional-cross}
Fix $\omega$ satisfying \eqref{eq:fixed-normalizer-condition} and $t_{\rm dev}\ge1$. If $N\ge d+t_{\rm dev}$, with failure at most $\mathsf{K}e^{-t_{\rm dev}}$,
\begin{align}
 \norm{E_Dw}&\le \mathsf{K}\{\nu+\sqrt{\overline V(d+t_{\rm dev})/N}\},
 \label{eq:directional-cross-norm}\\
 |w^\top E_Dw|&\le \mathsf{K}\{\nu+\sqrt{\overline Vt_{\rm dev}/N}
                                  +\sqrt{\overline V}t_{\rm dev}/N\}.
 \label{eq:directional-cross-scalar}
\end{align}
\end{lemma}

\begin{proof}
Let $\theta$ be a deterministic unit vector in $\R^d$. For each coordinate $1\le i\le N$, let $Z_{is}$ denote coordinate $i$ of the observation $z_s$, and define the scalar variables
\begin{equation}
 \xi_i(\theta)=\sum_{s=1}^d\theta_s Z_{is},\qquad
 \psi_i(\theta)=\xi_i(\theta)f_i.
 \label{eq:cross-coordinate-products}
\end{equation}
Since $f=Uw$ and $\widehat D=Z^\top U/N$, the empirical cross moment in direction $\theta$ is
\begin{equation}
 \theta^\top\widehat Dw=\frac1N\sum_{i=1}^N\psi_i(\theta),\qquad
 \E[\psi_i(\theta)]=\theta^\top Dw.
 \label{eq:cross-coordinate-average}
\end{equation}
The expectation identity follows from the definition of $D$ in \eqref{eq:empirical-matrices} and $f=Uw$. Proposition~\ref{prop:gaussian-law} implies that the variables $\psi_i(\theta)$ are independent and identically distributed over $i$.

For each fixed $i$, the entries $Z_{i1},\ldots,Z_{id}$ are independent standard Gaussians and $\sum_s\theta_s^2=1$, so $\xi_i(\theta)$ is standard Gaussian. The other factor $f_i$ has the law of the scalar terminal magnetization $M_K^\omega$, whose tail satisfies the second bound in \eqref{eq:predictable-tail}. Explicitly, for every $t_{\rm tail}>0$,
\[
 \Prob\{|\xi_i(\theta)|>\sqrt{2t_{\rm tail}}\}\le2e^{-t_{\rm tail}},
 \qquad
 \Prob\{|f_i|>\sqrt{2Vt_{\rm tail}}\}\le2e^{-t_{\rm tail}}.
\]
Thus $|\xi_i(\theta)|\le\sqrt{2t_{\rm tail}}$ and $|f_i|\le\sqrt{2Vt_{\rm tail}}$ hold simultaneously with failure probability at most $4e^{-t_{\rm tail}}$, by a union bound. The factors need not be independent. Integrating the separate tail bounds and applying Cauchy--Schwarz, we obtain, for every integer $p\ge2$,
\begin{equation}
 \norm{\psi_i(\theta)}_p
 \le\norm{\xi_i(\theta)}_{2p}\norm{f_i}_{2p}
 \le \mathsf{K}\sqrt p\sqrt{p\overline V}=\mathsf{K}p\sqrt{\overline V}.
 \label{eq:cross-coordinate-moments}
\end{equation}
Summing the exponential series with these moment bounds, we find that the centered variable $\psi_i(\theta)-\theta^\top Dw$ satisfies \eqref{eq:moment-bernstein-mgf}, with variance parameter $\mathsf{K}\overline V$ and scale $\mathsf{K}\sqrt{\overline V}$. Applying \eqref{eq:moment-bernstein-tail} to the independent summands in \eqref{eq:cross-coordinate-average}, we bound $|\theta^\top(\widehat D-D)w|$ by
\begin{equation}
 \mathsf{K}\sqrt{\overline V}\{\sqrt{t/N}+t/N\}
 \label{eq:cross-bernstein-radius}
\end{equation}
with failure probability at most $2e^{-t}$.

Use the deterministic net $\mathcal N_d$ from \eqref{eq:euclidean-net-bound} and apply \eqref{eq:cross-bernstein-radius} at every point with $t=t_{\rm dev}+d\log5$. By a union bound, the probability that at least one net point violates its bound is at most $2|\mathcal N_d|e^{-t}\le2e^{-t_{\rm dev}}$. Thus, with probability at least $1-2e^{-t_{\rm dev}}$,
\[
 |\theta^\top(\widehat D-D)w|
 \le \mathsf{K}\sqrt{\overline V}\{\sqrt{t/N}+t/N\}
 \quad\text{for every }\theta\in\mathcal N_d.
\]
On this event, applying the norm estimate in \eqref{eq:euclidean-net-bound} yields
\[
 \norm{(\widehat D-D)w}
 \le2\max_{\theta\in\mathcal N_d}|\theta^\top(\widehat D-D)w|
 \le \mathsf{K}\sqrt{\overline V}
       \left\{\sqrt{\frac{d+t_{\rm dev}}N}+\frac{d+t_{\rm dev}}N\right\}.
\]
Since $N\ge d+t_{\rm dev}$, the last linear term is at most the square-root term, which proves the bound \eqref{eq:directional-cross-norm} for $(\widehat D-D)w$. To bound $|w^\top(\widehat D-D)w|$, we apply \eqref{eq:cross-bernstein-radius} once, with $\theta=w/\norm w$ and $t=t_{\rm dev}$, and use $\norm w\le1$. Finally, $E_D=(\widehat D-D)+(D-D^0)$, and $\norm{D-D^0}\le \mathsf{K}\nu$ by \eqref{eq:population-mismatch}, which adds at most $\mathsf{K}\nu$ to both bounds.
\end{proof}

The auxiliary matrix $J$ enters the expansion through $w^\top Jw$ and through products with $Pw$. The next lemma uses only the law \eqref{eq:auxiliary-j} of $J$.

\begin{lemma}[Bounds on the auxiliary matrix $J$]
\label{lem:auxiliary-events}
With failure at most $\mathsf{K}\beta$, for $0<\beta<1/2$ and $t=\log(\mathsf{K}/\beta)$,
\begin{equation}
 \norm J\le \mathsf{K}\sqrt{(d+t)/N},\qquad |w^\top Jw|\le \mathsf{K}\sqrt{t/N}.
 \label{eq:aux-norm-event}
\end{equation}
\end{lemma}

\begin{proof}
The entries of $J$ have the law \eqref{eq:auxiliary-j}, so the GOE norm bound \eqref{eq:gaussian-matrix-norms} of Lemma~\ref{lem:probability-tools}, with $s=1/\sqrt N$, $m=d$, and $u=t$, proves the first bound. The vector $w$ is deterministic, so $w^\top Jw$ is a centered Gaussian variable with variance $2\norm w^4/N=2T^2/N$, which proves the second bound. Both bounds use only the marginal law of $J$; we condition neither on $L_0$ nor on events defined by the empirical matrices. When these bounds are intersected with events involving $L_0$, we use a union bound, so no independence between $J$ and $L_0$ is needed.
\end{proof}

We now combine Lemmas~\ref{lem:directional-qr}, \ref{lem:directional-cross}, and~\ref{lem:auxiliary-events} in the exact energy identity \eqref{eq:exact-energy}.

\begin{lemma}[Energy before clipping]
\label{lem:finite-energy}
Fix $\omega$ satisfying \eqref{eq:fixed-normalizer-condition} and $t_{\rm dev}\ge1$. Suppose $d\le N$, $N\ge d+t_{\rm dev}$, \eqref{eq:directional-dimension} holds, and $r_N$ in \eqref{eq:directional-radius} is sufficiently small. With failure at most $\mathsf{K}e^{-t_{\rm dev}}$,
\begin{gather}
 \left|\frac{H_N(f)}N-(\mathcal E-c_1)\right|
 \le \mathsf{K}\{\nu+\sqrt{\overline Vt_{\rm dev}/N}+\overline V(d+t_{\rm dev})/N\},
 \label{eq:uncut-energy-bound}\\
 \norm f/\sqrt N\le5/4.
 \label{eq:soft-output-norm}
\end{gather}
\end{lemma}

\begin{proof}
We apply the preceding bounds to \eqref{eq:energy-expansion}--\eqref{eq:energy-remainder}.
By Cauchy--Schwarz, $|(Pw)^\top E_Dw|\le\norm{Pw}\norm{E_Dw}$. Lemma~\ref{lem:directional-qr} bounds $\norm{Pw}$ by $\mathsf{K}r_N$, and Lemma~\ref{lem:directional-cross} bounds $\norm{E_Dw}$ by $\mathsf{K}\{\nu+\sqrt{\overline V(d+t_{\rm dev})/N}\}\le \mathsf{K}r_N$. Hence the first product in \eqref{eq:energy-remainder} is at most $\mathsf{K}r_N^2$. We apply Lemma~\ref{lem:auxiliary-events} with $\beta$ a sufficiently small fixed multiple of $e^{-t_{\rm dev}}$, so that $t=\log(\mathsf{K}/\beta)$ is at most a constant multiple of $t_{\rm dev}$. Its bounds then read $\norm J\le \mathsf{K}\sqrt{(d+t_{\rm dev})/N}$ and $|w^\top Jw|\le \mathsf{K}\sqrt{t_{\rm dev}/N}$. Since $\overline V\ge1$,
\begin{equation}
 (r_N+r_N^2)\sqrt{(d+t_{\rm dev})/N}\le \mathsf{K}r_N^2
\end{equation}
when $r_N$ is small, because $\sqrt{(d+t_{\rm dev})/N}\le r_N$. Since $|(Pw)^\top Jw|\le\norm{Pw}\norm J\norm w$ and $|(Pw)^\top J(Pw)|\le\norm{Pw}^2\norm J$, with $\norm{Pw}\le \mathsf{K}r_N$ and $\norm w\le1$, the last two products in \eqref{eq:energy-remainder} are also at most $\mathsf{K}r_N^2$. By \eqref{eq:population-cross}, the first term in \eqref{eq:energy-expansion} is $w^\top D^0w=\mathcal E-c_1$. We bound the next three terms by \eqref{eq:directional-cross-scalar}, \eqref{eq:directional-qr-scalar}, and the bound on $|w^\top Jw|$, and we use $|\mathcal R_N|\le \mathsf{K}r_N^2=\mathsf{K}\{\nu^2+\overline V(d+t_{\rm dev})/N\}$. After absorbing $\nu^2$ into $\nu$, and $\sqrt{\overline V}t_{\rm dev}/N$ into $\overline V(d+t_{\rm dev})/N$, this proves \eqref{eq:uncut-energy-bound}. Finally $f=QL_0w$ and $Q^\top Q=NI$, so
\begin{equation}
 \norm f/\sqrt N=\norm{L_0w}\le\norm w+\norm{Pw}\le1+\mathsf{K}r_N\le5/4,
\end{equation}
which proves \eqref{eq:soft-output-norm}. We intersect the events of Lemmas~\ref{lem:directional-qr}, \ref{lem:directional-cross}, and~\ref{lem:auxiliary-events} by a union bound. We never condition the Gaussian law of $J$, or its bounds, on the matrix $P$, which depends on $J$ through the observations.
\end{proof}

In \eqref{eq:uncut-energy-bound}, the term $\sqrt{\overline Vt_{\rm dev}/N}$ bounds the fluctuations of the scalars $w^\top E_Dw$, $x^\top Pw$, and $w^\top Jw$, and the term $\overline V(d+t_{\rm dev})/N$ bounds the products of two error terms, such as those in \eqref{eq:energy-remainder}. With $\overline V=O(L_\varepsilon^2)$ and $d=O(\varepsilon^{-1})$, both terms are $O(\varepsilon)$ when $N\ge \mathsf{K}\varepsilon^{-2}L_\varepsilon^2t_{\rm dev}$.

\subsection{Clipping}
\label{sec:matrix-clipping}

Clipping replaces the magnetization vector $f$ by $v=\clip(f,[-1,1]^N)$. We bound the average squared distance $\norm{f-v}^2/N$ in terms of the scalar terminal norm $\mathfrak s=\norm{M_K^\omega-m(T,Y_K)}_2$, without restricting its value in Lemma~\ref{lem:empirical-clipping}. We first decompose this terminal error into a continuous martingale and a bounded remainder. If the remainder is bounded by $b_0$, the martingale's second moment is at most $(\mathfrak s+b_0)^2$, to which we apply Lemma~\ref{lem:bracket-moments}. For the application in Section~\ref{sec:main-proofs}, we use \eqref{eq:scalar-final} and the chosen parameters to make both $\mathfrak s$ and $b_0$ small. The Brownian bridges are used only in the proof and are not generated by the algorithm.

\begin{lemma}[A martingale decomposition of the terminal error]
\label{lem:terminal-error-decomposition}
\label{app:clipping-concentration}
Fix deterministic $\omega$ satisfying \eqref{eq:fixed-normalizer-condition}, and let $(Y_k,M_k^\omega)$ follow \eqref{eq:scalar-y}--\eqref{eq:scalar-m} with the quadrature coefficients \eqref{eq:drift-quadrature}. Suppose $h\Lambda(T)\le1$, and put
\begin{equation}
 b_0=\mathsf{K}\left\{\frac{hL_\tau^2}{\tau}
       +\eta_{\rm q}\left(1+\frac{hL_\tau^{3/2}}{\sqrt\tau}\right)\right\}.
 \label{eq:clipping-remainder-size}
\end{equation}
On a probability space enlarged by independent Brownian bridges inside the time-grid cells,
\begin{equation}
 M_K^\omega-m(T,Y_K)=Z_T+\mathcal D_T,\qquad
 \langle Z\rangle_T\le \mathsf{K}\overline V,\qquad |\mathcal D_T|\le b_0,
 \label{eq:clipping-martingale-decomposition}
\end{equation}
where $Z$ is a continuous square-integrable martingale starting at zero. For independent coordinate paths, the bridges and resulting martingales and remainders can be constructed independently across coordinates.
\end{lemma}

\begin{proof}
Use the Brownian motion $W$ in \eqref{eq:brownian-bridge-interpolation}, whose grid increments are $G_k$. Let $k(r)$ satisfy $t_{k(r)}\le r<t_{k(r)+1}$. Put
\begin{equation}
 \delta_k^{\rm q}=(\Gamma_k^{\rm q}-\Gamma_k)/h,
 \qquad |\delta_k^{\rm q}|\le\eta_{\rm q}/2,
\end{equation}
and interpolate on a cell by
\begin{equation}
 \dd Y_t=\{\gamma(t)m_k+\delta_k^{\rm q}m_k\}\dd t+\dd W_t,
 \qquad m_k=m(t_k,Y_{t_k}).
 \label{eq:implemented-interpolation}
\end{equation}
Integrating \eqref{eq:implemented-interpolation} over cell $k$ gives $Y_{t_{k+1}}=Y_{t_k}+\Gamma_k^{\rm q}m_k+G_k$, which is the implemented recursion \eqref{eq:scalar-y}. Abbreviate $a_t=a(t,Y_t)$, $b_t=a_x(t,Y_t)$, and $\delta_t=m(t,Y_t)-m_k$. Lemma~\ref{lem:parisi-ito}, applied with $\mathfrak d_t=\gamma(t)\delta_t-\delta_k^{\rm q}m_k$, gives
\begin{align}
 \dd\delta_t&=a_t\dd W_t-\gamma a_t\delta_t\dd t
                         +\delta_k^{\rm q}a_tm_k\dd t,\nonumber\\
 \dd a_t&=b_t\dd W_t-\gamma(a_t^2+b_t\delta_t)\dd t
                         +\delta_k^{\rm q}b_tm_k\dd t.
\end{align}
Define
\begin{align}
 \Xi_t&=a_t^2+b_t\delta_t,\nonumber\\
 \Psi_t&=a_tb_t-2\gamma a_t^2\delta_t-\gamma b_t\delta_t^2
                                      +\delta_k^{\rm q}m_k\Xi_t.
\end{align}
The product rule gives $\dd(a_t\delta_t)=\Xi_t\dd W_t+\Psi_t\dd t$, and this product is zero at the left endpoint of every cell. Since $m(0,0)=0$, summing the equation for $\delta$ over the cells yields $m(T,Y_K)=\int_0^Ta_t\dd W_t-\int_0^T\gamma a\delta\dd t+\int_0^T\delta_{k(t)}^{\rm q}a_tm_{k(t)}\dd t$. With $R_k(t)=\int_t^{t_{k+1}}\gamma(s)\dd s$, stochastic Fubini shows that $\int_0^T\gamma a\delta\dd t=\int_0^TR_{k(t)}(t)(\Xi_t\dd W_t+\Psi_t\dd t)$. Subtracting $m(T,Y_K)$ from $M_K^\omega=\int_0^Ta_{t_{k(t)}}\omega_{k(t)}^{-1}\dd W_t$, we obtain the exact decomposition
\begin{align}
 Z_T&=\int_0^T\left\{\frac{a_{t_{k(t)}}}{\omega_{k(t)}}-a_t
                                      +R_{k(t)}(t)\Xi_t\right\}\dd W_t,
 \label{eq:clipping-martingale}\\
 \mathcal D_T&=\int_0^TR_{k(t)}(t)\Psi_t\dd t
                  -\int_0^T\delta_{k(t)}^{\rm q}a_tm_{k(t)}\dd t,
 \label{eq:clipping-remainder}\\
 M_K^\omega-m(T,Y_K)&=Z_T+\mathcal D_T.\nonumber
\end{align}
All coefficients are bounded on the fixed strip, so these stochastic integrals and the stochastic Fubini calculation are justified. Since $|\delta_t|\le2$, Proposition~\ref{prop:regularity} and \eqref{eq:euler-cell-weight} imply
\begin{equation}
 |\Xi_t|\le \mathsf{K}\Lambda(t),\qquad
 |\Psi_t|\le \mathsf{K}\Lambda(t)^{3/2}+\mathsf{K}\eta_{\rm q}\Lambda(t),\qquad
 R_k(t)\le \mathsf{K}h\sqrt{\Lambda(t)}.
\end{equation}
Since $\omega_k\ge1/2$ by \eqref{eq:fixed-normalizer-condition}, these bounds imply, when $h\Lambda(T)\le1$,
\begin{equation}
 \langle Z\rangle_T
 \le \mathsf{K}\int_0^T\Lambda(t)\dd t+\mathsf{K}h^2\int_0^T\Lambda(t)^3\dd t
 \le \mathsf{K}\overline V.
\end{equation}
For the last inequality, $h\le\tau$ makes $\Lambda$ comparable across each cell, so its integral is at most a constant times its left sum in \eqref{eq:clipping-coefficient-sum}. The time integrals in \eqref{eq:clipping-remainder} satisfy
\begin{equation}
 |\mathcal D_T|\le \mathsf{K}\left\{h\int_0^T\Lambda^2
        +\eta_{\rm q}h\int_0^T\Lambda^{3/2}
        +\eta_{\rm q}\int_0^T\sqrt\Lambda\right\}\le b_0,
\end{equation}
by \eqref{eq:singular-weights}--\eqref{eq:integrable-weights}. This proves the decomposition and its deterministic bounds. Each continuous process and remainder is a function of one coordinate's increments and bridges, so independent coordinate paths with independent bridges yield independent copies.
\end{proof}

We now apply \eqref{eq:clipping-martingale-decomposition} and Lemma~\ref{lem:bracket-moments} to the independent coordinates of $f$.

\begin{lemma}[The empirical clipping loss]
\label{lem:empirical-clipping}
For fixed $\omega$ satisfying \eqref{eq:fixed-normalizer-condition}, put $\overline V=1+V$ and $\mathfrak s=\norm{M_K^\omega-m(T,Y_K)}_2$, and let $b_0$ be as in \eqref{eq:clipping-remainder-size}.
If $h\Lambda(T)\le1$, then, for $t_{\rm dev}\ge1$, with failure at most $e^{-t_{\rm dev}}$,
\begin{equation}
 \frac{\norm{f-\clip(f,[-1,1]^N)}^2}{N}
 \le4(\mathfrak s+b_0)^2+2b_0^2+\mathsf{K}\overline Vt_{\rm dev}/N.
 \label{eq:empirical-clip}
\end{equation}
At the chosen scalar parameters, $b_0=o(\varepsilon)$.
\end{lemma}

\begin{proof}
Since $m(T,Y_K)\in[-1,1]$, the scalar clipping distance is at most $|M_K^\omega-m(T,Y_K)|$. By Proposition~\ref{prop:gaussian-law}, the coordinates of $f$ are independent copies of $M_K^\omega$, so
\[
 \E\!\left[\frac{\norm{f-\clip(f,[-1,1]^N)}^2}{N}\right]
 =\E[\dist(M_K^\omega,[-1,1])^2]\le\mathfrak s^2.
\]
This bounds the expected empirical loss. To prove \eqref{eq:empirical-clip} with its stated failure probability, we also control the fluctuations of the realized average.

By Lemma~\ref{lem:terminal-error-decomposition}, the terminal error has the decomposition \eqref{eq:clipping-martingale-decomposition}. Since $Z_T=M_K^\omega-m(T,Y_K)-\mathcal D_T$ and $|\mathcal D_T|\le b_0$, the triangle inequality shows that $\mu_Z:=\E Z_T^2\le(\mathfrak s+b_0)^2$. The scalar clipping distance is at most $|Z_T|+b_0$, and its square is at most $2Z_T^2+2b_0^2$. The coordinatewise construction in Lemma~\ref{lem:terminal-error-decomposition} supplies independent martingales with terminal values $Z_{T,i}$ distributed as $Z_T$ and quadratic variation at most $\mathsf{K}\overline V$. Applying \eqref{eq:bracket-bernstein} with $v=\mathsf{K}\overline V$, we obtain, with failure at most $e^{-t_{\rm dev}}$,
\[
 \begin{aligned}
 \frac{\norm{f-\clip(f,[-1,1]^N)}^2}{N}
 &\le\frac2N\sum_{i=1}^N Z_{T,i}^2+2b_0^2\\
 &\le4\mu_Z+2b_0^2+\mathsf{K}\overline Vt_{\rm dev}/N\\
 &\le4(\mathfrak s+b_0)^2+2b_0^2+\mathsf{K}\overline Vt_{\rm dev}/N.
 \end{aligned}
\]
The first inequality holds on every outcome; the second is the concentration estimate for the squared martingales, and the third uses the bound on $\mu_Z$. This proves \eqref{eq:empirical-clip}. Finally,
\begin{equation}
 hL_\tau^2/\tau=O(\varepsilon^{4/3}L_\varepsilon^{1/3})=o(\varepsilon),
\end{equation}
and the two terms with $\eta_{\rm q}$ in \eqref{eq:clipping-remainder-size} are smaller because $\eta_{\rm q}=c_{\rm q}\varepsilon^2/R$ by \eqref{eq:quadrature-tolerance}. This proves $b_0=o(\varepsilon)$.
\end{proof}

The bound in Lemma~\ref{lem:empirical-clipping} holds for any $\mathfrak s$ under its stated hypotheses. In the application, we have $\mathfrak s\le\varepsilon/1000$ by \eqref{eq:scalar-final} and $b_0=o(\varepsilon)$ by \eqref{eq:clipping-remainder-size}. The distance to the specified $m(T,Y_K)\in[-1,1]$ is stronger than a bound on distance to the interval: through \eqref{eq:clipping-martingale-decomposition}, it implies $\E[Z_T^2]\le(\mathfrak s+b_0)^2=O(\varepsilon^2)$. The right side of \eqref{eq:empirical-clip} is therefore $O(\varepsilon^2)$ once $N\ge \mathsf{K}\varepsilon^{-2}\overline Vt_{\rm dev}$.

The following deterministic inequality converts the distance $\norm{f-v}$ into a bound on the energy loss from clipping.

\begin{lemma}[Energy loss from clipping, adapted from {\cite[Section~5.3]{AMS}}]
\label{lem:clipping-energy}
For symmetric $A$ and $v=\clip(f,[-1,1]^N)$,
\begin{equation}
 |H_N(f)-H_N(v)|/N\le\frac{\norm A}{2}
       \frac{\norm{f-v}}{\sqrt N}
       \left(\frac{\norm f}{\sqrt N}+\frac{\norm v}{\sqrt N}\right).
 \label{eq:clip-energy}
\end{equation}
\end{lemma}

\begin{proof}
Since $A$ is symmetric, $f^\top Af-v^\top Av=(f-v)^\top A(f+v)$; dividing by $2N$ and applying Cauchy--Schwarz proves \eqref{eq:clip-energy}.
\end{proof}

\subsection{Sequential rounding}
\label{sec:matrix-rounding}

Sequential rounding converts $v\in[-1,1]^N$ to a spin configuration $\sigma\in\{-1,1\}^N$. The possible decrease in energy is bounded by the diagonal contribution of $A$.

\begin{lemma}[Sequential rounding, adapted from {\cite[Section~5.3]{AMS}}]
\label{lem:rounding}
For symmetric $A$ and $v\in[-1,1]^N$, sequential rounding costs $O(N^2)$ and returns spins satisfying
\begin{equation}
 H_N(\sigma)/N\ge H_N(v)/N-\frac1{2N}\sum_i|A_{ii}|.
 \label{eq:round-energy}
\end{equation}
For GOE input, with probability at least $1-3e^{-u}$,
\begin{equation}
 \norm A\le2+2\sqrt{u/N},\qquad
 N^{-1}\sum_i|A_{ii}|\le(2+2\sqrt u)/\sqrt N.
 \label{eq:diagonal-bound}
\end{equation}
\end{lemma}

\begin{proof}
With the diagonal of $A$ removed, the Hamiltonian is affine in each coordinate. Replacing one coordinate by the sign of its current local field therefore never decreases the zero-diagonal Hamiltonian, and sequential rounding makes $N$ such replacements. Because $\sigma_i^2=1$, the diagonal terms contribute $\tfrac12\sum_iA_{ii}(1-v_i^2)$ to $H_N(\sigma)-H_N(v)$. This contribution is at least $-\tfrac12\sum_i|A_{ii}|$ because $0\le1-v_i^2\le1$, which proves \eqref{eq:round-energy}. One initial product and $N$ column updates cost $O(N^2)$. The norm bound in \eqref{eq:diagonal-bound} is \eqref{eq:gaussian-matrix-norms} with $s=1/\sqrt N$ and $m=N$. For the diagonal, write $A_{ii}=\sqrt{2/N}g_i$ with independent standard Gaussian $g_i$. The average $N^{-1}\sum_i|A_{ii}|$ has mean $2/\sqrt{\pi N}$ and is $\sqrt2/N$-Lipschitz as a function of $g$. Gaussian concentration and a union bound with the norm event prove \eqref{eq:diagonal-bound}.
\end{proof}

\subsection{Comparison with the optimum}
\label{sec:matrix-optimum}

To compare $H_N(\sigma)$ with $\OPT_N$, we bound $\OPT_N/N$ above in terms of $\Par$. The following bound holds at every finite $N$ and uses no rate of convergence of $\E\OPT_N/N$ to $\Par$.

\begin{proposition}[Finite-size optimum bound {\cite{GuerraToninelli,BLM}}]
\label{prop:optimum}
For every $N$, $\E\OPT_N/N\le\Par$, and
\begin{equation}
                  \Prob\{\OPT_N/N>\Par+u\}\le e^{-Nu^2},\qquad u>0.
 \label{eq:optimum-tail}
\end{equation}
\end{proposition}

\begin{proof}
For the expectation bound, following the superadditivity argument of \cite[Theorem~3]{GuerraToninelli}, we compare a $kN$-spin Hamiltonian $X_\sigma$ with the sum $Y_\sigma$ of $k$ independent $N$-spin Hamiltonians, one for each block of $N$ consecutive coordinates. If $r_j$ is the overlap of $\sigma$ and $\tau$ on block $j$, then
\begin{equation}
 \E X_\sigma X_\tau=\frac{kN}{2}\left(\frac1k\sum_jr_j\right)^2
       \le\frac N2\sum_jr_j^2=\E Y_\sigma Y_\tau,
\end{equation}
and the variances agree because $r_j=1$ for every $j$ when $\sigma=\tau$. For $F_b(x)=b^{-1}\log\sum_\sigma e^{bx_\sigma}$, Gaussian integration by parts shows that the derivative of $\E F_b(\sqrt sX+\sqrt{1-s}Y)$ in $s$ is
\begin{equation}
 \frac12\sum_{\sigma,\tau}(\E X_\sigma X_\tau-\E Y_\sigma Y_\tau)
                      \E\partial_\sigma\partial_\tau F_b\ge0.
\end{equation}
The terms with $\sigma=\tau$ vanish because $X_\sigma$ and $Y_\sigma$ have equal variances. For $\sigma\ne\tau$, the covariance differences are nonpositive and the Hessian entries are $-bp_\sigma p_\tau\le0$, where $p_\sigma$ are the Gibbs weights proportional to $e^{bx_\sigma}$. Hence the derivative is nonnegative, and $\E F_b(X)\ge\E F_b(Y)$. Letting $b\to\infty$, we obtain $\E\OPT_{kN}\ge k\E\OPT_N$, because the maximum of $Y$ is the sum of the maxima of its $k$ independent parts. Thus $\E\OPT_N/N\le\E\OPT_{kN}/(kN)$ for every $k$, and letting $k\to\infty$ with $\E\OPT_{kN}/(kN)\to\Par$ proves the expectation bound. For the tail bound, represent $A=(G+G^\top)/\sqrt{2N}$, where $G$ has independent standard Gaussian entries. Each $H_N(\sigma)/N$ is a $1/\sqrt{2N}$-Lipschitz function of the entries of $G$, and so is their maximum $\OPT_N/N$. Gaussian concentration above the mean, together with the expectation bound, proves \eqref{eq:optimum-tail}.
\end{proof}
\section{The relative guarantee and complete work bound}
\label{sec:main-proofs}

We prove Theorem~\ref{thm:main}, including its $O(N^2/\varepsilon)$ complete work bound above the sufficient dimension, and then Corollary~\ref{thm:afptas}. The main step, Lemma~\ref{lem:eg-energy}, applies for $0<\varepsilon\le\varepsilon_0$ above \eqref{eq:main-size}: with probability at least $1-\alpha$, Algorithm~\ref{alg:egiamp} satisfies $H_N(\sigma)\ge(\Par-\varepsilon/2)N$, while $\OPT_N\le(\Par+\varepsilon/2)N$. We fix the algorithmic constants so that \eqref{eq:weighted-smallness}, the smallness condition on $\sqrt R\eta_{\rm q}$ in Lemma~\ref{lem:euler-quadrature}, and the accuracy range of Lemma~\ref{lem:partition} hold. We verify the matrix-analysis hypotheses and combine the energy, clipping, rounding, and optimum events by a union bound. After counting all work, the positive lower bound on $\Par$ converts additive accuracy into relative accuracy. Returning $\one$ below the threshold then proves the AFPRAS corollary.

\subsection{Energy bounds above the dimension threshold}

\begin{lemma}[Energy estimates for GEG-IAMP]
\label{lem:eg-energy}
There are universal $\mathsf{K},\varepsilon_0>0$ and fixed algorithmic constants with the following property. For $0<\varepsilon\le\varepsilon_0$ and $0<\alpha<1/2$, let $L_\varepsilon,L_\alpha$ be as in \eqref{eq:accuracy-logarithms}. If
\begin{equation}
                 N\ge \mathsf{K}\varepsilon^{-2}L_\varepsilon^2L_\alpha
 \label{eq:main-size}
\end{equation}
then, with probability at least $1-\alpha$, the output $\sigma$ of Algorithm~\ref{alg:egiamp} at internal additive accuracy per spin $\varepsilon$ and the optimum of the instance satisfy
\begin{equation}
 H_N(\sigma)\ge(\Par-\varepsilon/2)N,\qquad
                         \OPT_N\le(\Par+\varepsilon/2)N.
 \label{eq:main-gap}
\end{equation}
In particular $0\le\OPT_N-H_N(\sigma)\le\varepsilon N$ on this event. Before simplification, its work is
\begin{equation}
 O\{N^2B+NB^2+NK+K+\varepsilon^{-3}L_\varepsilon^2+N^2\},
 \label{eq:main-work}
\end{equation}
where $B=\Theta(\varepsilon^{-1})$ and $K=\Theta(\varepsilon^{-2}L_\varepsilon^2)$ by \eqref{eq:partition-size}. Under \eqref{eq:main-size}, this work is $O(N^2/\varepsilon)$. There are $B+1$ executed dense products and storage is $O(N^2)$ on this range.
\end{lemma}

\begin{proof}
We fix the constants in the following order. First choose $c_\tau$ and $c_h$ in Lemma~\ref{lem:partition} so that the right sides of \eqref{eq:scalar-clipping} and \eqref{eq:scalar-energy-bound} are small fixed multiples of $\varepsilon$. Next choose $c_{\rm q}$ so that the additional terms in Lemma~\ref{lem:euler-quadrature} keep these bounds small fixed multiples of $\varepsilon$. Then choose $c_\nu$ so that the changes of at most $\mathsf{K}\nu$ in Lemma~\ref{lem:fixed-normalizers} preserve \eqref{eq:scalar-final}, and so that the term $\mathsf{K}\nu$ in \eqref{eq:uncut-energy-bound} is a small fixed multiple of $\varepsilon$. Choose $c_B\le1/16$, so that $\Delta\le\varepsilon/16$. With these constants fixed, decrease $\varepsilon_0$ so that Proposition~\ref{prop:unit-normalizers} applies, the smallness conditions of Section~\ref{sec:scalar-analysis} hold, and $b_0\le\varepsilon/1000$ in \eqref{eq:clipping-remainder-size}. In particular, \eqref{eq:weighted-smallness} holds; since $L_\tau\ge1$, it implies $h\Lambda(T)=hL_\tau/\tau\le hL_\tau^3/\tau\le1$, as required in Lemma~\ref{lem:empirical-clipping}. From now on $\omega_j=1$ in all matrix estimates, and the dimension constant is chosen last. By Proposition~\ref{prop:virtual-final-query}, the executed output equals the output of the extended query construction, so it suffices to prove the energy bounds for that construction.

The extended construction has $d=B+1$ query directions, including the final proof-only query $q_d$. Execution instead uses $B$ query products and the distinct rounding product $Av$; their total count is also $B+1$.

Set $L=L_\varepsilon$ and $t_{\rm dev}=\log(\mathsf{K}/\beta)$ with $\beta=\alpha/64$, so $t_{\rm dev}\asymp L_\alpha\ge1$. We first verify the hypotheses of Section~\ref{sec:matrix-analysis} at the dimension
\begin{equation}
                 N_*=\mathsf{K}_*\varepsilon^{-2}L^2L_\alpha.
 \label{eq:dimension-test}
\end{equation}
Here $\mathsf{K}_*$ is a large constant chosen last. At the end of the dimension checks, we explain why they also hold for every $N\ge N_*$. By \eqref{eq:partition-size}, \eqref{eq:clipping-coefficient-sum}, \eqref{eq:quadrature-tolerance}, and \eqref{eq:normalizer-tolerance},
\begin{equation}
 d\le \mathsf{K}\varepsilon^{-1},\quad \overline V\le \mathsf{K}L^2,\quad
 R\le \mathsf{K}\varepsilon^{-2/3}L^{4/3},\quad
 \nu\le c_\nu\varepsilon,\quad L_\tau\asymp L.
 \label{eq:dimension-basic-bounds}
\end{equation}

\paragraph{The energy before clipping.}
The conditions $d\le N$ and $N\ge d+t_{\rm dev}$ of Lemmas~\ref{lem:directional-qr}--\ref{lem:finite-energy} hold at $N_*$, because $d\le \mathsf{K}\varepsilon^{-1}$ and $t_{\rm dev}\le \mathsf{K}L_\alpha$. The remaining condition $N\ge \mathsf{K}R\{t_{\rm dev}+\log(\mathsf{K}d)\}$ in \eqref{eq:directional-dimension} also holds, because
\begin{equation}
 \frac{R\{t_{\rm dev}+\log(\mathsf{K}d)\}}{N_*}
       \le\frac{\mathsf{K}}{\mathsf{K}_*}\varepsilon^{4/3}L^{1/3}.
 \label{eq:dimension-matrix-check}
\end{equation}
Substituting \eqref{eq:dimension-basic-bounds} into $r_N^2$ in \eqref{eq:directional-radius} and into the right side of \eqref{eq:uncut-energy-bound}, we obtain
\begin{align}
 r_{N_*}^2&\le \mathsf{K}c_\nu^2\varepsilon^2
              +\frac{\mathsf{K}}{\mathsf{K}_*}(\varepsilon/L_\alpha+\varepsilon^2),\nonumber\\
 \sqrt{\overline Vt_{\rm dev}/N_*}&\le \mathsf{K}\varepsilon/\sqrt{\mathsf{K}_*},\qquad
 \overline V(d+t_{\rm dev})/N_*\le\frac{\mathsf{K}}{\mathsf{K}_*}
                              (\varepsilon/L_\alpha+\varepsilon^2).
 \label{eq:dimension-scalar-check}
\end{align}
Hence, once $c_\nu$ and $\varepsilon_0$ are small and $\mathsf{K}_*$ is large, $r_{N_*}$ satisfies the smallness condition of Lemma~\ref{lem:finite-energy}, and the right side of \eqref{eq:uncut-energy-bound} is at most $\varepsilon/8$. The bounds in \eqref{eq:dimension-scalar-check} hold for every $t_{\rm dev}\ge1$; they do not require $t_{\rm dev}\le d$.

\paragraph{Clipping and the remaining Gaussian events.}
By \eqref{eq:scalar-final}, $\mathfrak s\le\varepsilon/1000$, and the constants were chosen so that $b_0\le\varepsilon/1000$. The first two terms on the right side of \eqref{eq:empirical-clip} are then at most $18\varepsilon^2/10^6$, which is less than one fifth of $(\varepsilon/100)^2$. Hence, by Lemma~\ref{lem:empirical-clipping}, $\norm{f-v}/\sqrt N\le\varepsilon/100$ when
\begin{equation}
                  N\ge \mathsf{K}\varepsilon^{-2}\overline Vt_{\rm dev},
 \label{eq:dimension-clipping-check}
\end{equation}
with a sufficiently large fixed constant. Since $\overline V\le \mathsf{K}L^2$ and $t_{\rm dev}\le \mathsf{K}L_\alpha$, this condition holds at $N_*$ after $\mathsf{K}_*$ is enlarged. The bounds on $\norm A$ and on the diagonal in \eqref{eq:diagonal-bound}, and the optimum tail \eqref{eq:optimum-tail}, require only $N\ge \mathsf{K}\varepsilon^{-2}t_{\rm dev}$, which also holds at $N_*$. In particular, $e^{-N\varepsilon^2/4}\le\beta$, and by Lemma~\ref{lem:rounding}, $\norm A\le4$ and the diagonal term $\frac1{2N}\sum_i|A_{ii}|$ in \eqref{eq:round-energy} is at most $\varepsilon/16$.

Every dimension-dependent term in these checks is a constant multiple of $N^{-1}$ or $N^{-1/2}$, with no logarithm of $N$, so it decreases as $N$ grows. The same checks therefore hold at every $N\ge N_*$, including when $\log(1/\alpha)>d$.

\paragraph{The common success event and the energy losses.}
We assign each of the following events failure probability at most a fixed multiple of $\beta$: the bounds on $Pw$ and $x^\top Pw$ in Lemma~\ref{lem:directional-qr}; the bounds on $E_Dw$ and $w^\top E_Dw$ in Lemma~\ref{lem:directional-cross}; the clipping bound of Lemma~\ref{lem:empirical-clipping}; the bounds on $\norm J$ and $w^\top Jw$ in \eqref{eq:aux-norm-event}; the bounds on $\norm A$ and on the diagonal in \eqref{eq:diagonal-bound}; and the optimum tail \eqref{eq:optimum-tail}. Condition~\eqref{eq:fixed-normalizer-condition} holds deterministically for $\omega_j=1$, so it contributes no failure event. We increase the fixed constants in the confidence parameters of these finitely many events so that their total failure probability is at most $\alpha$. A union bound controls their intersection, so no independence between the events is needed.

On the intersection of these events, the right side of \eqref{eq:uncut-energy-bound} is at most $\varepsilon/8$ by the choice of $\mathsf{K}_*$ above. Since $\mathcal E\ge\Par-\varepsilon/16$ by \eqref{eq:scalar-final} and $c_1\le\Delta\le\varepsilon/16$,
\begin{equation}
 H_N(f)/N\ge\Par-\varepsilon/16-\varepsilon/16-\varepsilon/8
                         =\Par-\varepsilon/4.
\end{equation}
On the same event, $\norm{f-v}/\sqrt N\le\varepsilon/100$ by the clipping check above. Since $\norm f/\sqrt N\le5/4$ by \eqref{eq:soft-output-norm}, $\norm v/\sqrt N\le1$, and $\norm A\le4$, \eqref{eq:clip-energy} bounds the clipping loss by
\begin{equation}
 \frac42\frac\varepsilon{100}\left(\frac54+1\right)
                         =\frac{9\varepsilon}{200}<\frac\varepsilon{16}.
\end{equation}
By \eqref{eq:round-energy} and the diagonal bound above, rounding loses at most $\varepsilon/16$. Hence $H_N(\sigma)/N\ge\Par-\varepsilon/4-\varepsilon/16-\varepsilon/16=\Par-3\varepsilon/8\ge\Par-\varepsilon/2$. On the same event, $\OPT_N\le(\Par+\varepsilon/2)N$ by \eqref{eq:optimum-tail} with $u=\varepsilon/2$, which completes the proof of \eqref{eq:main-gap}. Since $\sigma\in\{-1,1\}^N$, $\OPT_N\ge H_N(\sigma)$, so the gap $\OPT_N-H_N(\sigma)$ is nonnegative.

\paragraph{Work and storage.}
The executed observation loop uses $B$ dense products, at cost $O(N^2B)$. At query $s$, projecting $F_s$ onto the earlier queries, selecting and projecting a coordinate vector by the rule after \eqref{eq:query-proposals} if the residual is zero, and forming the Gaussian corrections in \eqref{eq:gaussian-coupling} cost $O(Ns)$, hence $O(NB^2)$ in total. Sampling the leading $B\times B$ block of $J$ costs $O(B^2)$. For each coordinate and group, generating the $m_{\rm f}-1$ side variables and applying \eqref{eq:householder} cost $O(m_{\rm f})$, and the per-step updates \eqref{eq:vector-scalar-update} cost the same. Over $N$ coordinates and $B$ groups, these operations cost $O(Nm_{\rm f}B)=O(NK)$. The point quadrature of Lemma~\ref{lem:drift-quadrature} is performed once, at cost $O(K+\varepsilon^{-3}L^2)$. Clipping and rounding cost $O(N^2)$, including the additional product $Av$. This proves \eqref{eq:main-work} and the executed count $B+1$. The final query $q_d$ and the entries $J_{rd}$ appear only in the proof and are never computed.

We now show that $N^2B$ dominates the other terms of \eqref{eq:main-work} when $N\ge N_*$. At $N=N_*$ we have $N_*^2B\asymp\varepsilon^{-5}L^4L_\alpha^2$, and direct comparison shows
\begin{equation}
 \begin{aligned}
 \frac{NK}{N^2B}&\le \mathsf{K}\varepsilon/L_\alpha,&
 \frac{NB^2}{N^2B}&\le \mathsf{K}\varepsilon/(L^2L_\alpha),\\
 \frac{K}{N^2B}&\le \mathsf{K}\varepsilon^3/(L^2L_\alpha^2),&
 \frac{\varepsilon^{-3}L^2}{N^2B}&\le \mathsf{K}\varepsilon^2/(L^2L_\alpha^2).
 \end{aligned}
 \label{eq:work-domination}
\end{equation}
All ratios in \eqref{eq:work-domination} are uniformly bounded because $L_\alpha\ge1$, and they decrease for larger $N$. The term $N^2$ for clipping and rounding is at most $N^2B$ because $B\ge1$. Thus the complete work is $O(N^2B)=O(N^2/\varepsilon)$. The algorithm stores $A$, the $B$ query vectors, the current states of all coordinates, the leading $B\times B$ block of $J$, and the $K$ quadrature coefficients. Per-step Gaussian increments can be processed one group and coordinate at a time, with $O(m_{\rm f})$ temporary storage. Since $B,K\le \mathsf{K}N$ on this range, the storage $O(N^2+NB+K)$ is $O(N^2)$.
\end{proof}

\subsection{Relative approximation and the AFPRAS corollary}

To obtain the relative guarantee, we bound $\OPT_N/N$ below by a constant independent of accuracy. On the event of Lemma~\ref{lem:eg-energy}, $\OPT_N\ge H_N(\sigma)\ge(\Par-\varepsilon/2)N$, so a lower bound on $\Par$ suffices.

\begin{lemma}[Bounds on the Parisi value]
\label{lem:positive-value}
Under Assumption~\ref{ass:frsb}, $1-\log2\le\Par\le1$.
\end{lemma}

\begin{proof}
Since $\gamma$ is nondecreasing and $\int_0^1\gamma=1$ by Proposition~\ref{prop:regularity}, we have $(1-t)\gamma(t)\le\int_t^1\gamma\le1$, so $\gamma(t)\le(1-t)^{-1}$. For $t\le1/2$, the formula for $g$ in \eqref{eq:g-identities} and $\int_0^t\gamma\le t\gamma(t)$ imply $g(t)=1-\int_0^t\gamma\ge1-t\gamma(t)\ge(1-2t)/(1-t)$. Since $g\ge0$, we may drop the interval $(1/2,1)$ from \eqref{eq:parisi-integral} and integrate this nonnegative lower bound over $[0,1/2]$:
\begin{equation}
 \Par\ge\int_0^{1/2}\frac{1-2t}{1-t}\dd t
       =[2t+\log(1-t)]_0^{1/2}=1-\log2.
\end{equation}
The upper bound follows from $g\le1$ in \eqref{eq:g-identities} and \eqref{eq:parisi-integral}.
\end{proof}

\begin{proof}[Proof of Theorem~\ref{thm:main}]
Put $p_0=1-\log2$ and fix $c\le\min(\varepsilon_0,p_0/4)$. Set $\etaacc=c\varepsilon$, and choose the threshold constant in \eqref{eq:all-dimension-threshold} so that every $N\ge N_0$ satisfies \eqref{eq:main-size} at accuracy $\etaacc$. For $N\ge N_0$, Lemma~\ref{lem:eg-energy} at accuracy $\etaacc$ and Lemma~\ref{lem:positive-value} imply that, on an event of probability at least $1-\alpha$,
\begin{equation}
 \OPT_N\ge H_N(\sigma)\ge(p_0-\etaacc/2)N\ge(p_0/2)N,
 \qquad \OPT_N-H_N(\sigma)\le\etaacc N\le\varepsilon\OPT_N.
\end{equation}
In the first chain, the first inequality holds because $\sigma\in\{-1,1\}^N$, the second follows from \eqref{eq:main-gap} and $\Par\ge p_0$, and the third from $\etaacc\le c\le p_0/4$. In the second chain, the first inequality is \eqref{eq:main-gap}, and the second follows from $\OPT_N\ge(p_0/2)N$ and $\etaacc=c\varepsilon\le p_0\varepsilon/4$. This proves \eqref{eq:main-relative}.

Lemma~\ref{lem:eg-energy} bounds complete work by $O(N^2/\etaacc)=O(N^2/\varepsilon)$ and storage by $O(N^2)$. The fixed factor $c$ implies $L_{c\varepsilon}\le \mathsf{K}L_\varepsilon$, so the grid scales and the ledger \eqref{eq:main-work} give the stated $K$, $B$, product count, and \eqref{eq:afptas-work}. Since $c\le\varepsilon_0$, Proposition~\ref{prop:unit-normalizers} applies at internal accuracy $\etaacc$ for every requested $0<\varepsilon\le1$.
\end{proof}

\begin{proof}[Proof of Corollary~\ref{thm:afptas}]
For $N\ge N_0$, Theorem~\ref{thm:main} proves the stronger relative guarantee \eqref{eq:main-relative}. Enlarge its threshold constant, if necessary, so that $N_0\ge\log(2/\alpha)$.

For $N<N_0$, Algorithm~\ref{alg:afptas} returns the all-plus vector $\one$ before it constructs a grid or computes quadrature coefficients. Deterministically, $|H_N(\one)|\le N\norm A/2$ and $|\OPT_N|\le N\norm A/2$. Since $0\le1-\varepsilon<1$, we have $(1-\varepsilon)\OPT_N\le|\OPT_N|$ whether or not $\OPT_N$ is negative, so
\begin{equation}
                 (1-\varepsilon)\OPT_N-H_N(\one)\le N\norm A.
\end{equation}
With $u=\log(2/\alpha)$, the GOE norm bound \eqref{eq:gaussian-matrix-norms} with $s=1/\sqrt N$ and $m=N$ shows that $\norm A\le2+2\sqrt{u/N}$ with probability at least $1-2e^{-u}=1-\alpha$. We enlarge the threshold constant so that $N_0\ge u$. Since $N<N_0$, we then have
\begin{equation}
                       N\norm A\le2N+2\sqrt{Nu}\le4N_0.
\end{equation}
This proves \eqref{eq:afptas-result} below the threshold. In this branch, computing $N_0$ and testing $N<N_0$ use arithmetic, logarithms, and integer rounding; reading the matrix and returning the fixed vector cost $O(N^2)$. Above the threshold, Theorem~\ref{thm:main} gives the stated work and storage. Finally, $L_{c\varepsilon}\le \mathsf{K}L_\varepsilon$ and the threshold definition imply \eqref{eq:eg-afptas-allowance}, so the additive term is independent of $N$ and the work satisfies Definition~\ref{def:afptas}.
\end{proof}
\section{Open questions}
\label{sec:open-questions}

We conclude with two kinds of questions: whether the bounds in our guarantees can be sharpened, namely those on the accuracy attainable at a given dimension, the number of matrix queries, and the sufficient dimension for AMS IAMP; and whether the guarantees extend to finite-precision computation and to non-Gaussian disorder.

\paragraph{Sharper finite-dimensional accuracy.}
Does \gegiamp, or another polynomial-time algorithm under the same assumptions and oracle model, achieve a per-spin deficit of $o(\log(eN)/\sqrt N)$ at fixed confidence? We compare the output energy and the optimum separately with the Parisi value, whereas a direct comparison might exploit cancellations lost in these separate bounds. The question concerns finite-accuracy approximation rather than exact optimization, discussed in Section~\ref{sec:exact}; our estimates give no lower bound on the achievable deficit.

\paragraph{Fewer matrix queries.}
Can one obtain an $O(\varepsilon N)$ energy deficit with prescribed confidence using $o(\varepsilon^{-1})$ dense matrix queries, while retaining polynomial total work and a polynomial sufficient dimension? In our proof, the omitted first-group contribution to the deterministic leading term is at most $\Delta$, which leads to $B=\Theta(\varepsilon^{-1})$. Avoiding this omission or changing how the updates are grouped could permit longer observation intervals. A higher-order scalar discretization alone would not settle the question, because we would also have to generate its increments by a matrix construction with the required energy correlation and a controlled clipping loss.

\paragraph{A sharper dimension bound for AMS IAMP.}
Does AMS IAMP, with the cutoff on field increments and the empirical coefficients of Appendix~\ref{sec:iamp-comparison}, attain its relative guarantee above a dimension polynomial in $1/\varepsilon$ and $\log(1/\alpha)$? The sufficient dimension in Theorem~\ref{thm:onsager-main} is exponential in a power of $1/\varepsilon$, because we bound the distance from the Onsager iteration to its Gaussian comparator and the effect of coefficient errors over many steps. A sharper stability estimate, or a direct energy estimate for the Onsager iteration, might avoid this exponential dependence. The comparator uses the AMS IAMP maps, rather than the grouped construction, so our guarantee for \gegiamp\ does not answer this question.

\paragraph{Finite precision and computation of the Parisi solution.}
Can the work bound in the oracle model of Definition~\ref{def:oracles} be replaced by a polynomial bound on bit complexity that includes computing the minimizing order parameter and the required Parisi derivatives? Appendix~\ref{app:preprocessing} assumes construction costs for exact evaluators. Approximate evaluation would require error bounds for the derivatives and quadrature, stability of the adaptive query projections when a residual is small, and a finite representation of the random input and Gaussian samples. These estimates remain to be established beyond the present exact real arithmetic model.

\paragraph{Beyond Gaussian disorder.}
Does a guarantee like Theorem~\ref{thm:main}, with polynomial dependence on $1/\varepsilon$ and $\log(1/\alpha)$, hold for SK-type matrices with independent non-Gaussian entries on and above the diagonal, matched first two moments, and suitable tail bounds? Bayati, Lelarge and Montanari prove universality for polynomial AMP updates at fixed iteration counts~\cite{BayatiLelargeMontanari}; Han proves quantitative universality for first-order methods under regularity assumptions~\cite{HanGFOM}. These results do not directly cover the adaptive projections in \gegiamp. The exact Gaussian law of its corrected query outputs rests on the conditional law of the remaining GOE block in Lemma~\ref{lem:conditional-goe-blocks}, which is specific to Gaussian disorder. An extension would require a replacement for that block law or a quantitative comparison preserving polynomial parameter dependence, in addition to an asymptotic formula for the optimum.

\section*{Acknowledgment}
We thank Rob Otter and Ruslan Shaydulin for executive support and valuable feedback on this project. We also acknowledge our colleagues at the Global Technology Applied Research (GTAR) Center of JPMorganChase. We thank Enrico Fontana and James Sud for numerical investigations conducted during J.S.'s summer internship at GTAR. Their investigations helped motivate many of the questions answered in this manuscript.

\section*{Disclaimer}
This paper was prepared for informational purposes by the Global Technology Applied Research center of JPMorgan Chase \& Co. This paper is not a product of the Research Department of JPMorgan Chase \& Co. or its affiliates. Neither JPMorgan Chase \& Co. nor any of its affiliates makes any explicit or implied representation or warranty and none of them accept any liability in connection with this paper, including, without limitation, with respect to the completeness, accuracy, or reliability of the information contained herein and the potential legal, compliance, tax, or accounting effects thereof. This document is not intended as investment research or investment advice, or as a recommendation, offer, or solicitation for the purchase or sale of any security, financial instrument, financial product or service, or to be used in any way for evaluating the merits of participating in any transaction.

\appendix
\section{GEG-IAMP pseudocode}
\label{app:gegiamp-pseudocode}

The following algorithms implement the specification in Section~\ref{sec:algorithm}. Algorithm~\ref{alg:egiamp} is the grouped iteration of Theorem~\ref{thm:main}; Algorithm~\ref{alg:afptas} adds its relative-accuracy rescaling and a dimension check, yielding Corollary~\ref{thm:afptas}.

Algorithm~\ref{alg:egiamp} requires $0<\varepsilon\le\varepsilon_0$, where $\varepsilon_0$ is the universal accuracy threshold of Lemma~\ref{lem:eg-energy}. It returns $\one$ when $d>N$, because the exact state-evolution argument, including the final query used in the proof, requires $d\le N$. During query formation, $\lambda_i$ stores $(P_{s-1})_{ii}$, the diagonal of the projection onto the preceding query span in \eqref{eq:query-proposals}.

All constants below are fixed in the proofs, not supplied as tuning parameters. The algorithm draws the auxiliary matrix once and generates the side variables one group at a time, all from independent standard Gaussian draws. We fix all side variables only when proving the conditional law in Proposition~\ref{prop:gaussian-law}. Algorithm~\ref{alg:egiamp} does not use a confidence parameter, while Algorithm~\ref{alg:afptas} uses it only to choose the dimension threshold.

\begingroup\normalsize
\begin{pseudocodebox}
\captionsetup{type=algorithm,hypcap=false,font=normalsize,labelfont=bf,justification=raggedright,singlelinecheck=false,skip=6pt}
\captionof{algorithm}{GEG-IAMP}
\label{alg:egiamp}
\begin{algorithmic}[1]
\Require Symmetric $A$, $0<\varepsilon\le\varepsilon_0$, and the oracles of Definition~\ref{def:oracles}.
\State Construct $\tau,T,\Delta,h,m_{\rm f},K,B,d$ by \eqref{eq:grouped-cutoff}--\eqref{eq:grid-parameters}.
\If{$d>N$}\State \Return $\one$.\EndIf
\State Compute $R,\eta_{\rm q}$ by \eqref{eq:quadrature-tolerance}; store $\Gamma_k^{\rm q}$ from \eqref{eq:drift-quadrature}.
\State Draw the leading $B\times B$ block of $J$ by \eqref{eq:auxiliary-j}; set $Y\gets0$, $\widehat M\gets\sqrt\Delta\one$, and $\lambda_i\gets0$ for all $i$.
\For{$s=1,\ldots,B$}
 \State $F_s\gets\widehat M$; $r_s\gets F_s-\sum_{r<s}(q_r^\top F_s/N)q_r$.
 \If{$\norm{r_s}=0$}
      \State $i_*\gets\min\{j:\lambda_j<1\}$; $r_s\gets\mathbf e_{i_*}-\sum_{r<s}(q_{r,i_*}/N)q_r$.
 \EndIf
 \State $q_s\gets\sqrt N r_s/\norm{r_s}$; $\lambda_i\gets\lambda_i+q_{s,i}^2/N$ for all $i$.
 \State Compute $v_s\gets Aq_s$ and $z_s$ by \eqref{eq:gaussian-coupling}.
 \For{$i=1,\ldots,N$}
  \State Draw $m_{\rm f}-1$ independent standard Gaussians; form $(G_{j,i})_{j\in I_s}$ by \eqref{eq:group-reconstruction}.
  \For{$j\in I_s$ in increasing order}
   \State $x\gets Y_i$; $\widehat M_i\gets\widehat M_i+a(t_j,x)G_{j,i}$.
   \State $Y_i\gets x+\Gamma_j^{\rm q}m(t_j,x)+G_{j,i}$.
  \EndFor
 \EndFor
\EndFor
\State $f\gets\widehat M-\sqrt\Delta\one$; $v\gets\clip(f,[-1,1]^N)$; $\widetilde A\gets A-\diag(A)$.
\State $\ell\gets Av-\diag(A)v$ using one additional product.
\For{$i=1,\ldots,N$}
 \State $s_i\gets+1$ if $\ell_i\ge0$, and $-1$ otherwise; $\delta_i\gets s_i-v_i$; $v_i\gets s_i$.
 \State $\ell\gets\ell+\delta_i\widetilde A_{:i}$.
\EndFor
\State \Return $v$.
\end{algorithmic}
\end{pseudocodebox}
\endgroup

Algorithm~\ref{alg:afptas} tests dimension before constructing the accuracy-dependent grid or computing quadrature coefficients. These computations cost $O(K+\varepsilon^{-3}L_\varepsilon^2)$ after the fixed accuracy rescaling, by Lemma~\ref{lem:drift-quadrature}, and can exceed $\mathsf{K}N^2/\varepsilon$ for $N<N_0$. Returning $\one$ before this preprocessing preserves the work bound at every dimension.

\begingroup\normalsize
\begin{pseudocodebox}
\captionsetup{type=algorithm,hypcap=false,font=normalsize,labelfont=bf,justification=raggedright,singlelinecheck=false,skip=6pt}
\captionof{algorithm}{The GEG-IAMP AFPRAS}
\label{alg:afptas}
\begin{algorithmic}[1]
\Require Symmetric $A$, $0<\varepsilon\le1$, $0<\alpha<1/2$, and point Parisi data.
\State $\etaacc\gets c\varepsilon$; $N_0\gets\lceil \mathsf{K}\etaacc^{-2}\log^2(e/\etaacc)\log(\mathsf{K}/\alpha)\rceil$.
\If{$N<N_0$}
 \State \Return $\one$.
\Else
 \State \Return Algorithm~\ref{alg:egiamp}$(A,\etaacc)$.
\EndIf
\end{algorithmic}
\end{pseudocodebox}
\endgroup
% Compile the AMS IAMP guarantee and construction only here, after \appendix.
% Supporting proofs follow; the complete algorithms form the final subsection.
\section{The AMS IAMP guarantee and its proof}
\label{sec:iamp-comparison}

We prove Theorem~\ref{thm:onsager-main}: under Assumption~\ref{ass:frsb} and Definition~\ref{def:oracles}, AMS IAMP attains a $(1-\varepsilon)$ approximation with probability at least $1-\alpha$, using $O(N^2\varepsilon^{-3}L_\varepsilon^2)$ total work above the sufficient dimension $\mathsf{K}\exp(\mathsf{K}\varepsilon^{-7/2}L_\varepsilon^{5/2})L_\alpha^4$. This exponential dimension bound is substantially worse than the polynomial bound proved for \gegiamp. Here AMS IAMP is the zero-temperature algorithm of El Alaoui, Montanari and Sellke~\cite[Sections~3 and~5]{AMS}, with each field increment clipped to $[-1,1]$ and normalization factors and Onsager coefficients estimated from the current iterates. We retain their left-endpoint updates, Onsager correction, nonzero initial magnetization, separate state, accumulated magnetization, final clipping, and sequential rounding~\cite[Section~5.3]{AMS}.

We first bound the scalar errors and the effect of the increment cutoff (Theorem~\ref{thm:uniform-scalar} and Lemma~\ref{lem:onsager-cutoff}), then verify the matching moments and Lipschitz constants (Lemma~\ref{lem:onsager-history}). The directional estimates of Section~\ref{sec:matrix-energy}, with their inputs verified in Appendix~\ref{app:onsager-comparator}, bound the realized Hamiltonian $H_N(F_d(W))/N$, not only a scalar correlation: the Gaussian comparator $W$ is coupled to the same matrix $A$ as the implemented iteration. Lemma~\ref{lem:reeves-onsager-bound} transfers this bound to the clipped output of the population Onsager iteration. Lemma~\ref{lem:onsager-calibration} then compares the clipped population and empirical outputs on this $A$, by concentrating comparator statistics and propagating coefficient errors deterministically. These energy comparisons and sequential rounding yield Lemma~\ref{lem:onsager-implementation}. Appendix~\ref{app:onsager-afptas} converts its energy bounds to relative accuracy and proves the secondary AFPRAS corollary. The trajectory and coefficient comparisons contain factors exponential in the number of steps, leading to the exponential sufficient dimension; this is a sufficient bound, not a dimension lower bound for AMS IAMP.

\subsection{The relative approximation guarantee}
\label{sec:onsager-results}

We use $L_\varepsilon=\log(e/\varepsilon)$ and $L_\alpha=\log(\mathsf{K}/\alpha)$ as in \eqref{eq:accuracy-logarithms}, and extend the accuracy notation by $L_u=\log(e/u)$ for $0<u\le1$; in particular, $L_\xi=\log(e/\xi)$. We use one common fixed sufficiently large $\mathsf{K}$ inside $L_\alpha$ and $\log(\mathsf{K}d/\alpha)$ throughout these guarantees. Event-tail logarithms retain their explicitly stated constants, and constants multiplying estimates may be increased separately.

\begin{theorem}[Relative approximation by AMS IAMP]
\label{thm:onsager-main}
Under Assumption~\ref{ass:frsb} and Definition~\ref{def:oracles}, there are $c,\mathsf{K}>0$
such that, for $0<\varepsilon\le1$ and $0<\alpha<1/2$, put
\begin{equation}
 \begin{gathered}
 \xi=c\varepsilon,\qquad
 \mathcal S_\xi=\xi^{-7/2}L_\xi^{5/2},\qquad
 m_{\rm AMS}=\lceil \mathsf{K}\mathcal S_\xi\rceil,\\
 N_0^{\rm AMS}=\left\lceil2^{m_{\rm AMS}}L_\alpha^4\right\rceil.
 \end{gathered}
 \label{eq:onsager-all-dimension-threshold}
\end{equation}
For every $N\ge N_0^{\rm AMS}$, Algorithm~\ref{alg:onsager-calibrated} at accuracy $\xi$ returns $\sigma\in\{-1,1\}^N$ satisfying
\begin{equation}
 \Prob\{H_N(\sigma)\ge(1-\varepsilon)\OPT_N\}\ge1-\alpha.
 \label{eq:onsager-relative}
\end{equation}
Total work is at most $\mathsf{K}N^2\varepsilon^{-3}L_\varepsilon^2$,
storage is $O(N^2)$,
$K=\Theta(\varepsilon^{-3}L_\varepsilon^2)$, and there are $K+2$ matrix products. Estimating the coefficients and updating the Onsager correction cost $O(NK)$ operations in total. The algorithm is deterministic given $A$, and the success probability is over the GOE input.
\end{theorem}

\begin{corollary}[An AFPRAS from AMS IAMP]
\label{thm:onsager-afptas}
Under the assumptions of Theorem~\ref{thm:onsager-main}, Algorithm~\ref{alg:onsager-afptas} returns all-plus for $N<N_0^{\rm AMS}$, before building a grid or estimating coefficients, and otherwise runs Algorithm~\ref{alg:onsager-calibrated} at accuracy $\xi$. For every $N\ge1$,
\begin{equation}
 \Prob\{H_N(\sigma)\ge(1-\varepsilon)\OPT_N-4N_0^{\rm AMS}\}
                                                  \ge1-\alpha,
 \label{eq:onsager-afptas-result}
\end{equation}
with total work and storage as in Theorem~\ref{thm:onsager-main}, and
\begin{equation}
 4N_0^{\rm AMS}\le
 \mathsf{K}\exp\!\left(\mathsf{K}\varepsilon^{-7/2}L_\varepsilon^{5/2}\right)
                                      L_\alpha^4.
 \label{eq:onsager-afptas-allowance}
\end{equation}
For $N\ge N_0^{\rm AMS}$, the stronger relative guarantee \eqref{eq:onsager-relative} holds.
\end{corollary}

The coefficient estimates require no independent scalar paths. The all-plus branch avoids grid construction below the sufficient dimension. Corollary~\ref{cor:onsager-earlier-stopping} improves the bounds for the same updates with an earlier terminal time, but the resulting sufficient dimension remains exponential in inverse accuracy.

\subsection{Projection in the implementation of El Alaoui and Montanari}
\label{sec:projected-iamp}

The query orthogonalization of \gegiamp\ differs from the projections that El Alaoui and Montanari propose for IAMP~\cite{EAM}: they project field increments, whereas we project matrix queries before multiplying by $A$. Their full-history construction projects each gradient increment off all preceding magnetizations~\cite[Eq.~(2.9)]{EAM}. Their numerical implementation instead retains the Onsager update, orthogonalizes each field increment against the previous five, and rescales the increment and magnetization to prescribed norms~\cite[Section~3.2, Eqs.~(3.6)--(3.8)]{EAM}.

After query orthogonalization, Reeves's corrections in \eqref{eq:gaussian-coupling} make the corrected query outputs $z_s$ exactly Gaussian under Proposition~\ref{prop:gaussian-law}; their norms are random, and the outputs need not be mutually orthogonal. For the $B$ matrix queries of \gegiamp, full-history query projection and correction cost $O(NB^2)$. AMS IAMP uses $K$ steps. Evaluating its full-history correction sum directly costs $O(NK^2)$, but the coefficient structure in \eqref{eq:estimated-onsager-entries} permits the equivalent $O(NK)$ recurrence \eqref{eq:onsager-streamed-correction}. Projection against a fixed number of preceding increments also costs $O(NK)$, in addition to matrix products. These counts concern the specified matrix operations, rather than an ordering of achieved energies. Our AMS iteration uses the streamed Onsager correction without the numerical projections or norm rescaling of El Alaoui and Montanari.

\subsection{Left-endpoint discretization}
\label{sec:uniform-onsager-scalar}

We first bound the errors of the scalar recursion of AMS IAMP, which uses $h\gamma(t_k)$ instead of the integral of $\gamma$ over a step. Following AMS, we denote the Gaussian increment over $[t_k,t_{k+1}]$ by $G_{k+1}$; this increment is denoted $G_k$ in Section~\ref{sec:scalar-analysis}.

For $T=1-\tau=Kh$, $t_k=kh$, and $h\le\tau$, let $W$ be standard Brownian motion on $[0,T+h]$ and set $G_{k+1}=W_{t_{k+1}}-W_{t_k}$ for $0\le k<K$. These increments are independent $\mathcal N(0,h)$ variables. Define
\begin{align}
 Y_{k+1}&=Y_k+h\gamma(t_k)m(t_k,Y_k)+G_{k+1},\qquad Y_0=0,\nonumber\\
 \rho_k^2&=\E a(t_k,Y_k)^2,\nonumber\\
 M_{k+1}&=M_k+\frac{a(t_k,Y_k)}{\rho_k}G_{k+1},\qquad M_0=\sqrt h.
 \label{eq:uniform-iamp-scheme}
\end{align}
These are the SK scalar updates and normalization of \cite[Sections~3 and 5]{AMS}; in particular $a(0,0)=\rho_0=1$. The initial value $M_0=\sqrt h$ remains in the final magnetization. Set $t_{K+1}=T+h$. Each increment of $M$ is a predictable multiple of the next Gaussian increment, so
\begin{equation}
 \E M_jM_k=(\min(j,k)+1)h,\qquad
 \mathcal E_h=\E[W_{T+h}M_K]
           =\sum_{k<K}h\frac{\E a(t_k,Y_k)}{\rho_k}.
 \label{eq:uniform-energy}
\end{equation}
We use $W_{T+h}$ because the matrix iteration computes a final field $z^{K+1}$, so the Gaussian comparator has $d=K+1$ columns. The extra increment $W_{T+h}-W_T$ is centered and independent of $M_K$, and therefore leaves the correlation $\mathcal E_h$ unchanged. We compare this correlation with the output energy in Lemma~\ref{lem:onsager-comparator-energy}.

\begin{theorem}[Errors of the uniform left-endpoint recursion]
\label{thm:uniform-scalar}
Use $\Lambda(t)$ and $L_\tau$ from \eqref{eq:derivative-weight}.
There are universal $c_0,\mathsf{K}>0$ such that, if
$h\Lambda(T)^2\le c_0$, then $3/4\le\rho_k\le5/4$ and
\begin{align}
 \norm{\dist(M_K,[-1,1])}_2
 &\le \mathsf{K}\left(\frac{\sqrt h\,L_\tau}{\sqrt\tau}
                +\sqrt h\,L_\tau^2+\frac{hL_\tau^{3/2}}\tau\right),
                                               \label{eq:uniform-clipping}\\
 \mathcal E_h&\ge\Par-\mathsf{K}\tau^{3/2}\sqrt{L_\tau}
                                      -\mathsf{K}\sqrt h\,L_\tau^2.
                                               \label{eq:uniform-energy-bound}
\end{align}
\end{theorem}

Appendix~\ref{app:scalar-grids} proves Theorem~\ref{thm:uniform-scalar}. The proof bounds the relative entropy between the true and interpolated path laws up to each time $s$, not only up to $T$; this yields the bound on normalization factors $|\rho_k-1|\le \mathsf{K}\sqrt h\,\Lambda(t_k)$. Since the scheme freezes $\gamma$ as well as $m$ at the left endpoint, the proof also keeps the variation $\gamma(t)-\gamma(t_k)$ in every within-cell drift calculation.

Algorithm~\ref{alg:onsager-calibrated} takes a target accuracy $u$ and uses the smaller scalar accuracy $\etaacc=c_{\rm AMS}u$ to choose its grid. The wrapper in Corollary~\ref{thm:onsager-afptas} calls it with $u=\xi=c\varepsilon$. Thus $u$ and $\xi$ play the role of the internal accuracy in Theorem~\ref{thm:main}, while $\etaacc$ in this appendix denotes the scalar tolerance. At scalar accuracy $\etaacc$, choose
$J_0=\lceil\log_2(32/\etaacc)\rceil$ and set
\begin{equation}
 \begin{gathered}
 \etaacc/64<\tau=2^{-J_0}\le\etaacc/32,\qquad T=1-\tau,
 \qquad L_\tau=\log(e/\tau),\\
 h_*=c_h\etaacc^2\tau/L_\tau^2,\qquad
 K=\lceil T/h_*\rceil,\qquad h=T/K,\qquad t_k=kh.
 \end{gathered}
 \label{eq:uniform-onsager-grid}
\end{equation}
Here $c_h>0$ is a sufficiently small fixed constant. Since $T\ge1/2$ and $h_*\to0$, the ceiling in $K$ gives $h_*/2\le h\le h_*$. Thus $h\le\tau$, $T+h\le1$, and $K=\Theta(\etaacc^{-3}\log^2(e/\etaacc))$. Substituting these parameters into the three terms of \eqref{eq:uniform-clipping}, we obtain the bounds
\begin{equation}
 \sqrt{c_h}\etaacc,\qquad
 \sqrt{c_h}\etaacc\sqrt\tau L_\tau,\qquad
 c_h\etaacc^2/\sqrt{L_\tau},
 \label{eq:uniform-parameter-errors}
\end{equation}
up to universal constants. The second and third are smaller than the first for sufficiently small $\etaacc$. Also $h\Lambda(T)^2\le c_h\etaacc^2/\tau\to0$, so the hypothesis $h\Lambda(T)^2\le c_0$ of Theorem~\ref{thm:uniform-scalar} holds for small $\etaacc$. In \eqref{eq:uniform-energy-bound}, the terminal loss is at most $\tau$, and the remaining correlation error is $\mathsf{K}\sqrt{c_h}\etaacc\sqrt\tau L_\tau$. Therefore, given any two prescribed fixed multiples of the target accuracy, choosing $\etaacc$ as a sufficiently small fixed multiple of the target accuracy makes the clipping error \eqref{eq:uniform-clipping} and the energy deficit $\Par-\mathcal E_h$ smaller than these multiples, as required in the proof of Lemma~\ref{lem:onsager-implementation}.

Stopping earlier means taking a larger $\tau=1-T$. This increases $h_*$ in \eqref{eq:uniform-onsager-grid} and reduces the number of steps; Corollary~\ref{cor:onsager-earlier-stopping} uses the following choice. Fix a sufficiently small universal $c_{\rm early}>0$.
Starting with $\tau=1/2$, halve it until
\begin{equation}
             \tau^3\log(e/\etaacc)\le c_{\rm early}^3\etaacc^2.
 \label{eq:uniform-earlier-cutoff}
\end{equation}
For sufficiently small $\etaacc$, the first accepted value lies
within a factor two of
\begin{equation*}
                  c_{\rm early}\etaacc^{2/3}\log^{-1/3}(e/\etaacc).
\end{equation*}
Keep $h_*,K,h$ as in \eqref{eq:uniform-onsager-grid}. Then $K=\Theta(\etaacc^{-8/3}\log^{7/3}(e/\etaacc))$. The tail bound \eqref{eq:terminal-g}, sharper than the bound $\tau$ used for \eqref{eq:uniform-onsager-grid}, bounds the discarded energy by $\mathsf{K}\tau^{3/2}\sqrt{L_\tau}\le \mathsf{K} c_{\rm early}^{3/2}\sqrt{\log(e/c_{\rm early})}\etaacc$. Given two prescribed fixed multiples of $\etaacc$, choose $c_{\rm early}$ first, then $c_h$, and finally restrict $\etaacc$ to a sufficiently small range, so that the two scalar errors are at most these multiples. The test \eqref{eq:uniform-earlier-cutoff} uses arithmetic, logarithms, and comparisons. Both choices of $\tau$ retain the updates \eqref{eq:uniform-iamp-scheme}.

\subsection{AMS IAMP with bounded increments}

Reeves's trajectory comparison in Lemma~\ref{lem:reeves-onsager-bound} requires maps that are globally Lipschitz in the field history. Differentiating the uncut magnetization increment $a(t_k,Y_k)G_{k+1}/\rho_k$ with respect to an earlier field introduces the unbounded factor $G_{k+1}$. We therefore clip each field increment before using it in the state and magnetization updates.

Let $\chi_R(x)=\clip(x,[-R,R])$ and
$v_{h,R}=\E\chi_R(G)^2$ for $G\sim\mathcal N(0,h)$.
The scalar recursion with cutoff increments is
\begin{align}
 Y^R_{k+1}&=Y^R_k+h\gamma(t_k)m(t_k,Y^R_k)+\chi_R(G_{k+1}),\qquad Y_0^R=0,\nonumber\\
 (\rho_k^R)^2&=(v_{h,R}/h)\E a(t_k,Y_k^R)^2,\nonumber\\
 M^R_{k+1}&=M^R_k+\frac{a(t_k,Y_k^R)}{\rho_k^R}\chi_R(G_{k+1}),\qquad M_0^R=\sqrt h.
 \label{eq:iamp-cutoff-scalar}
\end{align}
Since $\chi_R$ is odd and $G_{k+1}$ is symmetric and independent of $G_1,\ldots,G_k$, the increments of $M^R$ have conditional mean zero. The factor $v_{h,R}/h$ in $(\rho_k^R)^2$ makes their unconditional second moment $\E[(M^R_{k+1}-M^R_k)^2]=h$. Because $a(0,0)=1$, the first normalization factor satisfies $(\rho_0^R)^2=v_{h,R}/h$.

For cumulative fields $z^0=0,z^1,\ldots,z^k$, run
\eqref{eq:iamp-cutoff-scalar} with $G_j=z^j-z^{j-1}$ and denote
the output by $f_k^R$. The algorithm uses $R=1$, and the superscript $1$ in $Y^1,M^1,\rho^1,f^1$ denotes this choice. The matrix iteration retains the Onsager form
of El Alaoui, Montanari and Sellke~\cite[Sections~3 and 5]{AMS}:
\begin{equation}
 z^{k+1}=A m^k-\sum_{j=1}^k b_{k,j}m^{j-1},\qquad
 m^k=f_k^1(z^1,\ldots,z^k),\quad m^0=\sqrt h\one.
 \label{eq:ordinary-iamp}
\end{equation}
The population coefficients $b_{k,j}$ are expectations of the partial derivatives of $f_k^1$ with respect to $z^j$, evaluated on the Gaussian comparator; Lemma~\ref{lem:onsager-history} identifies them with entries of the matrix $B^1$ in \eqref{eq:onsager-population-B}. Section~\ref{sec:onsager-calibration} estimates the corresponding moments from the actual fields. These estimates need not equal the empirical derivatives of the resulting maps, whose coefficients depend on all coordinates; we compare their correction directly with the population correction.

For the scalar recursion \eqref{eq:iamp-cutoff-scalar} with $R=1$, let $\mathcal E_h^1=\E[W_{T+h}M_K^1]$.

\begin{lemma}[Effect of the increment cutoff]
\label{lem:onsager-cutoff}
Let $\kappa_a(T)=\max\{1,\mathsf{K}\sqrt{\Lambda(T)}\}$, where the fixed constant
bounds $a(t,x)$ for $t\le T$. For the grids above, sufficiently
small scalar accuracy gives $1/2\le\rho_k^1\le2$ and
\begin{gather}
 \norm{M_K^1-M_K}_2+|\mathcal E_h^1-\mathcal E_h|
      \le \mathsf{K}\kappa_a(T)^2(K+1)e^{-c/h},
 \label{eq:onsager-cutoff-bias}\\
 \mathcal E_h^1:=\sum_{k<K}\frac{\E a(t_k,Y_k^1)}{\rho_k^1}
                                      \E[G\chi_1(G)].
 \label{eq:cutoff-correlation}
\end{gather}
The coupled scalar paths use the same Gaussian increments.
\end{lemma}

Fix a universal $\eta_0>0$ small enough that, on either grid above with $0<\etaacc\le\eta_0$, the cutoff factors satisfy $1/2\le\rho_k^1\le2$, $v_{h,1}/h\ge1/2$, $h\Lambda(T)^2\le1$, and $dh\le1$. The scalar theorem, the grid estimates, and \eqref{eq:onsager-normalized-tail} ensure that such a choice exists. All subsequent population, directional, and empirical-coefficient lemmas in this appendix use this common accuracy range. The fixed rescaling in Theorem~\ref{thm:onsager-main} enforces it for every requested accuracy.

\subsection{Gaussian trajectory comparison and population moments}

To compare the AMS IAMP trajectory with a Gaussian process at finite $N$, we use Reeves's GOE comparison theorem, whose bound depends explicitly on the Lipschitz constants of the maps that send past fields to the next query and correction. Related finite-sample analyses concern Gaussian regression~\cite{RushVenkataramanan}, spiked matrix estimation~\cite{LiWeiSpiked}, and general first-order dynamics under regularity assumptions~\cite{HanGFOM}.

\begin{lemma}[Comparison of a GOE iteration with its Gaussian comparator, specialization of {\cite[Theorem~5]{Reeves}}]
\label{lem:reeves-onsager-bound}
Consider $d\le N$ steps $x_t=A F_t(x_{<t})+g_t(x_{<t})$,
where $F_t,g_t:\R^{N\times(t-1)}\to\R^N$ have common global
Lipschitz constants $L_F,L_g$, using the Frobenius norm on histories
and the Euclidean norm on outputs.
Choose deterministic upper triangular $\Omega$ with positive diagonal.
Construct $W=Z\Omega$ on the same $A$ by
Definition~\ref{def:reeves-coupling}, using $F_t$ and the identically zero
mean map. The entries of $Z$ are iid standard Gaussians, so the rows of
$W$ are independent centered Gaussian vectors with covariance
$\Sigma=\Omega^\top\Omega$; $W$ need not be independent of $A$.
For $W_{<t}=[W_{:,1}|\cdots|W_{:,t-1}]$, define
\begin{equation*}
 \begin{gathered}
 F(W)=[F_1(W_{<1})|\cdots|F_d(W_{<d})]\in\R^{N\times d},\\
 G(W)=[g_1(W_{<1})|\cdots|g_d(W_{<d})]\in\R^{N\times d}.
 \end{gathered}
\end{equation*}
Suppose the deterministic matrix $B$ is strictly upper triangular and
\begin{equation}
 \begin{gathered}
 \E[F(W)^\top F(W)]/N=\Sigma,\qquad
 B=\Sigma^{-1}\E[W^\top F(W)]/N,\\
 g_t(x_{<t})=-\sum_{s<t}B_{st}F_s(x_{<s})
                   \quad\text{for every history }x_{<t}.
 \end{gathered}
 \label{eq:onsager-matching}
\end{equation}
For $0\le r\le N$,
with probability at least $1-5e^{-r}$,
\begin{equation}
 \frac{\norm{X-W}_{\rm F}}{\sqrt N}
 \le \mathsf{K}(1+4L_F+L_g)^{d-1}\mathcal L^3
       \norm\Omega_{2,1}\sqrt{\frac{d+r}{N}},
 \label{eq:reeves-onsager-bound}
\end{equation}
where $X=[x_1|\cdots|x_d]$ and
\begin{equation}
 \mathcal L=\log_2(2d)+\sqrt d(L_F+L_g)
                       \operatorname{cond}(\Sigma)^{1/2},\qquad
 \norm\Omega_{2,1}=\sum_{j=1}^d\norm{\Omega\mathbf e_j}.
 \label{eq:reeves-onsager-constants}
\end{equation}
Here $\mathbf e_j$ is the $j$th standard coordinate vector in $\R^d$.
\end{lemma}

We quote this specialization of \cite[Theorem~5]{Reeves} with the zero mean map. The first matching condition uses raw second moments, not covariances centered at the mean, and gives $S=N^{-1}\Omega^{-\top}\E[F(W)^\top F(W)]\Omega^{-1}=I_d$. The last condition gives the pointwise identity $G(W)+F(W)B=0$, while the middle condition fixes the population cross moments to this same $B$. Thus the normalized second-moment and correction residuals in the general theorem vanish; no other population mismatch remains in \eqref{eq:reeves-onsager-bound}. Proposition~\ref{prop:energy-identity} is a deterministic identity for the same coupled $A,W$, valid on every outcome, and therefore applies to the realized $H_N(F_d(W))/N$.

Set $d=K+1$, $\Omega_{st}=\sqrt h\one_{\{s\le t\}}$,
$F_t=f_{t-1}^1$, and
\begin{equation}
 U^1=F(W)\Omega^{-1},\qquad D^1=\E[Z_i^\top U_i^1],\qquad
 B^1=\Omega^{-1}D^1\Omega.
 \label{eq:onsager-population-B}
\end{equation}
The population iteration means \eqref{eq:ordinary-iamp} with the normalization factors $\rho_k^1$ and coefficient matrix $B^1$ defined by scalar expectations. With these choices, the iteration of Lemma~\ref{lem:reeves-onsager-bound} is \eqref{eq:ordinary-iamp}, with $x_t=z^t$, $F_t=m^{t-1}$, and $b_{k,j}=B^1_{j,k+1}$. Its comparator $W$ is the ungrouped specialization of Reeves's construction using the cutoff AMS maps, one query per scalar step, and the left-endpoint drift. We compare $W$ with the population iteration by \eqref{eq:reeves-onsager-bound} and handle estimated coefficients separately in Lemma~\ref{lem:onsager-calibration}. In contrast, \gegiamp\ uses Reeves's construction with grouped query and scalar maps, and we bound its energy directly.

To bound the amplification factor $(1+4L_F+L_g)^{d-1}$ in \eqref{eq:reeves-onsager-bound}, write
\begin{equation}
 b_{h,\tau}=1+\sqrt{\Lambda(T)}+\log(d+h^{-1}+2),\qquad
 \mathfrak S_{h,\tau}=d\,b_{h,\tau}.
 \label{eq:onsager-history-quantity}
\end{equation}
By Lemma~\ref{lem:onsager-history}, this factor is at most $e^{\mathsf{K}\mathfrak S_{h,\tau}}$. Thus $\mathfrak S_{h,\tau}$ determines the coefficient accuracy and a sufficient dimension for the comparison.

\begin{lemma}[Population moments and Lipschitz constants]
\label{lem:onsager-history}
For the scalar recursion \eqref{eq:iamp-cutoff-scalar} with $R=1$ and
fixed population normalization factors $\rho_k^1$, take $d=K+1$ and the
Gaussian comparator $W=Z\Omega$ with
$\Omega_{st}=\sqrt h\one_{\{s\le t\}}$.
Run the scalar updates along its rows using
$G_{k+1,i}=W_{i,k+1}-W_{i,k}$, with $W_{i,0}=0$, so that
$F_t(W_{<t})=(M_{t-1,i}^1)_{i=1}^N$.
Put $U^1=F(W)\Omega^{-1}$ and $D^1=\E[Z_i^\top U_i^1]$,
where $Z_i,U_i^1$ denote rows. Then, for $0\le k<K$ and
$G\sim\mathcal N(0,h)$,
\begin{equation}
 \E[(U_i^1)^\top U_i^1]=I_d,\qquad
 D^1_{k+1,k+2}=\frac{\E a(t_k,Y_k^1)}{\rho_k^1}
                  \frac{\E[G\chi_1(G)]}{h},\qquad
 \norm{D^1}\le1,
 \label{eq:cutoff-cross-moments}
\end{equation}
with all other cross entries zero. Equations~\eqref{eq:onsager-matching}
hold with $B=B^1$, the population derivative coefficients.
Furthermore
\begin{equation}
 \log(1+4L_F+L_g)\le \mathsf{K}b_{h,\tau},\qquad
 \operatorname{cond}(\Sigma)\le4d/h,\qquad
 \norm\Omega_{2,1}\le d.
 \label{eq:onsager-history-bounds}
\end{equation}
The constants include the final magnetization map.
\end{lemma}

\subsection{Estimation of the coefficients from the iterates}
\label{sec:onsager-calibration}

The current state and field increment determine two empirical averages from which we estimate both coefficients. For the scalar comparator, let $a_k^1=a(t_k,Y_k^1)$, and put
\begin{gather}
 q_k=(\rho_k^1)^2=\E[(a_k^1)^2\chi_1(G_{k+1})^2/h],\qquad
 v_k=\E[a_k^1G_{k+1}\chi_1(G_{k+1})/h],\nonumber\\
 c_k=v_k/\rho_k^1=D^1_{k+1,k+2}.
 \label{eq:onsager-calibration-targets}
\end{gather}
Here $G_{k+1}\sim\mathcal N(0,h)$ is independent of $Y_k^1$. Stein's identity states that $\E[G\chi_1(G)]/h=\Prob\{|G|\le1\}$, so $v_k$ is the numerator of the cross moment in \eqref{eq:cutoff-cross-moments}. The function $x\mapsto x\chi_1(x)$ is globally $2$-Lipschitz. We use this product instead of an indicator, so transferring the statistic from the Gaussian comparator to the actual fields requires no anti-concentration estimate.

Let $\widehat z^k,\widehat Y_k,\widehat m^k$ denote the implemented fields, states, and magnetizations. After forming $\widehat z^{k+1}$ and before updating the state, set $\widehat G_{k+1,i}=\widehat z_i^{k+1}-\widehat z_i^k$ and $\widehat a_{k,i}=a(t_k,\widehat Y_{k,i})$. Define
\begin{align}
 \widehat q_k&=\frac1{Nh}\sum_{i=1}^N
                 \widehat a_{k,i}^2\chi_1(\widehat G_{k+1,i})^2,\nonumber\\
 \widehat v_k&=\frac1{Nh}\sum_{i=1}^N
                 \widehat a_{k,i}\widehat G_{k+1,i}\chi_1(\widehat G_{k+1,i}).
 \label{eq:onsager-calibration-means}
\end{align}
Set
\begin{equation}
 \widehat\rho_k^{\rm AMS}=\sqrt{\clip(\widehat q_k,[1/4,4])},\qquad
 \widehat c_k=\clip(\widehat v_k/\widehat\rho_k^{\rm AMS},[0,1]),\qquad
 \widehat D^{\rm AMS}_{k+1,k+2}=\widehat c_k.
 \label{eq:estimated-onsager-coefficients}
\end{equation}
All other entries of $\widehat D^{\rm AMS}$ are zero. Since $q_k\in[1/4,4]$ and $c_k\in[0,1]$, the projections do not increase errors relative to these population targets. In particular, the first normalization factor is estimated rather than set to one: $(\rho_0^1)^2=v_{h,1}/h$.

The corresponding correction matrix $\widehat B=\Omega^{-1}\widehat D^{\rm AMS}\Omega$ has entries
\begin{equation}
 \widehat B_{ij}=\begin{cases}
 \widehat c_{i-1},&j=i+1,\\
 \widehat c_{i-1}-\widehat c_i,&j\ge i+2,\\
 0,&j\le i.
 \end{cases}
 \label{eq:estimated-onsager-entries}
\end{equation}
The second case occurs only for $i\le d-2$; the last row is zero. Each column uses only coefficients already available when its field is computed. The matrix need not be stored. Define $\widehat m^{-1}=0$ and let $\mathcal O^k$ be the Onsager correction subtracted from $A\widehat m^k$. Summing \eqref{eq:estimated-onsager-entries} yields
\begin{equation}
 \mathcal O^0=0,\qquad
 \mathcal O^{k+1}=\mathcal O^k+
       \widehat c_k(\widehat m^k-\widehat m^{k-1}),\qquad
 \widehat z^{k+1}=A\widehat m^k-\mathcal O^k.
 \label{eq:onsager-streamed-correction}
\end{equation}
Thus $\widehat c_k$, computed from field $k+1$, first enters field $k+2$. Both empirical averages and the correction update cost $O(N)$ per step.

For the error bounds only, fix $\mathsf{K}_\kappa$ so that $\kappa^2\ge\max\{1,4\sup_{t\le T,x}a(t,x)^2\}$, and write
\begin{equation}
 \begin{gathered}
 \kappa^2=\mathsf{K}_\kappa L_\tau/\tau,\qquad
 L_{\alpha,d}:=L_\alpha+\log d=\log(\mathsf{K}d/\alpha),\\
 \eta_N=\mathsf{K}\left\{\kappa\sqrt{L_{\alpha,d}/N}
                               +\kappa^2L_{\alpha,d}/N\right\}.
 \end{gathered}
 \label{eq:onsager-calibration-parameters}
\end{equation}
The moment bound $\kappa$ is chosen separately from the pointwise curvature envelope $\kappa_a(T)$. These quantities are not inputs to the empirical estimates. For any field history $\mathsf Z$, let $Q_k(\mathsf Z)$ and $V_k(\mathsf Z)$ be the averages in \eqref{eq:onsager-calibration-means}, with the state reconstructed from that history. In particular, $Q_k(W)$ and $V_k(W)$ are evaluated on the fixed Gaussian comparator.

\begin{lemma}[Coefficient accuracy and its effect on the matrix iteration]
\label{lem:onsager-calibration}
For $N\ge\mathsf{K}L_{\alpha,d}$, with probability at least $1-\alpha/8$,
\begin{equation}
 \max_{k<K}\max\{|Q_k(W)-q_k|,|V_k(W)-v_k|\}\le\eta_N,\qquad
 \max_{k<K}\frac{\norm{W_{:,k+1}-W_{:,k}}}{\sqrt N}\le2\sqrt h,
 \quad W_{:,0}=0.
 \label{eq:onsager-calibration-event}
\end{equation}
On this event and $\norm A\le3$, let $X$ be the population iteration and $\widehat X$ the empirical iteration on the same $A$. Then
\begin{equation}
 \frac{\norm{\widehat m^K-m^K}}{\sqrt N}
 \le e^{\mathsf{K}_{\rm stab}\mathfrak S_{h,\tau}}
       \left\{\frac{\norm{X-W}_{\rm F}}{\sqrt N}+\eta_N\right\}.
 \label{eq:onsager-calibration-stability}
\end{equation}
Computing the empirical coefficients and updating \eqref{eq:onsager-streamed-correction} cost $O(NK)$ work and $O(N)$ storage.
\end{lemma}

\begin{lemma}[Energy of the implemented AMS IAMP iteration]
\label{lem:onsager-implementation}
There are universal $\mathsf{K},u_{\rm AMS}>0$ such that, for
$0<u\le u_{\rm AMS}$ and $0<\alpha<1/2$, the following holds
for either stopping rule. Take scalar accuracy
$\etaacc=c_{\rm AMS}u$ with a sufficiently small fixed $c_{\rm AMS}$.
Choose the terminal time by the proportional rule
\eqref{eq:uniform-onsager-grid} or the earlier deterministic rule
\eqref{eq:uniform-earlier-cutoff}; these choices of $T=1-\tau$ do not
change the field-increment cutoff $\chi_1$.
Then
\begin{equation}
                 N\ge e^{\mathsf{K}\mathfrak S_{h,\tau}}L_\alpha^4
 \label{eq:onsager-implementation-size}
\end{equation}
implies, with probability at least $1-\alpha$,
\begin{equation}
 H_N(\widehat\sigma_{\rm AMS})\ge(\Par-u/2)N,\qquad
                         \OPT_N\le(\Par+u/2)N.
 \label{eq:onsager-implementation-energy}
\end{equation}
The total work, including empirical coefficient estimation, is $O(N^2K)$,
storage is $O(N^2)$, and there are $K+2$ matrix products.
\end{lemma}

The proofs of the cutoff, directional, coefficient, and finite-dimensional energy estimates are in Appendices~\ref{app:matrix-onsager-regularity}--\ref{app:matrix-onsager-finite}. We concentrate the two statistics only on the Gaussian comparator, whose rows are independent. Lipschitz estimates transfer them to the actual field history, and a causal deterministic recurrence controls their feedback into later fields. The energies of the empirical and population iterations are compared after clipping and before rounding, because sequential rounding need not map nearby magnetization vectors to nearby spin vectors.
\subsection{Scalar estimates for the left-endpoint AMS IAMP recursion}
\label{app:scalar-grids}

We prove the clipping and correlation bounds of Theorem~\ref{thm:uniform-scalar} by comparing the path laws through relative entropy, transferring moments of $a$ and $a_x$ with Lemma~\ref{lem:entropy}, and bounding the normalization, coefficient-variation, and drift terms in the terminal error. Unlike the integrated-drift recursion \eqref{eq:reference-euler}, the AMS recursion \eqref{eq:uniform-iamp-scheme} freezes $\gamma$ as well as $m$, so its drift discrepancy includes $\gamma(t)-\gamma(t_k)$.

In this proof, $\mathcal M_t=m(t,X_t)$ denotes the magnetization of the Parisi diffusion \eqref{eq:true-diffusion}, $k(t)=k$ for $t_k\le t<t_{k+1}$, with $k(T)=K-1$, and $a_k=a(t_k,Y_k)=a(t_k,Y_{t_k})$ for the interpolation of \eqref{eq:uniform-iamp-scheme}.

\begin{proof}[Proof of Theorem~\ref{thm:uniform-scalar}]
\par\smallskip\noindent\textit{Path laws and relative entropy.}\quad
We interpolate \eqref{eq:uniform-iamp-scheme} by
\begin{equation}
 \dd Y_t=\gamma(t_k)m(t_k,Y_{t_k})\dd t+\dd W_t,
 \qquad t_k\le t<t_{k+1},\qquad Y_0=0.
 \label{eq:ams-interpolation}
\end{equation}
Integrating over $[t_k,t_{k+1}]$ recovers the node update, so $Y_{t_k}=Y_k$. Fix $s\le T$. On $C([0,s];\R)$, let $Z_t$ be the coordinate process, let $\mathcal F_t^Z=\sigma(Z_u:0\le u\le t)$ be its natural filtration, and let $\mathsf P_s$ and $\mathsf Q_s$ be the laws of the Parisi diffusion \eqref{eq:true-diffusion} and the interpolation \eqref{eq:ams-interpolation}. We use the usual augmentation of the filtration under the law in question. The two drift functionals, evaluated on the same coordinate path, are
\begin{equation}
 \mu_t(Z)=\gamma(t)m(t,Z_t),\qquad
 \mu_t^{\mathrm{AMS}}(Z)=\gamma(t_{k(t)})m(t_{k(t)},Z_{t_{k(t)}}).
 \label{eq:ams-path-drifts}
\end{equation}
Both are progressively measurable, since they use only coordinates at times at most $t$. On cell $k$, their difference is
\begin{align}
 \Delta\mu_t&=\mu_t(Z)-\mu_t^{\mathrm{AMS}}(Z)\nonumber\\
 &=\gamma(t)\{m(t,Z_t)-m(t_k,Z_{t_k})\}
       +[\gamma(t)-\gamma(t_k)]m(t_k,Z_{t_k}).
 \label{eq:ams-path-drift-difference}
\end{align}
Under $\mathsf Q_s$, the process $W_t^Q=Z_t-\int_0^t\mu_u^{\mathrm{AMS}}(Z)\dd u$ is Brownian in the natural filtration. Monotonicity of $\gamma$ and $|m|\le1$ imply $|\Delta\mu_t|\le2\gamma(t)$, so Novikov's condition follows from
\begin{equation}
 \E_{\mathsf Q_s}\exp\left(\frac12\int_0^s(\Delta\mu_t)^2\dd t\right)
 \le\exp\left(2\int_0^s\gamma(t)^2\dd t\right)<\infty.
 \label{eq:ams-novikov}
\end{equation}
The last integral is finite because $T<1$ and $\gamma$ is bounded on $[0,T]$. Girsanov's theorem changes the drift from $\mu^{\mathrm{AMS}}$ to $\mu$. The true drift is uniformly Lipschitz in space on this strip, since $\partial_x\mu_t=\gamma(t)a(t,\cdot)$ is bounded. Uniqueness in law for \eqref{eq:true-diffusion} therefore identifies the changed measure with $\mathsf P_s$, and
\begin{equation}
 \mathcal L_s:=\frac{\dd\mathsf P_s}{\dd\mathsf Q_s}
 =\exp\left(\int_0^s\Delta\mu_t\dd W_t^Q
                    -\frac12\int_0^s(\Delta\mu_t)^2\dd t\right).
 \label{eq:ams-girsanov-density}
\end{equation}
Under $\mathsf P_s$, the process $B_t^P=W_t^Q-\int_0^t\Delta\mu_u\dd u$ is Brownian. Substituting $\dd W_t^Q=\dd B_t^P+\Delta\mu_t\dd t$ in \eqref{eq:ams-girsanov-density}, we obtain
\begin{align}
 \log\mathcal L_s&=\int_0^s\Delta\mu_t\dd B_t^P
                         +\frac12\int_0^s(\Delta\mu_t)^2\dd t,\nonumber\\
 \KL(\mathsf P_s\Vert\mathsf Q_s)
       &=\frac12\int_0^s\E_{\mathsf P_s}[(\Delta\mu_t)^2]\dd t.
 \label{eq:ams-entropy-identity}
\end{align}
The stochastic integral is square-integrable under $\mathsf P_s$ because its integrand is bounded on this fixed interval, and its expectation is zero. This justifies the entropy identity and the interchange of expectation and time integration.

Under the true law, $\mathcal M_t=m(t,Z_t)$ is the bounded martingale in \eqref{eq:martingale-identities}. Its increments have conditional mean zero, so, for $t_k\le t\le s$ in cell $k$,
\begin{align}
 \E_{\mathsf P_s}[(\mathcal M_t-\mathcal M_{t_k})\mathcal M_{t_k}]
 &=\E_{\mathsf P_s}\bigl[\mathcal M_{t_k}
       \E_{\mathsf P_s}[\mathcal M_t-\mathcal M_{t_k}\mid\mathcal F_{t_k}^Z]\bigr]=0,\nonumber\\
 \E_{\mathsf P_s}[(\mathcal M_t-\mathcal M_{t_k})^2]
 &=\int_{t_k}^t\E_{\mathsf P_s}[a(u,Z_u)^2]\dd u=t-t_k,
 \qquad \E_{\mathsf P_s}[\mathcal M_{t_k}^2]=t_k.
 \label{eq:ams-true-increment-moments}
\end{align}
The increment second moment follows from It\^o isometry, and the last identity is \eqref{eq:frsb}. Thus the two terms in \eqref{eq:ams-path-drift-difference} are orthogonal in $L^2(\mathsf P_s)$. Substituting their second moments in \eqref{eq:ams-entropy-identity}, we obtain
\begin{equation}
 \KL(\mathsf P_s\Vert\mathsf Q_s)
 =\frac12\sum_k\int_{[t_k,t_{k+1}]\cap[0,s]}
 \{\gamma(t)^2(t-t_k)+t_k[\gamma(t)-\gamma(t_k)]^2\}\dd t.
 \label{eq:uniform-entropy}
\end{equation}
For each cell with $t_k<s$, define the prefix-dependent increment $\Delta_{k,s}\gamma=\gamma(\min(t_{k+1},s))-\gamma(t_k)$. These increments are nonnegative, and their sum telescopes:
\begin{equation}
 \sum_{k:t_k<s}\Delta_{k,s}\gamma=\gamma(s)-\gamma(0)\le\gamma(s),
 \qquad
 \sum_{k:t_k<s}(\Delta_{k,s}\gamma)^2
       \le\left(\sum_{k:t_k<s}\Delta_{k,s}\gamma\right)^2\le\gamma(s)^2.
 \label{eq:ams-prefix-gamma-increments}
\end{equation}
On each intersected cell, $0\le\gamma(t)-\gamma(t_k)\le\Delta_{k,s}\gamma$, and its length is at most $h$. Consequently,
\begin{equation}
 \sum_k\int_{[t_k,t_{k+1}]\cap[0,s]}[\gamma(t)-\gamma(t_k)]^2\dd t
       \le h\sum_{k:t_k<s}(\Delta_{k,s}\gamma)^2\le h\gamma(s)^2.
 \label{eq:uniform-gamma-variation}
\end{equation}
The endpoint differences include any jumps of $\gamma$, and the calculation also covers a partially intersected last cell. When $s=0$, both sums are empty and the bound is immediate. Since $t-t_k\le h$, $t_k\le1$, and $\int_0^1\gamma=1$, equations \eqref{eq:uniform-entropy} and \eqref{eq:uniform-gamma-variation} imply
\begin{align}
 \KL(\mathsf P_s\Vert\mathsf Q_s)
 &\le\frac h2\left\{\int_0^s\gamma(t)^2\dd t+\gamma(s)^2\right\}\nonumber\\
 &\le\frac h2\{\gamma(s)+\gamma(s)^2\}\le \mathsf{K}h\Lambda(s).
 \label{eq:ams-prefix-entropy-bound}
\end{align}
Here $\int_0^s\gamma(t)^2\dd t\le\gamma(s)\int_0^s\gamma(t)\dd t\le\gamma(s)$, and the final inequality uses \eqref{eq:sharp-regularity} and $\Lambda(s)\ge1$.

\par\smallskip\noindent\textit{Moment transfer and normalization factors.}\quad
For each fixed $t\le T$, Lemma~\ref{lem:entropy} and \eqref{eq:ams-prefix-entropy-bound} imply $d_{\rm Hel}(\mathsf P_t,\mathsf Q_t)\le \mathsf{K}\sqrt{h\Lambda(t)}$. We apply the lemma to the fixed path observables $a(t,Z_t)$ and $a_x(t,Z_t)$. Their supremum bounds are $\mathsf{K}\sqrt{\Lambda(t)}$ and $\mathsf{K}\Lambda(t)$ by \eqref{eq:sharp-regularity} and \eqref{eq:sharp-spatial-derivative}, while $\E[a(t,X_t)^2]=1$. For the derivative moment under the true law, \eqref{eq:weighted-identity} and $a\ge0$ imply, initially for almost every $t$,
\begin{align}
 \E[a_x(t,X_t)^2]&=2\gamma(t)\E[a(t,X_t)^3]\nonumber\\
 &\le2\gamma(t)\norm{a(t,\cdot)}_\infty\E[a(t,X_t)^2]
       \le \mathsf{K}\gamma(t)\sqrt{\Lambda(t)}.
 \label{eq:ams-true-derivative-moment}
\end{align}
The function $t\mapsto\E[a_x(t,X_t)^2]$ is continuous by joint continuity of $a_x$, continuity of $X$, and dominated convergence on every fixed strip below time one. For any $t\le T$, choose $t_n\downarrow t$ outside the exceptional null set, using a strip ending strictly after $T$ if $t=T$. Right-continuity of $\gamma$ and continuity of $\Lambda$ then extend the upper bound in \eqref{eq:ams-true-derivative-moment} to $t$, including every grid node. No derivative of $\gamma$ is used.

The Hellinger norm comparison for $a(t,Z_t)$ bounds its norm error by $\mathsf{K}\sqrt h\Lambda(t)$. Choosing $c_0$ sufficiently small in $h\Lambda(T)^2\le c_0$ therefore ensures $3/4\le\norm{a(t,Y_t)}_2\le5/4$ for every $t\le T$. For the derivative observable, the same comparison yields
\begin{align}
 \norm{a_x(t,Y_t)}_2
 &\le\norm{a_x(t,X_t)}_2
       +\norm{a_x(t,\cdot)}_\infty d_{\rm Hel}(\mathsf P_t,\mathsf Q_t)\nonumber\\
 &\le \mathsf{K}\{\gamma(t)\sqrt{\Lambda(t)}\}^{1/2}+\mathsf{K}\sqrt h\Lambda(t)^{3/2}.
 \label{eq:ams-derivative-norm-transfer}
\end{align}
Squaring with $(u+v)^2\le2u^2+2v^2$, and using the second moments under both laws for the mean comparison, we obtain
\begin{align}
 |\norm{a(t,Y_t)}_2-1|
 &\le\norm{a(t,\cdot)}_\infty d_{\rm Hel}(\mathsf P_t,\mathsf Q_t)
       \le \mathsf{K}\sqrt h\Lambda(t),\nonumber\\
 \E[a_x(t,Y_t)^2]
 &\le \mathsf{K}\gamma(t)\sqrt{\Lambda(t)}+\mathsf{K}h\Lambda(t)^3,\nonumber\\
 |\E[a(t,Y_t)]-g(t)|
 &\le d_{\rm Hel}(\mathsf P_t,\mathsf Q_t)
       \{1+\norm{a(t,Y_t)}_2\}\le \mathsf{K}\sqrt{h\Lambda(t)}.
 \label{eq:uniform-moment-transfer}
\end{align}
In particular, $3/4\le\rho_k\le5/4$ by the definition in \eqref{eq:uniform-iamp-scheme}. The fourth moment needed below satisfies
\begin{equation}
 \E[a(t,Y_t)^4]
 \le\norm{a(t,\cdot)}_\infty^2\E[a(t,Y_t)^2]
 \le\frac{25}{16}\norm{a(t,\cdot)}_\infty^2\le \mathsf{K}\Lambda(t).
 \label{eq:ams-curvature-fourth-moment}
\end{equation}
To sum the normalization errors, square the first estimate in \eqref{eq:uniform-moment-transfer} at $t_k$ and multiply by $h$. Since $\Lambda(t_k)^2\le\Lambda(t)^2$ on $[t_k,t_{k+1}]$, its left-endpoint sum is at most its integral. Thus \eqref{eq:singular-weights} implies
\begin{equation}
 \sum_kh(1-\rho_k)^2
 \le \mathsf{K}h\sum_kh\Lambda(t_k)^2
 \le \mathsf{K}h\int_0^T\Lambda(t)^2\dd t
       \le \mathsf{K}hL_\tau^2/\tau.
 \label{eq:uniform-normalization-integral}
\end{equation}

\par\smallskip\noindent\textit{It\^o identities and the within-cell drift discrepancy.}\quad
We now work with the Brownian motion $W$ in \eqref{eq:uniform-iamp-scheme} and its completed natural filtration $\mathcal F_t=\sigma(W_u:0\le u\le t)$, with the usual right-continuous augmentation. All expectations and norms below refer to this construction; the path $Y$ through time $T$ has law $\mathsf Q_T$. The interpolation is adapted, and $a_k$ is $\mathcal F_{t_k}$-measurable. Its use on $(t_k,t_{k+1}]$ is predictable, and $G_{k+1}=W_{t_{k+1}}-W_{t_k}$ is independent of $\mathcal F_{t_k}$.

Write $U_t=m(t,Y_t)$ and, on cell $k$, set
\begin{equation}
 \mathfrak d_t=\gamma(t)U_t-\gamma(t_k)U_{t_k}
       =\gamma(t)(U_t-U_{t_k})+[\gamma(t)-\gamma(t_k)]U_{t_k}.
 \label{eq:ams-interpolated-drift-difference}
\end{equation}
Lemma~\ref{lem:parisi-ito} applies to \eqref{eq:ams-interpolation}: its drift is bounded and progressively measurable in this Brownian filtration, and the lemma uses weak time derivatives of $m$ and $a$. It implies
\begin{equation}
 \begin{aligned}
 \dd U_t&=a(t,Y_t)\dd W_t-a(t,Y_t)\mathfrak d_t\dd t,\\
 \dd a(t,Y_t)&=a_x(t,Y_t)\dd W_t
       -\gamma(t)a(t,Y_t)^2\dd t-a_x(t,Y_t)\mathfrak d_t\dd t.
 \end{aligned}
 \label{eq:uniform-ito}
\end{equation}
All stochastic coefficients are bounded on $[0,T]$, so the stochastic integrals are square-integrable martingales in $(\mathcal F_t)$. In particular, they have conditional mean zero over each cell. Integrating \eqref{eq:uniform-ito} from $t_k$ to $t\le t_{k+1}$, we obtain
\begin{align}
 U_t-U_{t_k}
 &=\int_{t_k}^t a(s,Y_s)\dd W_s
       -\int_{t_k}^t a(s,Y_s)\mathfrak d_s\dd s,\nonumber\\
 a(t,Y_t)-a_k
 &=\int_{t_k}^t a_x(s,Y_s)\dd W_s
       -\int_{t_k}^t\gamma(s)a(s,Y_s)^2\dd s
       -\int_{t_k}^t a_x(s,Y_s)\mathfrak d_s\dd s.
 \label{eq:ams-cell-integrated-ito}
\end{align}
These are integral identities across jumps of $\gamma$.

For the within-cell estimates, use the full-cell increment $\Delta_k\gamma=\gamma(t_{k+1})-\gamma(t_k)$, rather than the prefix-dependent increment in \eqref{eq:ams-prefix-gamma-increments}. Put $A_k=\sup_{[t_k,t_{k+1}]\times\R}a$ and $f(s)=\norm{U_s-U_{t_k}}_2$. It\^o isometry, Minkowski's integral inequality, and $\norm{U_{t_k}}_2\le1$ imply
\begin{align}
 f(s)
 &\le\left(\int_{t_k}^s\E[a(v,Y_v)^2]\dd v\right)^{1/2}
       +\int_{t_k}^s\norm{a(v,Y_v)\mathfrak d_v}_2\dd v\nonumber\\
 &\le2\sqrt{s-t_k}+A_k\int_{t_k}^s\gamma(v)f(v)\dd v
       +A_k\int_{t_k}^s[\gamma(v)-\gamma(t_k)]\dd v\nonumber\\
 &\le2\sqrt{s-t_k}+A_k\int_{t_k}^s\gamma(v)f(v)\dd v
       +A_kh\Delta_k\gamma.
 \label{eq:ams-cell-gronwall}
\end{align}
The bound on the first term uses $\norm{a(v,Y_v)}_2\le5/4$. By \eqref{eq:sharp-regularity} and monotonicity of $\Lambda$,
\begin{align}
 A_k\int_{t_k}^{t_{k+1}}\gamma(v)\dd v&\le \mathsf{K}h\Lambda(T),\nonumber\\
 A_kh\Delta_k\gamma&\le \mathsf{K}h\Lambda(T)
       =\mathsf{K}\sqrt h\sqrt{h\Lambda(T)^2}\le \mathsf{K}\sqrt{c_0}\sqrt h.
 \label{eq:ams-cell-gronwall-parameters}
\end{align}
Since $h\Lambda(T)\le c_0$, the exponential factor in Gr\"onwall's inequality is universally bounded. Applying it to \eqref{eq:ams-cell-gronwall} with $2\sqrt{s-t_k}\le2\sqrt h$, we obtain
\begin{align}
 f(s)&\le(2\sqrt h+A_kh\Delta_k\gamma)
       \exp\left(A_k\int_{t_k}^s\gamma(v)\dd v\right)\nonumber\\
 &\le \mathsf{K}(\sqrt h+A_kh\Delta_k\gamma)\le \mathsf{K}\sqrt h,
 \qquad t_k\le s\le t_{k+1}.
 \label{eq:ams-cell-increment-bound}
\end{align}
Substitution in \eqref{eq:ams-interpolated-drift-difference} proves
\begin{equation}
 \norm{\mathfrak d_t}_2
 \le\gamma(t)f(t)+[\gamma(t)-\gamma(t_k)]\norm{U_{t_k}}_2
 \le \mathsf{K}\gamma(t)\sqrt h+\gamma(t)-\gamma(t_k).
 \label{eq:uniform-defect-bound}
\end{equation}

\par\smallskip\noindent\textit{Accumulated drift.}\quad
We estimate the two terms in \eqref{eq:uniform-defect-bound} with their deterministic weights. Minkowski's integral inequality and \eqref{eq:sharp-regularity} imply
\begin{align}
 \left\|\int_0^T a(t,Y_t)\mathfrak d_t\dd t\right\|_2
 &\le\int_0^T\norm{a(t,\cdot)}_\infty\norm{\mathfrak d_t}_2\dd t\nonumber\\
 &\le \mathsf{K}\sqrt h\int_0^T\gamma(t)\sqrt{\Lambda(t)}\dd t
       +\mathsf{K}\sum_k\int_{t_k}^{t_{k+1}}\sqrt{\Lambda(t)}
                    [\gamma(t)-\gamma(t_k)]\dd t.
 \label{eq:ams-accumulated-drift-weights}
\end{align}
The first integral is at most $\mathsf{K}\int_0^T\Lambda(t)\dd t\le \mathsf{K}L_\tau^2$. For the second contribution, each cell has length $h$ and $\gamma(t)-\gamma(t_k)\le\Delta_k\gamma$, so
\begin{equation}
 \sum_k\int_{t_k}^{t_{k+1}}\sqrt{\Lambda(t)}[\gamma(t)-\gamma(t_k)]\dd t
 \le h\sqrt{\Lambda(T)}\sum_k\Delta_k\gamma
 \le h\sqrt{\Lambda(T)}\gamma(T)\le \mathsf{K}h\Lambda(T).
 \label{eq:ams-accumulated-gamma-variation}
\end{equation}
The sum of the increments is $\gamma(T)-\gamma(0)$. Combining these estimates, without any independence assumption between cells, we obtain
\begin{align}
 \left\|\int_0^T a(t,Y_t)\mathfrak d_t\dd t\right\|_2
 &\le \mathsf{K}\sqrt h\int_0^T\gamma\sqrt\Lambda\dd t
       +\mathsf{K}h\sqrt{\Lambda(T)}\sum_k\Delta_k\gamma\nonumber\\
 &\le \mathsf{K}(\sqrt hL_\tau^2+h\Lambda(T)).
 \label{eq:uniform-cumulative-drift}
\end{align}

\par\smallskip\noindent\textit{Variation of the curvature coefficient.}\quad
Define $V_h=\sum_k\int_{t_k}^{t_{k+1}}\E[(a(t,Y_t)-a_k)^2]\dd t$. The second identity in \eqref{eq:ams-cell-integrated-ito} separates the coefficient change into one stochastic integral and two drift integrals. For the stochastic term, It\^o isometry and Tonelli's theorem yield
\begin{align}
 &\sum_k\int_{t_k}^{t_{k+1}}
       \E\left[\left(\int_{t_k}^t a_x(s,Y_s)\dd W_s\right)^2\right]\dd t\nonumber\\
 &\quad=\sum_k\int_{t_k}^{t_{k+1}}\int_{t_k}^t
                    \E[a_x(s,Y_s)^2]\dd s\dd t\nonumber\\
 &\quad=\sum_k\int_{t_k}^{t_{k+1}}(t_{k+1}-s)\E[a_x(s,Y_s)^2]\dd s\nonumber\\
 &\quad\le \mathsf{K}h\int_0^T\gamma(s)\sqrt{\Lambda(s)}\dd s
                    +\mathsf{K}h^2\int_0^T\Lambda(s)^3\dd s.
 \label{eq:ams-curvature-stochastic}
\end{align}
The last line uses \eqref{eq:uniform-moment-transfer} and $t_{k+1}-s\le h$, and accounts for the first two integrals below.

For the drift involving $\gamma a^2$, Cauchy--Schwarz on $[t_k,t]$ implies
\begin{equation}
 \E\left[\left(\int_{t_k}^t\gamma(s)a(s,Y_s)^2\dd s\right)^2\right]
 \le(t-t_k)\int_{t_k}^t\gamma(s)^2\E[a(s,Y_s)^4]\dd s.
 \label{eq:ams-curvature-first-drift-cell}
\end{equation}
Using $t-t_k\le h$ and then changing the order of integration, we obtain
\begin{align}
 &\sum_k\int_{t_k}^{t_{k+1}}
       \E\left[\left(\int_{t_k}^t\gamma(s)a(s,Y_s)^2\dd s\right)^2\right]\dd t\nonumber\\
 &\quad\le h\sum_k\int_{t_k}^{t_{k+1}}(t_{k+1}-s)
                   \gamma(s)^2\E[a(s,Y_s)^4]\dd s\nonumber\\
 &\quad\le h^2\int_0^T\gamma(s)^2\E[a(s,Y_s)^4]\dd s
       \le \mathsf{K}h^2\int_0^T\gamma(s)^2\Lambda(s)\dd s.
 \label{eq:ams-curvature-first-drift}
\end{align}
Here \eqref{eq:ams-curvature-fourth-moment} bounds the fourth moment. One factor of $h$ comes from Cauchy--Schwarz, and the other comes from the remaining cell length $t_{k+1}-s$.

For the drift involving $a_x\mathfrak d$, the corresponding cellwise inequality is
\begin{equation}
 \E\left[\left(\int_{t_k}^t a_x(s,Y_s)\mathfrak d_s\dd s\right)^2\right]
 \le(t-t_k)\int_{t_k}^t\E[(a_x(s,Y_s)\mathfrak d_s)^2]\dd s.
 \label{eq:ams-curvature-second-drift-cell}
\end{equation}
By \eqref{eq:uniform-defect-bound},
\begin{equation}
 \E[\mathfrak d_s^2]\le \mathsf{K}h\gamma(s)^2
                +\mathsf{K}[\gamma(s)-\gamma(t_{k(s)})]^2.
 \label{eq:ams-defect-second-moment}
\end{equation}
The supremum bound on $a_x$ in \eqref{eq:sharp-spatial-derivative}, followed by the same two factors of $h$, therefore implies
\begin{align}
 &\sum_k\int_{t_k}^{t_{k+1}}
       \E\left[\left(\int_{t_k}^t a_x(s,Y_s)\mathfrak d_s\dd s\right)^2\right]\dd t\nonumber\\
 &\quad\le h\sum_k\int_{t_k}^{t_{k+1}}(t_{k+1}-s)
                   \E[(a_x(s,Y_s)\mathfrak d_s)^2]\dd s\nonumber\\
 &\quad\le h^2\int_0^T\norm{a_x(s,\cdot)}_\infty^2\E[\mathfrak d_s^2]\dd s\nonumber\\
 &\quad\le \mathsf{K}h^3\int_0^T\gamma(s)^2\Lambda(s)^2\dd s
       +\mathsf{K}h^2\Lambda(T)^2\int_0^T[\gamma(s)-\gamma(t_{k(s)})]^2\dd s.
 \label{eq:ams-curvature-second-drift}
\end{align}
Applying $(x+y+z)^2\le3(x^2+y^2+z^2)$ to \eqref{eq:ams-cell-integrated-ito}, and combining \eqref{eq:ams-curvature-stochastic}, \eqref{eq:ams-curvature-first-drift}, and \eqref{eq:ams-curvature-second-drift}, we conclude that
\begin{align}
 V_h\le \mathsf{K}\bigg\{&h\int_0^T\gamma\sqrt\Lambda\dd t
 +h^2\int_0^T\Lambda^3\dd t+h^2\int_0^T\gamma^2\Lambda\dd t\nonumber\\
 &+h^3\int_0^T\gamma^2\Lambda^2\dd t
 +h^2\Lambda(T)^2\int_0^T[\gamma(t)-\gamma(t_{k(t)})]^2\dd t\bigg\}.
 \label{eq:uniform-curvature-integrals}
\end{align}
We bound these five terms separately. Equations~\eqref{eq:sharp-regularity}, \eqref{eq:singular-weights}, and \eqref{eq:uniform-gamma-variation} with $s=T$ imply
\begin{align}
 h\int_0^T\gamma\sqrt\Lambda\dd t
 &\le \mathsf{K}h\int_0^T\Lambda\dd t\le \mathsf{K}hL_\tau^2,\nonumber\\
 h^2\int_0^T\Lambda^3\dd t
 &\le \mathsf{K}h^2L_\tau^3/\tau^2,\nonumber\\
 h^2\int_0^T\gamma^2\Lambda\dd t
 &\le \mathsf{K}h^2\int_0^T\Lambda^2\dd t\le \mathsf{K}h^2L_\tau^2/\tau,\nonumber\\
 h^3\int_0^T\gamma^2\Lambda^2\dd t
 &\le \mathsf{K}h^3\int_0^T\Lambda^3\dd t\le \mathsf{K}h^3L_\tau^3/\tau^2,\nonumber\\
 h^2\Lambda(T)^2\int_0^T[\gamma(t)-\gamma(t_{k(t)})]^2\dd t
 &\le \mathsf{K}h^3\Lambda(T)^2\gamma(T)^2\le \mathsf{K}h^3\Lambda(T)^3.
 \label{eq:ams-curvature-weight-estimates}
\end{align}
Since $h\le\tau\le1$ and $\Lambda(T)=L_\tau/\tau$, the last three bounds satisfy
\begin{gather}
 h^2L_\tau^2/\tau=(h/\tau)hL_\tau^2\le hL_\tau^2,\nonumber\\
 h^3L_\tau^3/\tau^2\le h^2L_\tau^3/\tau^2,\qquad
 h^3\Lambda(T)^3=(h/\tau)h^2L_\tau^3/\tau^2\le h^2L_\tau^3/\tau^2.
 \label{eq:ams-curvature-absorption}
\end{gather}
Consequently, $V_h\le \mathsf{K}(hL_\tau^2+h^2L_\tau^3/\tau^2)$.

\par\smallskip\noindent\textit{Terminal error.}\quad
The initial values are $U_0=m(0,0)=0$ and $M_0=\sqrt h$. Integrating the first equation in \eqref{eq:uniform-ito} over $[0,T]$ and subtracting it from the magnetization recursion in \eqref{eq:uniform-iamp-scheme}, we obtain
\begin{align}
 M_K-m(T,Y_T)&=\sqrt h+
 \int_0^T\left(\frac{a_{k(t)}}{\rho_{k(t)}}-a(t,Y_t)\right)\dd W_t\nonumber\\
 &\quad+\int_0^T a(t,Y_t)\mathfrak d_t\dd t.
 \label{eq:ams-terminal-error-decomposition}
\end{align}
On cell $k$, split the stochastic integrand as
\begin{equation}
 \frac{a_k}{\rho_k}-a(t,Y_t)
       =a_k(1/\rho_k-1)+a_k-a(t,Y_t).
 \label{eq:ams-terminal-stochastic-splitting}
\end{equation}
For the normalization term, put $\zeta_k=(1/\rho_k-1)a_kG_{k+1}$. The normalization factors are deterministic, and $a_k$ is measurable at time $t_k$. Conditional centering and the Gaussian increment variance imply
\begin{equation}
 \E[\zeta_k\mid\mathcal F_{t_k}]=0,\qquad
 \E[\zeta_k^2\mid\mathcal F_{t_k}]
       =h(1/\rho_k-1)^2a_k^2,\qquad
 \E[\zeta_k^2]=h(1-\rho_k)^2.
 \label{eq:ams-normalization-increment-moments}
\end{equation}
The last equality uses $\E[a_k^2]=\rho_k^2$. For $j<k$, the variable $\zeta_j$ is $\mathcal F_{t_k}$-measurable, so $\E[\zeta_j\zeta_k]=\E[\zeta_j\E[\zeta_k\mid\mathcal F_{t_k}]]=0$. Hence the normalization increments are orthogonal, and
\begin{equation}
 \left\|\sum_k\zeta_k\right\|_2^2=\sum_kh(1-\rho_k)^2,
 \qquad
 \left\|\int_0^T\{a_{k(t)}-a(t,Y_t)\}\dd W_t\right\|_2^2=V_h.
 \label{eq:ams-terminal-stochastic-moments}
\end{equation}
The second identity is It\^o isometry. The two stochastic terms in \eqref{eq:ams-terminal-stochastic-splitting} need not be orthogonal to each other; the squared-sum inequality suffices:
\begin{align}
 &\left\|\int_0^T\left(\frac{a_{k(t)}}{\rho_{k(t)}}-a(t,Y_t)\right)\dd W_t\right\|_2^2\nonumber\\
 &\quad\le2\sum_kh(1-\rho_k)^2+2V_h
       \le \mathsf{K}\{hL_\tau^2/\tau+hL_\tau^2+h^2L_\tau^3/\tau^2\}.
 \label{eq:ams-terminal-stochastic-bound}
\end{align}
Taking square roots and using \eqref{eq:uniform-cumulative-drift} in \eqref{eq:ams-terminal-error-decomposition}, we conclude that
\begin{align}
 \norm{M_K-m(T,Y_T)}_2
 &\le\sqrt h+\mathsf{K}\left\{\frac{\sqrt hL_\tau}{\sqrt\tau}
       +\sqrt hL_\tau+\frac{hL_\tau^{3/2}}\tau
       +\sqrt hL_\tau^2+h\Lambda(T)\right\}\nonumber\\
 &\le \mathsf{K}\left\{\frac{\sqrt hL_\tau}{\sqrt\tau}
       +\sqrt hL_\tau^2+\frac{hL_\tau^{3/2}}\tau\right\}.
 \label{eq:ams-terminal-error-bound}
\end{align}
For the last inequality, $L_\tau\ge1$ absorbs the initial $\sqrt h$ and the term $\sqrt hL_\tau$ into $\sqrt hL_\tau^2$, and $h\Lambda(T)=hL_\tau/\tau\le hL_\tau^{3/2}/\tau$. Since $m(T,Y_T)\in[-1,1]$, the pointwise inequality $\dist(M_K,[-1,1])\le|M_K-m(T,Y_T)|$ proves \eqref{eq:uniform-clipping}.

\par\smallskip\noindent\textit{Correlation and the terminal integral.}\quad
By \eqref{eq:uniform-energy}, $\mathcal E_h=\sum_kh\E[a_k]/\rho_k$. The additional Brownian increment from $T$ to $T+h$ is centered and independent of $M_K$, so it does not change this correlation. Nonnegativity of $a$ and Cauchy--Schwarz imply $0\le\E[a_k]\le\rho_k$, and therefore
\begin{equation}
 \left|\frac{\E[a_k]}{\rho_k}-\E[a_k]\right|
       =\frac{\E[a_k]}{\rho_k}|1-\rho_k|\le|1-\rho_k|.
 \label{eq:ams-correlation-normalization-pointwise}
\end{equation}
Multiplying this bound by $h$ and summing the first estimate in \eqref{eq:uniform-moment-transfer} at the grid nodes, we obtain
\begin{align}
 \left|\mathcal E_h-\sum_kh\E[a_k]\right|
 &\le\sum_kh|1-\rho_k|
       \le \mathsf{K}\sqrt h\sum_kh\Lambda(t_k)\nonumber\\
 &\le \mathsf{K}\sqrt h\int_0^T\Lambda(t)\dd t\le \mathsf{K}\sqrt hL_\tau^2.
 \label{eq:ams-correlation-normalization-sum}
\end{align}
The mean comparison in \eqref{eq:uniform-moment-transfer}, with the same weights, implies separately
\begin{align}
 \left|\sum_kh\E[a_k]-\sum_khg(t_k)\right|
 &\le\sum_kh|\E[a_k]-g(t_k)|
       \le \mathsf{K}\sqrt h\sum_kh\sqrt{\Lambda(t_k)}\nonumber\\
 &\le \mathsf{K}\sqrt h\int_0^T\sqrt{\Lambda(t)}\dd t\le \mathsf{K}\sqrt h.
 \label{eq:ams-correlation-mean-sum}
\end{align}
Both left-endpoint sums are bounded by their integrals because $\Lambda$ and $\sqrt\Lambda$ are increasing. The first integral is bounded by \eqref{eq:singular-weights}, and the second is uniformly finite by \eqref{eq:integrable-weights} with $p=j=1/2$.

Finally, $g$ is nonincreasing by \eqref{eq:g-identities}, so $hg(t_k)\ge\int_{t_k}^{t_{k+1}}g(t)\dd t$ on every cell. Using \eqref{eq:parisi-integral} and \eqref{eq:terminal-g}, we obtain
\begin{equation}
 \sum_khg(t_k)\ge\int_0^Tg(t)\dd t
       =\Par-\int_T^1g(t)\dd t
       \ge\Par-\mathsf{K}\tau^{3/2}\sqrt{L_\tau}.
 \label{eq:ams-correlation-terminal-integral}
\end{equation}
Combining \eqref{eq:ams-correlation-normalization-sum}, \eqref{eq:ams-correlation-mean-sum}, and \eqref{eq:ams-correlation-terminal-integral}, and absorbing $\mathsf{K}\sqrt h$ into $\mathsf{K}\sqrt hL_\tau^2$, proves \eqref{eq:uniform-energy-bound}.
\end{proof}
\subsection{Increment cutoff, matching moments, and Lipschitz constants for AMS IAMP}
\label{app:matrix-onsager-regularity}

We prove Lemmas~\ref{lem:onsager-cutoff} and~\ref{lem:onsager-history} for the population iteration defined after \eqref{eq:onsager-population-B}. We first compare the scalar recursions with and without the increment cutoff, then verify the matching conditions \eqref{eq:onsager-matching} and the Lipschitz bounds required by Lemma~\ref{lem:reeves-onsager-bound}.

\begin{proof}[Proof of Lemma~\ref{lem:onsager-cutoff}]
\par\smallskip\noindent\textit{Gaussian tails and normalization factors.}\quad
We couple the two scalar recursions with the same increments $G_1,\ldots,G_K$. Let $\varphi$ be the standard Gaussian density and put $x=h^{-1/2}\ge1$. Since $\varphi'(z)=-z\varphi(z)$, integration by parts implies
\begin{align}
 p:=\Prob\{\max_{k<K}|G_{k+1}|>1\}
 &\le2K\int_x^\infty\varphi(z)\dd z\le2Ke^{-1/(2h)},\nonumber\\
 b:=h^{-1}\E[G^2\one_{\{|G|>1\}}]
 &=2\int_x^\infty z^2\varphi(z)\dd z
   =2x\varphi(x)+2\int_x^\infty\varphi(z)\dd z
   \le \mathsf{K}e^{-1/(4h)}.
 \label{eq:onsager-normalized-tail}
\end{align}
For the last inequality, we absorb the factor $1+h^{-1/2}$ into the exponential. On $\{|G|>1\}$, the differences $G^2-\chi_1(G)^2$ and $G^2-G\chi_1(G)$ are nonnegative and at most $G^2$, and $|G-\chi_1(G)|^2\le G^2$. Therefore
\begin{equation}
 0\le1-v_{h,1}/h\le b,\qquad
 0\le1-\E[G\chi_1(G)]/h\le b,\qquad
 h^{-1}\E[(G-\chi_1(G))^2]\le b.
\end{equation}

Let $\mathcal A_k=\{\max_{1\le j\le k}|G_j|>1\}$, with $\mathcal A_0=\varnothing$. Induction in the state updates shows that $Y_j^1=Y_j$ for every $j\le k$ on $\mathcal A_k^c$. Write $a_k^1=a(t_k,Y_k^1)$ and $a_k=a(t_k,Y_k)$. Since both curvatures are bounded by $\kappa_a(T)$,
\begin{equation}
 \E[|a_k^1-a_k|^2]\le4\kappa_a(T)^2p,\qquad
 |\E[(a_k^1)^2]-\E[a_k^2]|\le2\kappa_a(T)^2p.
\end{equation}
Put $\nu=v_{h,1}/h\le1$. The defining identity $(\rho_k^1)^2=\nu\E[(a_k^1)^2]$ and $\E[a_k^2]=\rho_k^2\le25/16$ imply
\begin{equation}
 |(\rho_k^1)^2-\rho_k^2|
 \le\nu|\E[(a_k^1)^2]-\E[a_k^2]|+(1-\nu)\rho_k^2
 \le \mathsf{K}\kappa_a(T)^2(p+b).
 \label{eq:cutoff-normalizer-bias}
\end{equation}
For both admissible grids, $\kappa_a(T)$ and $K$ are polynomially bounded in inverse accuracy, whereas $p+b$ decays exponentially in $1/h$. By Theorem~\ref{thm:uniform-scalar}, $\rho_k\in[3/4,5/4]$, so sufficiently small accuracy ensures $\rho_k^1\in[1/2,2]$. Factoring the difference of the squares and then the reciprocals, we obtain
\begin{equation}
 |\rho_k^1-\rho_k|+
 |(\rho_k^1)^{-1}-\rho_k^{-1}|\le \mathsf{K}\kappa_a(T)^2(p+b).
\end{equation}
In particular, we retain $(\rho_0^1)^2=\nu$, since $a(0,0)=1$; the cutoff does not leave the first normalization factor equal to one.

\par\smallskip\noindent\textit{Magnetization increments and correlation.}\quad
Let $\mathcal F_k=\sigma(G_1,\ldots,G_k)$ be the common past of the coupled recursions, and put $\xi_{k+1}=(M_{k+1}^1-M_k^1)-(M_{k+1}-M_k)$. The coefficients below are $\mathcal F_k$-measurable, and
\begin{equation}
 \xi_{k+1}=c_kG_{k+1}
   +\frac{a_k^1}{\rho_k^1}\{\chi_1(G_{k+1})-G_{k+1}\},\qquad
 c_k=\frac{a_k^1-a_k}{\rho_k^1}
   +a_k\{(\rho_k^1)^{-1}-\rho_k^{-1}\}.
\end{equation}
The current increment is independent of $\mathcal F_k$, and both functions of it in this display have mean zero by oddness. Thus $\E[\xi_{k+1}\mid\mathcal F_k]=0$. The coefficient estimate uses the second moment of $a_k$:
\begin{equation}
 \E[c_k^2]\le \mathsf{K}\E[|a_k^1-a_k|^2]
   +\mathsf{K}\kappa_a(T)^4(p+b)^2\E[a_k^2]
 \le \mathsf{K}\{\kappa_a(T)^2p+\kappa_a(T)^4(p+b)^2\}.
\end{equation}
The two functions of $G_{k+1}$ in $\xi_{k+1}$ are not independent. Applying $(s+t)^2\le2s^2+2t^2$ before conditioning, we obtain
\begin{align}
 \E[\xi_{k+1}^2\mid\mathcal F_k]
 &\le2hc_k^2+2hb(a_k^1/\rho_k^1)^2,\nonumber\\
 h^{-1}\E[\xi_{k+1}^2]
 &\le \mathsf{K}\{\kappa_a(T)^2p+\kappa_a(T)^4(p+b)^2+\kappa_a(T)^2b\}.
\end{align}
For $j<k$, the tower property implies $\E[\xi_{j+1}\xi_{k+1}]=\E[\xi_{j+1}\E[\xi_{k+1}\mid\mathcal F_k]]=0$. Both magnetizations start at $\sqrt h$, so their difference is the sum of these orthogonal increments. Since $Kh\le1$,
\begin{equation}
 \norm{M_K^1-M_K}_2^2=\sum_{k<K}\E[\xi_{k+1}^2],\qquad
 \norm{M_K^1-M_K}_2
 \le \mathsf{K}\{\kappa_a(T)\sqrt p+\kappa_a(T)^2(p+b)+\kappa_a(T)\sqrt b\}.
\end{equation}

To compare the correlations, put $\lambda=\E[G\chi_1(G)]/h$, so $0\le\lambda\le1$ and $|1-\lambda|\le b$. Conditioning on successive increments, including the independent last increment of $W_{T+h}$, proves the sum in \eqref{eq:cutoff-correlation}. Subtracting \eqref{eq:uniform-energy}, we find
\begin{align}
 \mathcal E_h^1-\mathcal E_h
 =\sum_{k<K}h\biggl\{
 &\lambda\frac{\E[a_k^1-a_k]}{\rho_k^1}
 +\lambda\E[a_k]\{(\rho_k^1)^{-1}-\rho_k^{-1}\}
 +(\lambda-1)\frac{\E[a_k]}{\rho_k}\biggr\}.
\end{align}
Here $|\E[a_k^1-a_k]|\le2\kappa_a(T)p$, whereas $|\E[a_k]|\le\rho_k\le5/4$ by Cauchy--Schwarz. This second-moment bound avoids a further factor $\kappa_a(T)$ in the denominator contribution. Summing with total weight $Kh\le1$, we obtain
\begin{equation}
 |\mathcal E_h^1-\mathcal E_h|
 \le \mathsf{K}\{\kappa_a(T)p+\kappa_a(T)^2(p+b)+b\}.
\end{equation}
Substituting \eqref{eq:onsager-normalized-tail} into the last two estimates and decreasing the universal $c>0$ proves \eqref{eq:onsager-cutoff-bias}. This comparison holds the scalar increments fixed; the matrix iteration retains the cutoff.
\end{proof}

For the row calculation in Lemma~\ref{lem:onsager-history}, $W=Z\Omega$ makes each field increment equal to $G_{k+1}=\sqrt h z_{k+1}$. Here $z_s=Z_{is}$ denotes the entry in one fixed coordinate $i$; within that calculation, $U_{k+2}^1$ denotes the entry $(U_i^1)_{k+2}$ of the row $U_i^1$.

\begin{proof}[Proof of Lemma~\ref{lem:onsager-history}]
\par\smallskip\noindent\textit{Raw second moments and matching.}\quad
We use the admissible grids at sufficiently small accuracy, so that $\rho_k^1\in[1/2,2]$, $h\Lambda(T)^2\le1$, and $(K+1)h\le1$. In one row, let $\mathcal F_k=\sigma(z_1,\ldots,z_k)$ and $G_{k+1}=\sqrt h z_{k+1}$. The state $Y_k^1$ is $\mathcal F_k$-measurable, and the current increment is independent of this sigma-algebra. For $\Delta M_{k+1}^1=M_{k+1}^1-M_k^1$, oddness and the definition of $\rho_k^1$ imply
\begin{equation}
 \E[\Delta M_{k+1}^1\mid\mathcal F_k]=0,\qquad
 \E[(\Delta M_{k+1}^1)^2]
 =\frac{\E[a(t_k,Y_k^1)^2]v_{h,1}}{(\rho_k^1)^2}=h.
\end{equation}
Distinct increments are orthogonal by conditioning on the past of the later one. Since $M_j^1=\sqrt h+\sum_{\ell<j}\Delta M_{\ell+1}^1$, we obtain
\begin{equation}
 \E[M_j^1]=\sqrt h,\qquad
 \E[M_j^1M_k^1]=h+\min(j,k)h.
\end{equation}
These are raw second moments, not covariances centered at the mean. Right multiplication by $\Omega^{-1}$ takes successive differences and divides by $\sqrt h$, so
\begin{equation}
 U_i^1=\left(1,
 \frac{a(t_0,Y_0^1)\chi_1(\sqrt h z_1)}{\rho_0^1\sqrt h},\ldots,
 \frac{a(t_{K-1},Y_{K-1}^1)\chi_1(\sqrt h z_K)}{\rho_{K-1}^1\sqrt h}\right).
 \label{eq:onsager-comparator-row}
\end{equation}
Every nonconstant entry has mean zero and second moment one, and different such entries are orthogonal. Their cross moments with the constant first entry vanish. Thus $\E[(U_i^1)^\top U_i^1]=I_d$, including its first diagonal entry.

We compute the cross moments with $z_s$ in three cases. If $s\le k$, then $z_s$ and $a(t_k,Y_k^1)$ are $\mathcal F_k$-measurable, and
\begin{equation}
 \E[z_sU_{k+2}^1]
 =\E\left[\frac{z_sa(t_k,Y_k^1)}{\rho_k^1\sqrt h}
   \E[\chi_1(\sqrt h z_{k+1})\mid\mathcal F_k]\right]=0.
\end{equation}
If $s>k+1$, then $U_{k+2}^1$ is $\mathcal F_{k+1}$-measurable, whereas $z_s$ is independent and centered, so the cross moment is again zero. If $s=k+1$, independence of the current increment from the past implies
\begin{equation}
 \E[z_{k+1}U_{k+2}^1]
 =\frac{\E[a(t_k,Y_k^1)]}{\rho_k^1}
   \frac{\E[G\chi_1(G)]}{h}.
\end{equation}
The first column has zero cross moments because $\E z_s=0$. Each remaining cross moment has magnitude at most one by Cauchy--Schwarz. For a matrix supported at positions $(s,s+1)$, its squared action on $x$ is $\sum_{s<d}|D^1_{s,s+1}|^2x_{s+1}^2$; hence $\norm{D^1}\le1$.

The same row law holds for each of the $N$ coordinates. Since $F=U^1\Omega$ and $W=Z\Omega$, matrix multiplication now yields
\begin{equation}
 \E[F^\top F]/N=\Omega^\top I_d\Omega,\qquad
 \E[W^\top F]/N=\Omega^\top D^1\Omega,\qquad
 (\Omega^\top\Omega)^{-1}\E[W^\top F]/N
 =\Omega^{-1}D^1\Omega.
\end{equation}
Conjugating a strictly upper triangular matrix by an invertible upper triangular matrix preserves strict upper triangularity, so this is the correction matrix required in \eqref{eq:onsager-matching}.

To identify the derivative coefficients, view one row $w$ of $W$ as a Gaussian column with covariance $\Sigma=\Omega^\top\Omega$, and write $F_t(w)=f_{t-1}^1(w_1,\ldots,w_{t-1})$ for the scalar coordinate function. Gaussian integration by parts in covariance-matrix form states that
\begin{equation}
 \E[wF_t(w)]=\Sigma\E[\nabla F_t(w)],\qquad
 \E[\nabla F_t(w)]=\Sigma^{-1}\E[wF_t(w)]=B^1_{:,t}.
\end{equation}
We verify global Lipschitz continuity below, so this identity is valid for weak derivatives. The possible cutoff boundaries are the hyperplanes where a field increment equals $\pm1$; the nondegenerate Gaussian law assigns them probability zero. Since $F_t=f_{t-1}^1$ depends only on the first $t-1$ fields, its derivatives in the remaining coordinates vanish. Thus $B^1$ is precisely the population Onsager coefficient matrix.

\par\smallskip\noindent\textit{Lipschitz continuity in cumulative fields.}\quad
In this part, $z=(z^1,\ldots,z^K)$ and $z'=(z'^1,\ldots,z'^K)$ are arbitrary cumulative field histories, with $z^0=z'^0=0$. They are not the standardized Gaussian entries used in the row calculation. Put $\delta_{k+1}=(z^{k+1}-z^k)-(z'^{k+1}-z'^k)$ and $\delta_z=\norm{z-z'}$. The difference operator is the identity minus a coordinate shift, so $\sum_{k<K}\delta_{k+1}^2\le4\delta_z^2$. The state update and the spatial Lipschitz bound $\norm{m_x}_\infty\le\kappa_a(T)$ imply
\begin{equation}
 |Y_{k+1}^1(z)-Y_{k+1}^1(z')|
 \le(1+\mathsf{K}\kappa_a(T)h\gamma(t_k))|Y_k^1(z)-Y_k^1(z')|+|\delta_{k+1}|.
\end{equation}
We unroll this recurrence. Monotonicity of $\gamma$ implies $\sum_{k<K}h\gamma(t_k)\le\int_0^T\gamma\le1$, and hence
\begin{align}
 |Y_j^1(z)-Y_j^1(z')|
 &\le\sum_{\ell<j}|\delta_{\ell+1}|
   \prod_{q=\ell+1}^{j-1}(1+\mathsf{K}\kappa_a(T)h\gamma(t_q))\nonumber\\
 &\le e^{\mathsf{K}\kappa_a(T)}\sum_{\ell<j}|\delta_{\ell+1}|
 \le2\sqrt K e^{\mathsf{K}\kappa_a(T)}\delta_z.
\end{align}
The last step is Cauchy--Schwarz. For the magnetization map, we subtract the updates and first change the curvature, then the cutoff increment. Since $\rho_k^1\ge1/2$, $|\chi_1|\le1$, and $\norm{a_x}_\infty\le \mathsf{K}\Lambda(T)$,
\begin{align}
 |f_j^1(z)-f_j^1(z')|
 &\le \mathsf{K}\Lambda(T)\sum_{k<j}|Y_k^1(z)-Y_k^1(z')|
   +2\kappa_a(T)\sum_{k<j}|\delta_{k+1}|\nonumber\\
 &\le \mathsf{K}\{K^{3/2}\Lambda(T)e^{\mathsf{K}\kappa_a(T)}+\sqrt K\kappa_a(T)\}\delta_z.
\end{align}
This estimate includes $j=K$, the final magnetization map. Applying the scalar estimate in each coordinate and summing its square proves the same bound, with no factor depending on $N$, for the vector maps with the Euclidean history norm. We may therefore take
\begin{equation}
 L_F\le \mathsf{K}\{K^{3/2}\Lambda(T)e^{\mathsf{K}\kappa_a(T)}+\sqrt K\kappa_a(T)\}.
\end{equation}

It remains to bound the deterministic matrices and the correction maps. Factor $\Omega=\sqrt h T_d$, where $(T_d)_{st}=\one_{\{s\le t\}}$. Its inverse is the identity minus the matrix with ones on the first diagonal above the main diagonal, so $\norm{T_d^{-1}}\le2$. Each column of $\Omega$ has squared norm $jh\le dh\le1$. Consequently,
\begin{equation}
 \norm{\Omega^{-1}}\le2/\sqrt h,\qquad
 \norm\Omega^2\le\operatorname{Tr}(\Omega^\top\Omega)\le d,\qquad
 \operatorname{cond}(\Sigma)\le4d/h,\qquad
 \norm\Omega_{2,1}=\sum_{j=1}^d\sqrt{jh}\le d.
\end{equation}
Also $\norm{B^1}\le\norm{\Omega^{-1}}\norm{D^1}\norm\Omega\le\sqrt{4d/h}$. A column of $B^1$ has absolute sum at most $\sqrt d\norm{B^1}$, so for $g_t=-\sum_{s<t}B^1_{st}F_s$ we may take $L_g\le\sqrt d\norm{B^1}L_F$. The logarithms of the polynomial factors in these bounds are $O(\log(d+h^{-1}+2))$, since $h\Lambda(T)^2\le1$ implies $\log\Lambda(T)\le\frac12\log(1/h)$. Finally $\kappa_a(T)\le \mathsf{K}\sqrt{\Lambda(T)}$, so taking logarithms proves every bound in \eqref{eq:onsager-history-bounds}.
\end{proof}

\subsection{Directional bounds and energy of the Gaussian comparator}
\label{app:onsager-comparator}
\label{app:matrix-concentration}

We apply the directional calculations of Lemmas~\ref{lem:directional-qr} and~\ref{lem:directional-cross} to the ungrouped Gaussian comparator $W=Z\Omega$. Its increments use $\chi_1$ and it retains the initial magnetization $\sqrt h\one$. We verify the conditional exponential moments, higher moments, query measurability, and output weights needed in those proofs. No estimate of the operator norm of $L_0-I$ is used.

For $F=F(W)=QR_0$ and $U^1=F\Omega^{-1}$, define
\begin{equation}
 \begin{gathered}
 L_0=R_0\Omega^{-1},\qquad P=L_0-I,\qquad
 \widehat D=Z^\top U^1/N,\qquad E_D=\widehat D-D^1,\\
 w_{\rm A}=\Omega\mathbf e_d=\sqrt h(1,\ldots,1)^\top,\qquad
 x=D^1w_{\rm A},\qquad f=U^1w_{\rm A}=F_d(W).
 \end{gathered}
 \label{eq:onsager-directional-objects}
\end{equation}
Choose $\lambda_k=\mathsf{K}_\lambda\Lambda(t_k)$ so that $a(t_k,y)^2/(\rho_k^1)^2\le\lambda_k$ for every $y$, and put, for $t\ge1$,
\begin{equation}
 V_{\rm A}=h\sum_{k<K}\lambda_k\le\mathsf{K}L_\tau^2,\qquad
 \overline V_{\rm A}=1+V_{\rm A},\qquad
 \mathfrak r_N^2=\overline V_{\rm A}(d+t)/N.
 \label{eq:onsager-comparator-radius}
\end{equation}
The bound on $V_{\rm A}$ follows because $\Lambda$ is increasing and its integral is bounded by \eqref{eq:singular-weights}.

\begin{lemma}[Directional bounds for the AMS Gaussian comparator]
\label{lem:onsager-comparator-moments}
Let $t\ge1$, $d=K+1\le N$, and suppose
\begin{equation}
 N\ge d+t,\qquad
 N\ge\mathsf{K}\Lambda(T)\{t+\log(\mathsf{K}d)\}.
 \label{eq:onsager-directional-dimension}
\end{equation}
With failure at most $\mathsf{K}e^{-t}$, simultaneously
\begin{align}
 \norm{Pw_{\rm A}}+\norm{E_Dw_{\rm A}}
 &\le\mathsf{K}\mathfrak r_N,\nonumber\\
 |x^\top Pw_{\rm A}|
 &\le\mathsf{K}\left\{\sqrt{\overline V_{\rm A}t/N}
                         +\overline V_{\rm A}(d+t)/N\right\},\nonumber\\
 |w_{\rm A}^\top E_Dw_{\rm A}|
 &\le\mathsf{K}\left\{\sqrt{\overline V_{\rm A}t/N}
                         +\sqrt{\overline V_{\rm A}}t/N\right\}.
 \label{eq:onsager-comparator-event}
\end{align}
\end{lemma}

\begin{proof}
\par\smallskip\noindent\textit{Rows, filtrations, and conditional exponential moments.}\quad
The population maps and their normalization factors are deterministic. Reeves's Gaussian law for Definition~\ref{def:reeves-coupling}, recalled in Section~\ref{sec:reeves-coupling}, makes the entries of $Z$ iid standard Gaussians. Equation~\eqref{eq:onsager-comparator-row} then makes the rows of $(U^1,Z)$ iid. Lemma~\ref{lem:onsager-history} proves
\begin{equation}
 \E[(U_i^1)^\top U_i^1]=I_d,\qquad
 \E[Z_i^\top U_i^1]=D^1,\qquad 0\le D^1_{j,j+1}\le1.
 \label{eq:onsager-row-moments}
\end{equation}
Let $\mathcal H_k=\sigma(Z_{i,s}:1\le i\le N,\ s\le k)$. For $j=k+2$, the entries of column $U_j^1$ are
\begin{equation}
 (U_j^1)_i=\frac{a(t_k,Y_{k,i}^1)}{\rho_k^1\sqrt h}
                         \chi_1(\sqrt hZ_{i,k+1}).
\end{equation}
Conditional on $\mathcal H_k$, their coefficients are fixed and their current Gaussian variables are independent across $i$. They are centered by oddness of $\chi_1$. Since $F_j=\sqrt h\sum_{r\le j}U_r^1$, orthogonalizing $F$ and $U^1$ produces the same directions, including at a zero residual. The query $q_i$ is a deterministic function of $F_1,\ldots,F_i$, so it is $\mathcal H_{i-1}$-measurable. Every earlier query $q_i$, $i<j$, is therefore $\mathcal H_{j-2}$-measurable.

Write $b_k=a(t_k,Y_k^1)/\rho_k^1$ in one coordinate. Conditional symmetry, $|\chi_1(G)|\le|G|$, and the Gaussian exponential moment imply
\begin{align}
 \E[e^{\theta U_{k+2}^1}\mid\mathcal H_k]
 &=\E[\cosh(\theta b_k\chi_1(G)/\sqrt h)\mid\mathcal H_k]\nonumber\\
 &\le\E[\cosh(\theta b_kG/\sqrt h)\mid\mathcal H_k]
 =e^{\theta^2b_k^2/2}\le e^{\theta^2\lambda_k/2}.
 \label{eq:onsager-cutoff-mgf}
\end{align}
Here $G\sim\mathcal N(0,h)$ is independent of the past. Thus the conditional exponential-moment input to the projection argument holds with $\upsilon_j=\lambda_{j-2}$.

\par\smallskip\noindent\textit{Higher moments of the column entries.}\quad
Let $\nu_h=v_{h,1}/h$. For $h\le1$ and standard Gaussian $z$,
\begin{equation}
 \nu_h=\E[\min\{z^2,h^{-1}\}]
 \ge\E[\min\{z^2,1\}]=:c_\chi>0.
\end{equation}
The population normalization implies $\nu_h\E[b_k^2]=1$. Independence of $b_k$ from $G$ and $b_k^2\le\lambda_k$ therefore imply, for each integer $p\ge2$,
\begin{align}
 \E|U_{k+2}^1|^{2p}
 &=h^{-p}\E[b_k^{2p}]\E[\chi_1(G)^{2p}]\nonumber\\
 &\le\lambda_k^{p-1}\E[b_k^2](2p-1)!!
 \le p!\mathsf{K}^p\lambda_k^{p-1}.
 \label{eq:onsager-column-moments}
\end{align}
The last step uses $(2p-1)!!\le2^pp!$ and $\E[b_k^2]\le c_\chi^{-1}$. These are the higher-moment inputs of \eqref{eq:group-variance-moments}, with second moment exactly one. A continuous-martingale representation of the cutoff increment is not required.

\par\smallskip\noindent\textit{Strict upper entries of the query factor.}\quad
For deterministic $a\in\R^{j-1}$, conditional independence across coordinates and $Q^\top Q=NI$ imply
\begin{equation}
 \E\left[\left.\exp\left\{\theta\sum_{i<j}a_i(L_0)_{ij}\right\}
                 \right|\mathcal H_{j-2}\right]
 \le e^{\theta^2\upsilon_j\norm a^2/(2N)},\qquad
 (L_0)_{ij}=q_i^\top U_j^1/N.
 \label{eq:onsager-projection-mgf}
\end{equation}
This is \eqref{eq:predictable-qr-mgf}, whose proof uses only the conditional moments and measurability just verified. Let $P_{\rm str}$ be the strict upper triangle of $P$, and set $w_0=w_{\rm A}-\sqrt h\mathbf e_1$. Iterating \eqref{eq:onsager-projection-mgf} over columns with weights $(w_0)_j=\sqrt h$, $j\ge2$, gives
\begin{equation}
 \E e^{\theta a^\top P_{\rm str}w_0}
 \le e^{\theta^2V_{\rm A}\norm a^2/(2N)}.
\end{equation}
For a deterministic unit vector $a$, Chernoff's inequality therefore has failure $2e^{-v}$ at $\sqrt{2V_{\rm A}v/N}$. A norm larger than twice this radius requires a larger projection at a point of the deterministic $1/2$-net in \eqref{eq:euclidean-net-bound}. Taking $v=t+d\log5$, the union bound gives failure at most $2|\mathcal N_d|e^{-v}\le2e^{-t}$. A single application in the deterministic direction $x$, whose norm is at most one, gives a second failure probability $2e^{-t}$. Hence
\begin{equation}
 \norm{P_{\rm str}w_0}^2\le\mathsf{K}V_{\rm A}(d+t)/N,\qquad
 |x^\top P_{\rm str}w_0|\le\mathsf{K}\sqrt{V_{\rm A}t/N}.
 \label{eq:onsager-strict-directions}
\end{equation}

\par\smallskip\noindent\textit{Projection squares and diagonal entries.}\quad
Put $s_j=\norm{U_j^1}^2/N$, $b_j=\sum_{i<j}(L_0)_{ij}^2$, $\lambda_j^{\rm diag}=(L_0)_{jj}$, and $t_d=t+\log(\mathsf{K}d)$. For a deterministic unit vector $a\in\R^{j-1}$, the conditional Chernoff bound from \eqref{eq:onsager-projection-mgf} reads
\begin{equation}
 \Prob\left\{\left|\sum_{i<j}a_i(L_0)_{ij}\right|
          >\sqrt{2\upsilon_jv/N}\,\middle|\,\mathcal H_{j-2}\right\}
 \le2e^{-v}.
\end{equation}
The vector $((L_0)_{ij})_{i<j}$ has squared norm $b_j$. The factor-two norm estimate in \eqref{eq:euclidean-net-bound} implies that $b_j>8\upsilon_jv/N$ requires a projection larger than $\sqrt{2\upsilon_jv/N}$ at some $a\in\mathcal N_{j-1}$. The conditional union bound therefore gives
\begin{equation}
 \Prob\{b_j>8\upsilon_jv/N\mid\mathcal H_{j-2}\}
 \le2|\mathcal N_{j-1}|e^{-v}\le2\cdot5^{j-1}e^{-v}.
\end{equation}
Choose $v=(j-1)\log5+t_d$ and denote this failure event by $E_j$. The tower property gives $\Prob(E_j)=\E[\Prob(E_j\mid\mathcal H_{j-2})]\le2e^{-t_d}$. A union bound over $2\le j\le d$, followed by multiplication of the individual bounds by $h$ and summation, implies
\begin{equation}
 h\sum_{j=2}^db_j
 \le\frac{\mathsf{K}(d+t)}N h\sum_{j=2}^d\upsilon_j
 =\mathsf{K}V_{\rm A}(d+t)/N
 \label{eq:onsager-projection-squares}
\end{equation}
with failure at most $\mathsf{K}e^{-t}$. We used $j+t_d\le\mathsf{K}(d+t)$ and $h\sum_{j=2}^d\upsilon_j=V_{\rm A}$.

By \eqref{eq:onsager-column-moments}, the centered squares of the independent entries in column $j$ satisfy the moment Bernstein condition with variance and scale at most $\mathsf{K}\upsilon_j$. A union bound over columns gives
\begin{equation}
 |s_j-1|\le\mathsf{K}\{\sqrt{\upsilon_jt_d/N}+\upsilon_jt_d/N\},\qquad
 h\sum_{j=2}^d(s_j-1)^2\le\mathsf{K}V_{\rm A}t_d/N.
 \label{eq:onsager-diagonal-squares}
\end{equation}
For the second inequality, $N\ge\mathsf{K}(\max_j\upsilon_j)t_d$ absorbs the squared linear term before the weighted summation. The nonnegative diagonal entry $\lambda_j^{\rm diag}$ equals the residual norm divided by $\sqrt{Nh}$ when the residual is nonzero and is zero otherwise. The factorization $L_0^\top L_0=(U^1)^\top U^1/N$ therefore implies the same identities as \eqref{eq:qr-diagonal-identity}:
\begin{equation}
 (\lambda_j^{\rm diag})^2=s_j-b_j,\qquad
 \lambda_j^{\rm diag}-1
 =\tfrac12(s_j-1)-\tfrac12b_j-\tfrac12(\lambda_j^{\rm diag}-1)^2.
\end{equation}
Since $\lambda_j^{\rm diag}\ge0$, also $|\lambda_j^{\rm diag}-1|\le|s_j-1|+\sqrt{b_j}$. Thus
\begin{equation}
 h\sum_{j=2}^d(\lambda_j^{\rm diag}-1)^2
 \le\mathsf{K}\overline V_{\rm A}(d+t)/N.
 \label{eq:onsager-factor-diagonal-squares}
\end{equation}
The first column is $U_1^1=q_1=\one$, so $P\mathbf e_1=0$ and $Pw_{\rm A}=Pw_0$. Combining \eqref{eq:onsager-strict-directions} and \eqref{eq:onsager-factor-diagonal-squares} proves the required bound on $\norm{Pw_{\rm A}}$, including zero-residual outcomes.

\par\smallskip\noindent\textit{The scalar projection $x^\top Pw_{\rm A}$.}\quad
For $j\ge2$, put $\pi_j=x_j\sqrt h$. The shifted nonnegative structure of $D^1$ implies
\begin{equation}
 0\le\pi_j\le h,\qquad \sum_{j=2}^d\pi_j\le1,\qquad
 \sum_{j=2}^d\pi_j\upsilon_j\le V_{\rm A}.
\end{equation}
Only $j\ge2$ occurs because the first column of $P$ is zero; $x_1(w_{\rm A})_1$ need not vanish. For one row, set $X_\pi=\sum_{j=2}^d\pi_j(U_i^1)_j^2$. Minkowski's inequality, \eqref{eq:onsager-column-moments}, and weighted H\"older imply
\begin{align}
 \norm{X_\pi}_p
 &\le\mathsf{K}(p!)^{1/p}\sum_{j=2}^d\pi_j\upsilon_j^{1-1/p}\nonumber\\
 &\le\mathsf{K}(p!)^{1/p}
       \left(\sum_{j=2}^d\pi_j\right)^{1/p}
       \left(\sum_{j=2}^d\pi_j\upsilon_j\right)^{1-1/p}
 \le\mathsf{K}(p!)^{1/p}\overline V_{\rm A}^{1-1/p}.
\end{align}
Its mean is $\sum_{j=2}^d\pi_j$. The centered moments and independence across rows allow moment Bernstein with variance and scale at most $\mathsf{K}\overline V_{\rm A}$, so
\begin{equation}
 \left|\sum_{j=2}^d\pi_j(s_j-1)\right|
 \le\mathsf{K}\{\sqrt{\overline V_{\rm A}t/N}+\overline V_{\rm A}t/N\}
 \label{eq:onsager-weighted-diagonal}
\end{equation}
with failure at most $2e^{-t}$. The diagonal identity yields
\begin{align}
 x^\top Pw_{\rm A}
 ={}&x^\top P_{\rm str}w_0+\frac12\sum_{j=2}^d\pi_j(s_j-1)
       -\frac12\sum_{j=2}^d\pi_jb_j\nonumber\\
 &-\frac12\sum_{j=2}^d\pi_j(\lambda_j^{\rm diag}-1)^2.
\end{align}
Equations~\eqref{eq:onsager-strict-directions}--\eqref{eq:onsager-weighted-diagonal}, with $\pi_j\le h$, prove the scalar assertion. These are the weighted projection and diagonal calculations in the proof of Lemma~\ref{lem:directional-qr}, with their inputs verified above.

\par\smallskip\noindent\textit{Empirical cross moments in the output direction.}\quad
Iterating \eqref{eq:onsager-cutoff-mgf} with the factor $\sqrt h$ shows that $\E e^{\theta(f_i-\sqrt h)}\le e^{\theta^2V_{\rm A}/2}$. Consequently $\norm{f_i}_{2p}\le\mathsf{K}\sqrt{p\overline V_{\rm A}}$, including the deterministic initial value. For a deterministic unit vector $\vartheta\in\R^d$, define $\xi_i(\vartheta)=\sum_s\vartheta_sZ_{is}$ and $\psi_i(\vartheta)=\xi_i(\vartheta)f_i$. Then
\begin{equation}
 \vartheta^\top\widehat Dw_{\rm A}=N^{-1}\sum_i\psi_i(\vartheta),\qquad
 \E\psi_i(\vartheta)=\vartheta^\top D^1w_{\rm A}.
\end{equation}
The Gaussian projection is standard normal. Cauchy--Schwarz gives
\begin{equation}
 \norm{\psi_i(\vartheta)}_p
 \le\norm{\xi_i(\vartheta)}_{2p}\norm{f_i}_{2p}
 \le\mathsf{K}p\sqrt{\overline V_{\rm A}}.
\end{equation}
The two factors need not be independent, but the products are independent across $i$. As in \eqref{eq:cross-coordinate-moments}--\eqref{eq:cross-bernstein-radius}, the centered moments imply the fixed-direction radius $\mathsf{K}\sqrt{\overline V_{\rm A}}(\sqrt{v/N}+v/N)$ with failure $2e^{-v}$. Take $v=t+d\log5$. The union bound over $\mathcal N_d$ has failure at most $2e^{-t}$; on its complement this radius holds for every point of the net. The factor-two norm estimate and $N\ge d+t$ prove the bound on $\norm{E_Dw_{\rm A}}$. Apply the same scalar bound once at $\vartheta=w_{\rm A}/\norm{w_{\rm A}}$ and $v=t$ to obtain the bound on $|w_{\rm A}^\top E_Dw_{\rm A}|$, since $\norm{w_{\rm A}}\le1$. Here the full weight $w_{\rm A}$ is retained because $E_D\mathbf e_1$ need not vanish. A union bound over the displayed events proves \eqref{eq:onsager-comparator-event}.
\end{proof}

\begin{lemma}[Energy and clipping for the Gaussian comparator]
\label{lem:onsager-comparator-energy}
Under \eqref{eq:onsager-directional-dimension}, if $\mathfrak r_N$ is sufficiently small, then, with failure at most $\mathsf{K}e^{-t}$,
\begin{equation}
 \left|\frac{H_N(f)}N-\mathcal E_h^1\right|
 \le\mathsf{K}\left\{\sqrt{\overline V_{\rm A}t/N}
                          +\overline V_{\rm A}(d+t)/N\right\},\qquad
 \norm f/\sqrt N\le5/4.
 \label{eq:onsager-comparator-energy}
\end{equation}
Moreover, let $0<\beta<1/2$, $t_\beta=\log(4/\beta)$, $\kappa^2\ge\max\{1,4\sup_{s\le T,x}a(s,x)^2\}$, and $\mu=\E[\dist(M_K^1,[-1,1])^2]$. Except with probability $\beta$,
\begin{equation}
 \frac{\norm{f-\clip(f,[-1,1]^N)}^2}{N}
 \le2\mu+\mathsf{K}\kappa^2\frac{\log(4N/\beta)t_\beta}{N}.
 \label{eq:onsager-comparator-clipping}
\end{equation}
\end{lemma}

\begin{proof}
\par\smallskip\noindent\textit{Energy of the coupled comparator.}\quad
The queries $F_t=f_{t-1}^1$ construct $W$ from the same $A$ that occurs in $H_N$, so Proposition~\ref{prop:energy-identity} applies with \eqref{eq:onsager-directional-objects}. Since $w_{\rm A}$ has all entries $\sqrt h$, \eqref{eq:cutoff-cross-moments} implies
\begin{equation}
 w_{\rm A}^\top D^1w_{\rm A}
 =h\sum_{k<K}D^1_{k+1,k+2}=\mathcal E_h^1.
\end{equation}
Substitute $L_0=I+P$ and $\widehat D=D^1+E_D$ into \eqref{eq:exact-energy}, and expand the auxiliary quadratic form:
\begin{align}
 H_N(f)/N-\mathcal E_h^1
 ={}&w_{\rm A}^\top E_Dw_{\rm A}
       +x^\top Pw_{\rm A}-\tfrac12w_{\rm A}^\top Jw_{\rm A}
       +\mathcal R_N^{\rm AMS},\nonumber\\
 \mathcal R_N^{\rm AMS}
 ={}&(Pw_{\rm A})^\top E_Dw_{\rm A}
       -(Pw_{\rm A})^\top Jw_{\rm A}
       -\tfrac12(Pw_{\rm A})^\top J(Pw_{\rm A}).
 \label{eq:ams-matrix-energy-expansion}
\end{align}
The marginal law of $J$ and \eqref{eq:gaussian-matrix-norms} imply $\norm J\le\mathsf{K}\sqrt{(d+t)/N}$ with failure at most $2e^{-t}$. The deterministic vector $w_{\rm A}$ has squared norm $T+h\le1$, so $w_{\rm A}^\top Jw_{\rm A}$ is centered Gaussian with variance $2(T+h)^2/N$. Its absolute value is at most $\mathsf{K}\sqrt{t/N}$ with failure at most $2e^{-t}$.

Intersect these events with Lemma~\ref{lem:onsager-comparator-moments}. Cauchy--Schwarz and $\sqrt{(d+t)/N}\le\mathfrak r_N$ imply
\begin{align}
 |\mathcal R_N^{\rm AMS}|
 &\le\norm{Pw_{\rm A}}\norm{E_Dw_{\rm A}}
 +\norm J\left\{\norm{Pw_{\rm A}}+\tfrac12\norm{Pw_{\rm A}}^2\right\}\nonumber\\
 &\le\mathsf{K}\mathfrak r_N^2
\end{align}
when $\mathfrak r_N$ is small. Bound the three linear terms in \eqref{eq:ams-matrix-energy-expansion} by \eqref{eq:onsager-comparator-event} and the scalar Gaussian bound, and absorb $\sqrt{\overline V_{\rm A}}t/N$ into $\overline V_{\rm A}(d+t)/N$. This proves the energy bound. Finally, $f=QL_0w_{\rm A}$ and $Q^\top Q=NI$ imply
\begin{equation}
 \norm f/\sqrt N=\norm{L_0w_{\rm A}}
 \le\norm{w_{\rm A}}+\norm{Pw_{\rm A}}
 \le1+\mathsf{K}\mathfrak r_N\le5/4.
\end{equation}
All failure probabilities are combined by a union bound. Neither Gaussian bound on $J$ is conditioned on $P$ or on a comparator event.

\par\smallskip\noindent\textit{Expected and realized clipping loss.}\quad
In one scalar row, let $\mathcal F_k=\sigma(G_1,\ldots,G_k)$ and $b_k=a(t_k,Y_k^1)/\rho_k^1$. These coefficients are measurable in the chronological past and satisfy $|b_k|\le\kappa$. The conditional cosh calculation in \eqref{eq:onsager-cutoff-mgf}, with the factor $\sqrt h$, implies
\begin{equation}
 \E[e^{\lambda b_k\chi_1(G_{k+1})}\mid\mathcal F_k]
 \le e^{\lambda^2\kappa^2h/2},\qquad
 \E e^{\lambda(M_K^1-\sqrt h)}\le e^{\lambda^2\kappa^2Kh/2}
 \le e^{\lambda^2\kappa^2/2}.
\end{equation}
The second bound follows by successive conditioning. Chernoff's inequality in both signs gives
\begin{equation}
 \Prob\{|M_K^1-\sqrt h|>x\}\le2e^{-x^2/(2\kappa^2)},\qquad
 \Prob\{\dist(M_K^1,[-1,1])^2>2\kappa^2s\}\le2e^{-s}.
\end{equation}
For the last bound, $\sqrt h\in[-1,1]$ implies $\dist(M_K^1,[-1,1])\le|M_K^1-\sqrt h|$.

Set $D_i=\dist(f_i,[-1,1])^2$. The $f_i$ are iid copies of $M_K^1$, so $\E[N^{-1}\sum_iD_i]=\mu$. To bound the realized average, set $R=2\kappa^2\log(4N/\beta)$ and $D_i'=D_i\one_{\{D_i\le R\}}$. These are independent functions of the Gaussian rows; we do not condition on the event that all rows are below the cutoff. With $\mu'=\E D_i'$, we have
\begin{equation}
 0\le D_i'\le R,\qquad \mu'\le\mu,\qquad
 \operatorname{Var}(D_i')\le\E[(D_i')^2]\le R\mu.
\end{equation}
Bounded scalar Bernstein at $t_\beta=\log(4/\beta)$ and Young's inequality imply, with failure at most $\beta/2$,
\begin{equation}
 N^{-1}\sum_iD_i'\le\mu'+\sqrt{2R\mu t_\beta/N}
                      +2Rt_\beta/(3N)
 \le2\mu+\mathsf{K}Rt_\beta/N.
\end{equation}
The row-tail union bound also gives $\Prob\{\max_iD_i>R\}\le\beta/2$. The event inclusion
\begin{equation}
 \{N^{-1}\sum_iD_i>2\mu+\mathsf{K}Rt_\beta/N\}
 \subseteq\{\max_iD_i>R\}
 \cup\{N^{-1}\sum_iD_i'>2\mu+\mathsf{K}Rt_\beta/N\}
\end{equation}
therefore proves \eqref{eq:onsager-comparator-clipping} with total failure at most $\beta$.
\end{proof}

\subsection{Empirical coefficients and their effect on the trajectory}
\label{app:onsager-calibration}

We prove concentration only for statistics evaluated on the fixed Gaussian comparator $W$. Their values on an arbitrary field history are compared with those on $W$ by deterministic inequalities. We then propagate the resulting coefficient errors through the empirical iteration on the same matrix as the population iteration. The actual field coordinates are not assumed independent or Gaussian.

\begin{proof}[Proof of Lemma~\ref{lem:onsager-calibration}]
\par\smallskip\noindent\textit{Concentration under the comparator law.}\quad
Put $\kappa_a^\star=\max\{1,\sup_{s\le T,x}a(s,x)\}$ and $L_* =\sup_{s\le T,x}|a_x(s,x)|$. The regularity bounds imply $\kappa_a^\star\le\mathsf{K}\sqrt{\Lambda(T)}$ and $L_*\le\mathsf{K}\Lambda(T)$. This exact curvature envelope satisfies $\kappa_a^\star\le\kappa_a(T)$ and is distinct from the moment bound $\kappa$ in \eqref{eq:onsager-calibration-parameters}. Let $r_k=\rho_k^1$ and $\nu_h=v_{h,1}/h$. For sufficiently small admissible accuracy, \eqref{eq:onsager-normalized-tail} ensures $\nu_h\ge1/2$. Thus
\begin{equation}
 \E[(a_k^1)^2]=q_k/\nu_h\le8,\qquad
 0\le v_k=r_kc_k\le2.
\end{equation}
In row $i$ of the comparator, let $G_{k+1,i}=W_{i,k+1}-W_{i,k}$, with $W_{i,0}=0$, and $a_{k,i}=a(t_k,Y_{k,i}^1(W))$. The state depends only on earlier increments, so $a_{k,i}$ is independent of the current $G_{k+1,i}$. Define
\begin{equation}
 T^q_{k,i}=a_{k,i}^2\chi_1(G_{k+1,i})^2/h,\qquad
 T^v_{k,i}=a_{k,i}G_{k+1,i}\chi_1(G_{k+1,i})/h.
\end{equation}
Their means are $q_k,v_k$. For $G=\sqrt h z$, $|\chi_1(G)|\le|G|$, so the summands are bounded above by $a_{k,i}^2z^2$ and $a_{k,i}z^2$, respectively. Independence and Gaussian moments yield, for integers $p\ge2$,
\begin{align}
 \E|T^q_{k,i}|^p
 &\le (\kappa_a^\star)^{2p-2}\E[a_{k,i}^2]\E[z^{2p}]
 \le p!\mathsf{K}^p(\kappa_a^\star)^{2p-2},\nonumber\\
 \E|T^v_{k,i}|^p
 &\le (\kappa_a^\star)^{p-2}\E[a_{k,i}^2]\E[z^{2p}]
 \le p!\mathsf{K}^p(\kappa_a^\star)^{p-2}.
 \label{eq:onsager-empirical-statistic-moments}
\end{align}
Centering changes only the universal constants. The rows are independent, although statistics at distinct times within a row need not be independent. Moment Bernstein in Lemma~\ref{lem:probability-tools} consequently gives, for $s\ge1$, each with failure at most $2e^{-s}$,
\begin{align}
 |Q_k(W)-q_k|&\le\mathsf{K}\{\kappa_a^\star\sqrt{s/N}+(\kappa_a^\star)^2s/N\},\nonumber\\
 |V_k(W)-v_k|&\le\mathsf{K}\{\sqrt{s/N}+\kappa_a^\star s/N\}.
 \label{eq:onsager-calibration-bernstein}
\end{align}
Take $s=L_{\alpha,d}=L_\alpha+\log d$. The common fixed confidence constant $\mathsf{K}\ge64$ inside $L_\alpha$ makes the union bound over the $2K$ statistics at most $\alpha/16$, and both deviations are at most $\eta_N$ in \eqref{eq:onsager-calibration-parameters}.

For the Gaussian increment norms, $\E\exp(\sum_i z_i^2/4)=2^{N/2}$ and Markov's inequality imply
\begin{equation}
 \Prob\{\norm{G_{k+1}}^2>4Nh\}
 \le\exp\{-N+(N/2)\log2\}\le e^{-cN}.
\end{equation}
If $N\ge\mathsf{K}L_{\alpha,d}$, a union bound over $k<K$ makes this failure at most $\alpha/16$. Intersecting the two events proves \eqref{eq:onsager-calibration-event}. No event involving the actual iterates enters either concentration calculation.

\par\smallskip\noindent\textit{Deterministic transfer of the statistics.}\quad
For a vector $y$, write $\norm y_N=\norm y/\sqrt N$. For histories $\mathsf Z,\mathsf Z'$ define $\delta(\mathsf Z,\mathsf Z')=\norm{\mathsf Z-\mathsf Z'}_{\rm F}/\sqrt N$, restricting to the columns used by the statistic in question. Put $g_{k+1}(\mathsf Z)=\mathsf Z_{:,k+1}-\mathsf Z_{:,k}$, with $\mathsf Z_{:,0}=0$, and let $Y_k(\mathsf Z)$ be the cutoff state reconstructed by \eqref{eq:iamp-cutoff-scalar}. The state uses no normalization factor or Onsager coefficient. Subtracting its updates and unrolling gives
\begin{align}
 \norm{Y_k(\mathsf Z)-Y_k(\mathsf Z')}_N
 &\le\sum_{j<k}\norm{g_{j+1}(\mathsf Z)-g_{j+1}(\mathsf Z')}_N
   \prod_{\ell=j+1}^{k-1}(1+\kappa_a^\star h\gamma(t_\ell))\nonumber\\
 &\le\mathsf{K}_Y\delta(\mathsf Z,\mathsf Z'),\qquad
 \mathsf{K}_Y=2\sqrt d\,e^{\kappa_a^\star}.
 \label{eq:onsager-empirical-state-lipschitz}
\end{align}
We used $\sum h\gamma(t_k)\le1$, the bound two on the norm of the difference operator, and Cauchy--Schwarz over the time indices. In particular, $\norm{g_{k+1}(\mathsf Z)-g_{k+1}(\mathsf Z')}_N\le2\delta(\mathsf Z,\mathsf Z')$.

Work now on \eqref{eq:onsager-calibration-event}, and compare an arbitrary history $\mathsf Z$ with $W$. Write $a_{\mathsf Z}=a(t_k,Y_k(\mathsf Z))$, $a_W=a(t_k,Y_k(W))$, and $G=g_{k+1}(W)$. The functions $\chi_1^2$ and $\psi(x)=x\chi_1(x)$ are globally $2$-Lipschitz; also $|\chi_1|\le1$ and $|\psi(G)|\le|G|$. Split the two statistic differences as
\begin{align}
 a_{\mathsf Z}^2\chi_1(g_{k+1}(\mathsf Z))^2-a_W^2\chi_1(G)^2
 ={}&(a_{\mathsf Z}^2-a_W^2)\chi_1(G)^2\nonumber\\
 &+a_{\mathsf Z}^2\{\chi_1(g_{k+1}(\mathsf Z))^2-\chi_1(G)^2\},\nonumber\\
 a_{\mathsf Z}\psi(g_{k+1}(\mathsf Z))-a_W\psi(G)
 ={}&(a_{\mathsf Z}-a_W)\psi(G)
   +a_{\mathsf Z}\{\psi(g_{k+1}(\mathsf Z))-\psi(G)\}.
\end{align}
Here all products are coordinatewise. Since an empirical absolute mean is at most the normalized Euclidean norm, the first identity and \eqref{eq:onsager-empirical-state-lipschitz} imply
\begin{equation}
 |Q_k(\mathsf Z)-Q_k(W)|\le\mathsf{K}_q\delta(\mathsf Z,W),\qquad
 \mathsf{K}_q=(2\kappa_a^\star L_*\mathsf{K}_Y+4(\kappa_a^\star)^2)/h.
\end{equation}
For the second identity, Cauchy--Schwarz bounds the empirical mean of $|(a_{\mathsf Z}-a_W)\psi(G)|$ by $\norm{a_{\mathsf Z}-a_W}_N\norm{\psi(G)}_N$. On the current event, $\norm{\psi(G)}_N\le\norm G_N\le2\sqrt h$. Hence
\begin{equation}
 |V_k(\mathsf Z)-V_k(W)|\le\mathsf{K}_v\delta(\mathsf Z,W),\qquad
 \mathsf{K}_v=2L_*\mathsf{K}_Y/\sqrt h+4\kappa_a^\star/h.
 \label{eq:onsager-statistic-transfer}
\end{equation}
The unbounded factor $\psi$ occurs only at the comparator increment $G$. No norm bound on $g_{k+1}(\mathsf Z)$ is needed. The constants satisfy
\begin{equation}
 \log(1+\mathsf{K}_q+\mathsf{K}_v)\le\mathsf{K}b_{h,\tau},
 \label{eq:onsager-statistic-lipschitz-size}
\end{equation}
because $\kappa_a^\star\le\mathsf{K}\sqrt{\Lambda(T)}$, $L_*\le\mathsf{K}\Lambda(T)$, and $h\Lambda(T)^2\le1$.

\par\smallskip\noindent\textit{Normalization factors and Onsager coefficients.}\quad
For a fixed history $\mathsf Z$, use \eqref{eq:estimated-onsager-coefficients} to define $\widehat r_k(\mathsf Z)$ and $\widehat c_k(\mathsf Z)$. Projection onto $[1/4,4]$ cannot increase distance from $q_k=r_k^2$. Both square roots lie in $[1/2,2]$, so
\begin{equation}
 |\widehat r_k(\mathsf Z)-r_k|\le|Q_k(\mathsf Z)-q_k|,\qquad
 |\widehat r_k(\mathsf Z)^{-1}-r_k^{-1}|
 \le4|Q_k(\mathsf Z)-q_k|.
 \label{eq:onsager-calibration-denominators}
\end{equation}
Likewise, $c_k=v_k/r_k\in[0,1]$ and projection onto $[0,1]$ cannot increase its error. Subtracting the ratios, using $v_k\le2$, gives
\begin{align}
 |\widehat c_k(\mathsf Z)-c_k|
 &\le\left|\frac{V_k(\mathsf Z)-v_k}{\widehat r_k(\mathsf Z)}
           +v_k\{\widehat r_k(\mathsf Z)^{-1}-r_k^{-1}\}\right|\nonumber\\
 &\le2|V_k(\mathsf Z)-v_k|+8|Q_k(\mathsf Z)-q_k|.
 \label{eq:onsager-calibration-matrix-errors}
\end{align}
Combining concentration and deterministic transfer, there is a deterministic $\mathsf{K}_{\rm coeff}\ge1$, with $\log\mathsf{K}_{\rm coeff}\le\mathsf{K}b_{h,\tau}$, such that
\begin{equation}
 |\widehat r_k(\mathsf Z)-r_k|+|\widehat c_k(\mathsf Z)-c_k|
 \le\mathsf{K}_{\rm coeff}\delta(\mathsf Z_{\le k+1},W_{\le k+1})
   +\mathsf{K}\eta_N.
 \label{eq:onsager-empirical-coefficient-transfer}
\end{equation}
This estimate is uniform over histories on the comparator event already established. It does not assert independence or conditional unbiasedness of the coefficients computed by the algorithm.

\par\smallskip\noindent\textit{Changes in the maps at a fixed history.}\quad
Let $F_s,g_s$ be the population maps. At a fixed history $\mathsf Z$, write $\widehat F_s(\mathsf Z),\widehat g_s(\mathsf Z)$ for the maps using empirical coefficients obtained from its prefixes. Let $\theta_{t-1}(\mathsf Z)$ be the largest error in $r_k$ or $c_k$ for $0\le k\le t-2$ at those prefixes; put $\theta_0(\mathsf Z)=0$. Since the state at a fixed history is unchanged when coefficients are replaced, subtraction of the magnetization sums gives, for $s\le t$,
\begin{equation}
 \norm{F_s(\mathsf Z)}_N\le1+2K\kappa_a^\star,\qquad
 \norm{\widehat F_s(\mathsf Z)-F_s(\mathsf Z)}_N
 \le4K\kappa_a^\star\theta_{t-1}(\mathsf Z).
 \label{eq:onsager-map-perturbations}
\end{equation}
The entry formula \eqref{eq:estimated-onsager-entries} gives absolute column sums at most $d$ for $\widehat B$ and at most $2d\theta_{t-1}(\mathsf Z)$ for $\widehat B-B^1$. The correction difference is exactly
\begin{equation}
 \widehat g_t(\mathsf Z)-g_t(\mathsf Z)
 =-\sum_{s<t}(\widehat B-B^1)_{st}F_s(\mathsf Z)
  -\sum_{s<t}\widehat B_{st}\{\widehat F_s(\mathsf Z)-F_s(\mathsf Z)\}.
\end{equation}
Thus
\begin{equation}
 \norm{\widehat g_t(\mathsf Z)-g_t(\mathsf Z)}_N
 \le\{2d(1+2K\kappa_a^\star)+4dK\kappa_a^\star\}\theta_{t-1}(\mathsf Z)
 \le\mathsf{K}d^2\kappa_a^\star\theta_{t-1}(\mathsf Z).
 \label{eq:onsager-correction-perturbation}
\end{equation}
These inequalities compare maps at one history; they do not hold the histories of the two matrix iterations equal.

\par\smallskip\noindent\textit{Causal propagation on the same matrix.}\quad
Let $X_t,\widehat X_t$ be the successive population and empirical fields on the same $A$, and set
\begin{equation}
 E_t=\frac{\norm{[\widehat X_1-X_1|\cdots|\widehat X_t-X_t]}_{\rm F}}{\sqrt N},
 \qquad \mathcal R=\frac{\norm{X-W}_{\rm F}}{\sqrt N},\qquad E_0=0.
\end{equation}
Set $\theta_{t-1}=\theta_{t-1}(\widehat X_{<t})$. Field $t$ uses only coefficients through index $t-2$. Each was computed from fields through $t-1$, before field $t$ was formed. Equation~\eqref{eq:onsager-empirical-coefficient-transfer} and the triangle inequality therefore imply
\begin{equation}
 \theta_{t-1}\le\mathsf{K}_{\rm coeff}(E_{t-1}+\mathcal R)+\mathsf{K}\eta_N.
\end{equation}
On $\norm A\le3$, compare the population maps at the two histories and then replace them at the empirical history:
\begin{align}
 \widehat X_t-X_t={}&A\{F_t(\widehat X_{<t})-F_t(X_{<t})\}
       +g_t(\widehat X_{<t})-g_t(X_{<t})\nonumber\\
 &+A\{\widehat F_t(\widehat X_{<t})-F_t(\widehat X_{<t})\}
       +\widehat g_t(\widehat X_{<t})-g_t(\widehat X_{<t}).
\end{align}
Let $L=3L_F+L_g$, with $L_F,L_g$ bounded in \eqref{eq:onsager-history-bounds}, and choose $\mathsf{K}_{\rm pert}=\mathsf{K}d^2\kappa_a^\star$ large enough to cover \eqref{eq:onsager-map-perturbations}--\eqref{eq:onsager-correction-perturbation}. Then
\begin{align}
 \norm{\widehat X_t-X_t}_N&\le LE_{t-1}+\mathsf{K}_{\rm pert}\theta_{t-1},\nonumber\\
 E_t&\le\Gamma E_{t-1}
   +\mathsf{K}_{\rm pert}\{\mathsf{K}_{\rm coeff}\mathcal R+\mathsf{K}\eta_N\},
 \qquad \Gamma=1+L+\mathsf{K}_{\rm pert}\mathsf{K}_{\rm coeff}.
 \label{eq:onsager-trajectory-perturbation}
\end{align}
The second inequality uses $E_t\le E_{t-1}+\norm{\widehat X_t-X_t}_N$. By \eqref{eq:onsager-history-bounds} and \eqref{eq:onsager-statistic-lipschitz-size}, $\log\Gamma\le\mathsf{K}b_{h,\tau}$. Unrolling from $E_0=0$ yields
\begin{equation}
 E_d\le d\mathsf{K}_{\rm pert}
   \{\mathsf{K}_{\rm coeff}\mathcal R+\mathsf{K}\eta_N\}\Gamma^{d-1}
 \le e^{\mathsf{K}\mathfrak S_{h,\tau}}(\mathcal R+\eta_N).
\end{equation}
Finally, the population terminal magnetization map has Lipschitz constant $L_F$. Its empirical counterpart uses normalization factors through index $K-1$, whose histories end at field $K$. With $\theta_K\le\mathsf{K}_{\rm coeff}(E_K+\mathcal R)+\mathsf{K}\eta_N$, we obtain
\begin{equation}
 \frac{\norm{\widehat m^K-m^K}}{\sqrt N}
 \le L_FE_K+4K\kappa_a^\star\theta_K
 \le e^{\mathsf{K}_{\rm stab}\mathfrak S_{h,\tau}}(\mathcal R+\eta_N).
\end{equation}
This proves \eqref{eq:onsager-calibration-stability}. The recurrence is causal and linear in the errors; no small-error bootstrap or matching assumption for the empirical maps is used. Exact matching is required only for the fixed population iteration to which Lemma~\ref{lem:reeves-onsager-bound} is applied.

\par\smallskip\noindent\textit{Correction identity, work, and storage.}\quad
For the zero-based coefficients $\widehat c_k$ of \eqref{eq:estimated-onsager-coefficients}, multiplication by $T_d^{-1}$ on the left and $T_d$ on the right gives, for $i<d$,
\begin{equation}
 (\widehat D^{\rm AMS}T_d)_{ij}
 =\widehat c_{i-1}\one_{\{j\ge i+1\}},\qquad
 \widehat B_{ij}
 =\widehat c_{i-1}\one_{\{j\ge i+1\}}
  -\widehat c_i\one_{\{j\ge i+2\}}.
\end{equation}
The second term is absent when $i=d-1$, and the last row is zero. In column $k+1$, the correction is
\begin{equation}
 \mathcal O^k
 =\sum_{j=1}^{k-1}(\widehat c_{j-1}-\widehat c_j)\widehat m^{j-1}
       +\widehat c_{k-1}\widehat m^{k-1},\qquad k\ge1.
\end{equation}
Subtracting this expression at consecutive values of $k$ proves \eqref{eq:onsager-streamed-correction}, including $k=0$ with $\widehat m^{-1}=0$. Each step evaluates two empirical sums in $O(N)$ work, computes two projected scalars, and updates the correction in $O(N)$ work. The complete coefficient and correction work is
\begin{equation}
 O(NK),\qquad\text{additional storage }O(N).
 \label{eq:onsager-calibration-work}
\end{equation}
The current and preceding fields and magnetizations, the state, and the vectors used in the two passes over coordinates suffice. No scalar simulation, stored matrix $\widehat B$, or full magnetization history is required.
\end{proof}

\subsection{Finite-dimensional energy for AMS IAMP}
\label{app:matrix-onsager-finite}

We start with the energy of the Gaussian comparator, transfer it to the clipped output of the population iteration by Lemma~\ref{lem:reeves-onsager-bound}, then to the empirical iteration by Lemma~\ref{lem:onsager-calibration}, and finally account for rounding.

\begin{proof}[Proof of Lemma~\ref{lem:onsager-implementation}]
\par\smallskip\noindent\textit{Scalar errors and constant choices.}\quad
First fix $c_{\rm early}$ for the earlier stopping rule, then $c_h$ in \eqref{eq:uniform-onsager-grid}. Next choose $c_{\rm AMS}\le1/100$ and set $\etaacc=c_{\rm AMS}u$. The parameter substitutions following Theorem~\ref{thm:uniform-scalar} make its clipping error and correlation deficit small fixed multiples of $u$. By Lemma~\ref{lem:onsager-cutoff}, the additional cutoff errors are smaller than any fixed power of $\etaacc$. Decreasing $c_{\rm AMS}$ and the fixed upper bound $u_{\rm AMS}$ ensures
\begin{equation}
 \norm{\dist(M_K^1,[-1,1])}_2\le u/1000,\qquad
 \mathcal E_h^1\ge\Par-u/100.
 \label{eq:onsager-scalar-allowances}
\end{equation}
Choose these fixed constants also so that $c_{\rm AMS}u_{\rm AMS}\le\eta_0$. On this range $h\Lambda(T)^2\le c_0$, $h\le\tau$, $dh\le1$, and all population normalization bounds hold. Fix the constants in the directional and empirical-coefficient estimates before the dimension constant.

\par\smallskip\noindent\textit{The comparator and the two Onsager iterations.}\quad
Let $X$ be the population iteration on $A$, using deterministic $\rho_k^1,B^1$, and $\widehat X$ the empirical iteration on the same $A$. Extend the probability space by an independent auxiliary $J$ with law \eqref{eq:auxiliary-j}. Definition~\ref{def:reeves-coupling}, with population queries $F_t=f_{t-1}^1$, constructs $W=Z\Omega$ from this $A$ and $J$. Its Gaussian rows produce independent cutoff scalar paths. Lemma~\ref{lem:onsager-history} verifies the exact matching \eqref{eq:onsager-matching}. The actual population and empirical fields are not assigned a Gaussian law.

Put $r=\log(\mathsf{K}_{\rm tail}/\alpha)\ge1$, with a sufficiently large fixed $\mathsf{K}_{\rm tail}$. We choose the common constant inside $L_\alpha$ at least $\mathsf{K}_{\rm tail}$, so $r\le L_\alpha$ and $r\asymp L_\alpha$. For $d,r\le N$, Lemma~\ref{lem:reeves-onsager-bound} implies
\begin{equation}
 \mathcal R:=\frac{\norm{X-W}_{\rm F}}{\sqrt N}
 \le e^{\mathsf{K}\mathfrak S_{h,\tau}}\sqrt{(d+r)/N},\qquad
 \frac{\norm{m^K(X)-m^K(W)}}{\sqrt N}\le L_F\mathcal R
 \label{eq:onsager-final-coupling}
\end{equation}
except with probability $5e^{-r}$. The logarithms of $\mathcal L^3$, $\norm\Omega_{2,1}$, $L_F$, and the trajectory amplification factor are bounded by $\mathsf{K}\mathfrak S_{h,\tau}$. We next verify the required bounds on dimension for every confidence level.

\par\smallskip\noindent\textit{Dimension requirements.}\quad
Write $\mathfrak S=\mathfrak S_{h,\tau}$ and test the bounds at $N_*=e^{\mathsf{K}_2\mathfrak S}r^4$. Both grids satisfy $d\le\mathfrak S$, while the logarithms of $h^{-1},\Lambda(T),\kappa^2,u^{-1}$ are $O(\mathfrak S)$. For $r\ge1$,
\begin{equation}
 \frac d{N_*}\le\mathfrak S e^{-\mathsf{K}_2\mathfrak S},\qquad
 \frac r{N_*}\le e^{-\mathsf{K}_2\mathfrak S},\qquad
 \frac{\mathfrak S+r}{r^2}\le\mathfrak S+1.
\end{equation}
Hence $d+r\le N_*$ for large $\mathsf{K}_2$. Also $r+\log(\mathsf{K}d)\le\mathsf{K}(r+\mathfrak S)$, so
\begin{equation}
 \frac{\Lambda(T)\{r+\log(\mathsf{K}d)\}}{N_*}
 \le\mathsf{K}\Lambda(T)(\mathfrak S+1)e^{-\mathsf{K}_2\mathfrak S}.
\end{equation}
This verifies \eqref{eq:onsager-directional-dimension}. From \eqref{eq:onsager-comparator-radius},
\begin{equation}
 \sqrt{\overline V_{\rm A}r/N_*}
 +\overline V_{\rm A}(d+r)/N_*
 \le\sqrt{\overline V_{\rm A}}e^{-\mathsf{K}_2\mathfrak S/2}
   +\overline V_{\rm A}(\mathfrak S+1)e^{-\mathsf{K}_2\mathfrak S}.
 \label{eq:onsager-concentration-check}
\end{equation}
Increasing $\mathsf{K}_2$ ensures that $\mathfrak r_{N_*}$ is sufficiently small and the energy error in \eqref{eq:onsager-comparator-energy}, with $t=r$, is at most $u/100$. This energy event includes both bounds on the auxiliary matrix used in its proof.

For the empirical statistics, the exact identity
$L_{\alpha,d}=r+\log d+\log(\mathsf{K}/\mathsf{K}_{\rm tail})$
gives $L_{\alpha,d}\le\mathsf{K}(\mathfrak S+r)$, and thus
$N_*\ge\mathsf{K}L_{\alpha,d}$. Equation~\eqref{eq:onsager-calibration-parameters} yields
\begin{equation}
 \eta_{N_*}
 \le\mathsf{K}\{\kappa\sqrt{\mathfrak S+1}\,e^{-\mathsf{K}_2\mathfrak S/2}
        +\kappa^2(\mathfrak S+1)e^{-\mathsf{K}_2\mathfrak S}\}.
\end{equation}
Substituting \eqref{eq:onsager-final-coupling} into \eqref{eq:onsager-calibration-stability}, increasing $\mathsf{K}_2$ after $\mathsf{K}_{\rm stab}$, and using $\log(\kappa u^{-1})=O(\mathfrak S)$ prove simultaneously
\begin{equation}
 L_F\mathcal R\le u/100,\qquad
 e^{\mathsf{K}_{\rm stab}\mathfrak S}(\mathcal R+\eta_{N_*})\le u/100.
 \label{eq:onsager-empirical-dimension-check}
\end{equation}
The comparator concentration and the coefficient-feedback error both tend to zero with $N$; neither contains a deterministic population mismatch.

For clipping, take $\beta=\alpha/16$ in \eqref{eq:onsager-comparator-clipping}. Then $t_\beta\le\mathsf{K}r$ and $\log(4N_*/\beta)\le\mathsf{K}(\mathsf{K}_2\mathfrak S+r)$, so
\begin{equation}
 \mathsf{K}\kappa^2\frac{\log(4N_*/\beta)t_\beta}{N_*}
 \le\mathsf{K}\kappa^2\mathsf{K}_2(\mathfrak S+1)e^{-\mathsf{K}_2\mathfrak S}
 \le u^2/10^6
\end{equation}
after enlarging $\mathsf{K}_2$. The input norm, diagonal, and optimum estimates require only $N_*\ge\mathsf{K}u^{-2}r$, which also follows from $\log(u^{-1})=O(\mathfrak S)$. Specifically, \eqref{eq:diagonal-bound} and \eqref{eq:optimum-tail} give
\begin{equation}
 \norm A\le3,\qquad \frac1{2N_*}\sum_i|A_{ii}|\le u/100,\qquad
 \OPT_{N_*}/N_*\le\Par+u/2,
\end{equation}
with failure at most $3e^{-r}$ for the first two bounds and $e^{-r}$ for the last.

These checks hold for every $N\ge N_*$. The trajectory, coefficient, and directional radii are sums of constants times $N^{-1/2}$ or $N^{-1}$. For fixed $u,\alpha$, the clipping radius has the form $(a+b\log N)/N$ with $a,b\ge0$, whose derivative is nonpositive for $N\ge e$. All dimension requirements remain valid as $N$ increases. Taking the constant in \eqref{eq:onsager-implementation-size} sufficiently large ensures $N\ge N_*$. The argument imposes no upper bound on $r$ in terms of $d$.

\par\smallskip\noindent\textit{The common event and the energy comparisons.}\quad
Choose $\mathsf{K}_{\rm tail}$ after the universal failure constant in Lemma~\ref{lem:onsager-comparator-energy}. The following events then have the displayed failure bounds:
\begin{center}
\begin{tabular}{lll}
 Event & Parameter & Failure bound\\\hline
 Comparator energy, including auxiliary bounds & $t=r$ & $\alpha/16$\\
 Comparator clipping & $\beta=\alpha/16$ & $\alpha/16$\\
 Population trajectory comparison & $r$ & $5e^{-r}\le\alpha/16$\\
 Input norm and diagonal & $r$ & $3e^{-r}\le\alpha/16$\\
 Optimum bound & $u/2$ & $e^{-Nu^2/4}\le\alpha/16$\\
 Comparator coefficient statistics & $L_{\alpha,d}$ & $\alpha/8$
\end{tabular}
\end{center}
A union bound puts their intersection at probability at least $1-\alpha$, since $5\alpha/16+\alpha/8<\alpha$. No independence between these success events is required. All Gaussian calculations used their stated marginal laws before the intersection was taken.

On this event, write $f=m^K(W)$, $v_W=\clip(f,[-1,1]^N)$, $v_X=\clip(m^K(X),[-1,1]^N)$, and $\widehat v=\clip(\widehat m^K,[-1,1]^N)$. Equations~\eqref{eq:onsager-scalar-allowances} and \eqref{eq:onsager-comparator-energy} imply $H_N(f)/N\ge\Par-u/50$. Also $\mu\le u^2/10^6$ and the clipping concentration term is at most $u^2/10^6$, so
\begin{equation}
 \norm{f-v_W}^2/N\le3u^2/10^6,\qquad
 \norm{f-v_W}/\sqrt N\le u/500.
\end{equation}
Since $\norm f/\sqrt N\le5/4$, $\norm{v_W}/\sqrt N\le1$, and $\norm A\le3$, the energy decrease from clipping is at most $(3/2)(u/500)(5/4+1)=27u/4000<u/100$.

For $v,v'\in[-1,1]^N$, symmetry and Cauchy--Schwarz imply
\begin{equation}
 \frac{|H_N(v)-H_N(v')|}{N}
 \le\frac{\norm A}{2}\frac{\norm{v-v'}}{\sqrt N}
                         \frac{\norm v+\norm{v'}}{\sqrt N}
 \le3\frac{\norm{v-v'}}{\sqrt N}.
 \label{eq:ams-matrix-clipped-energy-comparison}
\end{equation}
Clipping is nonexpansive. Equations~\eqref{eq:onsager-final-coupling} and \eqref{eq:onsager-empirical-dimension-check} therefore bound $\norm{v_X-v_W}/\sqrt N$ and $\norm{\widehat v-v_X}/\sqrt N$ by $u/100$. The successive energy bounds before rounding are
\begin{equation}
 H_N(v_W)/N\ge\Par-3u/100,\qquad
 H_N(v_X)/N\ge\Par-6u/100,\qquad
 H_N(\widehat v)/N\ge\Par-9u/100.
\end{equation}
We compare these clipped vectors, not their rounded configurations. Sequential rounding and the diagonal bound cost at most another $u/100$. Consequently
\begin{equation}
 H_N(\widehat\sigma_{\rm AMS})/N\ge\Par-u/10\ge\Par-u/2,
 \qquad \OPT_N/N\le\Par+u/2.
\end{equation}

\par\smallskip\noindent\textit{Marginalization, work, and storage.}\quad
The implemented output is a deterministic function of $A$; $J$ and $W$ are proof variables. Let $\mathcal Q(A)$ be the event in \eqref{eq:onsager-implementation-energy}, and $\mathcal G(A,J)$ the common event just used. Since $\mathcal G\subseteq\mathcal Q$,
\begin{equation}
 \Prob_A(\mathcal Q^c)
 =\E_A\E_J[\one_{\mathcal Q^c}]
 =\Prob_{A,J}(\mathcal Q^c)
 \le\Prob_{A,J}(\mathcal G^c)\le\alpha.
\end{equation}
Marginalizing over the auxiliary variables proves the probability guarantee over the GOE input alone.

The loop computes $K+1$ dense products, including the final field that changes no magnetization. These cost $O(N^2K)$. The two empirical averages, the state and magnetization updates, and \eqref{eq:onsager-streamed-correction} cost $O(NK)$. Grid construction costs at most $O(K+\log(1/\etaacc))$, and final clipping and sequential rounding cost $O(N^2)$. Rounding includes one more dense product, for exactly $K+2$ products. Since $d\le N$ and $K\ge1$, total work is $O(N^2K)$. Keeping $A$, the current and preceding fields and magnetizations, the state, the correction, the temporary coordinate vectors, and the grid uses $O(N^2+N+K)=O(N^2)$ storage on this range. Neither the Gaussian comparator nor any independent scalar paths are computed.
\end{proof}
\subsection{Earlier stopping and the AMS IAMP guarantees}
\label{app:onsager-afptas}

We prove the relative guarantee of Theorem~\ref{thm:onsager-main} from Lemma~\ref{lem:onsager-implementation}, then the all-dimension consequence in Corollary~\ref{thm:onsager-afptas}. We also state the smaller sufficient bounds obtained by stopping earlier. Here $\tau=1-T$ measures the distance of the terminal time from one, while $\chi_1$ clips each field increment inside the updates. Replacing $\tau\asymp\xi$ by the larger value $\tau\asymp\xi^{2/3}L_\xi^{-1/3}$ in \eqref{eq:uniform-earlier-cutoff} preserves an $O(\xi)$ discarded correlation by \eqref{eq:terminal-g}. Since $h_*\propto\etaacc^2\tau/L_\tau^2$ and $\etaacc=c_{\rm AMS}\xi$, the power of $1/\xi$ in the step count becomes $8/3=2+2/3$. The uniform mesh, update equations, and empirical coefficient formulas remain the same.

\begin{corollary}[Earlier stopping for AMS IAMP]
\label{cor:onsager-earlier-stopping}
There are fixed $c,\mathsf{K}>0$ for which the following holds.
Put $\xi=c\varepsilon$ and replace the terminal choice $\tau\asymp\xi$ in \eqref{eq:uniform-onsager-grid} by the dyadic choice
\eqref{eq:uniform-earlier-cutoff}, keeping all uniform left-endpoint
updates. Use $\mathcal S_\xi^{\rm early}=\xi^{-3}L_\xi^3$
in Algorithm~\ref{alg:onsager-afptas}, with the corresponding
earlier-stopping matrix iteration and threshold
\begin{equation}
 N_0^{\rm early}=\left\lceil
       2^{\lceil \mathsf{K}\mathcal S_\xi^{\rm early}\rceil}
                                  L_\alpha^4\right\rceil.
 \label{eq:earlier-afptas-threshold}
\end{equation}
For every $0<\varepsilon\le1$, $0<\alpha<1/2$, and $N\ge N_0^{\rm early}$, the output satisfies $H_N(\sigma)\ge(1-\varepsilon)\OPT_N$ with probability at least $1-\alpha$. Returning all-plus below the threshold extends this to \eqref{eq:afptas-definition} for every $N\ge1$, with $\mathfrak{E}_{\varepsilon,\alpha}^{\rm early}=4N_0^{\rm early}$ and
\begin{equation}
 \mathfrak{E}_{\varepsilon,\alpha}^{\rm early}
       \le \mathsf{K}\exp(\mathsf{K}\varepsilon^{-3}L_\varepsilon^3)L_\alpha^4,\qquad
 \text{total work}\le \mathsf{K}N^2\varepsilon^{-8/3}L_\varepsilon^{7/3}.
 \label{eq:earlier-afptas-resources}
\end{equation}
The approximation is purely relative for $N\ge N_0^{\rm early}$.
On that range $K=\Theta(\varepsilon^{-8/3}L_\varepsilon^{7/3})$,
with $K+2$ products.
\end{corollary}

\begin{proof}[Proof of Theorem~\ref{thm:onsager-main}]
\par\smallskip\noindent\textit{The sufficient dimension.}\quad
Put $p_0=1-\log2$ and choose $c\le\min\{u_{\rm AMS},p_0/4\}$, so $\xi=c\varepsilon$ is a valid target in Lemma~\ref{lem:onsager-implementation}. Its scalar accuracy is $\etaacc=c_{\rm AMS}\xi$. Both rescalings are fixed, and therefore $L_{\etaacc}\asymp L_\xi$. By \eqref{eq:uniform-onsager-grid}, the selected dyadic value has $\tau\asymp\xi$ and the integer-rounded mesh has $h\asymp\xi^3L_\xi^{-2}$. Hence
\begin{equation}
 d=\Theta(\xi^{-3}L_\xi^2),\qquad
 \sqrt{\Lambda(T)}=\Theta(\xi^{-1/2}L_\xi^{1/2}),\qquad
 \log(d+h^{-1}+2)=O(L_\xi).
\end{equation}
The logarithmic term is absorbed by the curvature term: $L_\xi/(\xi^{-1/2}L_\xi^{1/2})=\sqrt{\xi L_\xi}$ is bounded on $0<\xi\le c$. Substituting these estimates in \eqref{eq:onsager-history-quantity}, we obtain
\begin{equation}
 \mathfrak S_{h,\tau}=d\{1+\sqrt{\Lambda(T)}+\log(d+h^{-1}+2)\}
 \le \mathsf{K}\xi^{-7/2}L_\xi^{5/2}=\mathsf{K}\mathcal S_\xi.
 \label{eq:ams-proportional-history-check}
\end{equation}
Let $\mathsf{K}_{\rm imp}$ be the fixed exponential constant required by Lemma~\ref{lem:onsager-implementation}, chosen after the directional and empirical-coefficient constants. Both dimension thresholds use the same common confidence factor $L_\alpha^4$. We choose the constant in $m_{\rm AMS}=\lceil \mathsf{K}\mathcal S_\xi\rceil$ last, so that \eqref{eq:ams-proportional-history-check} gives
\begin{equation}
 2^{m_{\rm AMS}}\ge e^{\mathsf{K}(\log2)\mathcal S_\xi}
                   \ge e^{\mathsf{K}_{\rm imp}\mathfrak S_{h,\tau}}.
\end{equation}
Thus $N_0^{\rm AMS}$ exceeds the implementation threshold. We enlarge $\mathsf{K}$ also so that $N_0^{\rm AMS}\ge\log(2/\alpha)$. This latter requirement is used only for the all-dimension corollary.

\par\smallskip\noindent\textit{Relative accuracy and work.}\quad
For $N\ge N_0^{\rm AMS}$, Lemma~\ref{lem:onsager-implementation} and Lemma~\ref{lem:positive-value} imply, on one common event of probability at least $1-\alpha$,
\begin{equation}
 \OPT_N\ge H_N(\sigma)\ge(p_0-\xi/2)N\ge(p_0/2)N,
 \qquad \OPT_N-H_N(\sigma)\le\xi N\le\varepsilon\OPT_N.
\end{equation}
The first chain uses $\sigma\in\{-1,1\}^N$, the implemented energy lower bound, and $\Par\ge p_0$. The gap estimate follows by subtracting the upper and lower bounds in \eqref{eq:onsager-implementation-energy}. Finally, $\xi=c\varepsilon\le p_0\varepsilon/4$ and $\OPT_N\ge(p_0/2)N$ imply $\xi N\le\varepsilon\OPT_N$. This proves \eqref{eq:onsager-relative} without allocating an additional failure probability.

Lemma~\ref{lem:onsager-implementation} counts the empirical coefficient estimation and gives $O(N^2K)$ work and $O(N^2)$ storage. Substituting $K=\Theta(\xi^{-3}L_\xi^2)$ and $\xi=c\varepsilon$ proves the stated bounds, because $L_{c\varepsilon}\le \mathsf{K}L_\varepsilon$. The $K+2$ executed products consist of $K+1$ field updates and one product to initialize rounding. The two empirical averages and the streamed Onsager correction cost $O(NK)$ work and need no independent simulation or stored coefficient matrix.
\end{proof}

\begin{proof}[Proof of Corollary~\ref{thm:onsager-afptas}]
For $N\ge N_0^{\rm AMS}$, invoke Theorem~\ref{thm:onsager-main}. Enlarge its threshold constant so that $N_0^{\rm AMS}\ge\log(2/\alpha)$.

For $N<N_0^{\rm AMS}$, the algorithm returns all-plus. Deterministically, $|\OPT_N|$ and $|H_N(\one)|$ are at most $N\norm A/2$. Because $0\le1-\varepsilon\le1$, we have $(1-\varepsilon)\OPT_N\le|\OPT_N|$ for either sign of the optimum. Thus
\begin{equation}
 (1-\varepsilon)\OPT_N-H_N(\one)
 \le|\OPT_N|+|H_N(\one)|\le N\norm A.
\end{equation}
With $u_\alpha=\log(2/\alpha)$, equation \eqref{eq:gaussian-matrix-norms} gives $\norm A\le2+2\sqrt{u_\alpha/N}$ with probability at least $1-2e^{-u_\alpha}=1-\alpha$. On this event,
\begin{equation}
 N\norm A\le2N+2\sqrt{Nu_\alpha}\le4N_0^{\rm AMS}.
\end{equation}
This proves \eqref{eq:onsager-afptas-result} below the threshold; above it, the main theorem proves a stronger inequality. The two regimes are determined by $N,\varepsilon,\alpha$, not by randomness, so their probability bounds need not be added.

To implement the dimension comparison, put $x=2^{m_{\rm AMS}}L_\alpha^4$. For an integer $N$, $N<\lceil x\rceil$ is equivalent to $N<x$. Monotonicity of the logarithm consequently implies
\begin{equation}
 N<N_0^{\rm AMS}
 \quad\Longleftrightarrow\quad
 \log N<m_{\rm AMS}\log2+4\log L_\alpha.
\end{equation}
The algorithm therefore computes only $m_{\rm AMS}$ and logarithms before this test, rather than evaluating $2^{m_{\rm AMS}}$ or constructing a grid below the threshold. Its work there is $O(N^2)$, including reading $A$. Above the threshold, Theorem~\ref{thm:onsager-main} includes the empirical estimates and gives $O(N^2\varepsilon^{-3}L_\varepsilon^2)$ work.

Since $\mathcal S_\xi\ge1$, the two ceilings change $N_0^{\rm AMS}$ by at most a universal factor. More explicitly, $2^{\lceil \mathsf{K}\mathcal S_\xi\rceil}\le2e^{\mathsf{K}(\log2)\mathcal S_\xi}$ and $\lceil x\rceil\le2x$ for $x\ge1$. Substituting $\xi=c\varepsilon$ and $L_{c\varepsilon}\le \mathsf{K}L_\varepsilon$ proves \eqref{eq:onsager-afptas-allowance}. The additive term is independent of $N$, and both branches have the jointly polynomial work required by Definition~\ref{def:afptas}.
\end{proof}

\begin{proof}[Proof of Corollary~\ref{cor:onsager-earlier-stopping}]
\par\smallskip\noindent\textit{The terminal time and mesh.}\quad
Let $\etaacc=c_{\rm AMS}\xi$ and $L_{\etaacc}=\log(e/\etaacc)\asymp L_\xi$. For sufficiently small fixed $c$, the initial value $1/2$ fails the dyadic test, so the accepted $\tau$ has a preceding rejected value $2\tau$. Acceptance and rejection imply
\begin{equation}
 \tau^3L_{\etaacc}\le c_{\rm early}^3\etaacc^2
 <(2\tau)^3L_{\etaacc},\qquad
 \frac{c_{\rm early}}2\etaacc^{2/3}L_{\etaacc}^{-1/3}
 <\tau\le c_{\rm early}\etaacc^{2/3}L_{\etaacc}^{-1/3}.
\end{equation}
Thus $\tau\asymp\xi^{2/3}L_\xi^{-1/3}$ and $L_\tau\asymp L_\xi$. Keeping $h_*=c_h\etaacc^2\tau/L_\tau^2$ and $K=\lceil T/h_*\rceil$ gives $h_*/2\le h=T/K\le h_*$ on this accuracy range. Substituting the powers of $\xi$ and $L_\xi$ separately, we obtain
\begin{equation}
 \begin{gathered}
 h\asymp\xi^{8/3}L_\xi^{-7/3},\qquad
 d\asymp\xi^{-8/3}L_\xi^{7/3},\\
 \sqrt{\Lambda(T)}\asymp\xi^{-1/3}L_\xi^{2/3},\qquad
 \mathfrak S_{h,\tau}\le \mathsf{K}\xi^{-3}L_\xi^3.
 \end{gathered}
\end{equation}
The smallness requirements hold because $h/\tau=O(\xi^2L_\xi^{-2})$ and $h\Lambda(T)^2=O(\xi^{4/3}L_\xi^{1/3})\to0$. The terminal correlation loss is $O(\tau^{3/2}\sqrt{L_\tau})=O(\xi)$, with its constant made small by fixing $c_{\rm early}$ first, as explained after \eqref{eq:uniform-earlier-cutoff}. The scalar and cutoff estimates used in Lemma~\ref{lem:onsager-implementation} therefore remain valid.

\par\smallskip\noindent\textit{Dimension, relative accuracy, and the extension.}\quad
The logarithmic term in $b_{h,\tau}$ is $O(L_\xi)$ and is dominated by $\sqrt{\Lambda(T)}$ at small $\xi$. Multiplying its leading factors gives
\begin{equation}
 \mathfrak S_{h,\tau}
 \le \mathsf{K}(\xi^{-8/3}L_\xi^{7/3})(\xi^{-1/3}L_\xi^{2/3})
 =\mathsf{K}\xi^{-3}L_\xi^3=\mathsf{K}\mathcal S_\xi^{\rm early}.
\end{equation}
Thus the new threshold controls the population trajectory comparison and the empirical-coefficient feedback by the same reasoning as \eqref{eq:ams-proportional-history-check}. Use $\mathcal S_\xi^{\rm early}$ in the logarithmic dimension test before constructing the grid. Above the threshold, Lemma~\ref{lem:onsager-implementation} and $\xi=c\varepsilon\le p_0\varepsilon/4$ give the relative guarantee by the positive-energy calculation in the proof of Theorem~\ref{thm:onsager-main}. There are still $K+2$ products, and total work is $O(N^2K)=O(N^2\varepsilon^{-8/3}L_\varepsilon^{7/3})$.

Below the threshold, the sign-safe Gaussian norm calculation in Corollary~\ref{thm:onsager-afptas} applies without change because $N_0^{\rm early}\ge\log(2/\alpha)$ after fixing its constant. It yields the additive term $4N_0^{\rm early}$, bounded as in \eqref{eq:earlier-afptas-resources}; this branch costs $O(N^2)$. The logarithmic test, ceilings, and dyadic grid use only the operations in Definition~\ref{def:oracles}.
\end{proof}

\subsection{Complete AMS IAMP algorithms}
\label{sec:implemented-onsager}

Algorithm~\ref{alg:onsager-calibrated} specifies the implementation with $\tau$ proportional to accuracy, as in \eqref{eq:uniform-onsager-grid}; its constants are those fixed in the proofs. After the last magnetization update, the loop computes one more field $z^{K+1}$, so that the iteration has the $d=K+1$ queries of the comparator in the energy argument. This field updates no magnetization, and omitting it would not change the returned configuration; the product count $K+2$ includes it.

\begingroup\normalsize
\begin{pseudocodebox}
\captionsetup{type=algorithm,hypcap=false,font=normalsize,labelfont=bf,
 justification=raggedright,singlelinecheck=false,skip=6pt}
\captionof{algorithm}{AMS IAMP with bounded increments and estimated coefficients}
\label{alg:onsager-calibrated}
\begin{algorithmic}[1]
\Require Symmetric $A$, $0<u\le u_{\rm AMS}$, and point Parisi data.
\State $\etaacc\gets c_{\rm AMS}u$; construct $\tau,T,h,K,t_k$ by
 \eqref{eq:uniform-onsager-grid}; $d\gets K+1$.
\If{$d>N$}\State \Return $\one$.\EndIf
\State $\widehat z^0\gets0$, $Y\gets0$, $\widehat m^{-1}\gets0$,
 $\widehat m^0\gets\sqrt h\one$, $\mathcal O^0\gets0$.
\For{$k=0,\ldots,K$}
 \State $\widehat z^{k+1}\gets A\widehat m^k-\mathcal O^k$.
 \If{$k<K$}
  \State $S^q,S^v\gets0$.
  \For{$i=1,\ldots,N$}
   \State $g_i\gets\widehat z_i^{k+1}-\widehat z_i^k$;
    $a_i\gets a(t_k,Y_i)$; $c_i\gets\clip(g_i,[-1,1])$.
   \State $S^q\gets S^q+a_i^2c_i^2/h$;
    $S^v\gets S^v+a_ig_ic_i/h$.
  \EndFor
  \State $\widehat\rho_k^{\rm AMS}\gets\sqrt{\clip(S^q/N,[1/4,4])}$;
   $\widehat c_k\gets\clip(S^v/(N\widehat\rho_k^{\rm AMS}),[0,1])$.
  \State $\mathcal O^{k+1}\gets\mathcal O^k+
        \widehat c_k(\widehat m^k-\widehat m^{k-1})$.
  \For{$i=1,\ldots,N$}
   \State $\widehat m_i^{k+1}\gets\widehat m_i^k+a_ic_i/\widehat\rho_k^{\rm AMS}$.
   \State $Y_i\gets Y_i+h\gamma(t_k)m(t_k,Y_i)+c_i$.
  \EndFor
 \EndIf
\EndFor
\State Clip $\widehat m^K$ to $v$; set $\widetilde A\gets A-\diag A$,
 $\ell\gets Av-\diag(A)v$ using one further product.
\For{$i=1,\ldots,N$}
 \State $s_i\gets+1$ if $\ell_i\ge0$, and $-1$ otherwise;
 $\Delta\gets s_i-v_i$; $v_i\gets s_i$;
 $\ell\gets\ell+\Delta\widetilde A_{:i}$.
\EndFor
\State \Return $v$.
\end{algorithmic}
\end{pseudocodebox}
\endgroup

Only the current and preceding fields and magnetizations, the state, and the correction vector need to be retained in the loop. For earlier stopping, choose $\tau$ by \eqref{eq:uniform-earlier-cutoff} instead; all other formulas remain the same. Algorithm~\ref{alg:onsager-afptas} tests the dimension first using $\mathcal S_\xi$ from \eqref{eq:onsager-all-dimension-threshold}. With $L_\xi=\log(e/\xi)$, we recall this quantity and define the corresponding one for earlier stopping:
\begin{equation}
 \mathcal S_\xi=\xi^{-7/2}L_\xi^{5/2},\qquad
 \mathcal S_\xi^{\rm early}=\xi^{-3}L_\xi^3.
 \label{eq:onsager-accuracy-quantities}
\end{equation}
\begingroup\normalsize
\begin{pseudocodebox}
\captionsetup{type=algorithm,hypcap=false,font=normalsize,labelfont=bf,
 justification=raggedright,singlelinecheck=false,skip=6pt}
\captionof{algorithm}{The AFPRAS based on AMS IAMP}
\label{alg:onsager-afptas}
\begin{algorithmic}[1]
\Require Symmetric $A$, $0<\varepsilon\le1$, $0<\alpha<1/2$, and point Parisi data.
\State $\xi\gets c\varepsilon$; compute $\mathcal S_\xi$ by
 \eqref{eq:onsager-accuracy-quantities}; $m_{\rm AMS}\gets\lceil \mathsf{K}\mathcal S_\xi\rceil$;
 $L_\alpha\gets\log(\mathsf{K}/\alpha)$.
\If{$\log N<m_{\rm AMS}\log2+4\log L_\alpha$}
 \State \Return $\one$.
\Else
 \State \Return Algorithm~\ref{alg:onsager-calibrated}$(A,\xi)$.
\EndIf
\end{algorithmic}
\end{pseudocodebox}
\endgroup

The earlier-stopping version uses $\mathcal S_\xi^{\rm early}$ in this test and, above the threshold, runs Algorithm~\ref{alg:onsager-calibrated} with \eqref{eq:uniform-earlier-cutoff}. All quantities in Algorithms~\ref{alg:onsager-calibrated} and~\ref{alg:onsager-afptas} are computed with the operations of Definition~\ref{def:oracles}. The threshold is tested by comparing logarithms before constructing the grid. The empirical coefficient estimates introduce no independent randomness.
\section{It\^o formulas and derivative bounds for the Parisi solution}
\label{app:scalar}

We prove the bounds on $\gamma,a,a_x,a_{xx}$ in Proposition~\ref{prop:regularity} and on $v_x,v_{xx}$ in Lemma~\ref{lem:euler-backward}, which are used in Section~\ref{sec:scalar-analysis}. We first justify It\^o's formula for $m,a,a_x$ along a diffusion with bounded drift (Lemma~\ref{lem:parisi-ito}). We then bound $\int_t^1g$ through the covariance of a Brownian increment with the terminal magnetization and apply a heat-semigroup estimate to a Burgers equation. Finally, we use the weighted-flow identity \eqref{eq:euler-weighted-flow} and another heat-semigroup estimate to bound $v_x$ and $v_{xx}$ for the weak error of $a^2$ in Lemma~\ref{lem:euler-weak}. The It\^o identities also apply to the left-endpoint AMS IAMP recursion in Appendix~\ref{app:scalar-grids}.

\subsection{It\^o identities under a bounded drift}
\label{app:scalar-ito}

We record the It\^o equations for $m,a,a_x$ under a bounded progressively measurable drift so that the discretization and quadrature calculations can use the same formulas with their respective drift discrepancies.

\begin{lemma}[It\^o identities for Parisi derivatives]
\label{lem:parisi-ito}
Fix $T<1$ and let $Y$ satisfy $\dd Y_t=\mu_t\dd t+\dd W_t$, where $W$ is Brownian motion with respect to a filtration $(\mathcal F_t)$ and $\mu$ is bounded and progressively measurable for that filtration. Define $\mathfrak d_t=\gamma(t)m(t,Y_t)-\mu_t$. Then, in integral form,
\begin{align}
 \dd m(t,Y_t)&=a(t,Y_t)\dd W_t-a(t,Y_t)\mathfrak d_t\dd t,
 \label{eq:general-m-ito}\\
 \dd a(t,Y_t)&=a_x(t,Y_t)\dd W_t
       -\{\gamma(t)a(t,Y_t)^2+a_x(t,Y_t)\mathfrak d_t\}\dd t,
 \label{eq:general-a-ito}\\
 \dd a_x(t,Y_t)&=a_{xx}(t,Y_t)\dd W_t
       -\{3\gamma(t)a(t,Y_t)a_x(t,Y_t)+a_{xx}(t,Y_t)\mathfrak d_t\}\dd t.
 \label{eq:general-ax-ito}
\end{align}
The stochastic integrals are square-integrable martingales on $[0,T]$.
\end{lemma}

\begin{proof}
Spatially differentiate \eqref{eq:pde}. The equations for $m,a,a_x$ are
\begin{align}
 m_t+\tfrac12m_{xx}+\gamma ma&=0,\nonumber\\
 a_t+\tfrac12a_{xx}+\gamma(a^2+ma_x)&=0,\nonumber\\
 (a_x)_t+\tfrac12(a_x)_{xx}+\gamma(3aa_x+ma_{xx})&=0.
 \label{eq:parisi-differentiated-pdes}
\end{align}
Proposition~\ref{prop:pde-facts} supplies bounded continuous spatial representatives and bounded weak time derivatives on the fixed strip. The generalized It\^o formula in \cite[Proposition~22]{JT} also requires a representative of each weak time derivative that is uniformly Lipschitz in space. The PDE representative of $\partial_t\partial_x^j\Phi$ has spatial derivative
\begin{equation}
 \partial_x(\partial_t\partial_x^j\Phi)
 =-\tfrac12\partial_x^{j+3}\Phi
       -\tfrac12\gamma(t)\partial_x^{j+1}(m^2),\qquad j=0,1,2,3.
 \label{eq:parisi-time-spatial-lipschitz}
\end{equation}
All terms are bounded on $[0,T]\times\R$, because $\gamma$ is bounded there and Proposition~\ref{prop:pde-facts} supplies spatial derivatives through order six. Thus the weak time derivatives of $\Phi,m,a,a_x$ have the required spatially Lipschitz representatives. Together with their at most linear growth and bounded continuous first two spatial derivatives, this verifies the hypotheses of the generalized It\^o formula along $Y$. For example,
\begin{align}
 \dd a(t,Y_t)
 &=a_x(t,Y_t)\dd W_t+
       \{a_t+\tfrac12a_{xx}+\mu_ta_x\}(t,Y_t)\dd t\nonumber\\
 &=a_x(t,Y_t)\dd W_t-
       \{\gamma(t)a(t,Y_t)^2+[\gamma(t)m(t,Y_t)-\mu_t]a_x(t,Y_t)\}\dd t.
\end{align}
This is \eqref{eq:general-a-ito}. The other two identities follow by the same substitution. On $[0,T]$ all stochastic coefficients are bounded, so the stochastic integrals are square-integrable. No derivative of $\gamma$ is taken; the identities remain valid across its jumps.
\end{proof}

\subsection{The terminal covariance and spatial smoothing}
\label{app:scalar-endpoint}

We prove Proposition~\ref{prop:regularity} in two steps: a Gaussian covariance bound controls the discarded correlation $\int_t^1g$, and a short Burgers interval then controls the spatial derivatives. The first step uses the following covariance estimate for a Gaussian variable and a bounded variable.

\begin{lemma}[Gaussian covariance with a bounded variable]
\label{lem:sign-covariance}
If $G$ is standard Gaussian and $V\in[-1,1]$ has mean $\mu$, with arbitrary dependence between $G$ and $V$, then
\begin{equation}
 |\E GV|\le \mathsf{K}(1-\mu^2)\sqrt{\log(e/(1-\mu^2))}.
 \label{eq:sign-covariance}
\end{equation}
The right side is interpreted as zero when $|\mu|=1$.
\end{lemma}

\begin{proof}
By symmetry suppose $\mu\ge0$ and put $p=\E[(1-V)/2]=(1-\mu)/2$. Since $\E G=0$, for every $u\ge0$,
\begin{equation}
 |\E GV|=|\E[G(V-1)]|
       \le2up+2\E[|G|\one_{\{|G|>u\}}]
       =2up+4\varphi(u),
\end{equation}
where $\varphi$ is the standard Gaussian density. For $p>0$, take $u=\sqrt{2\log(1/p)}$. Then $\varphi(u)=p/\sqrt{2\pi}$, and $2p\le1-\mu^2\le4p$ proves \eqref{eq:sign-covariance}. If $p=0$, the variable $V$ is constant and the covariance is zero. This proof uses the Gaussian density only analytically, not as an oracle operation.
\end{proof}

\begin{lemma}[Derivative bounds on a short Burgers interval]
\label{lem:heat-derivatives}
Let $P_s$ denote the heat semigroup with generator $\partial_{xx}/2$. Let $u$ solve $u_s=u_{xx}/2+c(s)uu_x$ on $[0,w]\times\R$ in the sense of Duhamel's formula \eqref{eq:burgers-mild}, with $|u|\le1$ and measurable $|c|\le D$. Suppose its spatial derivatives through order three are bounded on this interval. If $D\sqrt w\le c_0$ for a sufficiently small universal constant, then
\begin{equation}
 \norm{u_x(w)}_\infty\le \mathsf{K}w^{-1/2},\quad
 \norm{u_{xx}(w)}_\infty\le \mathsf{K}w^{-1},\quad
 \norm{u_{xxx}(w)}_\infty\le \mathsf{K}w^{-3/2}.
 \label{eq:heat-derivatives}
\end{equation}
The constants do not depend on the initial derivative bounds.
\end{lemma}

\begin{proof}
For the heat semigroup $P_s$ with generator $\partial_{xx}/2$,
\begin{equation}
 \norm{\partial_xP_sf}_\infty\le \mathsf{K}_hs^{-1/2}\norm f_\infty,
 \qquad \mathsf{K}_h=\sqrt{2/\pi}.
 \label{eq:heat-kernel-bound}
\end{equation}
Duhamel's formula is
\begin{equation}
 u(s)=P_su(0)+\int_0^sP_{s-r}[c(r)u(r)u_x(r)]\dd r.
 \label{eq:burgers-mild}
\end{equation}
The finite quantity $D_1=\sup_{0<s\le w}\sqrt s\norm{u_x(s)}_\infty$ satisfies, after differentiating \eqref{eq:burgers-mild},
\begin{equation}
 D_1\le \mathsf{K}_h+\pi \mathsf{K}_hD\sqrt wD_1,
\end{equation}
because $\int_0^s(s-r)^{-1/2}r^{-1/2}\dd r=\pi$. Choose $c_0$ so the coefficient of $D_1$ on the right is at most $1/2$. Thus $D_1\le2\mathsf{K}_h$, and $\norm{u_x}_\infty\le \mathsf{K}w^{-1/2}$ on $[w/4,w]$.

For $v=u_x$, $v_s=v_{xx}/2+c(uv_x+v^2)$. Apply Duhamel's formula to this equation starting at $a_1=w/4$. The finite quantity $D_2=\sup_{a_1<s\le w}\sqrt{s-a_1}\norm{v_x(s)}_\infty$ satisfies
\begin{equation}
 D_2\le \mathsf{K}w^{-1/2}+\mathsf{K}D+\mathsf{K}D\sqrt wD_2.
\end{equation}
The initial heat term uses $\norm{v(a_1)}_\infty\le \mathsf{K}w^{-1/2}$. The $v^2$ term is bounded by
\begin{equation}
 \mathsf{K} D\sqrt{s-a_1}\int_{a_1}^s(s-r)^{-1/2}w^{-1}\dd r
       \le \mathsf{K}D(s-a_1)/w\le \mathsf{K}D.
\end{equation}
The drift term $cuv_x$ is bounded with the same integral, $\int_{a_1}^s(s-r)^{-1/2}(r-a_1)^{-1/2}\dd r=\pi$, as for $D_1$. Absorbing the term $\mathsf{K}D\sqrt wD_2$ and using $D\le c_0w^{-1/2}$, we obtain $D_2\le \mathsf{K}w^{-1/2}$, hence $\norm{u_{xx}}_\infty\le \mathsf{K}w^{-1}$ on $[w/2,w]$.

For $v_2=u_{xx}$, the equation is $(v_2)_s=(v_2)_{xx}/2+c(u(v_2)_x+3u_xv_2)$. Restart at $a_2=w/2$ and set $D_3=\sup_{a_2<s\le w}\sqrt{s-a_2}\norm{(v_2)_x(s)}_\infty<\infty$. The initial term is $\mathsf{K}w^{-1}$; the product $u_xv_2$ contributes $\mathsf{K}Dw^{-1/2}$; the drift contributes $\mathsf{K}D\sqrt wD_3$. Therefore
\begin{equation}
 D_3\le \mathsf{K}w^{-1}+\mathsf{K}Dw^{-1/2}+\mathsf{K}D\sqrt wD_3.
\end{equation}
Absorbing again proves the last bound in \eqref{eq:heat-derivatives}. The boundedness assumptions make each weighted supremum finite before absorption, but their sizes never enter the final estimates. The coefficient $c$ was used only as a measurable multiplier.
\end{proof}

\begin{proof}[Proof of Proposition~\ref{prop:regularity}]
The martingale representation in Proposition~\ref{prop:identities} extends to time one in $L^2$. Bilinear It\^o isometry gives
\begin{equation}
 F(t):=\int_t^1g(s)\dd s
       =\E[(B_1-B_t)(\mathcal M_1-\mathcal M_t)]
       =\E[(B_1-B_t)\mathcal M_1].
 \label{eq:endpoint-covariance}
\end{equation}
Conditional on the Brownian path up to time $t$, the variable $(B_1-B_t)/\sqrt{r_t}$ is standard Gaussian, and $\mathcal M_1\in\{-1,1\}$ has conditional mean $\mathcal M_t$. Lemma~\ref{lem:sign-covariance}, applied conditionally, bounds the conditional expectation of $(B_1-B_t)\mathcal M_1$ in absolute value by $\mathsf{K}\sqrt{r_t}\,\psi(1-\mathcal M_t^2)$, where $\psi(v)=v\sqrt{\log(e/v)}$ is extended continuously at zero. This function is concave on $[0,1]$, because
\begin{equation}
 \psi''(v)=-\frac1{2v\sqrt{\log(e/v)}}-
                 \frac1{4v\log^{3/2}(e/v)}<0.
\end{equation}
Since $\E(1-\mathcal M_t^2)=r_t$, Jensen's inequality gives $\E\psi(1-\mathcal M_t^2)\le\psi(r_t)$, hence $F(t)\le \mathsf{K}\sqrt{r_t}\,\psi(r_t)=\mathsf{K}r_t^{3/2}\sqrt{\ell_t}$. For $t\ge1/2$, since $g$ is nonincreasing,
\begin{equation}
 r_tg(t)\le\int_{t-r_t}^tg(s)\dd s\le F(t-r_t)
       \le \mathsf{K}(2r_t)^{3/2}\sqrt{\log(e/(2r_t))}.
\end{equation}
For earlier times use $g\le1$. This proves \eqref{eq:terminal-g}, so $g(1-)=0$ and \eqref{eq:g-identities} yields $\int_0^1\gamma=1$. Monotonicity now implies $r_t\gamma(t)\le\int_t^1\gamma=g(t)$.

Fix $t<1$ and choose $w=c/\Lambda(t)\le r_t/2$. Set $u(s,x)=m(t+w-s,x)$. Because $a=m_x$, the first equation in \eqref{eq:parisi-differentiated-pdes} is a Burgers equation for $m$; thus $u$ has the form in Lemma~\ref{lem:heat-derivatives}, with coefficient $c(s)=\gamma(t+w-s)$. Since $r_{t+w}\ge r_t/2$, the bounds $r_t\gamma(t)\le g(t)\le \mathsf{K}\sqrt{r_t\ell_t}$, applied at time $t+w$, imply $|c(s)|\le \mathsf{K}\sqrt{\Lambda(t)}$. Choosing the constant in $w$ small makes $D\sqrt w$ small. Proposition~\ref{prop:pde-facts} supplies qualitative boundedness of all needed derivatives and justifies Duhamel's formula on this strip. At $s=w$, the three derivatives of $u$ are $a,a_x,a_{xx}$ at time $t$, so \eqref{eq:heat-derivatives} proves \eqref{eq:sharp-regularity}--\eqref{eq:sharp-spatial-derivative}. Finally substitute $r=1-t$ in the time integrals. Direct integration proves the first identity in \eqref{eq:singular-weights}; bounding the logarithm by $L_\tau$ proves its $p>1$ inequality. The substitution $r=e^{-y}$ proves \eqref{eq:integrable-weights}.
\end{proof}

\subsection{The Kolmogorov backward equation and the weighted flow}
\label{app:euler-backward}

To prove Lemma~\ref{lem:euler-backward}, we first verify that the expected value $v$ of \eqref{eq:diffusion-curvature-expectation} is regular enough for It\^o's formula. We then bound $p=v_x$ through \eqref{eq:euler-weighted-flow} and bound $p_x$ by a heat-semigroup estimate on a short interval.

\begin{lemma}[Regularity of the expected value of $a^2$]
\label{lem:euler-backward-regularity}
For fixed $t\le T<1$, let $v(s,x)$ be the expectation in \eqref{eq:diffusion-curvature-expectation}, with $\E_{s,x}$ and $P_{s,t}$ defined in \eqref{eq:diffusion-transition-operator}. Every fixed finite spatial derivative is bounded and continuous on $[0,t]\times\R$ and is Lipschitz in $s$. Equation~\eqref{eq:diffusion-kolmogorov-equation} and its spatially differentiated equations hold in integral form and almost everywhere in time. For $p=v_x$ and $z=p_x$,
\begin{align}
 p_s+\tfrac12p_{xx}+\gamma mp_x+\gamma ap&=0,
 \label{eq:euler-backward-p}\\
 z_s+\tfrac12z_{xx}+\gamma mz_x+2\gamma az+\gamma a_xp&=0.
 \label{eq:euler-backward-z}
\end{align}
Duhamel's formula and the generalized It\^o formulas are valid.
\end{lemma}

\begin{proof}
On the fixed strip the drift $\mu(s,x)=\gamma(s)m(s,x)$ and each fixed finite number of its spatial derivatives are bounded. The additive-noise integral equation for the flow can be differentiated in its initial position. The first derivative solves a linear integral equation with bounded coefficient; each later derivative solves a linear equation with bounded lower-order forcing. Successive Gr\"onwall estimates supply finite deterministic bounds. Dominated differentiation under expectation proves bounded continuous spatial derivatives of $v$. These qualitative constants may be large and will not be used quantitatively.

For $h_0>0$ with $s+h_0\le t$, the Markov property and spatial It\^o applied to the fixed function $v(s+h_0,\cdot)$ imply
\begin{equation}
 v(s)-v(s+h_0)=\int_s^{s+h_0}
 P_{s,r}\bigl[\tfrac12v_{xx}(s+h_0)+\mu(r,\cdot)v_x(s+h_0)\bigr]\dd r.
 \label{eq:backward-markov}
\end{equation}
The qualitative bounds make this $O(h_0)$ uniformly. Differentiating \eqref{eq:backward-markov} in space proves the same for every fixed finite derivative. Lebesgue differentiation followed by integration gives
\begin{equation}
 v(s,x)=a(t,x)^2+\int_s^t
       \{\tfrac12v_{xx}(r,x)+\gamma(r)m(r,x)v_x(r,x)\}\dd r.
\end{equation}
Spatial differentiation proves \eqref{eq:euler-backward-p}--\eqref{eq:euler-backward-z}; Duhamel's formula for each equation follows by integration against the heat kernel. The PDE representative of $v_s$ satisfies
\begin{equation}
 \partial_x v_s=-\tfrac12v_{xxx}
                  -\gamma(s)(a v_x+m v_{xx}).
 \label{eq:backward-time-spatial-lipschitz}
\end{equation}
These terms are bounded on the fixed strip, and the same is true after any fixed finite number of spatial differentiations. Hence the weak time derivatives of $v$ and its spatial derivatives have uniformly spatially Lipschitz representatives. This verifies the remaining hypothesis of \cite[Proposition~22]{JT}, so generalized It\^o applies along any diffusion with bounded progressively measurable drift and Brownian noise in the same filtration, as in Lemma~\ref{lem:parisi-ito}. No time derivative of $\gamma$ occurs.
\end{proof}

\begin{proof}[Proof of Lemma~\ref{lem:euler-backward}]
The flow derivative $\mathcal J_{s,u}=\partial_xX_u^{s,x}$ starts at one and satisfies $\dd\mathcal J_{s,u}=\gamma(u)a(u,X_u)\mathcal J_{s,u}\dd u$. The product rule and \eqref{eq:true-curvature-sde} cancel its drift:
\begin{equation}
 \dd\{a(u,X_u)\mathcal J_{s,u}\}=a_x(u,X_u)\mathcal J_{s,u}\dd B_u.
\end{equation}
The boundedness in Lemma~\ref{lem:euler-backward-regularity} makes this a true martingale. Taking expectations yields
\begin{equation}
 \E_{s,x}[a(t,X_t)\mathcal J_{s,t}]=a(s,x),\qquad
 \mathcal J_{s,t}=\partial_xX_t^{s,x}>0.
 \label{eq:euler-weighted-flow}
\end{equation}
By differentiation under the expectation, $p(s,x)=2\E_{s,x}[a(t,X_t)a_x(t,X_t)\mathcal J_{s,t}]$, and therefore
\begin{equation}
 |p(s,x)|\le2\norm{a_x(t,\cdot)}_\infty
       \E_{s,x}[a(t,X_t)\mathcal J_{s,t}]
       \le \mathsf{K}\Lambda(t)a(s,x).
\end{equation}
This step uses positivity of $a$ and $\mathcal J$, not a quantitative bound on $\mathcal J$ alone.

Put $Q_s=\Lambda(s)$ and $w=c/Q_s\le r_s/2$. On $[s,s+w]$, the weight $\Lambda$ is comparable to $Q_s$. If $t\ge s+w$, consider \eqref{eq:euler-backward-p} on $[s,s+w]$ in reversed time. On this strip, $|p|\le M=\mathsf{K}\Lambda(t)\sqrt{Q_s}$ by the preceding bound on $p$ and \eqref{eq:sharp-regularity}, the drift coefficient $\gamma m$ is at most $\mathsf{K}\sqrt{Q_s}$, and the coefficient $\gamma a$ of the zero-order term $\gamma ap$ is at most $\mathsf{K}Q_s$. For the weighted gradient supremum $D=\sup_{0<r\le w}\sqrt r\norm{\partial_xp(s+w-r)}_\infty$, which is finite by Lemma~\ref{lem:euler-backward-regularity}, the heat-kernel estimate \eqref{eq:heat-kernel-bound} gives
\begin{equation}
 D\le \mathsf{K}M+\mathsf{K}MQ_sw+\mathsf{K}\sqrt{Q_sw}\,D.
\end{equation}
The initial heat term is $\mathsf{K}M$, the zero-order term contributes $\mathsf{K}MQ_sw$, and the drift term is bounded with the same integral equal to $\pi$ as in the proof of Lemma~\ref{lem:heat-derivatives}. A small fixed $c$ permits absorption of the last term, so $D\le \mathsf{K}M$. Taking $r=w$ and dividing by $\sqrt w$ gives $|p_x(s,x)|\le \mathsf{K}\Lambda(t)Q_s$.

If $t<s+w$, then $\Lambda(t)\asymp Q_s$. The terminal value of $z$ is $2a_x(t,\cdot)^2+2a(t,\cdot)a_{xx}(t,\cdot)$, bounded by $\mathsf{K}\Lambda(t)^2$. The source term $\gamma a_xp$ is bounded by $\mathsf{K}\Lambda(t)Q_s^2$. The coefficient $2\gamma a$ of the zero-order term in \eqref{eq:euler-backward-z} is at most $\mathsf{K}Q_s$, so its exponential over this short strip is bounded by a universal constant. The maximum principle or the Feynman--Kac formula therefore implies
\begin{equation}
 |z(s,x)|\le \mathsf{K}\{\Lambda(t)^2+(t-s)\Lambda(t)Q_s^2\}
       \le \mathsf{K}\Lambda(t)Q_s.
\end{equation}
This proves \eqref{eq:euler-backward-bounds}.
\end{proof}
\section{Total work including the Parisi functions}
\label{app:preprocessing}

Proposition~\ref{prop:conditional-total} bounds the total work when exact evaluators for the minimizing order parameter $\gamma$ and the functions $m=\partial_x\Phi$, $a=\partial_{xx}\Phi$ can be constructed at an assumed cost, independently of the sampled matrix. This is a conditional accounting result in the exact real arithmetic model of Definition~\ref{def:oracles}. El Alaoui and Montanari describe a numerical construction through piecewise constant controls, Cole--Hopf updates, and projected gradient descent~\cite[Section~3.1 and Appendix~A]{EAM}. If $\gamma$ equals a constant $c>0$ on $[s,t]$, the solution of \eqref{eq:pde} satisfies
\begin{equation}
 \Phi(s,x)=c^{-1}\log\E e^{c\Phi(t,x+\sqrt{t-s}\,G)},\qquad G\sim\mathcal N(0,1),
 \label{eq:cole-hopf}
\end{equation}
with the heat-semigroup formula for $c=0$.

\begin{proposition}[Conditional total work]
\label{prop:conditional-total}
Suppose evaluators returning the exact point values of the fixed minimizing $\gamma$ and its associated $m,a$ in \eqref{eq:oracle-data} can be constructed, independently of $A$ and the subsequent algorithmic draws, at cost $P_{\rm data}(\varepsilon,\alpha)$, with cost at most $C_{\rm eval}(\varepsilon)$ per point evaluation. If construction is randomized, assume these exactness guarantees hold with failure probability at most $\alpha/16$, included in the confidence budget and in the stated construction cost. For $N\ge N_0$ in Theorem~\ref{thm:main}, complete work is bounded conditionally by
\begin{equation}
 P_{\rm data}+\mathsf{K}\{N^2/\varepsilon
             +(NK+K+\varepsilon^{-3}L_\varepsilon^2)C_{\rm eval}\}.
 \label{eq:total-cost-simple}
\end{equation}
The evaluators return exact values at every point permitted by Definition~\ref{def:oracles}; their construction and the stated costs are hypotheses of this proposition.
\end{proposition}

\begin{proof}
The per-step updates in all $N$ coordinates use $O(NK)$ point queries, and quadrature uses $O(K+\varepsilon^{-3}L_\varepsilon^2)$. Setting every normalization factor to one requires no additional point queries. Lemma~\ref{lem:eg-energy} bounds all remaining exact-real work by $O(N^2/\varepsilon)$. Adding the costs of evaluating and constructing the data proves \eqref{eq:total-cost-simple}. If constructing the data is randomized, allocate a fixed fraction of the failure probability to it. Below the threshold, the algorithm returns without requesting these data.
\end{proof}

The hypothesis of exact evaluation is not a finite-bit complexity guarantee. A treatment of approximate evaluators would have to prove, under the recursion they generate, both terminal and correlation estimates in \eqref{eq:scalar-final}, with the correlation compared to the true Parisi value $\Par$, and the normalization condition \eqref{eq:fixed-normalizer-condition} with $\omega_k=1$. It would also have to verify the pointwise coefficient bounds used in \eqref{eq:sharp-regularity}--\eqref{eq:sharp-spatial-derivative}, the martingale representations and terminal-error decomposition of Lemmas~\ref{lem:martingale-representations} and~\ref{lem:terminal-error-decomposition}, the exponential and moment bounds used in Lemmas~\ref{lem:gaussian-sum} and~\ref{lem:bracket-moments}, and the quadrature error and work estimates of Lemmas~\ref{lem:drift-quadrature} and~\ref{lem:euler-quadrature}. A bound on the error of $\Phi$ alone does not bound the errors of $m$ and $a$. Constructing the evaluators, proving these approximation estimates, and controlling finite-precision matrix operations remain outside the conditional work calculation.
\section{Energy gaps and exact optimization}
\label{app:gap-lower-tail}

We prove the bound $\Prob\{\Delta_N^{\mathrm{gap}}\le t\}\le Nt/2$ for $N\ge2$ and $t\ge0$ on the gap between the two largest distinct energies of a GOE instance (Lemma~\ref{lem:gap-lower-tail}). We then deduce that an approximation scheme with polynomial work at every accuracy, without a dimension restriction or an additional additive error term, would yield an exact solver in the same computational model. We apply Gaussian integration by parts to a Lipschitz vector field in the Euclidean space of the independent Gaussian entries $g=(\sqrt N A_{ij})_{i<j}$; the argument depends only on the Gaussian law of $A$. We work with the off-diagonal maximum so it is nonnegative; the full optimum can be negative because of its common diagonal shift.

\begin{lemma}[Probability of a small first distinct gap]
\label{lem:gap-lower-tail}
For GOE input, $N\ge2$, and $t\ge0$, let $\Delta_N^{\mathrm{gap}}$ be the difference between the largest and second-largest distinct energies. Then
\begin{equation}
 \Prob\{\Delta_N^{\mathrm{gap}}\le t\}
 \le\frac{N\,\E\OPT_N}{2(N-1)}t
 \le\frac{\sqrt{\log2}}2Nt\le\frac N2t.
 \label{eq:gap-lower-tail}
\end{equation}
\end{lemma}

\begin{proof}[Proof of Lemma~\ref{lem:gap-lower-tail}]
Let $\Sigma_N=\{\sigma\in\{-1,1\}^N:\sigma_1=1\}$ contain one representative of each global-spin-reversal class. For $g=(\sqrt N A_{ij})_{i<j}$, a vector of independent standard Gaussians, put
\begin{equation}
 u_\sigma=N^{-1/2}(\sigma_i\sigma_j)_{i<j},\qquad
 \widetilde H_N(\sigma)=\langle g,u_\sigma\rangle
       =H_N(\sigma)-\tfrac12\operatorname{tr}A,\qquad
 M_N=\max_{\sigma\in\Sigma_N}\widetilde H_N(\sigma).
 \label{eq:gap-gaussian-representation}
\end{equation}
The diagonal term cancels from all energy differences, and $\E M_N=\E\OPT_N$. Averaging $\widetilde H_N$ over all spin configurations gives zero, hence $M_N\ge0$ on every realization.

Each $u_\sigma$ has squared norm $(N-1)/2$. If two different classes have representatives at Hamming distance $k$, then $1\le k\le N-1$, and counting the $k(N-k)$ edges on which their spin products differ gives
\begin{equation}
 \langle u_\sigma,u_\sigma-u_\tau\rangle
       =\tfrac12\norm{u_\sigma-u_\tau}^2
       =\frac{2k(N-k)}N\ge\frac{2(N-1)}N.
 \label{eq:gap-covariance-separation}
\end{equation}
The common norm is important for the first equality. In particular, each difference $\langle g,u_\sigma-u_\tau\rangle$ is a nondegenerate Gaussian. The finitely many energies in \eqref{eq:gap-gaussian-representation} therefore have no ties almost surely, and their two largest values differ by $\Delta_N^{\mathrm{gap}}$.

Fix $t>0$. On an outcome with a unique maximizing class $\sigma(g)$, define the vector field
\begin{equation}
 \mathcal V_t(g)=\min\{\Delta_N^{\mathrm{gap}}(g),t\}\,u_{\sigma(g)},
 \label{eq:gap-vector-field}
\end{equation}
and define it to be zero at a tie for the maximum. This field is globally Lipschitz. To see this, cover coupling space by the finitely many closed convex regions where a specified pair $\sigma,\tau$ has the largest and second-largest values. On such a region,
\begin{equation}
 \mathcal V_t(g)=\min\{\langle g,u_\sigma-u_\tau\rangle,t\}\,u_\sigma.
\end{equation}
The formula is Lipschitz with a constant bounded uniformly over the finite collection of pairs. At a change in the maximizing class, the top two values tie and the field vanishes on both sides. At a change only in the second-largest class, the maximizing vector and gap agree. Thus these formulas agree on all boundaries. A line segment meets each convex region in an interval, so summing the common Lipschitz bound along the resulting finite partition proves global Lipschitz continuity.

Gaussian integration by parts is therefore valid coordinate by coordinate, with no boundary contribution. Away from the finitely many ordering and truncation hyperplanes,
\begin{equation}
 \sum_r\partial_{g_r}(\mathcal V_t)_r(g)
 =\langle u_{\sigma(g)},u_{\sigma(g)}-u_{\tau(g)}\rangle
       \mathbf1_{\{\Delta_N^{\mathrm{gap}}<t\}},
\end{equation}
where $\tau(g)$ is the second-largest class. The exceptional hyperplanes have Gaussian measure zero. Combining integration by parts, \eqref{eq:gap-covariance-separation}, and $M_N\ge0$ gives
\begin{align}
 \frac{2(N-1)}N\Prob\{\Delta_N^{\mathrm{gap}}<t\}
 &\le\E\sum_r\partial_{g_r}(\mathcal V_t)_r(g)\nonumber\\
 &=\E\langle g,\mathcal V_t(g)\rangle
   =\E[M_N\min\{\Delta_N^{\mathrm{gap}},t\}]
   \le t\E M_N.
 \label{eq:gap-integration-by-parts}
\end{align}
Letting $t$ decrease to any prescribed nonnegative threshold proves the first inequality in \eqref{eq:gap-lower-tail}, including its non-strict event.

Finally, there are $2^{N-1}$ centered Gaussian energies in \eqref{eq:gap-gaussian-representation}, each of variance $(N-1)/2$. Independence between them is not required: for any $\lambda>0$,
\begin{equation}
 \E M_N\le\frac1\lambda\log\sum_{\sigma\in\Sigma_N}
          \E e^{\lambda\widetilde H_N(\sigma)}
       =\frac{(N-1)\log2}{\lambda}+\frac{(N-1)\lambda}{4}.
\end{equation}
Choosing $\lambda=2\sqrt{\log2}$ yields $\E M_N\le(N-1)\sqrt{\log2}$. Substitute this in \eqref{eq:gap-integration-by-parts} to obtain the last inequalities in \eqref{eq:gap-lower-tail}.
\end{proof}

\paragraph{Consequence for exact optimization.}
Suppose a method returns spins with $\OPT_N-H_N(\sigma)\le\varepsilon N$ with probability at least $1-\alpha$ for every $N\ge2$, $0<\varepsilon\le1$, and $0<\alpha<1/2$, with work polynomial in $N$, $1/\varepsilon$, and $\log(1/\alpha)$ and with no dimension restriction or additional term $\mathfrak{E}_{\varepsilon,\alpha}$. Fix $0<\delta<1$ and run it with $\varepsilon=\delta/N^2$ and $\alpha=\delta/2$. Its total deficit is then at most $\delta/N$ except with probability $\delta/2$, while \eqref{eq:gap-lower-tail} bounds $\Prob\{\Delta_N^{\mathrm{gap}}\le\delta/N\}$ by $\delta/2$. Any total deficit strictly below the first distinct gap is zero. A union bound therefore makes the output exact with probability at least $1-\delta$, with work polynomial in $N$ and $1/\delta$. This conditional implication holds in the same computational model, including its supplied oracles; it does not establish a hardness result for Gaussian SK optimization.

\clearpage
\bibliographystyle{alpha}
\bibliography{references}

\vfill
\begin{center}
    \normalsize © 2026 JPMorgan Chase \& Co. All rights reserved.
\end{center}
\end{document}